\documentclass[aps,rmp,amssymb,amsmath,twocolumn,floatfix]{revtex4-2}

\usepackage{graphicx,epsfig,xcolor}
\usepackage[T1]{fontenc}
\usepackage[latin1]{inputenc}
\usepackage[english]{babel}
\usepackage{amsmath}
\usepackage{mathtools}
\usepackage{amsfonts}
\usepackage{amssymb}
\usepackage{dsfont}

\numberwithin{equation}{section}

\begin{document}

\newcommand{\cP}{\ensuremath{\mathcal{P}}}
\newcommand{\cL}{\ensuremath{\mathcal{L}}}
\newcommand{\cH}{\ensuremath{\mathcal{H}}}
\newcommand{\cD}{\ensuremath{\mathcal{D}}}
\newcommand{\cT}{\ensuremath{\mathcal{T}}}
\newcommand{\cC}{\ensuremath{\mathcal{C}}}
\newcommand{\cE}{\ensuremath{\mathcal{E}}}
\newcommand{\cPT}{\ensuremath{\mathcal{PT}}}
\newcommand{\cCPT}{\ensuremath{\mathcal{CPT}}}
\newcommand{\half}{\mbox{$\textstyle{\frac{1}{2}}$}}
\newcommand{\fourth}{\mbox{$\textstyle{\frac{1}{4}}$}}
\newcommand{\vep}{\varepsilon}

\title{$\cPT$-symmetric quantum mechanics}
\author{Carl M.~Bender}
\affiliation{Department of Physics, Washington University, St. Louis, Missouri 63130, USA}
\email{cmb@wustl.edu}
\author{Daniel W. Hook}
\affiliation{Centre for Complexity Science, Imperial College London, London SW7 2AZ, UK}
\email{d.hook@imperial.ac.uk}

\begin{abstract}
It is generally assumed that a Hamiltonian for a physically acceptable quantum system (one that has a positive-definite spectrum and obeys the requirement of unitarity) must be Hermitian. However, a $\cPT$-symmetric Hamiltonian can also define a physically acceptable quantum-mechanical system even if the Hamiltonian is not Hermitian. The study of $\cPT$-symmetric quantum systems is a young and extremely active research area in both theoretical and experimental physics. The purpose of this Review is to provide established scientists as well as graduate students with a compact, easy-to-read introduction to this field that will enable them to understand more advanced publications and to begin their own theoretical or experimental research activity. The ideas and techniques of $\cPT$ symmetry have been applied in the context of many different branches of physics. This Review introduces the concepts of $\cPT$ symmetry by focusing on elementary one-dimensional $\cPT$-symmetric quantum and classical mechanics and relies in particular on oscillator models to illustrate and explain the basic properties of $\cPT$-symmetric quantum theory.
\end{abstract}
\vskip 0.2cm
\today

\maketitle
\tableofcontents
\newpage

{\small
\leftline{In realms unseen, where quantum whispers speak,}
\leftline{A novel symmetry doth enter stage,}
\leftline{With parity and time, a dance unique,}
\leftline{To bind the quantum actors, new age.}
\vspace{0.15cm}
\leftline{The $\cPT$-symmetry, a twofold guise,}
\leftline{Where $\cP$ reflects in space, yet $\cT$, in time,}
\leftline{Together sing a tale of compromise,} 
\leftline{In quantum world, a balance, most sublime.}
\vspace{0.15cm}
\leftline{Unbroken, this domain of spectral dance,} 
\leftline{The loss and gain of energy entwined,} 
\leftline{Their quantum states, a harmonizing trance,}
\leftline{In eerie beauty, to each other bind.}
\vspace{0.15cm}
\leftline{And thus, in deep embrace of mystic schemes,}
\leftline{$\cPT$-symmetric quantum magic gleams.}
\vspace{0.10cm}
\rightline{---Sonnet from ChatG{$\cPT$}} }

\section{Introduction to $\cPT$ symmetry}\label{s1}
The study of $\cPT$-symmetric quantum mechanics began in 1998 with the observation that a quantum-mechanical Hamiltonian that is invariant under combined space reflection (parity) $\cP$ and time reversal $\cT$ can have a spectrum that is entirely real and bounded below even if the Hamiltonian is not Hermitian \cite{pt1}. This letter was followed shortly thereafter by a second which demonstrated that when the spectrum of a $\cPT$-symmetric Hamiltonian is real and bounded below, the Hamiltonian defines a {\it unitary} (probability-conserving) quantum-mechanical theory \cite{pt90}. Thus, the mathematical constraint that a Hamiltonian operator be Hermitian in the Dirac sense $H^\dag=H$, where $\dag$ represents combined complex conjugation and matrix transposition, can be weakened to the more physically intuitive constraint of spacetime reflection invariance ($\cPT$ symmetry):
\begin{equation}
[H,\cPT]=0.
\label{e1.1}
\end{equation}

\begin{figure}
\center
\includegraphics[scale=0.25]{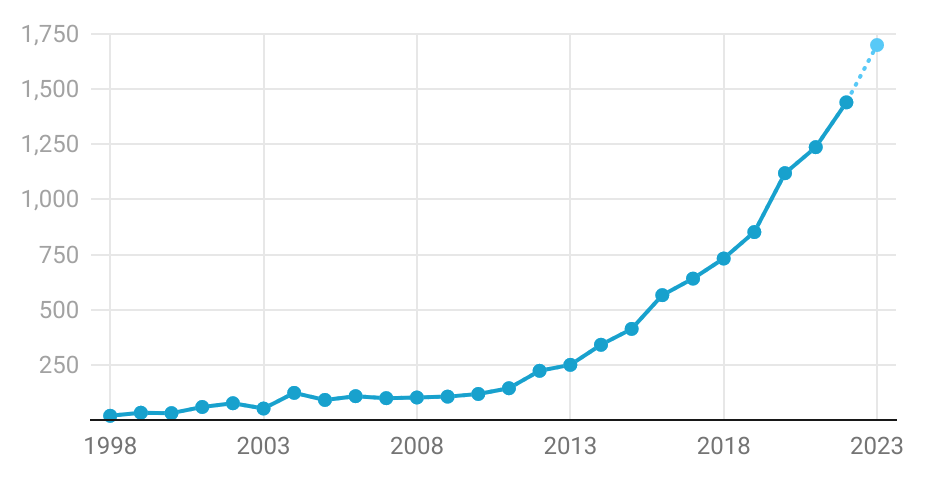}
\caption{Annual publications referencing $\cPT$ symmetry since its inception in 1998 (not including papers on the arXiv). The total number of research publications is growing rapidly and currently exceeds 10,000. Source: Dimensions from Digital Science \cite{pt578}.}
\label{f1}
\end{figure}

Because the condition of $\cPT$ invariance is a {\it weaker} constraint than Hermiticity, it allows one to study vastly many new non-Hermitian quantum theories, some of which exhibit surprising physical properties. Thus, $\cPT$ symmetry offers a rich area for theoretical and experimental investigation. There are already over 10,000 papers published and numerous reviews, books, and theses written on the subject of $\cPT$ symmetry. Many research grants have been awarded and numerous conferences and symposia have been held. Research activity on $\cPT$ symmetry is accelerating (see Fig.~\ref{f1}).

\subsection{Mathematical background}\label{ss1A} The study of $\cPT$-symmetric quantum mechanics draws heavily on complex analysis because the condition of $\cPT$ symmetry arises when we deform a Hermitian theory, such as the quantum harmonic oscillator, into the non-Hermitian {\it complex} domain. We will see that {\it $\cPT$ symmetry marries the fundamental concepts of quantum physics with the mathematical theory of complex variables.}

While complex numbers have been studied by mathematicians for many centuries, complex numbers did not play an essential role in the formulation of physical theories until the twentieth century when the study of quantum mechanics began. The complex number $i$ does not even appear in early studies of special and general relativity. It was not until many years later, when the representations of the homogeneous Lorentz group were analyzed in depth, did it become clear that fermionic spinor representations, which require complex numbers, were physically relevant. In particular, the {\it fundamental} spin-$\tfrac{1}{2}$ representation of the Lorentz group is complex \cite{pt599,pt600}.

The homogeneous Lorentz group is defined as the set of all {\it real} $4\times4$ matrices (matrices whose 16 elements are all real numbers) that map the spacetime vector $x^\mu$ into another spacetime vector $x'^\mu$ in such a way that $x^\mu x_\mu$ remains invariant:
\begin{equation}
x^2+y^2+z^2-t^2=x'^2+y'^2+z'^2-t'^2.
\label{e1.2}
\end{equation}
This condition guarantees that the speed of light is the same in all inertial frames of reference.

In the context of the Lorentz group, $\cPT$ symmetry appears naturally and at a fundamental level. The parity matrix $\cP$ is an element of the Lorentz group because it flips the sign of the spatial coordinates, $$\cP:\,(x,y,z,t)\to(-x,-y,-z,t),$$ and thus leaves $x^2+y^2+z^2-t^2$ invariant. The time-reversal matrix $\cT$ is also an element of the Lorentz group because it flips the sign of the time coordinate, $$\cT:\,(x,y,z,t)\to(x,y,z,-t),$$ and again leaves $x^2+y^2+z^2-t^2$ invariant. The $\cPT$ matrix changes the sign of all four spacetime coordinates, $$\cPT:\,(x,y,z,t)\to(-x,-y,-z,-t),$$ and thus it too is an element of the Lorentz group. 

\begin{figure}
\center
\includegraphics[scale = 0.30]{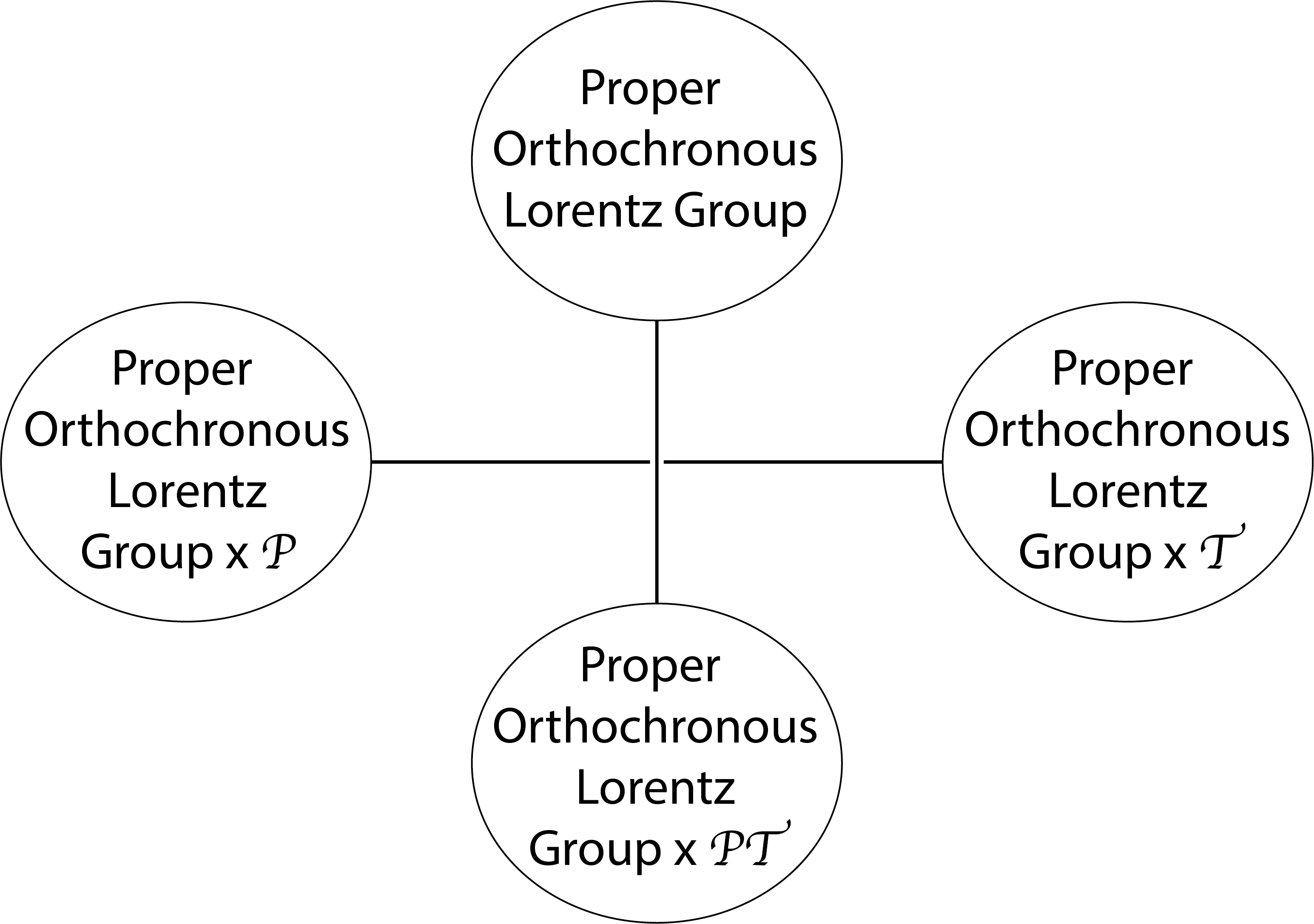}
\caption{Schematic picture of the group space of the homogeneous real Lorentz group. This continuous group consists of four disconnected parts: At the top of the diagram is the proper orthochronous Lorentz group (POLG); the POLG contains the identity element and is a subgroup of the Lorentz group. The other three components of the homogeneous real Lorentz group are the POLG multiplied by the parity (space-reflection) operator $\cP$, the time-reversal operator $\cT$, and the spacetime reflection operator $\cPT$. If we extend the real Lorentz group to the {\it complex} Lorentz group \cite{pt599}, there are now only {\it two} disconnected parts. Continuous paths through complex group space (indicated as vertical and horizontal lines) connect the top and bottom components and the left and right components of the real Lorentz group.}
\label{f2}
\end{figure}

Mathematically, the real Lorentz group is a six-parameter continuous group consisting of four {\it disconnected} components, as indicated schematically in Fig.~\ref{f2}. The first part, as shown at the top of Fig.~\ref{f2}, is called the {\it proper orthochronous Lorentz group} (POLG). The POLG contains all the elements of the homogeneous Lorentz group that are continuously connected in group space to the {\it identity} element, which is the transformation that leaves the vector $(x,y,z,t)$ unchanged. The POLG is a subgroup of the homogeneous Lorentz group. The second component of the Lorentz group (the left part of Fig.~\ref{f2}) consists of all of the Lorentz transformations in the POLG multiplied by $\cP$. This part of the Lorentz group is not a subgroup because it does not contain an identity element. The third component of the Lorentz group (right part of Fig.~\ref{f2}) consists of all of the Lorentz transformations in the POLG multiplied by $\cT$. Again, this part of the Lorentz group is not a subgroup. The fourth part of the Lorentz group consists of all of the elements of the POLG multiplied by the spacetime reflection operator $\cPT$ and it too is not a subgroup.

For many years after the discovery of special relativity, the homogeneous Lorentz group pictured in Fig.~\ref{f2} was believed to be the fundamental symmetry group of the natural world. However, Lee and Yang theorized that space reflection (parity) $\cP$ is not a fundamental symmetry of the universe \cite{pt580}; the 1957 Nobel Prize was awarded for this work. The work of Lee and Yang implies that an experiment can be designed that achieves different experimental results in left-handed and right-handed laboratories; that is, two laboratories that are spatial reflections (mirror images) of one another. Such an experiment was performed by Wu \cite{pt581}, for which she was awarded the Wolf Prize in 1978.

Subsequently, Cronin and Fitch were awarded the 1980 Nobel Prize for demonstrating that time reversal $\cT$ is also not a symmetry of nature \cite{pt582}. Their experiment demonstrates, in principle, that a laboratory traveling forward in time is experimentally distinguishable from a physically identical laboratory that is traveling backward in time. The implication of all this work is that the Lorentz transformations in the left and the right portions of the Lorentz group illustrated in Fig.~\ref{f2} are not fundamental symmetries of the universe.

These discoveries regarding parity and time reversal raise an obvious question
regarding the Lorentz transformations in the lower portion of Fig.~\ref{f2}. If we have two laboratories, one traveling forward in time and the other space-reflected and traveling backward in time, is there an experiment that will yield different results in these two laboratories? Are these two laboratories experimentally distinguishable or is combined $\cPT$ reflection (spacetime reflection) a fundamental discrete symmetry of nature?

To examine this question, it helps to extend the real homogeneous Lorentz group to the {\it complex} homogeneous Lorentz group, which is the set of all {\it complex} $4\times4$ matrices that leave $x^2+y^2+z^2-t^2$ invariant. The complex homogeneous Lorentz group consists of two, and not four, disconnected parts. Figure~\ref{f2} shows that there is a continuous path (horizontal line) through complex-Lorentz-group space connecting the left and right portions of the real Lorentz group. There is also a continuous path (vertical line) connecting the upper and lower parts of the real Lorentz group \cite{pt566}. This complex group structure suggests that since it is natural to extend the real number system to the complex number system and since complex numbers appear commonly in modern physics, one should consider $\cPT$ reflection as a possible symmetry of some physical theories.

\subsection{$\cCPT$ theorem}\label{ss1B}
Extending the real homogeneous Lorentz group to the complex Lorentz group is the key ingredient in the proof of a profound theorem in quantum field theory and particle physics known as the $\cCPT$ theorem \cite{pt566}. This theorem states that two identical physical measurements, where the first measurement is done in a laboratory traveling forward in time and consisting of ordinary matter and the second is done in a space-reflected laboratory traveling backward in time and consisting of antimatter, will return identical results.

The proof of the $\cCPT$ theorem requires several crucial assumptions made in axiomatic quantum field theory about the nature of particle interactions, such as unitarity, locality, causality, positivity of the particle spectrum, and Hermiticity of the Hamiltonian. These assumptions allow one to extend the real Lorentz group to the complex Lorentz group and to perform an analytic continuation in group space from the upper to the lower portion of Fig.~\ref{f2}. Along this path the state of an elementary particle undergoes a transition from the original particle to its corresponding antiparticle. [The proof of the $\cCPT$ theorem fills an entire book \cite{pt566}.]

We do not discuss quantum field theory, antiparticles, and the charge-conjugation operator in particle physics here; this Review is limited to a discussion of quantum mechanics. However, we will see that there is a fascinating connection between the $\cCPT$ theorem and the results on $\cPT$ symmetry presented in this Review. The development in this Review can be viewed as establishing the quantum-mechanical {\it converse} of the $\cCPT$ theorem. 

We begin by explaining how to construct various non-Hermitian quantum-mechanical theories (not field theories) that are $\cPT$ invariant and we examine their properties. The spectra of such theories are often entirely real and bounded below. We then show that when the spectrum of a $\cPT$-symmetric Hamiltonian is real, the Hamiltonian has a {\it hidden invariance}; there exists a new quantum-mechanical operator $\cC$ that commutes with the $\cPT$-symmetric Hamiltonian. (We call this operator $\cC$ because its properties closely resemble those of the charge-conjugation operator $\cC$ used in the $\cCPT$ theorem of particle physics.) We use this $\cC$ operator to construct an inner product endowed with a positive-definite norm and establish that the $\cPT$-symmetric Hamiltonian defines a {\it unitary} quantum-mechanical theory.

The organization of this Review is straightforward: Section \ref{s2} discusses the construction and properties of $\cPT$-symmetric Hamiltonians with special emphasis on the calculation of eigenvalues. It is shown how to obtain $\cPT$-symmetric quantum-mechanical theories as complex deformations of Hermitian theories. These examples emphasize the intimate connection between the physical constraint of $\cPT$ symmetry and the mathematical theory of complex variables. Section~\ref{s3} examines the process of complex analytical deformation in detail and Sec.~\ref{s4} gives a quantum and classical analysis of $\cPT$-symmetric upside-down potentials. Section~\ref{s5} describes the remarkable properties of the $\cPT$-symmetric quartic anharmonic oscillator at both the quantum-mechanical and the classical-mechanical levels.

Having discussed the properties of $\cPT$-symmetric eigenvalue problems for a variety of quantum-mechanical models, we then address some crucial theoretical issues regarding $\cPT$-symmetric quantum mechanics. In Sec.~\ref{s6} we construct the $\cC$ operator, which in turn is used to define the $\cPT$-symmetric inner product and the Hilbert space. (In conventional Hermitian quantum mechanics we already know what the inner product is even before we know the Hamiltonian, but in $\cPT$-symmetric quantum mechanics the inner product is determined by the Hamiltonian itself.) Section~\ref{s7} briefly summarizes some of the current research on $\cPT$-symmetric quantum-mechanical systems and suggests possibilities for future research.

Because the study of $\cPT$-symmetric quantum theory involves heavy use of complex-variable theory, we include an Appendix that briefly summarizes the historical interest in extending real mathematics to complex mathematics, the study of which began in earnest in the sixteenth century. Complex calculus is powerful and can often solve problems that the calculus of real variables is incapable of solving. Physicists have long understood the power of complex analysis as a calculational tool, but complex numbers played no role in formulating physical theories until the 20th century. Since its appearance in quantum theory, the number $i$ has played an increasingly important role in physics.

\section{Examples of $\cPT$-symmetric Hamiltonians}\label{s2}
How does one construct a $\cPT$-symmetric quantum-mechanical theory? An elementary recipe for creating a $\cPT$-symmetric Hamiltonian is to begin with a quantum-mechanical Hamiltonian that is both Hermitian and $\cPT$ symmetric, and then to perform a $\cPT$-symmetric {\it deformation} of this Hamiltonian \cite{pt1,pt175,pt579}. The resulting Hamiltonian, which is a complex analytic continuation of the original Hermitian Hamiltonian, is still $\cPT$ symmetric but it is no longer Hermitian. Nevertheless, the deformed Hamiltonian operator may continue to have an entirely real spectrum whose energy levels remain well separated and do not cross. [For further examples see \cite{pt51,pt61,pt128,pt129}].

\subsection{An elementary $\cPT$-symmetric Hamiltonian}\label{ss2A}
To illustrate the construction of a $\cPT$-symmetric Hamiltonian we perform a simple $\cPT$ deformation of the quantum harmonic-oscillator Hamiltonian
\begin{equation}
H=\half p^2+\half x^2.
\label{e2.1}
\end{equation}
This Hamiltonian is Hermitian $H^\dag=H$ and the eigenvalues $E_n$ of this Hamiltonian are all real and positive:
\begin{equation}
E_n=n+\half\quad(n=0,\,1,\,2,\,...).
\label{e2.2}
\end{equation}

The $\cP$ and $\cT$ operators are conventionally defined as follows: The parity operator $\cP$ is a linear Hermitian operator that performs space reflection, so $\cP^2=I$ ($I$ is the unity operator) and $\cP^{-1}=\cP$. Under parity reflection both the position operator $x$ and the momentum operator $p$ change sign:
\begin{equation}
\cP x\cP^{-1}=-x,\qquad \cP p\cP^{-1}=-p.
\label{e2.3}
\end{equation}
The time-reversal operator $\cT$ is also a reflection operator and for bosonic systems $\cT^2=I$, so $\cT^{-1}=\cT$. Time reversal leaves $x$ invariant but changes the sign of $p$:
\begin{equation}
\cT x\cT^{-1}=x,\qquad \cT p\cT^{-1}=-p.
\label{e2.4}
\end{equation}
However, unlike the $\cP$ operator, the $\cT$ operator is not linear; $\cT$ is said to be {\it antilinear} because it changes the sign of the complex number $i$:
\begin{equation}
\cT i\cT^{-1}=-i.
\label{e2.5}
\end{equation}
The operations of reversing space and reversing time are independent and thus $\cP$ and $\cT$ commute $[\cP,\cT]=0$.

According to the definitions above we see immediately that the Hamiltonian (\ref{e2.1}) satisfies (\ref{e1.1}) and thus this Hamiltonian is $\cPT$ symmetric as well as Hermitian. (In this case, $H$ is {\it independently} invariant under $\cP$ and $\cT$ reflections.) The combination $ix$ is not Hermitian, but it is $\cPT$ symmetric. Therefore, we can use the $ix$ operator to perform a $\cPT$-symmetric deformation of the Hamiltonian $H$: Let $\vep$ be a {\it real} parameter called a {\it deformation parameter}. The simplest kind of deformation is an {\it additive} deformation, so we use an additive deformation to construct the deformed Hamiltonian
\begin{equation}
H=\half p^2+\half x^2+\vep ix.
\label{e2.6}
\end{equation}
The Hamiltonian (\ref{e2.6}) is not Hermitian when $\vep\neq0$ because $ix$ is not Hermitian, but $H$ {\it is} $\cPT$ symmetric because $H$ commutes with the $\cPT$ operator.

A brief calculation in which we replace $x$ by $x-i\vep$ in the time-independent Schr\"odinger equation associated with this Hamiltonian verifies the surprising result that the eigenvalues of $H$ in (\ref{e2.6}) are all {\it real}:
\begin{equation}
E_n=n+\half\big(1+\vep^2)\quad(n=0,\,1,\,2,\,...).
\label{e2.7}
\end{equation}

Although the eigenvalues of $H$ in (\ref{e2.6}) are real, the corresponding eigenfunctions $\psi_n(x)$ are complex. However, they still satisfy the original undeformed boundary condition on $\psi_n(x)$; namely, that $\psi_n(x)$ vanishes exponentially as $x\to\pm\infty$. An important property of these eigenfunctions is that they are all $\cPT$ symmetric; that is, $\psi_n(x)$ remains invariant (apart from an overall multiplicative phase) if we replace $i$ by $-i$ (that is, take the complex conjugate) and replace $x$ by $-x$. Furthermore, the deformed eigenfunctions continue to be orthogonal, $$\int_{-\infty}^{\infty}dx\,\psi_m(x)\psi_n(x)=0\quad(m\neq n),$$ for all real $\vep$.

This example suggests that there are infinitely many more $\cPT$-symmetric Hamiltonians than Hermitian Hamiltonians. In place of the single Hermitian harmonic-oscillator Hamiltonian in (\ref{e2.1}), we now have a continuously infinite class of non-Hermitian Hamiltonians in (\ref{e2.6}) parametrized by $\vep$. All of these Hamiltonians have entirely real positive spectra and they define perfectly acceptable quantum-mechanical theories. The Hermitian Hamiltonian corresponding to $\vep=0$ is just one special case that is not fundamentally different from the other Hamiltonians. This shows that while Hermiticity is a sufficient condition for having a real spectrum, it is not a necessary condition. This also suggests that we might adopt the view that $\cPT$ symmetry is a fundamental property of a quantum Hamiltonian with Hermiticity being a mere special case.

\subsection{Some $\cPT$-symmetric Hamiltonians have complex spectra}
\label{ss2B}
The example above might lead one to conjecture that the property of $\cPT$ symmetry in (\ref{e1.1}) is strong enough to establish that the eigenvalues of {\it any} $\cPT$-symmetric Hamiltonian are all real. This conjecture is false, but it is nevertheless instructive to examine the following incorrect attempt to prove this conjecture \cite{pt175,pt579}.

Let $\psi$ be a (time-independent) eigenstate of the Hamiltonian operator $H$ with eigenvalue $E$:
\begin{equation}
H\psi=E\psi.
\label{e2.8}
\end{equation}
Assume that $H$ is $\cPT$ symmetric; that is, $H$ commutes with $\cPT$: $[H,\cPT]=0$. Next, recall a theorem from linear algebra that is often quoted in elementary quantum-mechanics courses: {\it If two operators $A$ and $B$ commute, then they are simultaneously diagonalizable}; that is, an eigenstate of $A$ is also an eigenstate of $B$. For example, if $H$ commutes with $\cP$, then the energy eigenstates of $H$ have definite parity. On the basis of this theorem, one might think that if $H$ commutes with $\cPT$, then $\psi$ must also be an eigenstate of $\cPT$:
\begin{equation}
\cPT\psi=\lambda\psi,
\label{e2.9}
\end{equation}
where $\lambda$ is an eigenvalue of the $\cPT$ operator. (Warning: This assumption is wrong!)

Let us examine the eigenvalue equation (\ref{e2.9}). Because $\cP^2=\cT^2=I$ and $[\cP,\cT]=0$ it follows that $(\cPT)^2=I$. Also, $\cT$ performs complex conjugation. 
Thus, if we multiply (\ref{e2.9}) on the left by $\cPT$, we obtain
\begin{equation}
\psi=\lambda^*\lambda\psi.
\label{e2.10}
\end{equation}
This proves that $|\lambda|=1$. Thus, {\it the eigenvalues of the $\cPT$ operator lie on the unit circle in the complex plane.} This is an interesting complex generalization of the result that the eigenvalues of the $\cP$ operator are $\pm1$; that is, that the eigenvalues of $\cP$ lie on a {\it one-dimensional} unit circle on the real axis (instead of on a two-dimensional circle in the complex plane).

Finally, we multiply the eigenvalue equation (\ref{e2.8}) on the left by the $\cPT$ operator and divide the resulting equation by $\lambda$, which is allowed because we have shown that $\lambda$ is nonzero. We are therefore tempted to conclude that
\begin{equation}
E=E^*.
\label{e2.11}
\end{equation}
While it is correct that $|\lambda|=1$, this conclusion is false!

The mistake in this argument is that we misquoted the elementary theorem above; we neglected to say that $A$ and $B$ must be {\it linear} operators. It is not necessarily true that an eigenstate of $H$ is also an eigenstate of $\cPT$ because while $\cPT$ commutes with $H$, $\cPT$ is not a {\it linear} operator; $\cPT$ is {\it antilinear} because $\cT$ performs complex conjugation. However, we have established an important result: {\bf If an eigenstate of a $\cPT$-symmetric Hamiltonian is also an eigenstate of $\cPT$, then the corresponding energy eigenvalue is real.}

Unfortunately, proving that an eigenstate of a $\cPT$-symmetric Hamiltonian is simultaneously an eigenstate of the $\cPT$ operator can be nontrivial; this must be done on a case-by-case basis for each $\cPT$ symmetric Hamiltonian. In some cases the proof may be difficult and one might have to resort to a convincing numerical analysis instead of a mathematical proof. Proving that the eigenvalues of $H$ in (\ref{e2.6}) are real is easy, but a general technique for proving spectral reality for a given $\cPT$ symmetric Hamiltonian is unlikely to be found.

Nevertheless, there are infinitely many more non-Hermitian $\cPT$-symmetric Hamiltonians that have entirely real spectra than Hermitian Hamiltonians. These $\cPT$-symmetric Hamiltonians often describe physical systems having novel and experimentally verifiable properties. This is why the theoretical and experimental study of $\cPT$ symmetry has become so active.

\subsection{Reality of eigenvalues}\label{ss2C}
An energy measurement performed on an isolated quantum system will always return a real number. This number is the expectation value of the Hamiltonian $H$ in the quantum state of the system and it is equal to one of the eigenvalues of $H$. Thus, the eigenvalues of a Hamiltonian that describes an isolated quantum system are always real.

Let us recall the standard proof that the energy eigenvalues of a Hermitian Hamiltonian are real. For simplicity, we assume that the quantum system is one-dimensional and defined on the whole real line and that the time-independent Schr\"odinger eigenvalue equation (\ref{e2.8}) has the simple form
\begin{equation}
-\psi''(x)+V(x)\psi(x)=E\psi(x)\quad(-\infty<x<\infty).
\label{e2.12}
\end{equation}
We then multiply (\ref{e2.12}) on the left by $\psi^*(x)$, integrate from $-\infty$ to $\infty$, and integrate the first term on the left side by parts. The resulting equation reads
\begin{eqnarray}
&&\int_{-\infty}^\infty dx\,|\psi'(x)|^2+\int_{-\infty}^\infty dx\,V(x)
|\psi(x)|^2\nonumber\\
&&\quad=E\int_{-\infty}^\infty dx\,|\psi(x)|^2.
\label{e2.13}
\end{eqnarray}
Finally, we note that the potential $V(x)$ is real if $H$ is Hermitian, so all three integrals in (\ref{e2.13}) are real. This implies that $E$ is real.

The key step in this argument is the integration by parts. Integration by parts is possible only if the eigenfunction $\psi(x)$ vanishes sufficiently fast as $|x|\to\infty$. Thus, the proof that the eigenspectrum is real requires {\it more} than just the Hermiticity of $H$. Hermiticity, like the condition of $\cPT$ symmetry, is a {\it local} condition that holds (or does not hold) at each value of $x$; thus, one can immediately see if a Hamiltonian is Hermitian or $\cPT$ symmetric by mere inspection. However, the condition that the eigenfunctions vanish fast enough as $|x|\to\infty$ to allow integration by parts is a nontrivial {\it global} (nonlocal) condition. These local and global conditions taken together are called {\it self-adjointness}. (Note that there is no difference between the condition of Hermiticity and self-adjointness for a {\it matrix} Hamiltonian because in that case the proof that the eigenvalues are real does not require an integration by parts.)

This condition of self-adjointness is crucial in quantum mechanics because it implies that the quantum theory is unitary (probability-conserving). In Sec.~\ref{s6} we show that if the eigenspectrum of a $\cPT$-symmetric Hamiltonian $H$ is entirely real, then $H$ defines a unitary theory of quantum mechanics. For many nontrivial $\cPT$-symmetric Hamiltonians, the spectrum has been proved to be entirely real, as we see in the next example.

\subsection{Multiplicative deformation}\label{ss2D}
Let us perform a {\it multiplicative} deformation of the harmonic-oscillator Hamiltonian (\ref{e2.1}) to obtain the non-Hermitian $\cPT$-symmetric class of Hamiltonians
\begin{equation}
H=p^2+x^2(ix)^\vep.
\label{e2.14}
\end{equation}
This nontrivial Hamiltonian is not Hermitian, but the spectrum of $H$ is entirely real and positive if $\vep\geq0$ \cite{pt1}. In addition, the eigenfunctions are all $\cPT$ symmetric if $\vep \geq0$. However, if $\vep<0$, only a finite number of the lowest-lying eigenvalues are real and the corresponding eigenfunctions are $\cPT$ symmetric. The rest of the eigenvalues are complex and the corresponding eigenfunctions are not $\cPT$ symmetric.

\begin{figure}
\center
\includegraphics[scale=0.245]{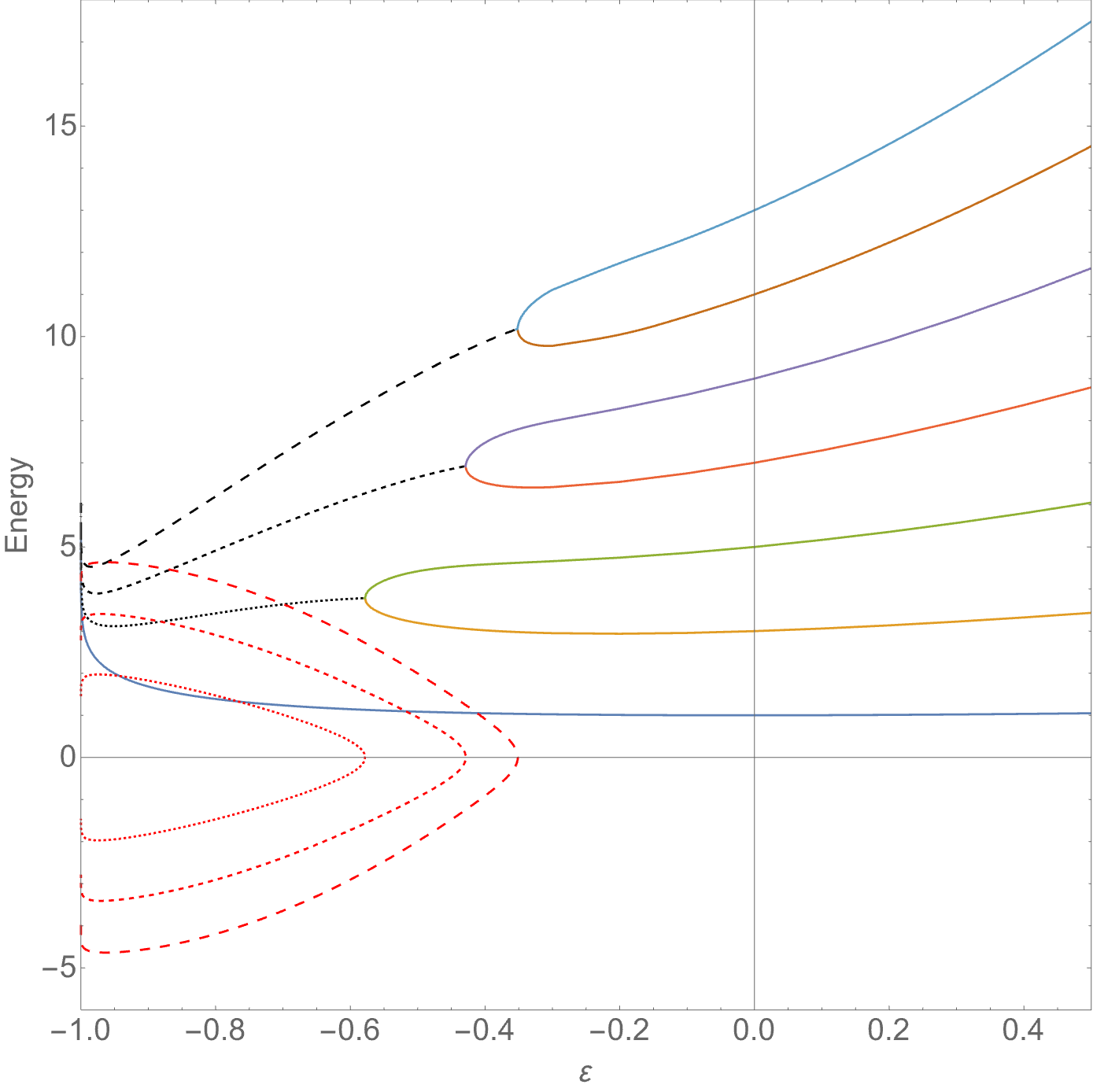}
\caption{Lowest seven eigenvalues of $H$ in (\ref{e2.14}) plotted for $-1<\vep<0.5$. These eigenvalues are real, positive, discrete, and monotonically increasing for $\vep\geq0$ and they are displayed as solid lines. As $\vep$ decreases below 0, the ground-state energy remains real and becomes singular at $\vep=-1$. The other eigenvalues merge pairwise in a regular and orderly fashion, starting with the highest eigenvalues, and immediately bifurcate sequentially into complex-conjugate pairs (whose real and imaginary parts are indicated by dashed/broken lines). The values of $\vep$ at which pairs of real eigenvalues merge are called {\it exceptional points}. These exceptional points are square-root singularities, sometimes referred to as {\it pitchfork bifurcations}, in the complex-$\vep$ plane.}
\label{f3}
\end{figure}

\begin{figure}
\center
\includegraphics[scale=0.245]{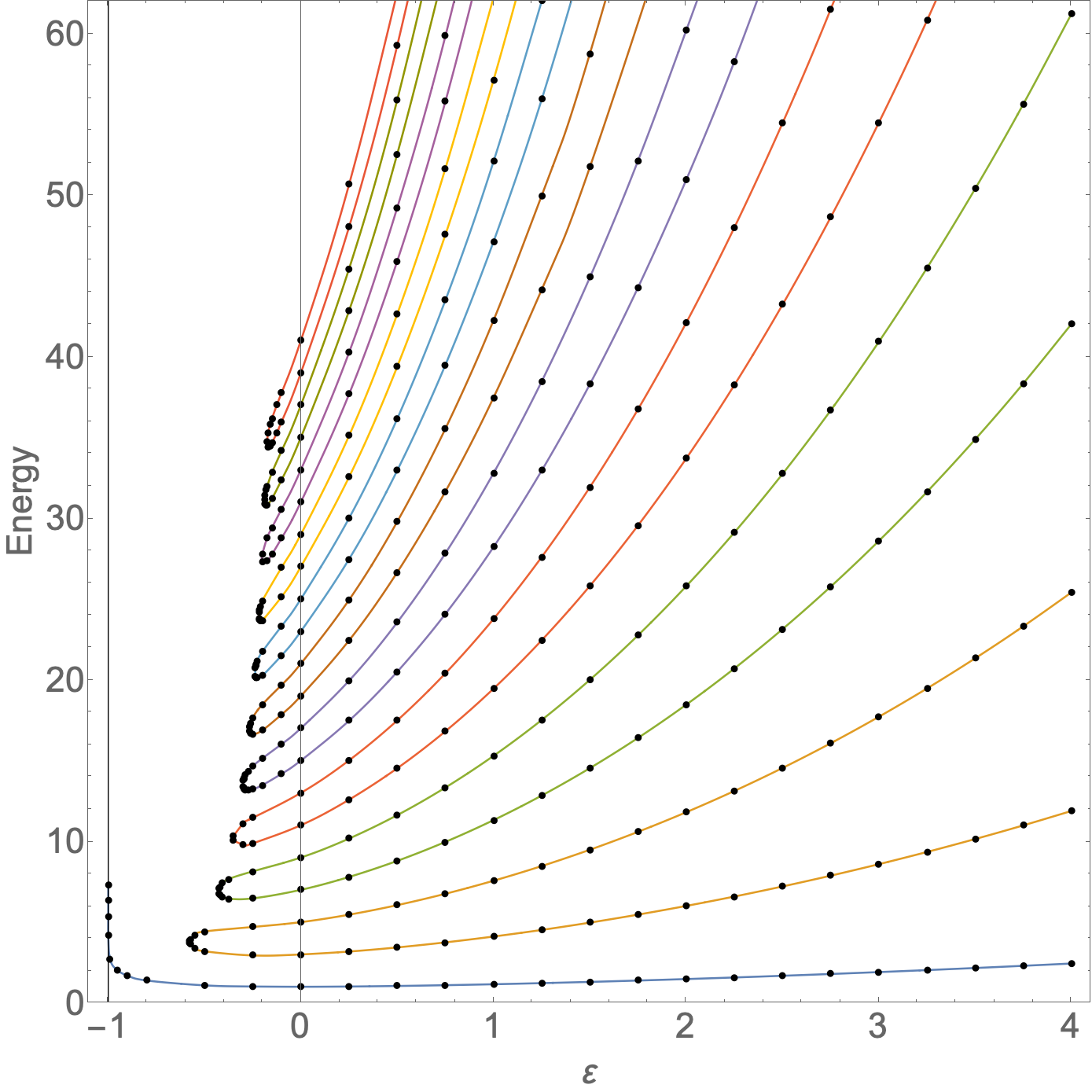}
\caption{First twenty-one eigenvalues of $H$ in (\ref{e2.14}) plotted for $-1<\vep\leq4$. The eigenvalues are all real, positive, discrete, and monotonically increasing for $\vep\geq0$. This plot only shows the eigenvalues when they are real. Note that as $\vep$ decreases below 0 the eigenvalues become degenerate pairwise and enter the complex plane as
complex-conjugate pairs, but the ground-state energy continues to remain real and becomes singular as $\vep$ approaches $-1$.}
\label{f4}
\end{figure}

A plot of the lowest seven eigenvalues for $-1<\vep<0.5$ is shown in Fig.~\ref{f3}. The range $\vep>0$, in which the eigenvalues are all real, is called the {\it region of unbroken $\cPT$ symmetry}. In this region the eigenvalues are positive, discrete, and increase monotonically with increasing $\vep$. At $\vep=0$, the low edge of the region of unbroken $\cPT$ symmetry, the eigenvalues are just the harmonic-oscillator eigenvalues given in (\ref{e2.2}). The range $\vep<0$ is called the {\it region of broken $\cPT$ symmetry}. The boundary between these two regions, which we call the {\it $\cPT$ transition}, occurs at the unique value of $\vep$ at which the Hamiltonian is Hermitian. For larger positive values of $\vep$ the energy levels all continue to rise and remain well separated. In Fig.~\ref{f4} the upper range of $\vep$ is extended from $0.5$ to $4$ and the first twenty-one real eigenvalues are shown.

In general, whenever a $\cPT$-symmetric deformation of a $\cPT$-symmetric Hermitian Hamiltonian $H$ is performed, the eigenvalues of the deformed Hamiltonian are either real or else they occur in complex-conjugate pairs. It is easy to see why this is true: The eigenvalues of a Hamiltonian are the roots of the secular-determinant equation
\begin{equation}
{\rm det}(H-IE)=0,
\label{e2.15}
\end{equation}
which is a polynomial in $E$ if $H$ is a finite-dimensional matrix. We insert the unit operator $I=\cPT\cPT$ in the determinant, commute one factor of $\cPT$ through $H-IE$, and then permute the other factor of $\cPT$ cyclically around so that the operator $\cPT\cPT$ reappears and can be replaced by the unit operator. Remember that if $H$ is $\cPT$ symmetric it commutes with $\cPT$, but when $\cPT$ commutes past $E$ it replaces $E$ with $E^*$. Thus, we get
\begin{equation}
{\rm det}(H-IE^*)=0.
\label{e2.16}
\end{equation}
Thus, if $E$ is a solution to (\ref{e2.15}), then so is $E^*$. Hence the secular equation is real, and as stated above, the eigenvalues $E$ are either real or come in complex-conjugate pairs, which is consistent with the plot in Fig.~\ref{f3}.

The entire spectrum of a $\cPT$-deformed Hamiltonian may remain real for all $\vep$ or for some region of $\vep$; there may also be more than one region of broken and of unbroken $\cPT$ symmetry. It was several years before it was established rigorously that the eigenvalues of the Hamiltonian $H$ in (\ref{e2.14}) are real and positive if $\vep>0$ \cite{pt274,pt294,pt295,pt297,pt298,pt312}. This proof requires the use of powerful theorems of complex analysis.

\subsection{$2\times2$ Matrix Hamiltonian with a $\cPT$ transition}\label{ss2E}
The $\cPT$ transition described in the previous subsection is a feature of many $\cPT$-symmetric Hamiltonians. As was demonstrated in a beautiful series of optics experiments done by Christodoulides {\it et al.} \cite{pt182, pt183, pt184, pt186, pt187, pt188}, one can use optical fibers to observe and measure the $\cPT$ transition in $\cPT$-symmetric matrix Hamiltonians. Much research activity on $\cPT$ symmetry was inspired by these pioneering and prize-winning experiments.\footnote{
Christodoulides was awarded the 2023 Schawlaw Laser Prize by the APS in part for his experimental studies of $\cPT$-symmetric optical fibers. His citation reads: ``For pioneering several areas in laser sciences, among them, the fields of parity-time non-Hermitian optics, accelerating Airy waves, and discrete solitons in periodic media.''} Following these early experiments, many more theoretical and experimental studies have been done on $\cPT$-symmetric systems involving lasers and photonics \cite{pt202,pt203,pt205,pt209,pt210,pt544,pt550,pt556,pt557,pt499,pt559,pt537,pt689}, atomic systems \cite{pt609}, quantum Zeno effect \cite{pt687}, optomechanics \cite{pt493,pt510}, phononics \cite{pt664,pt685,pt686}, acoustics \cite{pt242,pt500,pt549,pt551}, metamaterials \cite{pt560,pt233,pt220}, optical lattices \cite{pt554,pt225,pt217,pt218,pt527}, exceptional points and enhanced sensing \cite{pt545,pt553,pt568}, graphene \cite{pt558, pt542,pt592}, critical phenomena \cite{pt555}, metasurfaces \cite{pt237,pt239,pt494,pt596}, nonlinear waves and solitons \cite{pt227,pt229,pt226}, and mechanical systems \cite{pt222,pt223,pt224,pt487}.

\begin{figure}
\center
\includegraphics[scale=1.10]{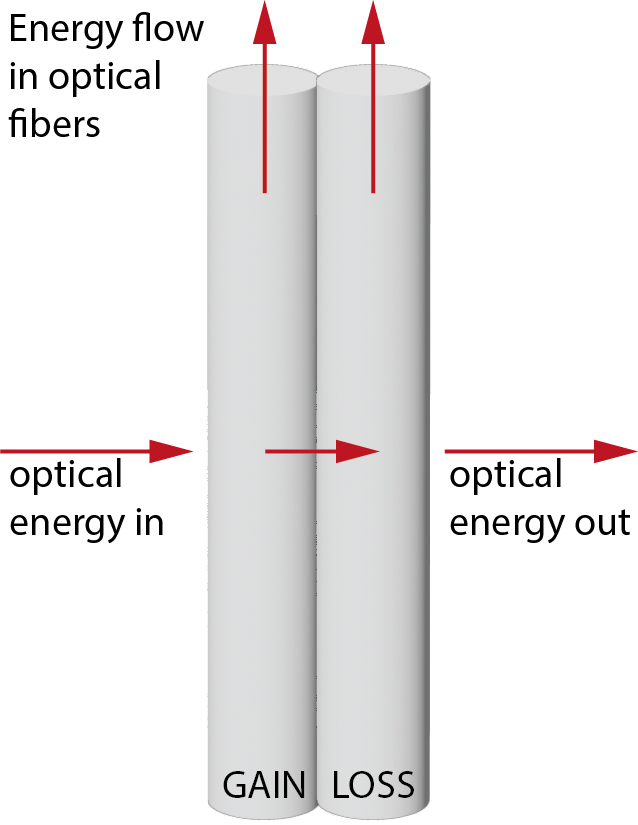}
\caption{Schematic picture of the light fiber experiment.}
\label{f5}
\end{figure}

The original optical-fiber experiments make use of two identical and parallel fibers \cite{pt186,pt188}. One fiber has {\it loss}; light traveling through this fiber is passively absorbed by the material in the fiber and it can also be actively absorbed by placing metallic substances in contact with the fiber to absorb some of the light energy traveling through the fiber. The other fiber has {\it gain}; light energy can be pumped into this fiber by shining a laser on it. The two fibers are placed in contact with one another so that they can exchange energy. Furthermore, the energy flow into the fiber with gain is adjusted so that it balances the energy flowing out of the fiber with loss (see Fig.~\ref{f5}).

There are two possibilities: (i) If the fiber system is strongly coupled, the light energy that flows into the fiber with gain readily flows across the boundary between the two fibers, where it then flows out of the fiber with loss. In this case the optical-fiber system is in {\it dynamic equilibrium}; that is, the total energy at each point along the fiber remains constant in time. (ii) If the fiber system is weakly coupled, the energy that is flowing into the fiber with gain is unable to flow easily across the boundary between the fibers and into the fiber with loss. In this case the optical-fiber system is not in dynamic equilibrium; the energy in the fiber with gain grows in time and the energy in the fiber with loss decays in time.

This optical-fiber system can be interpreted as being $\cPT$ symmetric if we define the $\cP$ operation as an exchange of the two fibers. The system is not $\cP$ symmetric because the fiber with loss becomes the fiber with gain, and the fiber with gain becomes the fiber with loss. However, if we also reverse time, the inflow of energy becomes outflow, outflow becomes inflow, and the original configuration is recovered.

The behavior of this classical optical-fiber system can be described mathematically by a $2\times2$ matrix that is $\cPT$-symmetric. This matrix contains a parameter $g$ that represents the strength of the optical coupling between the two fibers. We will {\it reinterpret} this matrix as a {\it Hamiltonian} that describes a two-state quantum-mechanical system in which the coupling parameter $g$ plays the role of a deformation parameter. An elementary study of this matrix predicts that the physical system will exhibit a $\cPT$ transition.

Let us consider first a simple quantum-mechanical system described by the $1\times1$ matrix Hamiltonian
\begin{equation}
H_1=[a+ib],
\label{e2.17}
\end{equation}
where $a$ and $b$ are real numbers. If the parameter $b$ is nonzero, this Hamiltonian is non-Hermitian. The system described by $H_1$ has no spatial extent; we can think of it as existing at just one point in space. The time-dependent Schr\"odinger equation associated with $H_1$ is the first-order differential equation
\begin{equation}
i\psi'(t)=(a+ib)\psi,
\label{2.18}
\end{equation}
whose solution is $$\psi(t)=Ce^{(-ai+b)t}\quad(C~{\rm constant}).$$

Note that the wave function $\psi(t)$ oscillates with frequency $a$, and if $b\neq 0$, $\psi(t)$ grows or decays in time so the system described by $H_1$ in (\ref{e2.17}) is not in equilibrium; $\psi(t)$ grows or decays exponentially when $b$ is positive or negative. (If $b$ is negative, we have an elementary model of radioactive decay, where $b$ is a measure of the half-life of the state.) The system described by $H_1$  is a simple quantum-mechanical analog of a single classical optical fiber with loss or gain.

Next, we construct a $2\times2$ quantum-mechanical matrix Hamiltonian that describes {\it two} noninteracting systems like that in (\ref{e2.17}), one whose wave function grows in time and the other whose wave function decays in time:
\begin{equation}
H_2=\left[\begin{array}{cc} a+ib & 0\\ 0 & a-ib \end{array}\right].
\label{e2.19}
\end{equation}
This matrix Hamiltonian describes two subsystems, one with loss and the other with gain. Thus, it serves as a quantum-mechanical analog of two independent optical fibers, one with gain and the other with loss, which are not in contact and thus cannot exchange energy.

While the physical system described by the Hamiltonian $H_2$ is not in equilibrium if $b\neq0$, we can achieve dynamic equilibrium by {\it coupling} the two subsystems. We model this coupling mathematically by replacing the zeros in the off-diagonal elements of $H_2$ with the real coupling parameter $g$:
\begin{equation}
H_{\rm coupled}=\left[\begin{array}{cc} a+ib & g \\ g &
a-ib\end{array}\right].
\label{e2.20}
\end{equation}
Now the subsystems with loss and with gain are symmetrically coupled and so this Hamiltonian models the situation in which the two optical fibers are in contact and light energy can leak from one fiber into the other.

The Hamiltonian matrix $H_{\rm coupled}$ is not Hermitian, but it is $\cPT$ symmetric. To see why this is so, we define the parity operator $\cP$, which interchanges the two subsystems, as the linear matrix operator
\begin{equation}
\cP=\left[\begin{array}{cc} 0 & 1 \\ 1 & 0 \end{array}\right].
\label{e2.21}
\end{equation}
Observe that $\cP$ behaves as a {\it reflection operator} because $\cP^2=I$ ($I$ is the identity matrix). Also, $\cP$ is conventionally Hermitian $\cP^\dag=\cP$. Next, we define the time-reversal operation as complex conjugation because this changes the sign of $t$ in the coupled Schr\"odinger equations for these matrix Hamiltonians. (As mentioned earlier in this section, $\cT$ is an antilinear operator.) With these definitions, both $\cP$ and $\cT$ are reflection operators:
$$\cP^2=I~{\rm and}~\cT^2=I.$$ Also, as discussed earlier, parity reflection and time reversal are independent operations, so $\cP$ and $\cT$ commute: $$[\cP,\cT]=0.$$ With these definitions of $\cP$ and $\cT$ it is clear that while the Hamiltonian $H_{\rm coupled}$ is not Hermitian, $$H_{\rm coupled}^\dag\neq H_{\rm coupled},$$ it is $\cPT$ symmetric:
\begin{equation}
\big[ H_{\rm coupled},\cPT\big]=0.
\label{e2.22}
\end{equation}

Even though $H_{\rm coupled}$ is not Hermitian, if $g$ is sufficiently large, the eigenvalues of $H_{\rm coupled}$ become real. The eigenvalues $E$ are the roots of the secular polynomial
\begin{equation}
\det\left(H_{\rm coupled}-IE\right)=E^2-2aE+a^2+b^2-g^2,
\label{e2.23}
\end{equation}
where $I$ is the identity matrix. As we expect from the discussion in 
Subsec.~\ref{ss2D}, this secular polynomial is {\it real}.

The roots of the polynomial (\ref{e2.23}) are
\begin{equation}
E_\pm=a\pm\sqrt{g^2-b^2}.
\label{e2.24}
\end{equation}
Hence, if the two subsystems are weakly coupled ($g^2<b^2$), then the eigenvalues are complex. In this case the total system is not in equilibrium because one eigenstate grows in time and the other decays in time. This is the region of {\it broken $\cPT$ symmetry}. However, if the subsystems are strongly coupled ($g>b$ or $g<-b$), then the eigenvalues in (\ref{e2.24}) become real and the entire composite system is in dynamic equilibrium; the eigenstates oscillate in time and do not grow or decay. These are the regions of {\it unbroken $\cPT$ symmetry}. So, as a function of coupling parameter $g$, there are {\it two} transition points and there are {\it two} regions of unbroken $\cPT$ symmetry with a region of broken $\cPT$ symmetry sandwiched between them.

Interestingly, the situation here is unlike that for the Hamiltonian (\ref{e2.14}) discussed in Subsec.~\ref{ss2D}. For that Hamiltonian there is just one transition point and at that transition point the Hamiltonian is Hermitian. In the current case, at each of the two transition points there is an abrupt transition between nonequilibrium and equilibrium states, but at these transition points the Hamiltonian does not become Hermitian.  

This matrix model illustrates an important distinction between Hermitian and $\cPT$-symmetric systems. An unbroken $\cPT$-symmetric system resembles a closed and isolated Hermitian system because its energy levels are real and it is in dynamic equilibrium. However, it is {\it not} closed because it is still in contact with the external environment. In this optical model light energy is flowing (in a balanced way) into and out of the system. In contrast, a broken $\cPT$-symmetric system resembles an open system because it is not in dynamic equilibrium; however, unlike a typical open system the net probability flux vanishes and the system has $\cPT$ symmetry. Thus, $\cPT$-symmetric systems are a new class of systems whose physical properties are intermediate between open and closed systems.

\subsection{Other multiplicative deformations}\label{ss2F}
The quantum harmonic-oscillator Hamiltonian $H=p^2+x^2$ is special in many ways. For example, the time-independent Schr\"odinger eigenvalue equation associated with this Hamiltonian can be solved exactly in terms of Hermite polynomials, the leading-order WKB approximation gives the exact values of its eigenvalues, the field-theoretic analog of the harmonic-oscillator Hamiltonian describes {\it free particles}, and so on. However, while the eigenvalue problems arising from a multiplicative deformation of the harmonic-oscillator Hamiltonian are not exactly solvable, the spectral behavior in the unbroken $\cPT$-symmetric region is generic. Indeed a plot of the eigenvalues of $H$ in (\ref{e2.14}) for $\vep\geq 0$, as shown in Fig.~\ref{f4}, resembles the spectral plots of infinite numbers of other $\cPT$-deformed Hermitian Hamiltonians. For example, if we perform multiplicative deformations of the huge class of Hermitian Hamiltonians $H=p^{2m}+x^{2n}$, where $m,\,n=1,\,2,\,3,\,\cdots$, we obtain the infinite class of $\cPT$-symmetric Hamiltonians \cite{pt579}
\begin{equation}
H=p^{2m}+x^{2n}(ix)^\vep\quad(m,\,n=1,\,2,\,3,\,\cdots).
\label{e2.25}
\end{equation}

\begin{figure}
\center
\includegraphics[scale=0.24]{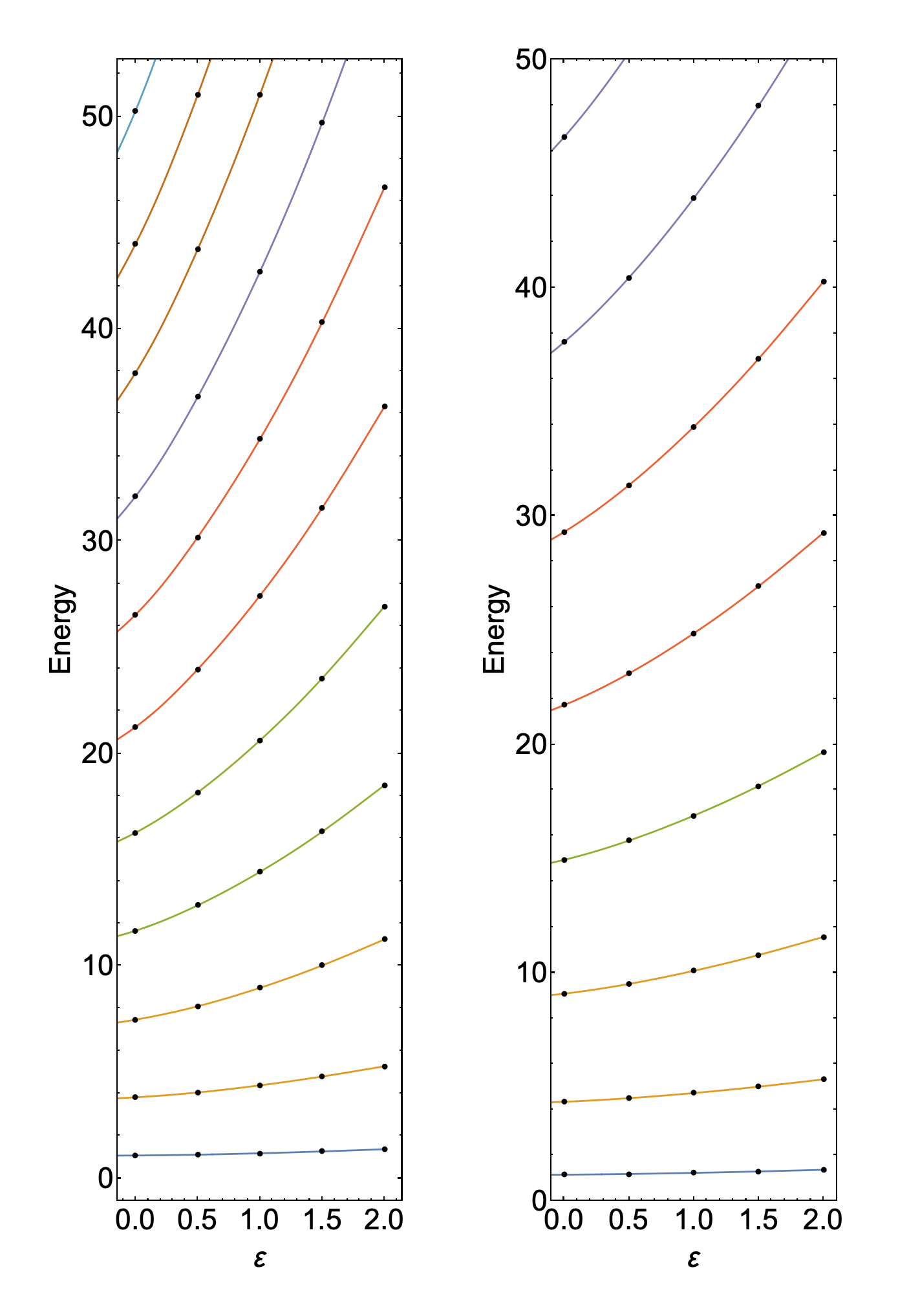}
\caption{Left panel: First eleven eigenvalues of $H$ in (\ref{e2.25}) with $m=1,\,n=2$ plotted for $0\leq\vep\leq2$. Right panel: First eight eigenvalues of $H$ in (\ref{e2.25}) with $m=1,\,n=3$ plotted for $0\leq\vep\leq2$. This figure emphasizes the generic features of the spectra of multiplicatively deformed Hermitian Hamiltonians in the unbroken $\cPT$-symmetric regions. In both cases the entire spectrum is real, positive, discrete, and rising monotonically for $\vep\geq0$. When $\vep$ decreases below 0 the eigenvalues begin to become complex, but the behavior of the eigenvalues in the broken-$\cPT$ symmetric region differs from potential to potential. For example, there is one new unbroken $\cPT$-symmetric region in which the entire spectrum of $H=p^2+x^6(ix)^\vep$ becomes real again when $-3 \leq\vep\leq-2$ (not shown). Other Hamiltonians may have several different unbroken regions for negative $\vep$ or no unbroken regions at all.}
\label{f6}
\end{figure}

\begin{figure}
\center
\includegraphics[scale=0.243]{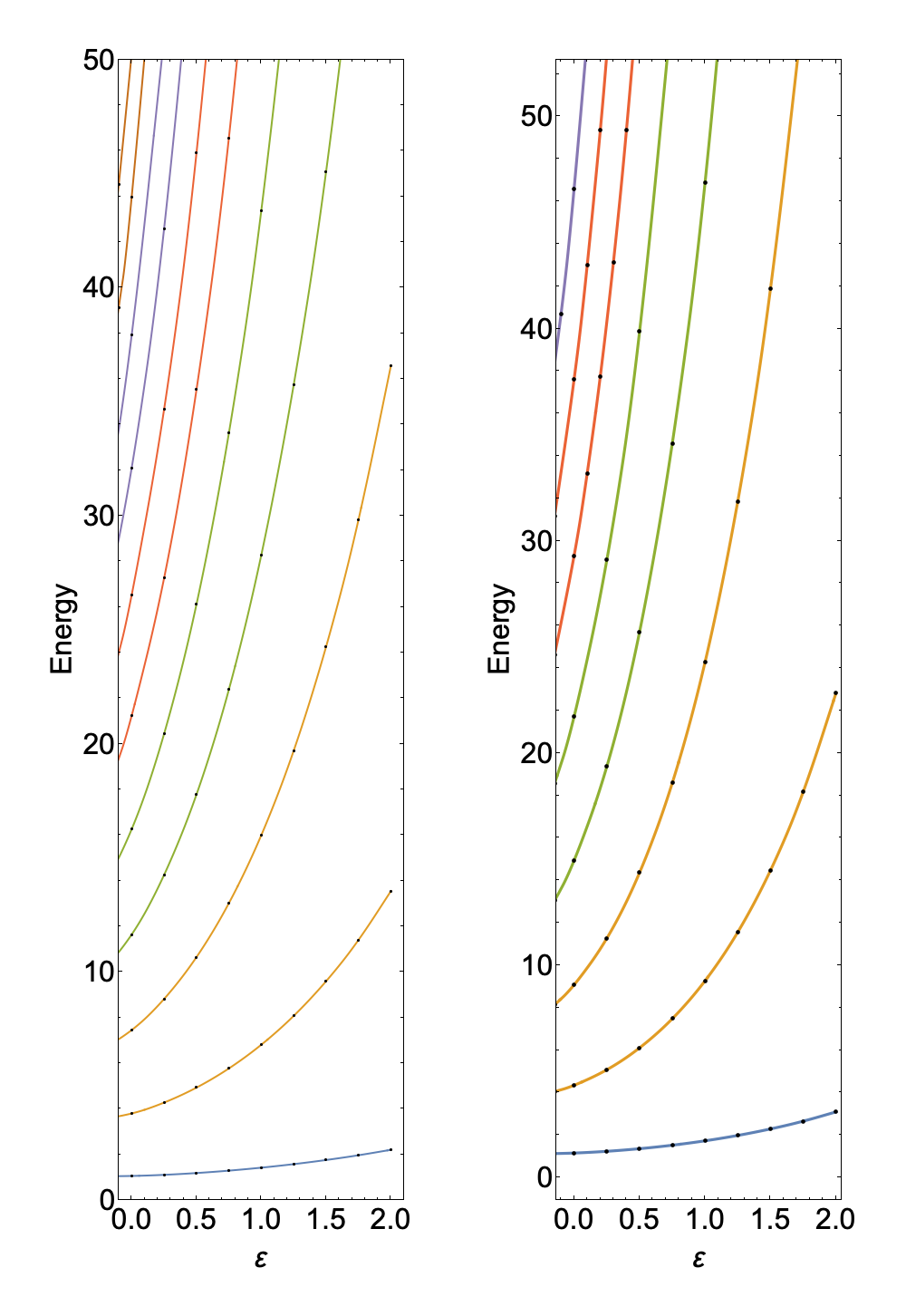}
\caption{Left panel: First eleven eigenvalues of $H$ in (\ref{e2.25}) with $m=2,\,n=1$ plotted for $0\leq\vep\leq2$. Right panel: First eight eigenvalues of $H$ in (\ref{e2.25}) for $m=3,\,n=1$ plotted for $0\leq\vep\leq2$. As in Fig.~\ref{f6}, the eigenvalues are all real, positive, discrete, and monotonically increasing in the unbroken $\cPT$-symmetric region $\vep\geq 0$. The eigenvalues become complex at exceptional points when $\vep<0$.}
\label{f7}
\end{figure}

Numerical studies indicate that there is universal behavior when $\vep$ is positive \cite{pt170}: For all $m$ and $n$ this is a region of unbroken $\cPT$ symmetry. The eigenvalues of these deformed Hamiltonians are all real, positive, discrete, and monotonically increasing in this region and there is a $\cPT$ transition at $\vep=0$. The low-lying eigenvalues for the cases $(m=1,\,n=2)$ and $(m=1,\,n=3)$ are plotted in Fig.~\ref{f6} and the low-lying eigenvalues for $(m=2,\,n=1)$ and $(m=3,\, n=1)$ are plotted in Fig.~\ref{f7} for $\vep\geq0$. However, in the broken $\cPT$-symmetric region $\vep<0$ these models exhibit a variety of interesting behaviors: While there are an infinite number of exceptional points just below $\vep=0$ at which high-lying energy levels become complex, as $\vep$ becomes more negative new finite-width regions of unbroken $\cPT$ symmetry may emerge. For example, for the case $m=2,\,n=6$ there is an additional region of unbroken $\cPT$ symmetry when $-3<\vep<-2$.

\section{Deformation of eigenvalue problems}\label{s3}
We have seen in Sec.~\ref{s2} that $\cPT$-symmetric deformations of Hermitian quantum-mechanical theories can lead to new non-Hermitian theories that exhibit remarkable and unexpected behaviors. Deformation is a powerful mathematical tool that can provide mathematical insights and establish relationships between seemingly unrelated mathematical systems. However, deforming an eigenvalue problem is a nontrivial global procedure because it requires a {\it simultaneous} deformation of the differential equation and the accompanying boundary conditions. This section discusses in depth the process of complex deformation of eigenvalue problems.

\subsection{Complex asymptotic approximations}\label{ss3A}
To explain complex deformation and to understand the differences between $\cPT$-symmetric quantum mechanics and Hermitian quantum mechanics it is essential to understand the nature of asymptotic approximations in the complex plane. An {\it asymptotic approximation} is a relation between two functions that is expressed as a limit \cite{pt16}. This relation approximates the behavior of a function $f(x)$ by a simpler function $g(x)$ as $x$ approaches a point $a$ (which may be $\infty$). The symbol $\sim$ is used to express an asymptotic approximation:
$$f(x)\sim g(x)\quad(x\to a).$$
This statement means that $\lim_{x\to a}[f(x)/g(x)]=1$.

In this Review we are specifically interested in the behavior of eigenfunctions as they approach the boundary of a region where the eigenfunctions are required to obey boundary conditions. The boundary is often the point at $|x|=\infty$. Usually, these eigenfunctions are so complicated that only very rarely are we able to solve for them in closed form and thus we need to use asymptotic methods to approximate the eigenfunctions.

Complex asymptotic relations are generally not valid as $x$ approaches $a$ in all complex directions; asymptotic approximations are typically valid in wedge-shaped regions of the complex plane called {\it Stokes sectors}. To describe a Stokes sector we specify two pieces of information: First, we specify the complex limit; for example, $|x|\to\infty$ or $|x|\to a$. Second, we give the {\it angular range} of $x$ in which this limit is valid: $\theta_1<{\rm arg}\,x<\theta_2$.

As an elementary example of a complex asymptotic approximation, let us approximate the function $\sinh(x)=\half \big(e^x-e^{-x}\big)$ for large $|x|$. For large positive $x$, $\sinh(x)$ is well approximated by the simpler function $\half e^x$ because $e^{-x}$ is negligible compared with $e^x$. This approximation continues to be valid for complex $x$ if ${\rm Re}\,x$ is large and positive; that is, if $|x|\to\infty$ in a complex Stokes sector that is centered about the positive-real-$x$ axis and that has an opening angle of $\pi$:
$$\sinh(x)\sim\half e^x\quad\big(|x|\to\infty,~-\tfrac{\pi}{2}<{\rm arg}\,x<\tfrac{\pi}{2}\big).$$

Outside this Stokes wedge $\sinh(x)$ has a {\it different} asymptotic behavior because for large $|x|$ with ${\rm Re}\,x<0$, $e^x$ is negligible compared with $e^{-x}$:
$$\sinh(x)\sim-\half e^{-x}\quad\big(|x|\to\infty,~\tfrac{\pi}{2}<{\rm arg}\, x<\tfrac{3\pi}{2}\big).$$
This abrupt change in the complex asymptotic behavior of $\sinh(x)$ for large $|x|$, which occurs at the edges ${\rm arg}\,x=\pm\tfrac{\pi}{2}$ of the Stokes sectors, is a general phenomenon that is typical of asymptotic approximations. The edges of Stokes sectors are called {\it Stokes lines}.

As we see in the next subsection, because the boundary conditions on solutions to the time-independent Schr\"odinger equation are valid inside Stokes sectors, these {\it boundary conditions distinguish $\cPT$-symmetric quantum theories from conventional Hermitian quantum theories}. In conventional quantum mechanics we are ordinarily interested in the behavior of solutions to the time-independent Schr\"odinger equation on the real axis subject to boundary conditions that are also imposed on the real axis, and we are usually not concerned with the behavior of solutions for complex $x$. However, for the case of $\cPT$-symmetric quantum mechanics the boundary conditions are often specified in the complex domain.

\subsection{Boundary conditions for the Sextic Hamiltonian}\label{ss3B}
Let us consider the Hamiltonian $H$ in (\ref{e2.14}) for the sextic case
$\vep=4$:
\begin{equation}
H=p^2+x^6.
\label{e3.1}
\end{equation}
The time-independent Schr\"odinger equation associated with $H$ in (\ref{e3.1})
is
\begin{equation}
-\psi''(x)+x^6\psi(x)=E\psi(x).
\label{e3.2}
\end{equation}
The exact solutions to this deceptively simple-looking differential equation are much too complicated to be expressed in closed form. However, we will show that elementary asymptotic analysis using WKB techniques easily gives the possible asymptotic behaviors of the solutions $\psi(x)$ to this Schr\"odinger equation for large $|x|$ \cite{pt16}.

For the Schr\"odinger differential equation $y''(x)=Q(x)y(x)$ where $\lim_{|x|\to\infty}|Q(x)|=\infty$, WKB theory provides a simple formula for the two possible exponential components of the asymptotic behavior: The solutions can either {\it grow} or {\it decay} exponentially like
\begin{equation}
\exp\left[\pm\int^x dt\,\sqrt{Q(t)}\right].
\label{e3.3}
\end{equation}
The expression in (\ref{e3.3}) is called a {\it geometrical-optics} approximation \cite{pt16}. We emphasize that (\ref{e3.3}) is {\it not} a leading-order WKB approximation (called a {\it physical-optics} approximation); it is just the {\it exponential component} of the WKB approximation. From (\ref{e3.3}) we see that the possible exponential behaviors of solutions to (\ref{e3.2}) for large $|x|$ are $\exp\big(\pm\tfrac{1}{4}x^4\big)$.

If we were to treat $H$ in (\ref{e3.1}) as a conventional Hermitian Hamiltonian, we would require that the eigenfunctions vanish as $|x|\to\infty$ on the real-$x$ axis. A numerical calculation would then show that the eigenvalues are all discrete, real, and positive. The approximate values of the lowest five of these eigenvalues are
\begin{eqnarray}
&&E_0=1.145...,~~E_1=4.339...,~E_2=9.073...,\nonumber\\
&&\quad E_3=14.935...,~~E_4=21.714...\,.
\label{e3.4}
\end{eqnarray}
However, if we treat $H$ as a $\cPT$-symmetric Hamiltonian, we find that while the eigenvalues are again all discrete, real, and positive, they are different from and larger than those in (\ref{e3.4}). The approximate values of the lowest five eigenvalues are
\begin{eqnarray}
&&E_0=2.439...,~E_1=11.882...,~E_2=25.412...,\nonumber\\
&&\quad E_3=42.024...,~~E_4=61.222...\,.
\label{e3.5}
\end{eqnarray}
These are the $\cPT$ eigenvalues that are shown in Fig.~\ref{f4}.

It is remarkable that one Hamiltonian can have two completely different spectra. The reason for this is that a Hamiltonian specifies only the {\it local} behavior but not the {\it global} behavior of the eigenfunctions. That is, the Hamiltonian  $H$ specifies the Schr\"odinger differential equation, which in turn determines the relationships between the eigenfunctions and their derivatives {\it at the point} $x$. However, $H$ does not specify the global {\it boundary conditions} on the solutions to the differential equation.

Treating (\ref{e3.1}) as a conventional Hermitian Hamiltonian, we require that the solutions to (\ref{e3.2}) be normalizable (square-integrable) on the real-$x$ axis. This means that the solutions must decay like $\exp\big(-\fourth x^4\big)$ as $x\to\infty$ and as $x\to-\infty$; we exclude the solutions to the Schr\"odinger equation that blow up like $\exp\big( \fourth x^4\big)$ as $x\to\infty$ or as $x\to-\infty$. Like the function ${\rm sinh}\,x$ discussed in Subsec.~\ref{ss3A}, the validity of these exponentially decaying behaviors is not limited to the real-$x$ axis; these behaviors continue to be valid in the complex-$x$ plane in two Stokes sectors, each having an angular opening of $\tfrac{\pi}{4}$. One sector is centered about the positive-real axis and the other is centered about the negative-real axis.

\begin{figure}
\center
\includegraphics[scale=0.42]{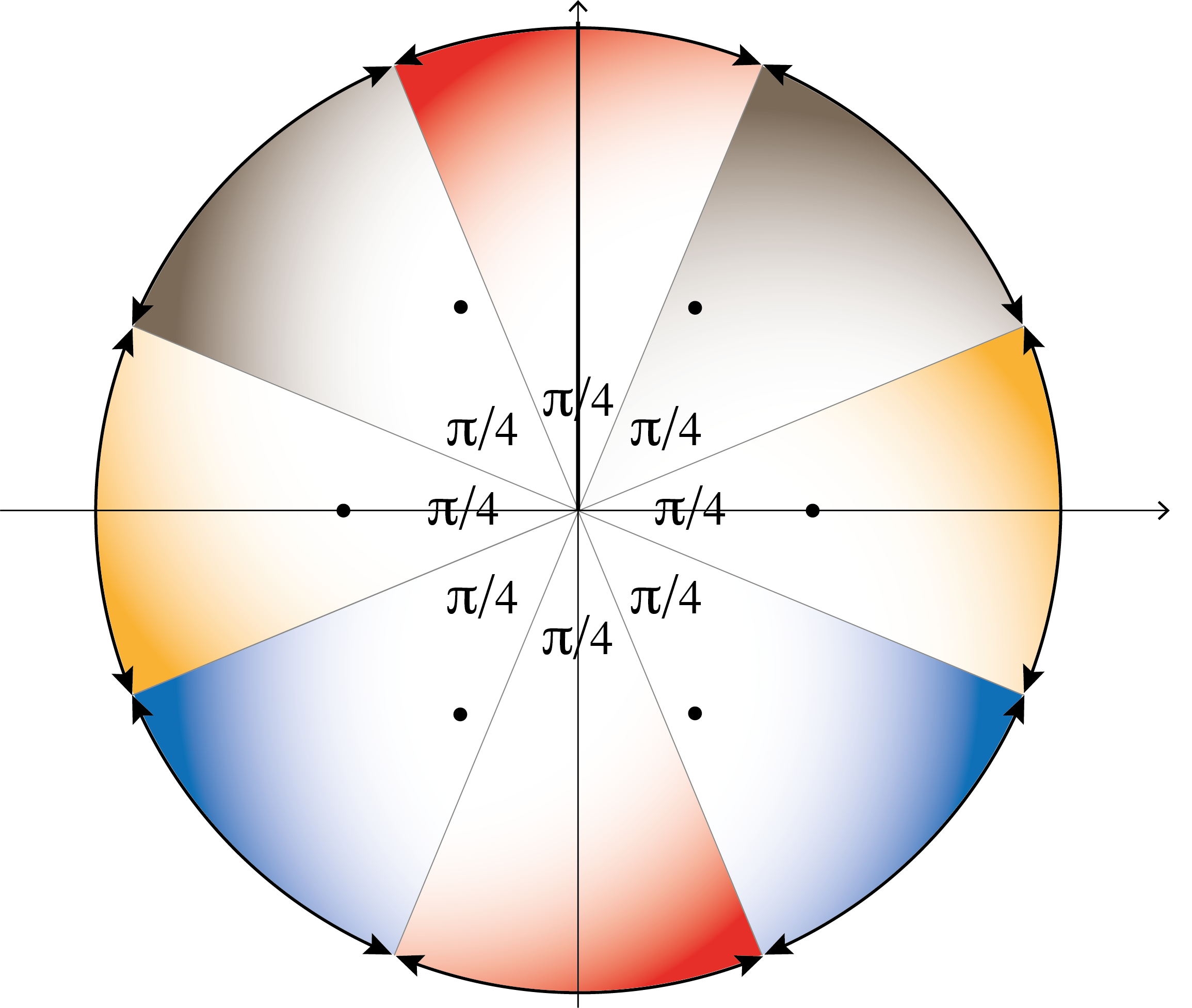}
\caption{Stokes sectors for the eigenfunctions of the sextic Hamiltonian (\ref{e3.1}). If the eigenfunctions are required to vanish exponentially in the yellow pair of Stokes sectors containing the positive-real axis and the negative-real axis (\ref{e3.6}), the associated eigenvalues are those of the conventional Hermitian $x^6$ potential that are listed in (\ref{e3.4}). However, if the eigenfunctions are required to vanish exponentially in the blue pair of Stokes sectors adjacent to and below the yellow Stokes sectors (\ref{e3.7}), the associated eigenvalues, which are listed in (\ref{e3.5}), are those of the $\cPT$-symmetric sextic theory in (\ref{e2.14}) with $\vep=4$. Since the sextic Hamiltonian is real, the Schr\"odinger eigenvalue problem is symmetric under complex conjugation, so the array of Stokes sectors is up-down symmetric. Hence the eigenvalues associated with the gray pair of Stokes sectors are identical to those associated with the blue pair. The dots indicate the locations of the classical turning points, which are the solutions to the equation $x^6=E$. Each horizontal (left-right symmetric) pair of turning points corresponds to a real spectrum. A fourth pair of Stokes sectors (red) is centered about the positive-imaginary and negative-imaginary axes. This pair of Stokes sectors, which does not contain turning points, is obtained by rotating the yellow pair by $90^\circ$. The eigenvalues associated with the eigenfunctions that vanish exponentially in the red Stokes sectors are the {\it negatives} of the Hermitian eigenvalues listed in (\ref{e3.4}).}
\label{f8}
\end{figure}

To summarize, the conventional quantum-mechanical eigenfunctions $\psi(x)$, which satisfy the differential equation (\ref{e3.2}) and vanish asymptotically as $x$ approaches $\pm\infty$, {\it continue} to vanish exponentially as $|x|\to\infty$ in two Stokes sectors whose opening angles are given by
\begin{equation}
-\tfrac{\pi}{8}<\rm{arg}\,x<\tfrac{\pi}{8}~~{\rm and}~~ -\tfrac{9\pi}{8}<{\rm
arg}\,x<-\tfrac{7\pi}{8}.
\label{e3.6}
\end{equation}
These Stokes sectors are displayed in Fig.~\ref{f8} and are colored yellow.

On the Stokes lines at the edges of the Stokes sectors (\ref{e3.6}) the character of the conventional Hermitian eigenfunctions, whose eigenvalues are given in (\ref{e3.4}), changes abruptly. If we cross the Stokes lines at
$${\rm arg}\,x=\pm\tfrac{\pi}{8}~~{\rm and}~~{\rm arg}\,x=-\pi\pm\tfrac{\pi}{8},$$
the conventional eigenfunctions no longer vanish exponentially for large $|x|$ and instead grow exponentially. However, if we impose {\it new} boundary conditions that require the solution to the Schr\"odinger equation (\ref{e3.2}) to vanish exponentially in two Stokes sectors of angular opening $\tfrac{\pi}{4}$ and centered about $-\tfrac{\pi}{4}$ and $-\tfrac{3\pi}{4}$,
\begin{equation}
-\tfrac{3\pi}{8}<{\rm arg}\,x<-\tfrac{\pi}{8}~~{\rm and}~~ -\tfrac{7\pi}{8}<{\rm arg}\,x<-\tfrac{5\pi}{8},
\label{e3.7}
\end{equation}
we find an entirely new spectrum of eigenvalues; the lowest five are listed in (\ref{e3.5}). The Stokes sectors in (\ref{e3.7}) are colored blue in Fig.~\ref{f8}. As will be explained in Subsec.~\ref{ss3D}, these Stokes sectors are obtained when we continuously deform the boundary conditions for the $\cPT$-symmetric Hamiltonian in (\ref{e2.14}) as $\vep\to4$.

There are two more pairs of Stokes sectors shown in Fig.~\ref{f8}, the gray pair and the red pair. The gray Stokes sectors are complex conjugates of the blue sectors. Note that the sextic Hamiltonian in (\ref{e3.1}) is special because it is {\it real}. Thus, in addition to $\cPT$ symmetry (left-right symmetry in the complex plane), $H$ exhibits another discrete symmetry, namely, complex-conjugation symmetry, which, as explained in Sec.~\ref{s2}, is time-reversal symmetry. The operator $\cT$ performs a reflection about the real axis, so Fig.~\ref{f8} is up-down symmetric. As a consequence of this symmetry, the eigenvalues associated with the gray pair of Stokes sectors are identical to those in the blue Stokes sectors. These eigenvalues are listed in (\ref{e3.5}).

The eigenvalues associated with the red pair of Stokes sectors are the {\it negatives} of the eigenvalues listed in (\ref{e3.4}). To see why this is so, we rotate the eigenvalue problem associated with the yellow Stokes wedges by $90^\circ$; that is, we replace $x$ in the eigenvalue equation (\ref{e3.2}) by $ix$. The new eigenvalue equation reads
$$-\psi''(x)+x^6\psi(x)=-E\psi(x).$$

We have recovered the original Hermitian eigenvalue problem except that $E$ is multiplied by $-1$ and $\psi(x)$ vanishes as $x\to\pm i\infty$. The eigenspectrum associated with this boundary-value problem is unbounded below. This sign change occurs for all Hamiltonians of the form
\begin{equation}
H=p^2+x^{2+4n}\quad(n=0,\,1,\,2,\,...).
\label{e3.8}
\end{equation}
The simplest case ($n=0$) of this spectral sign-flip phenomenon is discussed in the next subsection.

\subsection{Complex deformation of the harmonic oscillator}\label{ss3C} 
As explained earlier in this section, when we perform a deformation of an eigenvalue problem associated with a Hamiltonian, we must be careful to deform the boundary conditions on the eigenfunctions in addition to deforming the Schr\"odinger equation. If we neglect to do so, we may encounter contradictions and paradoxes.

For example, let us consider the quantum harmonic-oscillator Hamiltonian
\begin{equation}
H=p^2+\omega^2x^2,
\label{e3.9}
\end{equation}
where $\omega$ is a positive parameter associated with the classical spring constant. The Schr\"odinger eigenvalue problem for this elementary Hamiltonian reads
\begin{equation}
-\psi''(x)+\omega^2x^2\psi(x)=E\psi(x)\,.
\label{e3.10}
\end{equation}
The conventional Hermitian boundary conditions on the eigenfunctions are
\begin{equation}
\lim_{x\to\pm\infty}\psi(x)=0\,.
\label{e3.11}
\end{equation}
This is an exactly solvable eigenvalue problem. As we saw in Subsec.~\ref{ss2A}, the eigenfunctions are Hermite functions (Hermite polynomials multiplied by Gaussians) and the eigenvalues are all positive:
\begin{equation}
E_n=(2n+1)\omega\quad(n=0,\,1,\,2,\,\dots)\,.
\label{e3.12}
\end{equation}

Let us now perform a smooth complex deformation of the Hamiltonian (\ref{e3.9}) by rotating the parameter $\omega$, which is initially real and positive, into the complex plane:
\begin{equation}
\omega\to\omega e^{i\theta}.
\label{e3.13}
\end{equation}
If we rotate $\omega$ by an angle $\pi$, that is, $\theta:\,0\to\pi$, then $\omega$ changes sign. Under this rotation the Hamiltonian (\ref{e3.9}) remains {\it unchanged}. However, its eigenvalues all change sign! Thus, this Hamiltonian possesses {\it two} different spectra; it has an entirely negative spectrum in addition to the usual positive spectrum (\ref{e3.12}). Given our previous discussion of the $x^6$ potential (Subsec.~\ref{ss3B}), this conclusion is not surprising.

The advantage of this harmonic-oscillator example is that the simplicity of the Hamiltonian allows us to understand clearly what happens as we continuously deform the boundary conditions (\ref{e3.11}) together with the differential equation (\ref{e3.10}). Initially, when $\theta=0$, the eigenfunctions $\psi_n(x)$ all vanish exponentially on the positive-$x$ and the negative-$x$ axes like 
\begin{equation}
\exp\big(-\half\omega x^2\big)\quad(x\to\pm\infty).
\label{e3.14}
\end{equation}

This exponentially-decaying Gaussian behavior persists into a pair of Stokes sectors of angular opening $\tfrac{\pi}{2}$ that are centered about the positive-$x$ and negative-$x$ axes. However, when we rotate $\omega$ into the complex plane according to (\ref{e3.13}), we must also replace the formula for the exponential Gaussian decay in (\ref{e3.14}) by  
\begin{equation}
\exp\big(-\half e^{i\theta}\omega x^2\big).
\label{e3.15}
\end{equation}
Thus, as $\theta$ increases from 0 to $\pi$ and $\omega$ rotates anticlockwise, the pair of Stokes sectors (in which the eigenfunctions decay exponentially) correspondingly rotates {\it clockwise} like a propeller by an angle of $\half\pi$. The opening angles of the two sectors remain constant during this rotation, but the orientation of the Stokes sectors changes and the wedges end up centered about the positive-imaginary-$x$ and negative-imaginary-$x$ axes when $\theta=\pi$.

This argument makes it clear that the Hamiltonian (\ref{e3.9}) has two independent spectra, one entirely positive and the other entirely negative, just like the energy levels associated with the red pair of Stokes sectors in Fig.~\ref{f8}. The positive-spectrum case is physically relevant. However, while a negative spectrum may be interesting from a mathematical point of view, a negative spectrum is physically unacceptable because the energy levels are not bounded below. This violates the basic physical requirement of stability, namely, that a physical system must have a lowest-energy (ground) state.

\subsection{Boundary conditions for $H=p^2+x^2(ix)^\vep$}\label{ss3D}
Let us now return to the sextic Hamiltonian discussed in Subsec.~\ref{ss3B}. The question that we must answer is, Why do we require that the eigenfunctions of the $\cPT$-symmetric Hamiltonian $H=p^2+x^2(ix)^\vep$ in (\ref{e2.14}) with $\vep=4$ satisfy the vanishing boundary conditions shown in Fig.~\ref{f8}? (And more generally for any Hamiltonian, What determines the appropriate pair of Stokes sectors in which to impose the boundary conditions on the eigenfunctions?) To address this question, we go back to the WKB formula in (\ref{e3.3}). This formula implies that the eigenfunctions of $H$ either grow or decay exponentially for large $|x|$ like
\begin{equation}
\exp\left(\pm\tfrac{2}{4+\vep}i^{\vep/2}x^{2+\vep/2}\right).
\label{e3.16}
\end{equation}

To determine the appropriate choice of signs for particular values of $\vep$, we begin by setting $\vep=0$, which reduces $H$ to the harmonic-oscillator Hamiltonian. As explained in Subsec.~\ref{ss3B}, we get a real positive spectrum if we require that the eigenfunctions of the harmonic-oscillator Hamiltonian vanish in Stokes sectors centered about the positive-real-$x$ and negative-real-$x$ axes. The angular opening of these sectors is $\tfrac{\pi}{2}$. From (\ref{e3.16}) we see that if we {\it continuously} deform $H$ by smoothly increasing the real parameter $\vep$, both Stokes sectors rotate {\it downward} towards the negative-imaginary-$x$ axis. 

Specifically, for $\vep\geq0$, the right Stokes sector rotates {\it clockwise} and is centered about the angle
\begin{equation}
\theta_{\rm center,\,right}=-\tfrac{\pi\vep}{8+2\vep}
\label{e3.17}
\end{equation}
and the left Stokes sector rotates {\it anticlockwise} and is centered about the angle
\begin{equation}
\theta_{\rm center,\,left}=-\pi+\tfrac{\pi\vep}{8+2\vep}.
\label{e3.18}
\end{equation}
In addition, the angular opening $\theta_{\rm opening\,angle}$ of each Stokes
sector is given by
\begin{equation}
\theta_{\rm opening\,angle}=\tfrac{2\pi}{4+\vep}.
\label{e3.19}
\end{equation}
Thus, the sectors become thinner ($\theta_{\rm opening\,angle}$ decreases) as $\vep$ increases.

\begin{figure}
\center
\includegraphics[scale = 0.65]{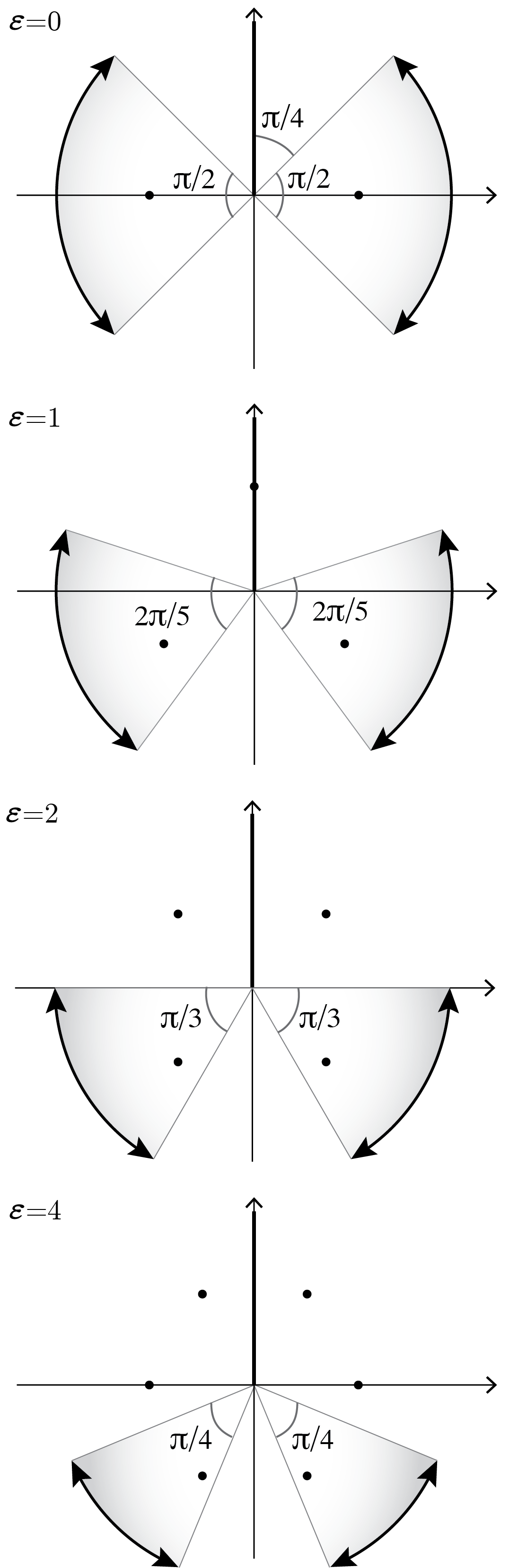}
\caption{Pairs of Stokes sectors in the complex-$x$ plane in which the eigenfunctions of the $\cPT$-symmetric Hamiltonian $H=p^2+x^2(ix)^\vep$ vanish as $|x|\to\infty$. The sectors are displayed for $\vep=0$ (harmonic
oscillator), $\vep=1$ (imaginary cubic oscillator), $\vep=2$ (upside-down quartic oscillator), and $\vep=4$ (right-side-up sextic oscillator). As $\vep$ increases smoothly, the sectors rotate downward and become narrower. The dots indicate the locations of classical turning points. Note that the turning points also rotate downward as $\vep$ increases (they rotate slightly faster than the sectors) and reach the positions shown in Fig.~\ref{f8}.}
\label{f9}
\end{figure}

This pair of Stokes sectors is shown in Fig.~\ref{f9} for four values of $\vep$. Note that at $\vep=4$, we recover the $\cPT$-symmetric
orientation of the blue Stokes sectors shown in Fig.~\ref{f8}, which was used in the discussion of the sextic $\cPT$-symmetric Hamiltonian in Subsec.~\ref{ss3A}. As $\vep\to\infty$, the sectors become infinitely thin and lie along the negative-imaginary axis. In this limit we obtain the $\cPT$-symmetric square well \cite{pt327}. For all $\vep>0$ the pair of Stokes sectors in Fig.~\ref{f9} is $\cPT$ symmetric (left-right symmetric), which enforces the global $\cPT$ symmetry of the eigenfunction problem.

Figure \ref{f9} also shows that as $\vep$ increases, there are more and more classical turning points. These turning points are solutions to the equation
$$x^2(ix)^\vep=E.$$
If $\vep$ is not an integer, the function $x^2(ix)^\vep$ is defined on a Riemann surface with a branch cut running from $x=0$ to $\infty$, and because we want the eigenvalue problem associated with the Hamiltonian (\ref{e2.14}) to respect $\cPT$ symmetry (left-right symmetry), we take this branch cut to lie along the positive-imaginary-$x$ axis. (The four configurations in Fig.~\ref{f9} correspond to integer values of $\vep$, so there are no branch cuts.) As $\vep$ increases smoothly through real values, the density of turning points increases and new pairs of turning points continue to emerge through the branch cut on the positive-imaginary-$x$ axis. One of the turning points in each new pair rotates clockwise and the other rotates anticlockwise as $\vep$ increases.

When $\vep$ ranges from 0 to 1, there are only two turning points on the principal sheet of the Riemann surface (top diagram in Fig.~\ref{f9}). Just as $\vep$ approaches 1, the first new pair of turning points appears on the positive-imaginary-$x$ axis. When $\vep$ is exactly 1, these turning points coincide (second diagram in Fig.~\ref{f9}). As $\vep$ increases above 1, the coincident turning points split with one moving east and the other moving west. When $\vep$ reaches 2, the four turning points on the principal sheet exhibit left-right and up-down symmetry (third diagram in Fig.~\ref{f9}). At $\vep=3$ another pair of turning points enters the principal sheet, and when $\vep=4$ the six turning points are now arranged in a regular hexagon (fourth diagram in Fig.~\ref{f9}).

We emphasize that the continuous deformation described by (\ref{e3.17}) and (\ref{e3.18}), in which we smoothly increase $\vep$, preserves the {\it global} $\cPT$ symmetry of the eigenvalue problem. The Hamiltonian $H$ is locally $\cPT$ symmetric for any real value of $\vep$. But the boundary conditions {\it also} must remain $\cPT$ symmetric (left-right symmetric) in order that the eigenfunctions vanish in the left and right sectors displayed in Fig.~\ref{f9}. 

Remember that $\cPT$ symmetry corresponds to left-right symmetry in the complex-$x$ plane because $\cP$ reflects $x$ through the origin,
$$\cP:\,x\to-x,$$
and $\cT$ performs complex conjugation, which reflects $x$ about the real axis:
$$\cT:\,x\to x^*.$$ 
Thus, the combined effect of a $\cPT$ reflection corresponds to a reflection about the imaginary axis. It is noteworthy that $\cPT$ (left-right) reflection and complex conjugation (up-down reflection) are an orthogonal pair of independent discrete symmetries. 

\subsection{Sextic potential revisited}\label{ss3E}
Let us return to the Hamiltonian (\ref{e3.1}) discussed in Subsec.~\ref{ss3A}. We have stated that this sextic Hamiltonian has two positive real spectra depending on the choice of boundary conditions; that is, depending on the pair of Stokes sectors inside of which the eigenfunctions are required to vanish as $|x|\to \infty$. The conventional Hermitian eigenvalues are associated with the left-right-symmetric pair of Stokes sectors that contain the real axis and the first five of these eigenvalues are listed in (\ref{e3.4}). The $\cPT$-symmetric eigenvalues are associated with the left-right-symmetric pair of Stokes sectors below and adjacent to the Hermitian pair and the first five eigenvalues associated with this pair of Stokes sectors are listed in (\ref{e3.5}). These two pairs of sectors are shown in Fig.~\ref{f8} and the second pair of Stokes sectors is also shown in the fourth diagram of Fig.~\ref{f9}.

There is a third pair of left-right-symmetric Stokes sectors not shown in Fig.~\ref{f9}. This pair (shown as gray in Fig.~\ref{f8}) lies above and adjacent to the Hermitian pair of Stokes sectors. The eigenvalues associated with this third pair of sectors are identical to those associated with the second pair of Stokes sectors. This is because the sextic Hamiltonian is {\it real}, and therefore in addition to $\cPT$ symmetry (left-right symmetry), it also has complex-conjugate (up-down) symmetry in the complex plane.

An easy way to verify that the sextic Hamiltonian has two different spectra is simply to calculate the eigenvalues. Of course, the Schr\"odinger equation (\ref{e3.2}) cannot be solved analytically. However, WKB theory gives a good semiclassical approximation to both spectra. The WKB procedure is straightforward \cite{pt16}: First, we find the turning points $x_{\pm}$ that separate the classically allowed and classically forbidden regions; these turning points satisfy the equation $V(x)=E$.
Then, we evaluate the integral in the WKB quantization condition
\begin{equation}
\int_{x_-}^{x_+} dx\,\sqrt{V(x)-E}\sim\big(n+\tfrac{1}{2}\big)\pi\quad(n\to\infty).
\label{e3.20}
\end{equation}
Finally, we solve the resulting equation algebraically for $E_n$. The accuracy of this WKB calculation increases rapidly with increasing $n$.

For the conventional Hermitian theory, the turning points are the real roots of the equation $x^6=E$: 
$$x_\pm=\pm E^{1/6}.$$
Thus, the quantization condition (\ref{e3.20}) for the Hermitian theory reads $$\int_{-E^{1/6}}^{E^{1/6}}dx\,\sqrt{E-x^6}=\big(n+\tfrac{1}{2}\big)\pi.$$
We evaluate this integral in terms of Gamma functions and solve for $E_n$:
\begin{equation}
E_n\sim\left[\frac{(4n+2)\sqrt{\pi}\,\Gamma(2/3)}{\Gamma(1/6)}\right]^{3/2}\quad
(n\to\infty).
\label{e3.21}
\end{equation}
From this formula we numerically read off the first five WKB approximations to the Hermitian eigenvalues:
\begin{eqnarray}
&&E_0=0.801...,~~E_1=4.161...,~~E_2=8.954...,\nonumber\\
&&\quad E_3=14.832...,~~E_4=21.622...\,.
\label{e3.22}
\end{eqnarray}
These approximations are in good agreement with the numerical results in (\ref{e3.4}) and the accuracy of the WKB calculation improves rapidly with increasing $n$.

For the $\cPT$-symmetric sextic theory in Fig.~\ref{f9} the turning points lie inside the lower pair of Stokes sectors at $x_-=E^{1/6}e^{-2i\pi/3}$ and $x_+=E^{1/6} e^{-i\pi/3}$. To evaluate the integral in the WKB quantization formula (\ref{e3.20}), it is simplest to integrate along a radial line from $x_-$ to the origin and then along a second radial line from the origin out to $x_+$. Evaluating these two integrals in terms of Gamma functions and then combining the results gives a formula that differs from that in (\ref{e3.21}) for the Hermitian case by a multiple of $2^{3/2}$:
\begin{equation}
E_n\sim\left[\frac{(8n+4)\sqrt{\pi}\,\Gamma(2/3)}{\Gamma(1/6)}\right]^{3/2}\quad
(n\to\infty).
\label{e3.23}
\end{equation}
From this formula we read off the first five WKB approximations to the $\cPT$-symmetric eigenvalues:
\begin{eqnarray}
&&E_0=2.265...,~~E_1=11.770...,~~E_2=25.954...,\nonumber\\
&&\quad E_3=41.950...,~~E_4=61.157...\,.
\label{e3.24}
\end{eqnarray}
Once again, these approximations are in good agreement with the numerical calculations reported in (\ref{e3.5}). As noted in Subsec.~\ref{ss3B}, the $\cPT$ eigenvalues are larger than the corresponding Hermitian eigenvalues.

As explained in Subsec.~\ref{ss3B}, {\it any} pair of Stokes sectors (regardless of whether they have a $\cPT$-symmetric orientation) provides a setting for a specific eigenvalue problem with its own tower of eigenvalues. However, we cannot impose exponentially vanishing boundary conditions on solutions in {\it adjacent} pairs of Stokes sectors. This is because an exponentially decaying solution becomes an exponentially growing solution as we cross the Stokes line separating the two sectors. Thus, we must cross at least two Stokes lines to formulate an eigenvalue problem in which the solutions decay exponentially in two directions.

There are {\it twenty} nonadjacent pairs of Stokes sectors in Fig.~\ref{f8} inside of which we can formulate eigenvalue problems. Three left-right pairs give rise to positive spectra (blue-blue, yellow-yellow, gray-gray), three up-down pairs give rise to negative spectra (gray-blue, red-red, gray-blue), and the fourteen other nonadjacent pairs give rise to complex spectra.

Having examined the multiple-spectrum phenomenon at the quantum and at the semiclassical (WKB) level, it intriguing to study it at the {\it classical} level. For the sextic Hamiltonian $H=p^2+x^6$, Hamilton's equations of motion read
\begin{eqnarray}
&&x'(t)=\frac{\partial H}{\partial p}=2p,\nonumber\\
&&p'(t)=-\frac{\partial H}{\partial x}= -6x^5\,.
\label{e3.25}
\end{eqnarray}

These equations imply that the classical energy $E=H$ is a constant of the motion, so from the first equation in (\ref{e3.25}) we obtain a first-order differential equation for the motion of a classical particle in the complex plane:
\begin{equation}
E=\sqrt{\tfrac{1}{4}[x'(t)]^2+[x(t)]^6}\,.
\label{e3.26}
\end{equation}
Without loss of generality we can fix the value for $E$, say $E=1$, and choose an initial position $x(0)$ for the particle. Then, (\ref{e3.26}) determines the classical path $x(t)$ of the particle for all times $t>0$.

\begin{figure}
\center
\includegraphics[scale=0.22]{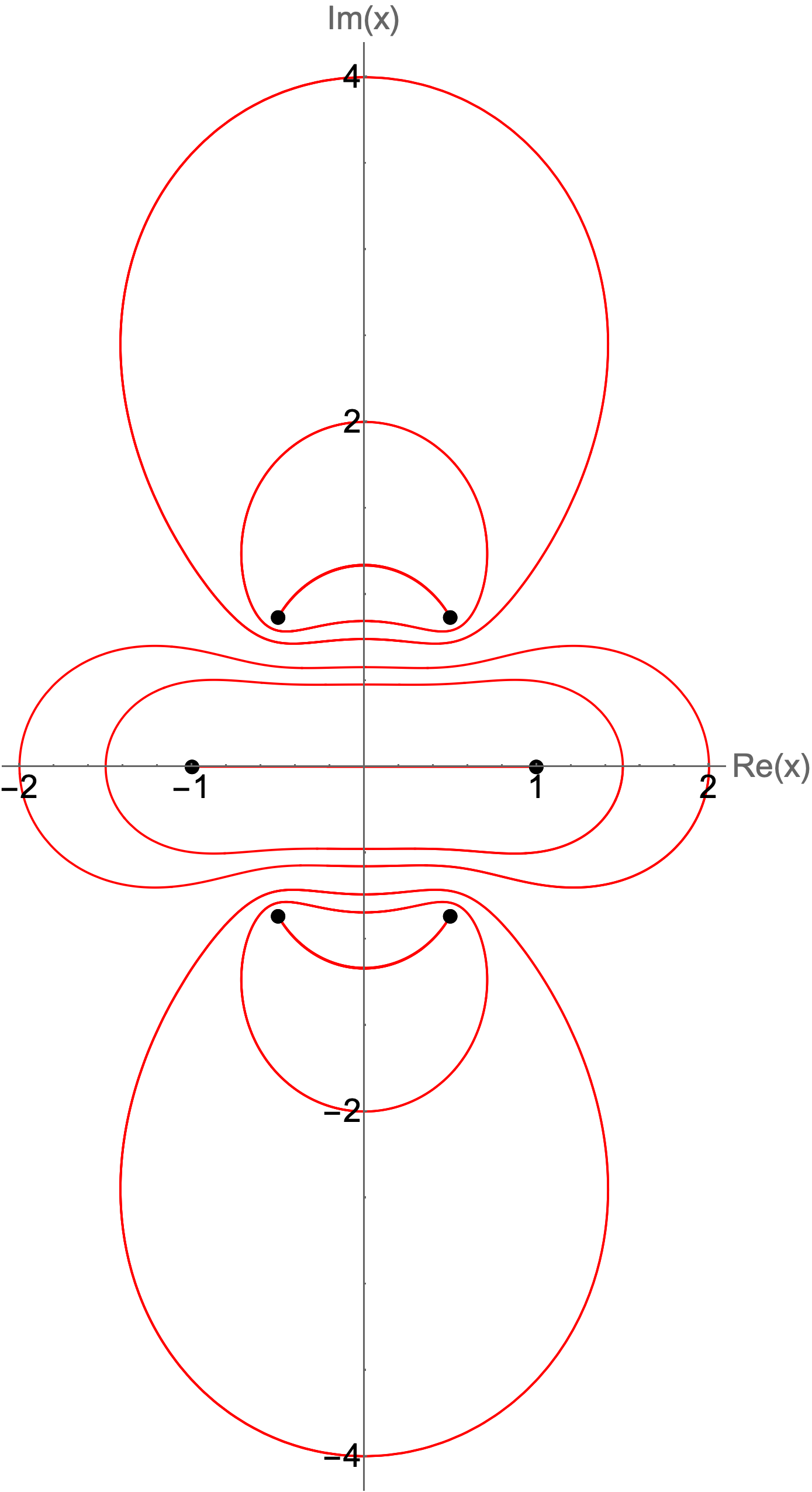}
\caption{Classical paths in the complex-$x$ plane for the sextic Hamiltonian $H=p^2+x^6$ for which the energy $E=1$. These paths are all closed and periodic. The classical turning points are indicated by black dots. The central group of paths is associated with the classical turning points at $\pm1$ in the yellow Stokes sectors in Fig.~\ref{f8}. All of these paths enclose the turning points at $\pm1$ except for the special path that oscillates between them; these paths all have the same period $P$, which is given in (\ref{e3.27}). The upper and lower groups of paths are associated with the classical turning points in the gray and blue Stokes sectors in Fig.~\ref{f8}. These paths all have the same period, which is given in (\ref{e3.28}), but this period is half as long as the period of the paths in the central region because the particles are moving faster. This is consistent with the quantized version of this Hamiltonian because at the quantum level the corresponding energies are higher. This figure emphasizes an important feature of $\cPT$-symmetric quantum theory: In the classical analog of the quantum theory the particles move through the complex plane. Thus, unlike the Hamiltonian operator, the operator $x(t)$, which represents the instantaneous position of a quantum particle, is not a quantum observable (see Subsec.~\ref{ss3F}).}
\label{f10}
\end{figure}

Figure~\ref{f10} displays three topological classes of paths for $E=1$. These positive-energy classical paths are all closed and periodic. Each class of paths is associated with one of the three pairs of Stokes sectors in Fig.~\ref{f8} for which the quantum theory has positive spectra and thus each class is associated with a pair of turning points.

The classical orbits associated with the yellow sectors all have the same period
\begin{equation}
P_1=\frac{\Gamma(1/6)\sqrt{\pi}}{3\Gamma(2/3)}.
\label{e3.27}
\end{equation}
However, the periods of the classical orbits associated with the gray-gray and blue-blue sectors are half as long:
\begin{equation}
P_2=\frac{\Gamma(1/6)\sqrt{\pi}}{6\Gamma(2/3)}.
\label{e3.28}
\end{equation}
Thus, these classical particles move faster than the particles whose orbits have period $P_1$. This is not surprising because in the associated quantum theories the energy eigenvalues (\ref{e3.5}) of the gray-gray and blue-blue sectors are larger than the corresponding energy eigenvalues (\ref{e3.4}) of the yellow-yellow sectors.

\begin{figure}
\center
\includegraphics[scale=0.23]{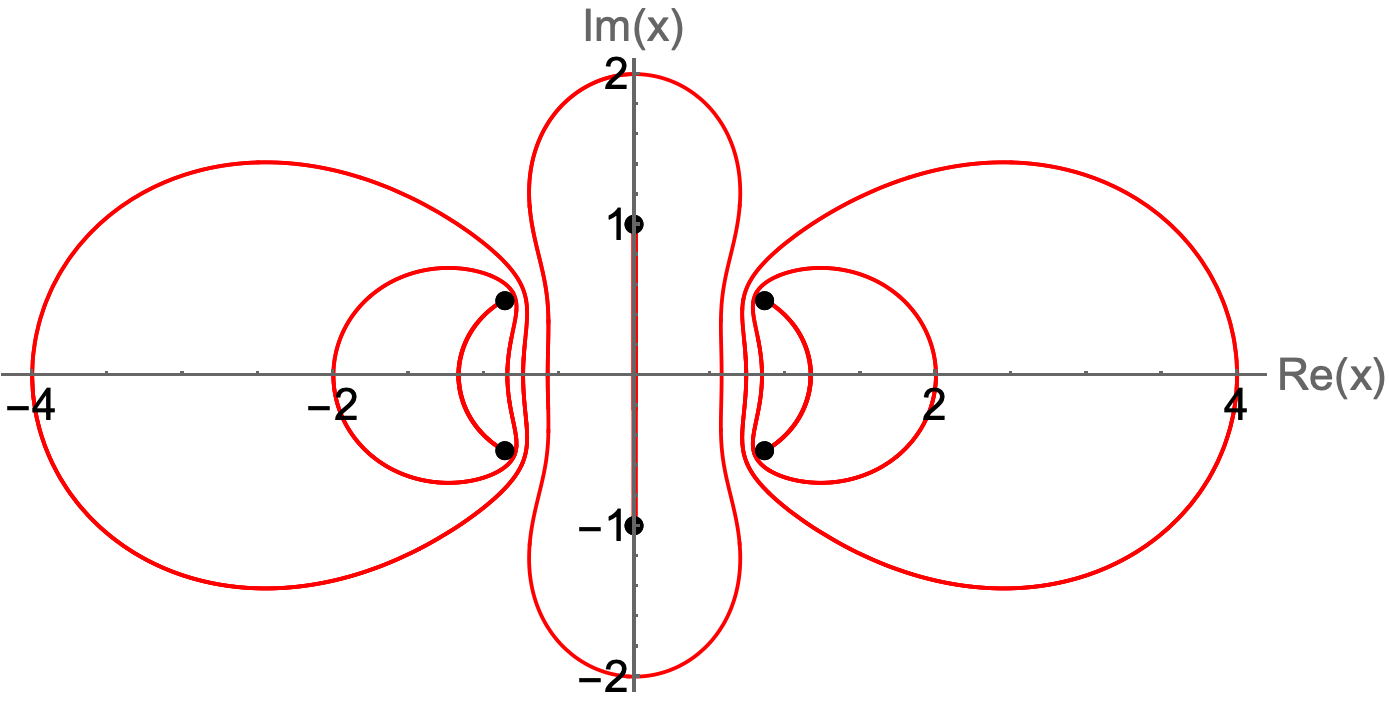}
\caption{Classical paths in the complex-$x$ plane for the sextic Hamiltonian $H=p^2+x^6$ for the energy $E=-1$. The black dots show the locations of the classical turning points. Like the paths in Fig.~\ref{f10}, there are three families of closed periodic orbits; the periods of these orbits are the same as those of the corresponding orbits in Fig.~\ref{f10}. The paths shown in this figure begin at $i$ (connecting the turning points) and $2i$ in the central region and at $\pm1.167$ (connecting the turning points) and $\pm2$ and $\pm4$ in the left and right regions.}
\label{f11}
\end{figure}

If the classical energy is negative, there are still three families of closed paths, but the orientation of these paths is rotated by $90^\circ$ compared to those in Fig.~\ref{f10}. The paths for $E=-1$ are shown in Fig.~\ref{f11}. These paths correspond to the Stokes-sector pairs in Fig.~\ref{f8} that are colored gray-blue, red-red, and blue-gray and for which the quantum energy levels are negative \cite{pt489}.

The classical-particle trajectories shown in Figs.~\ref{f10} and \ref{f11} fall into three different families in which each orbit associated with a distinct initial condition. The three families of orbits together cover the entire complex plane with limiting separatrix curves at the edges of the regions. A classical path that originates in one of the three regions never crosses into another region. 

The six regions of particle trajectories displayed in Figs.~\ref{f10} and \ref{f11} correspond to the six {\it real} quantum spectra that one obtains by imposing vanishing boundary conditions in the six pairs of Stokes sectors (three horizontal and three vertical) in Fig.~\ref{f8}. However, there are 14 more quantum-mechanical eigenvalue problems corresponding to other nonadjacent pairs of Stokes sectors. All of these eigenvalue problems have {\it complex energies}, which suggests that the quantum states corresponding to these eigenvalue problems grow or decay in time and tunnel to different regions of the complex domain.

\begin{figure*}
\center
\includegraphics[scale = 0.50]{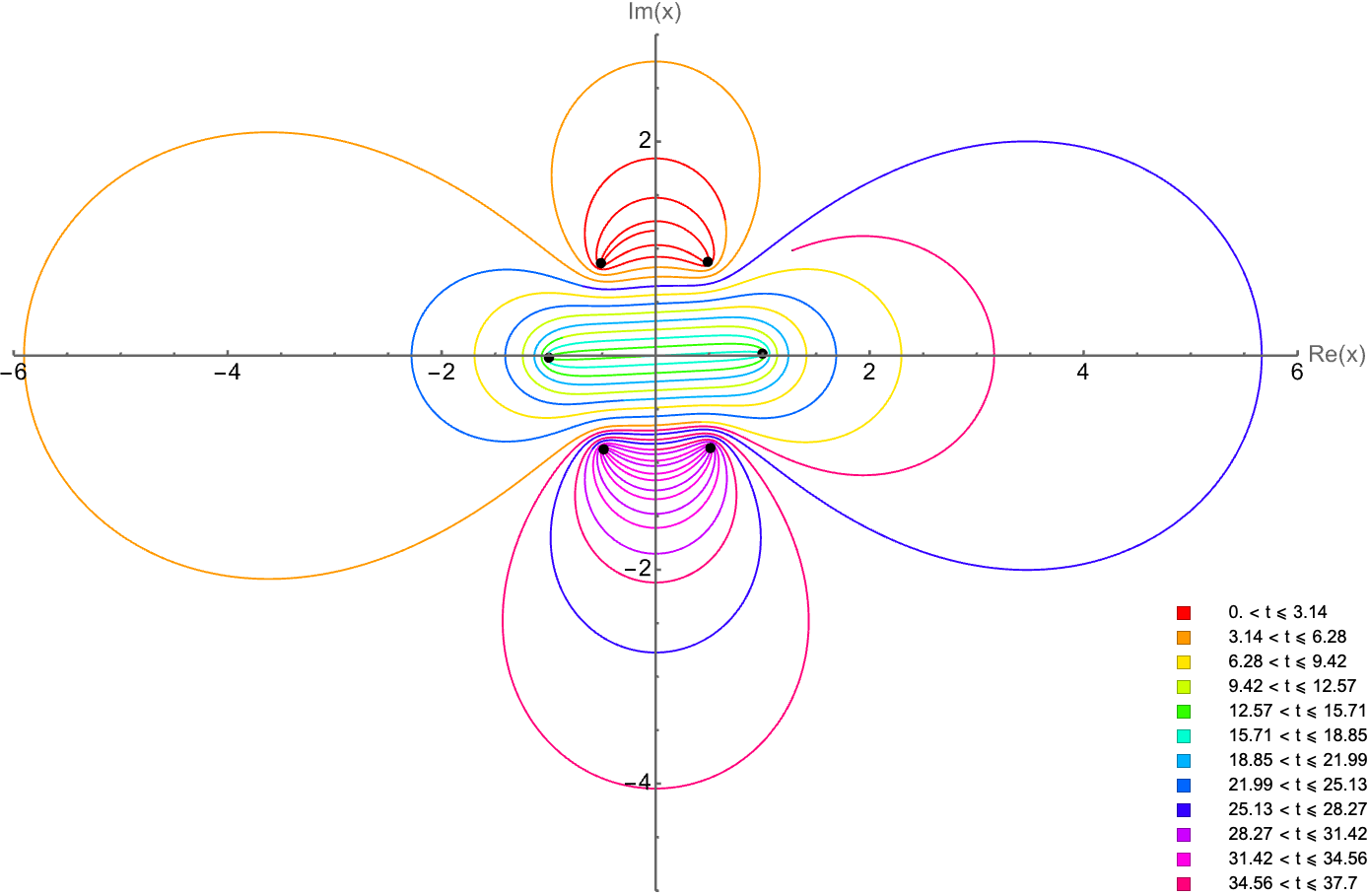}
\caption{Complex-energy version of Fig.~\ref{f10}. If the energy of the classical particle is complex, the particle is no longer confined to one of the three regions in Fig.~\ref{f10}. In this figure the energy of the particle is taken to be $E=1+0.2\,i$. As a result, the pairs of turning points are no longer horizontal (left-right symmetric). The upper pair of turning points is located at $-0.514743+0.858312\,i$ and $0.485949+0.874936\,i$, the middle pair is located at $\pm(1.00069+0.0166245\,i)$, and the lower pair is located at $-0.485949-0.874936\,i$ and $0.514743-0.858312\,i$. The initial position of the particle is $x(0)=1.167\,i$. The particle exhibits deterministic random behavior as it tunnels from region to region. The trajectory is uniquely determined by its initial position and the path of the particle never crosses itself. Each time the particle enters a region, it spirals inward and then outward, and then enters an adjacent region in a deterministic random fashion. The trajectory of the particle is plotted from $t=0$ until $t=37.70$. The color scheme below the plot enables one to trace the path of the particle from region to region. The particle in this plot goes from the upper to the middle to the lower and back to the middle region. The detailed motion of this particle up to the longer time $t=250$ is shown in Fig.~\ref{f13}.}
\label{f12}
\end{figure*}

The {\it classical} analog of this quantum behavior is shown in Fig.~\ref{f12}, which is a complex-energy version of Fig.~\ref{f10}. For this figure we choose a complex energy $E=1+0.2\,i$ that has a relatively small imaginary part and we plot the trajectory of a classical particle that begins on the positive imaginary axis at $x(0)=1.167\,i$. This trajectory remains in the upper region for several cycles but eventually spirals outward and enters the middle region. It then spirals inward as it encircles the central pair of turning points. The particle then goes {\it between} the two turning points and spirals outward. At this point the particle enters another region, spirals inward and outward again, and continues in this manner to enter and leave one region after another. The path of this classical motion is uniquely determined by the initial condition $x(0)$ and this trajectory does not cross itself. The plot in Fig.~\ref{f12} suggests that the path is ergodic and space-filling.

\begin{figure*}
\center
\includegraphics[scale = 0.57]{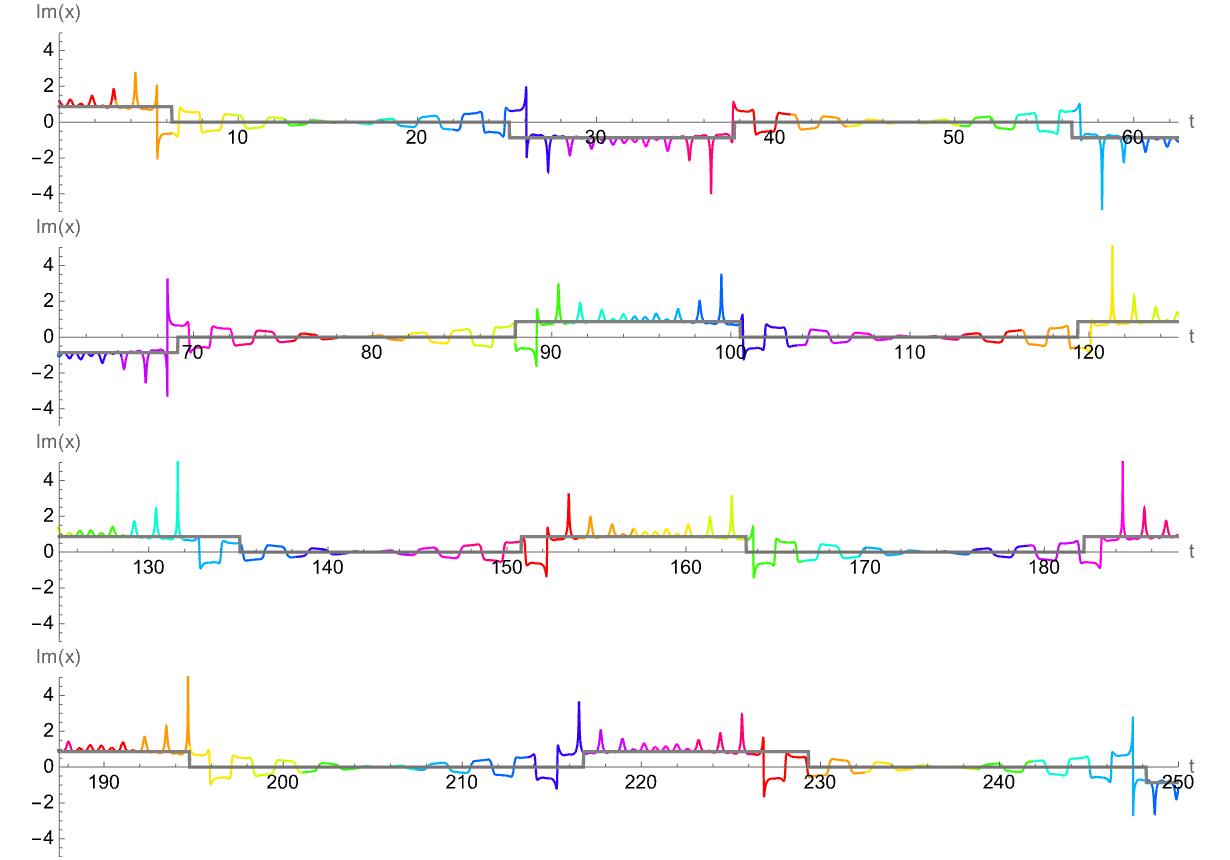}
\caption{A plot of the classical particle in Fig.~\ref{f12} for $t=0$ until $t=250$. The path cycles through the color scheme that is used in Fig.~\ref{f12}. To provide additional assistance with tracking the complex motion of the particle, only the imaginary part of the position of the particle is plotted. Also, the horizontal gray lines are provided to identify the region that the particle occupies. The height of the gray lines is the average of the imaginary parts of the turning points in each region. Note that the particle exhibits deterministic random behavior as it goes from region to region; when it leaves the middle region, it makes a deterministic random choice of whether to enter the upper or the lower region. If we label the upper, middle, and lower regions by U, M, and L, we see that as $t$ ranges from 0 to 250, the particle is in 17 regions: U, M, L, M, L, M, U, M, U, M, U, M, U, M, U, M, L, $...\,.$ The behavior pictured this figure can be interpreted as being a classical analog of the time-energy quantum uncertainty principle. If we increase (or decrease) the imaginary part of the energy ${\rm Im}\,E$ of the particle, the average time $T$ spent in each of the three regions decreases (or increases) in such a way that the product $({\rm Im}\, E)T$ remains constant.}
\label{f13}
\end{figure*}

The path plotted in Fig.~\ref{f12} exhibits {\it deterministic random behavior} as it travels from region to region. To verify this, we plot in Fig.~\ref{f13} the path of the particle from $t=0$ to $t=250$. The horizontal gray lines indicate the region the particle visits as a function of time $t$. As is typical of chaotic behavior, this pattern is exquisitely sensitive to the choice of initial condition: For an arbitrarily small change in $x(0)$ the pattern eventually becomes completely different. 

\subsection{Measuring the position of a particle}\label{ss3F}
The central region in Fig.~\ref{f10} contains a continuously infinite number of classical paths that are all up-down and left-right symmetric. At the quantum level, the expectation value of the $x$ operator is the functional average over all classical paths, so by symmetry we know immediately that this expectation value is 0. This result is correct: The particle spends as much time to the right of the origin as to the left of the origin and as much time above the real axis as below the real axis, so if we could perform many experiments in which we measure the instantaneous complex position of the particle and then average these measurements, we would find that $\langle x\rangle=0$. Indeed, for the Hermitian $x^6$ theory the expectation value of the $x$ operator is 0.

However, if we were to average the $\cPT$-symmetric paths in the upper-half or lower-half plane in Fig.~\ref{f10}, we would find that the expectation value of the $x$ operator for these $\cPT$-symmetric theories is a positive-imaginary or a negative-imaginary number. How can the expectation value of an operator be imaginary? The short answer to this question is simply that in $\cPT$-symmetric quantum theory the {\it position operator is not an observable}. Indeed, when we extend a Hermitian quantum theory to a non-Hermitian $\cPT$-symmetric quantum theory, particle trajectories are no longer confined to the real-$x$ axis and particles travel through the complex-$x$ plane. Since the outcome of a position measurement in nonrelativistic quantum mechanics should be a real number, we can no longer say that the $x$ operator is an observable.

This statement should not be surprising because in general the position of a quantum particle is not an observable. In quantum electrodynamics, for example, the expectation value of the electron field is not measurable because the result would be complex. We can, in principle, measure the instantaneous position of the {\it energy density} or the {\it charge density} of the electron even though we cannot measure the actual position of the electron.

In general, in relativistic particle physics the location of a particle cannot be an observable because if we attempt to trap the particle in an arbitrarily small box, the uncertainty principle implies that the momentum of the particle, and thus the energy of the particle, would be arbitrarily large. The particle would therefore pair-produce and not be in the box at all! The center of energy of a system, such as an atom, is measurable, but not the positions of the components of the system.

Even in elementary nonrelativistic Hermitian quantum mechanics, when a particle tunnels it is a {\it virtual particle} and its instantaneous position or its tunneling trajectory cannot be observed or measured. One can view the process of tunneling as one in which the particle is following a complex trajectory \cite{pt513,pt591,pt585}. Only in elementary quantum mechanics, which is a nonrelativistic approximation to relativistic quantum mechanics, can the $x$ operator be treated as an observable. (We examine these issues further in Sec.\ref{s6} when we discuss the $\cPT$-symmetric inner product and expectation values.)

Nevertheless, in $\cPT$-symmetric quantum mechanics there are other measurable observables, the most important being the Hamiltonian operator. In any physical theory the ability to make an energy measurement is a fundamental requirement. In Hermitian quantum mechanics when we try to observe the position of a particle, we are actually performing an energy measurement. Energy absorbed and emitted from the particle {\it if the particle is on the real axis} enters a detector and is transferred to a film (or a nerve cell in a retina). We then infer from this energy transfer where the particle is located. In any viable physical theory, whether or not it is possible to measure the position of a particle, it is crucial that we can still perform an energy measurement, and the outcome of such a measurement must be a real number. (Observables are discussed further in Subsec.~\ref{ss6H}.)

 Because energy is a measurable observable, one may wonder how an {\it upside-down} potential such as $-x^4$ can have a positive-energy spectrum (see Fig.~\ref{f4} for $\vep=2$, for example). One might think that an upside-down potential would be unstable and could not bind positive-energy bound states. We will see in Sec.~\ref{s4} that an upside-down potential can indeed have stable positive-energy states. This is because extending the real numbers to the complex numbers gives up an important property of the real numbers, namely, the {\it ordering principle}. That is, if $a$ and $b$ are complex numbers, inequalities such as $a>b$ or $a<b$ no longer make sense. When we say that a potential is unstable because it is unbounded below, we are implicitly using a {\it real} inequality that, as we emphasize in Sec.~\ref{s4}, no longer applies in the context of $\cPT$-symmetric quantum theory!

\subsection{$x^{10}$ potential}\label{ss3G}
This section has focused on Hamiltonians of the form (\ref{e3.8}) whose right-side-up potentials $x^{2+4n}$ ($n\geq0$ integer) are real and positive. In the complex-$x$ domain these Hamiltonians all possess two independent symmetries, left-right symmetry ($\cPT$ symmetry) and up-down symmetry (complex-conjugation symmetry). For $n=0$ (the quantum harmonic oscillator) there is one positive real (Hermitian) spectrum; for $n=1$ (the sextic oscillator) there are two different positive real spectra, the Hermitian spectrum associated with the yellow-yellow pair of Stokes sectors and the $\cPT$-symmetric spectrum associated with the gray-gray and blue-blue pairs of Stokes sectors in Fig.~\ref{f8}. There is no positive spectrum associated with the red-red pair of sectors; this is because there are no turning points inside these sectors (see Fig.~\ref{f8}).

The pattern is clear: For a $x^{2+4n}$ potential there are $2n+1$ left-right symmetric Stokes sectors, but because of up-down (complex-conjugation) symmetry there are $n+1$ different real positive discrete spectra. The central pair of sectors gives the Hermitian energy levels, which are the smallest levels of the multiple-eigenvalue spectrum. Moving downward or upward, the energy levels associated with each successive pair of left-right sectors continue to be real and become correspondingly larger.

\begin{figure}
\center
\includegraphics[scale = 0.42]{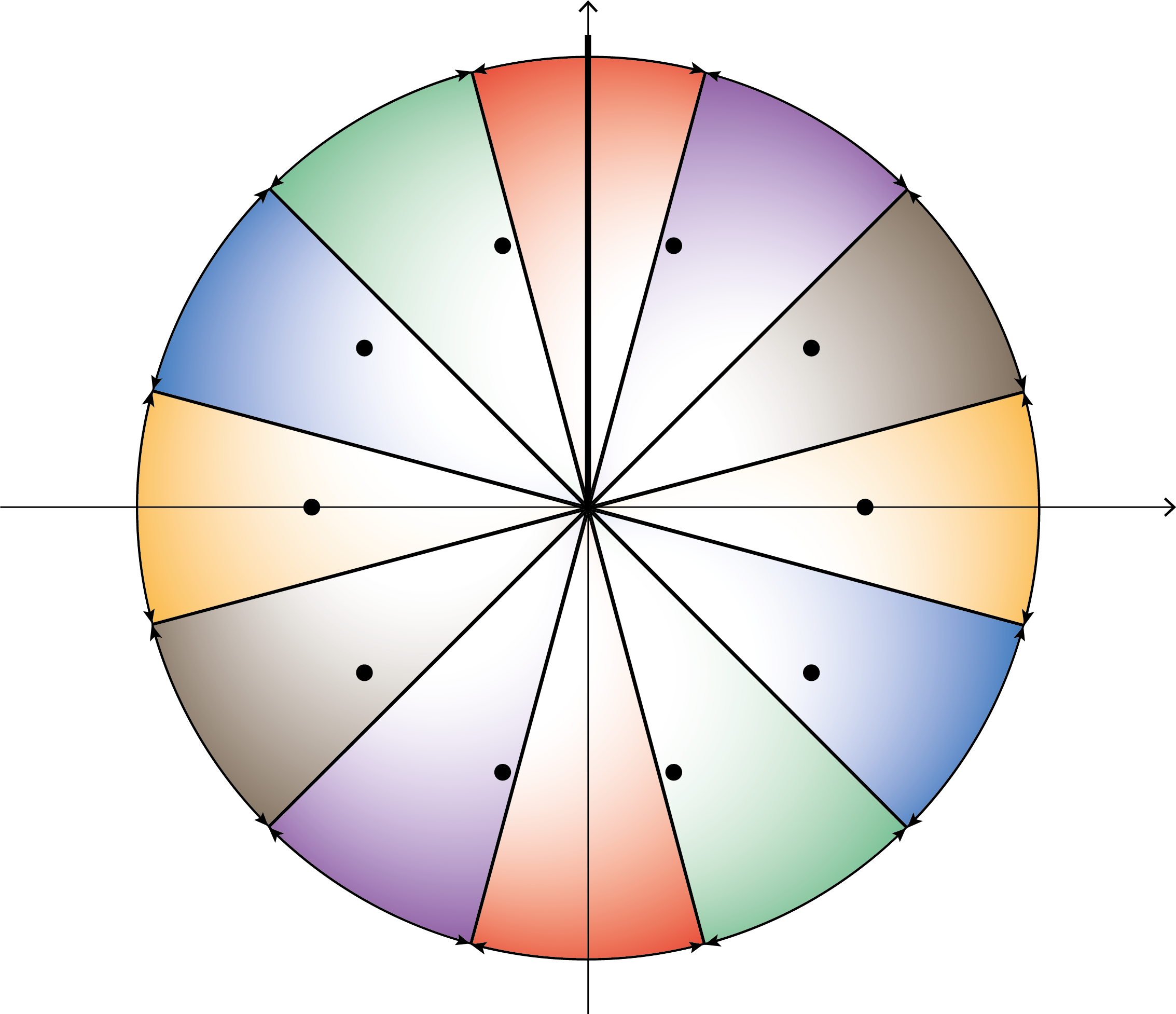}
\caption{Stokes sectors for the $x^{10}$ theory. There are six pairs of Stokes sectors, each having an opening angle of $\tfrac{\pi}{6}$. There are also ten classical turning points (dots) separated by angles of $\tfrac{\pi}{5}$. Each left-right-symmetric ($\cPT$-symmetric) pair of turning points is associated with a positive discrete quantum-mechanical spectrum. There are ten turning points (black dots) and twelve Stokes sectors, and as a consequence, the red up-down-symmetric pair of Stokes sectors does not contain any turning points. These sectors are not associated with a positive spectrum. This figure is the analog of Fig.~\ref{f8} for the $x^6$ theory.}
\label{f14}
\end{figure}

To illustrate this pattern, we consider briefly the case $n=2$, which corresponds to the $x^{10}$ potential. There are now five horizontal pairs of Stokes sectors in the complex-$x$ plane for the Schr\"odinger equation $$-y''(x)+x^{10}y(x)=Ey(x)$$ and each sector has an opening angle of $\tfrac{\pi}{6}$ (see Fig.~\ref{f14}). Because of up-down symmetry there are now three distinct positive-energy spectra. There are always two fewer turning points than Stokes sectors, and as with Fig.~\ref{f8}, the red up-down pair of Stokes sectors is not associated with a positive spectrum because there are no turning points in this sector.

As we saw in the sextic case, the WKB quantization condition (\ref{e3.20}) gives good approximations for each of the three eigenspectra. There are five pairs of turning points, which are the roots of the equation $x^{10}=E$. The central (Hermitian) turning points lie on the real axis at
$$x_\pm=\pm E^{1/10}$$
and the next two $\cPT$-symmetric pairs of turning points are situated at 
$$x_\pm=e^{-4i\pi/5}E^{1/10}~~{\rm and}~~e^{-i\pi/5}E^{1/10}$$
and at
$$x_\pm=e^{-3i\pi/5}E^{1/10}~~{\rm and}~~e^{-2i\pi/5}E^{1/10}.$$ 

The semiclassical WKB approximation to the Hermitian energy levels is $$E_n\sim\left[\frac{(6n+3)\sqrt{\pi}\,\Gamma(3/5)}{\Gamma(1/10)}\right]^{5/3}
\quad(n\to\infty).$$ From this formula we can read off approximations to the first five Hermitian eigenvalues:
\begin{eqnarray}
&&E_0=0.737...,~~E_1=4.596...,~~E_2=10.768...,\nonumber\\
&&\quad E_3=18.866...,~~E_4=28.680...\,.
\label{e3.29}
\end{eqnarray}

The middle pair of turning points leads to the WKB approximation to the $\cPT$-symmetric energy spectrum $$E_n\sim\left[\frac{(6n+3)\sqrt{\pi}\,\Gamma(3/5)}{\cos(\pi/5)\Gamma(1/10)}
\right]^{5/3}\quad(n\to\infty)$$ for which the first five eigenvalues are approximately
\begin{eqnarray}
&&E_0=1.049...,~~E_1=6.543...,~~E_2=15.330...,\nonumber\\
&&\quad E_3=26.858...,~~E_4=40.831...\,.
\label{e3.30}
\end{eqnarray}
These eigenvalues are larger than the corresponding Hermitian eigenvalues in
(\ref{e3.29}) by a factor of $1.424$.

The lower pair of turning points leads to the WKB approximation
$$E_n\sim\left[\frac{(6n+3)\sqrt{\pi}\,\Gamma(3/5)}{\cos(2\pi/5)
\Gamma(1/10)}\right]^{5/3}\quad(n\to\infty).$$ These $\cPT$-symmetric eigenvalues are larger than the corresponding
eigenvalues in (\ref{e3.30}) by a factor of $7.780$:
\begin{eqnarray}
&&E_0=5.214...,~~E_1=32.539...,~~E_2=76.235...,\nonumber\\
&&\quad E_3=133.568...,~~E_4=203.053...\,.
\label{e3.31}
\end{eqnarray}

We obtain the spectrum in (\ref{e3.30}) by analytically continuing the Hamiltonian $H=p^2+ x^6(ix)^\vep$, whose eigenvalues are shown in Fig.~\ref{f6} (right panel), from the Hermitian case ($\vep=0$) to the non-Hermitian $\cPT$-symmetric case ($\vep=4$). The $\cPT$-symmetric spectrum (\ref{e3.31}) is obtained by analytically continuing the Hamiltonian (\ref{e2.14}) in $\vep$, as shown in Fig.~\ref{f9}, from the Hermitian case ($\vep=0$) to the non-Hermitian $\cPT$-symmetric case ($\vep=8$).

The appearance of three real spectra suggests a plausible particle-physics interpretation. As of now, there are three known families of leptons (the electron, the muon, and the tau), which all have very similar physical properties except for one obvious feature, namely, that the masses of each successive family are greater. It is interesting that by merely extending a Hermitian theory into the complex domain, we obtain similarly structured families \cite{pt369}. Furthermore, as we saw in Subsec.~\ref{ss3F}, if the energies of the states are slightly complex, the particles become unstable and can decay into one another.

\begin{figure}
\center
\includegraphics[scale = 0.23]{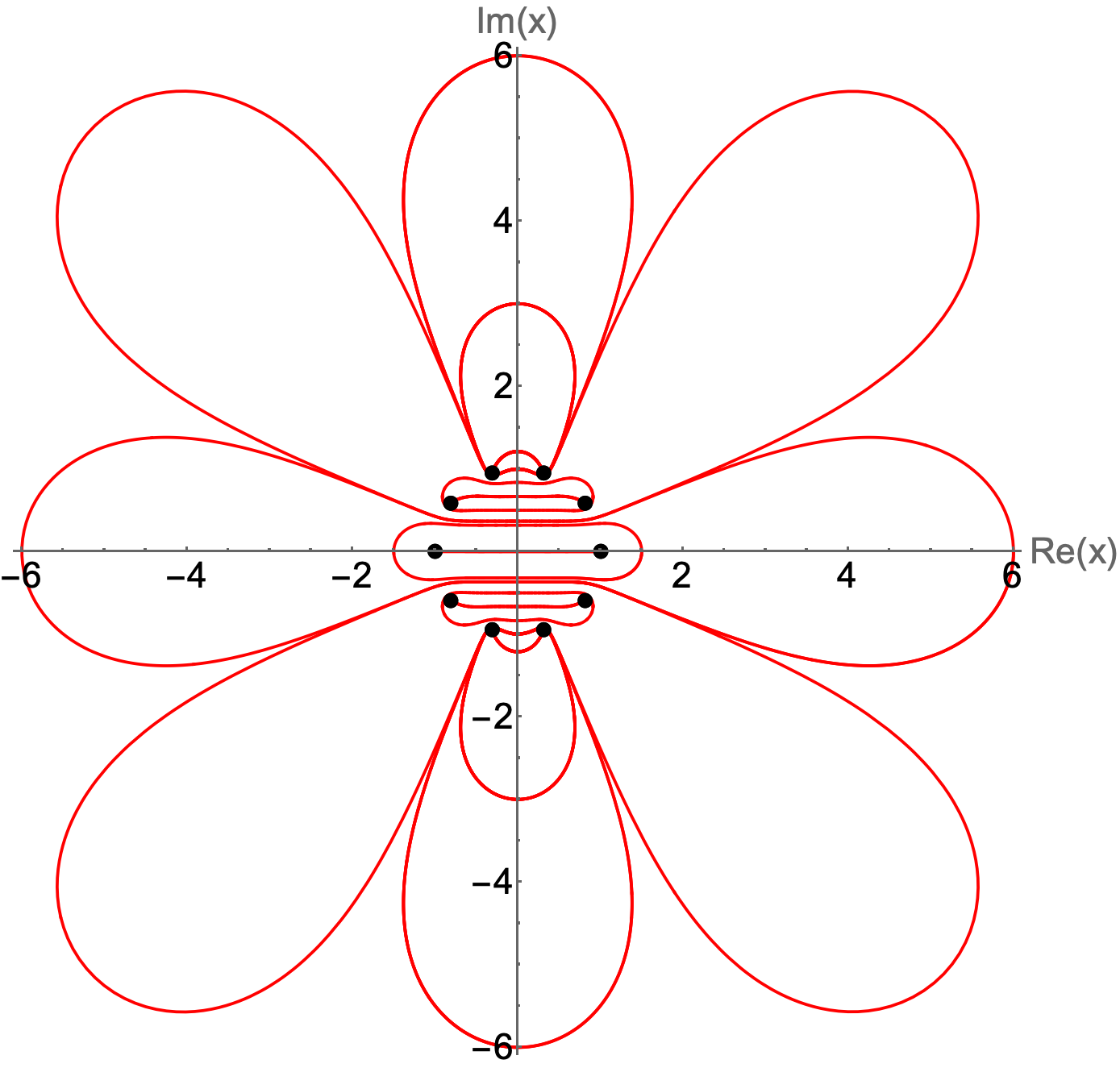}
\caption{Complex classical trajectories for the Hamiltonian $H=p^2+x^{10}$ corresponding to the energy $E=1$. There are five regions of periodic trajectories and all trajectories in each region have the same period. Three trajectories are shown in each region, one trajectory joining the turning points, which represents a particle that oscillates between the turning points, and two others, which represent a particle that encircles the turning points. The periods in the central region are the longest and those in the outer regions are shorter. The regions are divided by separatrices that run off to $\infty$. These regions along with the separatrices cover the entire complex plane. The black dots indicate the locations of the classical turning points. The classical paths shown in this figure begin at $\pm 6\,i$, $\pm 3\,i$, $\pm 1.21\,i$, $\pm 0.368\,i$, $\pm 0.5\,i$, $\pm 0.665\,i$, 3/2, and 1.}
\label{f15}
\end{figure}

The five real quantum spectra of the $x^{10}$ potential, one Hermitian spectrum and two identical up-down symmetric pairs of spectra, correspond to the five left-right-symmetric pairs of turning points in Fig.~\ref{f14}. These spectra have clear classical analogs. Figure~\ref{f15} shows all classical trajectories corresponding to a real positive energy $E=1$. These trajectories are all closed and periodic and they fill five distinct regions in the complex plane. There are two left-right pairs of regions that are symmetric under complex conjugation (up-down reflection) and which correspond to $\cPT$-symmetric theories, and a central region, which corresponds to the Hermitian quantum theory. The trajectories in each region all have the same periods. The trajectories in the central region have the longest periods and the trajectories in the upper and lower regions have shorter periods. These five regions along with the separatrices that divide the regions cover the entire complex plane.

\begin{figure}
\center
\includegraphics[scale = 0.23]{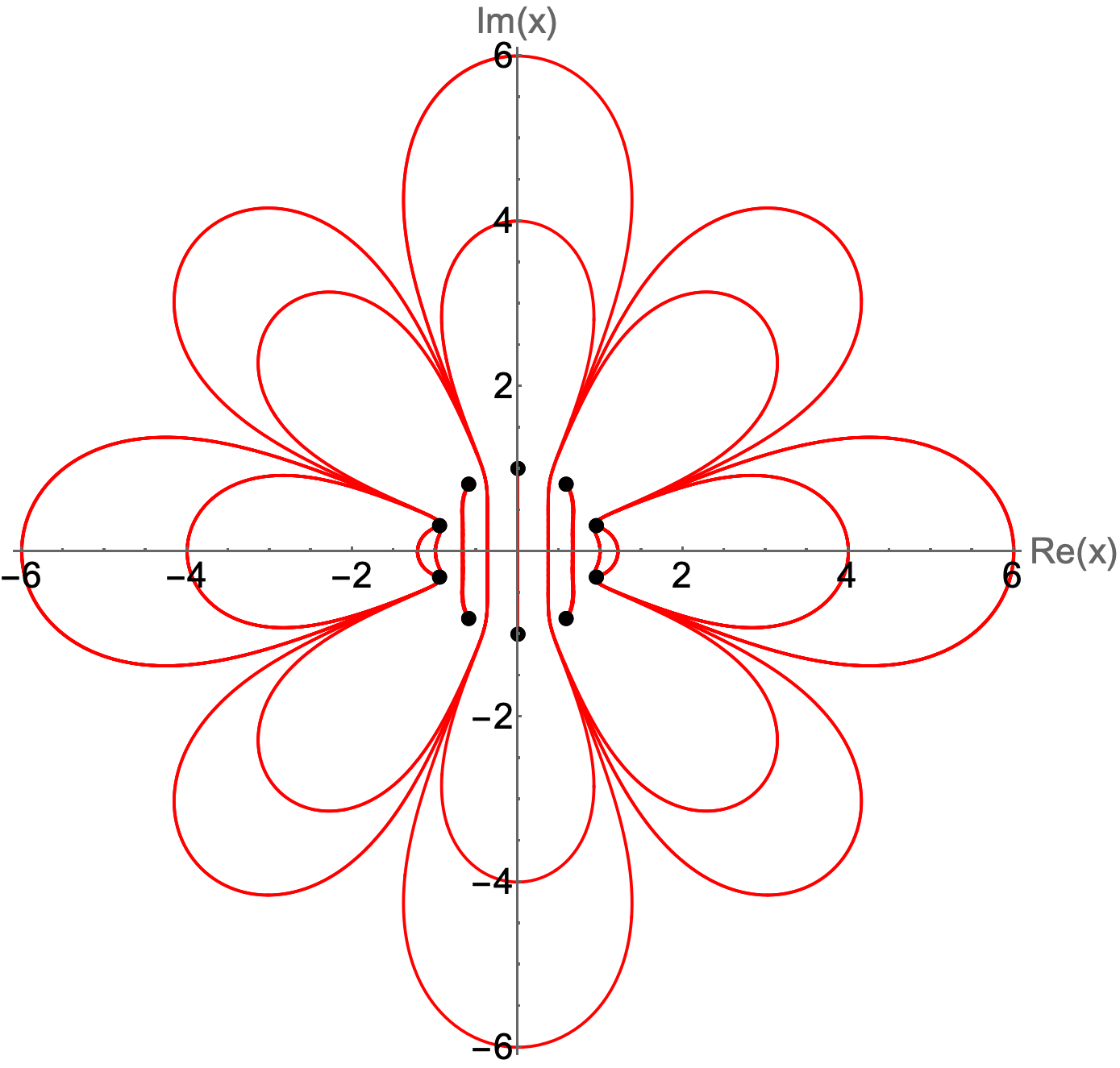}
\caption{Complex classical trajectories for the $x^{10}$ theory
with $E=-1$. This picture is identical to that in Fig.~\ref{f15} except that it is rotated by $90^\circ$. The paths shown in this figure begin at $i$, $\pm4\, i$, $\pm6\,i$, $\pm6$, $\pm4$, $\pm0.66$, $\pm0.3682$, $\pm0.3688$, and $\pm1.21$.}
\label{f16}
\end{figure}

Replacing $x$ by $ix$ in the Schr\"odinger equation $-y''(x)+x^{10}y(x)=Ey(x)$ corresponds to changing the sign of $E$. Correspondingly, at the classical level, if we plot the classical trajectories for $E=-1$, the picture in Fig.~\ref{f15} becomes rotated by $90^\circ$ (see Fig.~\ref{f16}).

In summary, there are 54 nonadjacent pairs of Stokes sectors in Fig.~\ref{f14} in which we can formulate eigenvalue problems. Five pairs of sectors lead to positive real spectra, of which three are distinct, and these are associated with the five regions of classical orbits plotted in Fig.~\ref{f15}. Five more pairs of sectors lead to negative real spectra, of which three are distinct, and these are associated with the five regions of classical orbits shown in Fig.~\ref{f16}. The 44 remaining pairs of Stokes sectors are associated with complex eigenvalues \cite{pt489}.

\section{Upside-down potentials}\label{s4}
This section examines Hamiltonians whose potentials have the general form
$$V(x)=x^{4n}\quad(n=1,\,2,\,3,\,...).$$
Like the potentials discussed in Sec.~\ref{s3}, these potentials are real if $x$ is real and they have an infinite number of positive-energy $\cPT$-symmetric bound states (see Figs.~\ref{f3} and \ref{f4}). We obtain these upside-down potentials by setting $\vep=2,\,6,\,10,\,...$ in (\ref{e2.14}). These potentials are profoundly different from the right-side-up $x^2$, $x^6$, and $x^{10}$ potentials discussed in Sec.~\ref{s3} because if the theories associated with these upside-down potentials are treated as Hermitian quantum theories, they do not have real energy levels. However, as $\cPT$-symmetric theories, their energy levels are all real, positive, and discrete.

The conventional (real-axis-only) treatment of upside-down potentials precludes the possibility that such potentials could have bound states. Rather, upside-down potentials are normally used as one-dimensional models of tunneling. One might think that an upside-down potential cannot confine a bound state simply because at the classical level the potential exerts a real force on a classical particle that tends to push the particle {\it away} from the peak of the potential. However, this argument fails in the context of $\cPT$-symmetric quantum mechanics because in the complex plane the intuitive concept of a potential being unbounded below does not apply. As explained in
Subsec.~\ref{ss3F}, the complex number system is not {\it ordered}; one cannot say that a complex number is greater than or less than another complex number. The notion of ``unbounded below'' relies on the ordering principle, which only holds in the real number system.

This section examines the subtle question of how an upside-down potential can actually confine particles and possess bound states and we explain this both physically and mathematically. We begin with a detailed examination of the $-x^4$ potential.

\subsection{Classical particle in a quartic upside-down potential}\label{ss4A}
Setting $\vep=2$ in (\ref{e2.14}) leads to the quartic Hamiltonian
\begin{equation}
H=p^2-x^4.
\label{e4.1}
\end{equation}
This real $\cPT$-symmetric Hamiltonian is special because it is {\it independently} symmetric under space-reflection $\cP$ and time reversal $\cT$. As discussed above, treating $H$ conventionally, we would say that a particle with positive energy and subject to forces provided by the upside-down $-x^4$ potential cannot be in a stable bound state. If a classical particle with positive energy and subject to this potential is initially located at a point $x$, the particle feels a positive force if $x>0$ and a negative force if $x<0$. Regardless of its initial velocity, the particle is pushed away from the origin, and as it moves further from the origin, the potential steepens so the force on the particle gets stronger. Thus, the particle accelerates to infinity. But is it gone forever?

Let us use complex analysis to examine this deceptively simple problem. At the classical level it is easy to solve the equations of motion and to calculate the time for a classical particle under the influence of an upside-down quartic potential to escape to infinity.\footnote{Classical studies of $\cPT$-symmetric systems include \cite{pt173,pt431,pt434,pt429,pt430,pt432,pt22,pt23,pt65,pt516,pt486,pt489,pt490,pt513,pt514,pt174,pt593,pt650,pt662,pt676,pt678,pt679,pt680}} Hamilton's equations of motion for $H$ in (\ref{e4.1}) read
\begin{equation}
\frac{dx}{dt}=\frac{\partial H}{\partial p}=2p,~~\quad\frac{dp}{dt}=-\frac{\partial H}{\partial x}=4x^3.
\label{e4.2}
\end{equation}

The Hamiltonian $H$ is a constant of the motion and the initial energy of the particle is $E$. Following the classical analysis in Subsec.~\ref{ss3E}, we use the first equation of motion to eliminate $p$ and then write $H=E$ as a nonlinear first-order differential equation:
$$\frac{dx}{dt}=2\sqrt{E+x^4},$$
where we have taken a square root and used the initial conditions $x(0)=0$ and $p(0)=\sqrt{E}$. This differential equation is {\it separable}, and thus the time $T=\int\!dt$ that it takes for the particle to reach $x=\infty$ is
\begin{equation}
T=\int_0^\infty\frac{dx}{2\sqrt{E+x^4}}.
\label{e4.3}
\end{equation}
We can evaluate (\ref{e4.3}) in terms of Gamma functions,
\begin{equation}
T=\frac{[\Gamma(1/4)]^2}{8\sqrt{\pi}}\,E^{-1/4},
\label{e4.4}
\end{equation}
but the point of this calculation is merely that this integral {\it exists}, so the particle reaches infinity in {\it finite} time.

If we repeat this calculation for a general upside-down $-|x|^N$ potential, we merely replace $x^4$ in (\ref{e4.3}) with $x^N$. Evidently, there is a {\it transition} at $N=2$, which corresponds to an upside-down harmonic (parabolic) potential. If $N>2$, an evaluation of the integral shows that the particle reaches infinity in finite time, but if $N\leq2$, the particle takes infinitely long to reach infinity. Thus, we have a peculiar version of the Zeno paradox: If $N\leq2$, the particle cannot reach infinity even though it is approaching infinity at an ever increasing speed, while if $N>2$, the particle gets to infinity in finite time $T$.

This result raises a fascinating question. If $N\leq2$, we know the location of the particle for all time; it is continuing forever on its journey to infinity. But if $N>2$, where is the particle when $t>T$? 

This is a question that is extremely difficult to answer using real analysis. There is a singularity on the real-$N$ axis, so we cannot keep $N$ real and examine what happens if we smoothly increase $N$ past 2. However, Fig.~\ref{f9} provides insight and suggests that we should use complex-variable theory to analyze this problem. This figure shows that at the quantum level the boundary conditions on the eigenvalue problem for $\vep=2$ are imposed in Stokes sectors whose upper edges lie on the real axes.

In general, the angular opening of a Stokes sector does {\it not} include the edges of the sector. In this case the opening angular ranges
$$-\tfrac{\pi}{3}<\theta_{\rm right}<0\,\quad{\rm and}\quad-\pi<\theta_{\rm left}<-\tfrac{2\pi}{3}$$
do not include the angles 0 and $-\pi$, so the Stokes sectors lie just below the real axis. Thus, at the quantum level one should not think of a particle as traveling to infinity exactly along the real axis; rather, the quantum particle travels infinitesimally {\it below} the real axis in the complex plane. We first study this behavior by understanding the classical dynamical motion in the complex plane and then approach the real axis as a limiting process. In effect, this analysis allows us to bypass and evade the singularity on the real axis at $N=2$. We then find that even at the {\it classical} level a complex upside-down quartic potential confines particles.

\begin{figure}
\center
\includegraphics[scale=0.242]{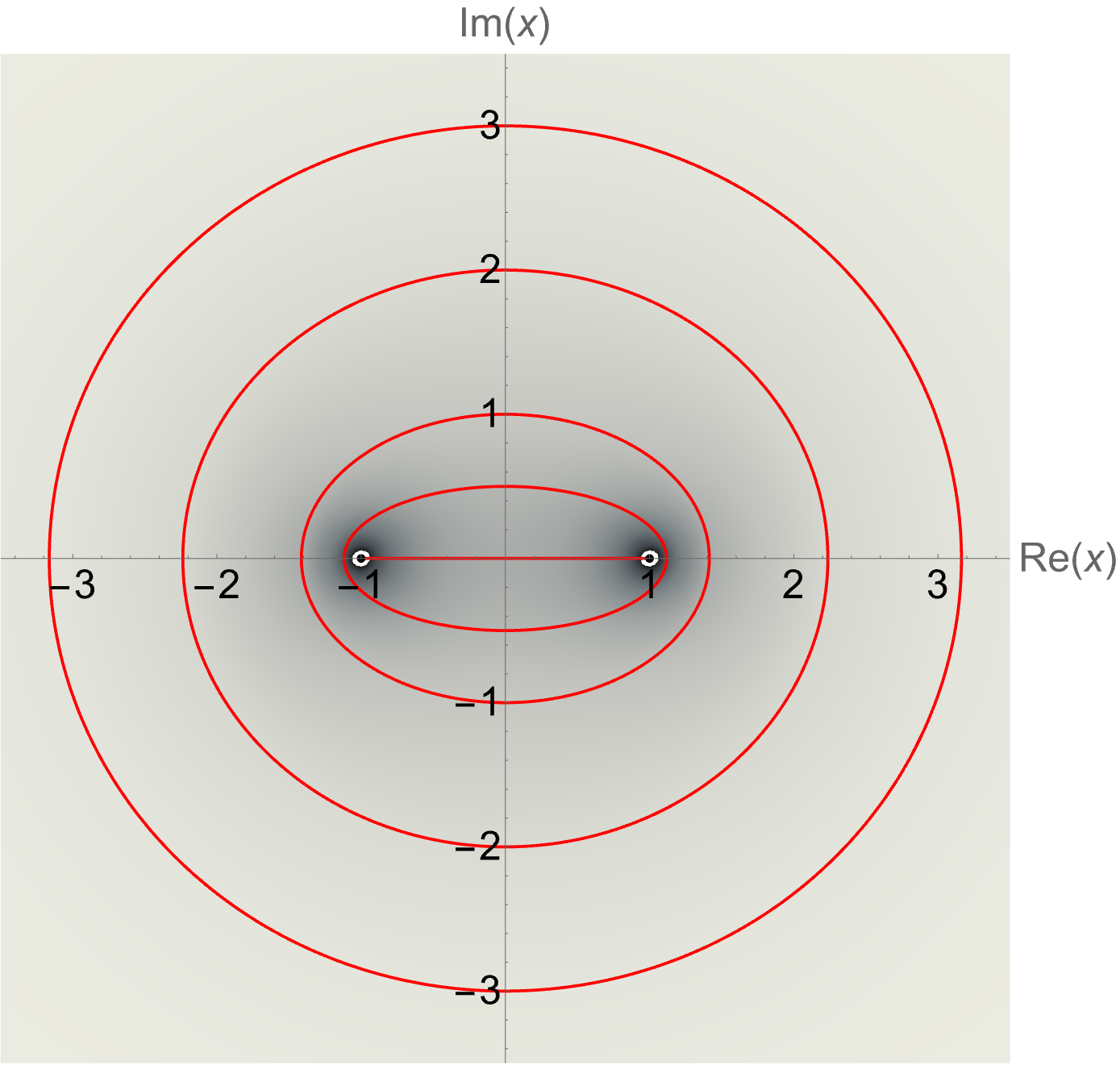}
\caption{Complex orbits for a classical particle in the harmonic-oscillator potential $x^2$. The particle has energy $E=1$. The possible paths of the particle are all ellipses whose foci are located at the classical turning points $x=\pm1$. Particles move slower in areas of heavier shading. (The classical turning points are indicated by dots, which are located at the centers of the shaded regions.) The main feature of these orbits is that because of Cauchy's integral theorem their periods are all the same. Therefore, particles in the outer orbits move faster than particles in the inner orbits. This is unlike the case of Keplerian gravitational orbits, which are also ellipses, because particles in the distant Keplerian orbits move more slowly than particles in the inner orbits. (A year on Pluto is much longer than an Earth year.)}
\label{f17}
\end{figure}

Let us examine the complex {\it classical} particle trajectories for the general $x^2(ix)^\vep$ potential in (\ref{e2.14}). At $\vep=0$ the complex trajectories for a classical particle of energy $E=1$ in the harmonic potential are pictured in Fig.~\ref{f17}. All but one of these trajectories are {\it ellipses} whose foci are the two turning points on the real axis at $x=\pm1$. The trajectory on the real axis between the turning points is a degenerate ellipse. The notable feature of this complex dynamical system is that the periods of all the ellipses are the same because of the Cauchy integral theorem. (Keplerian orbits are also ellipses, but particles traveling in outer orbits go faster than particles in inner orbits.)

\begin{figure}
\center
\includegraphics[scale=0.22]{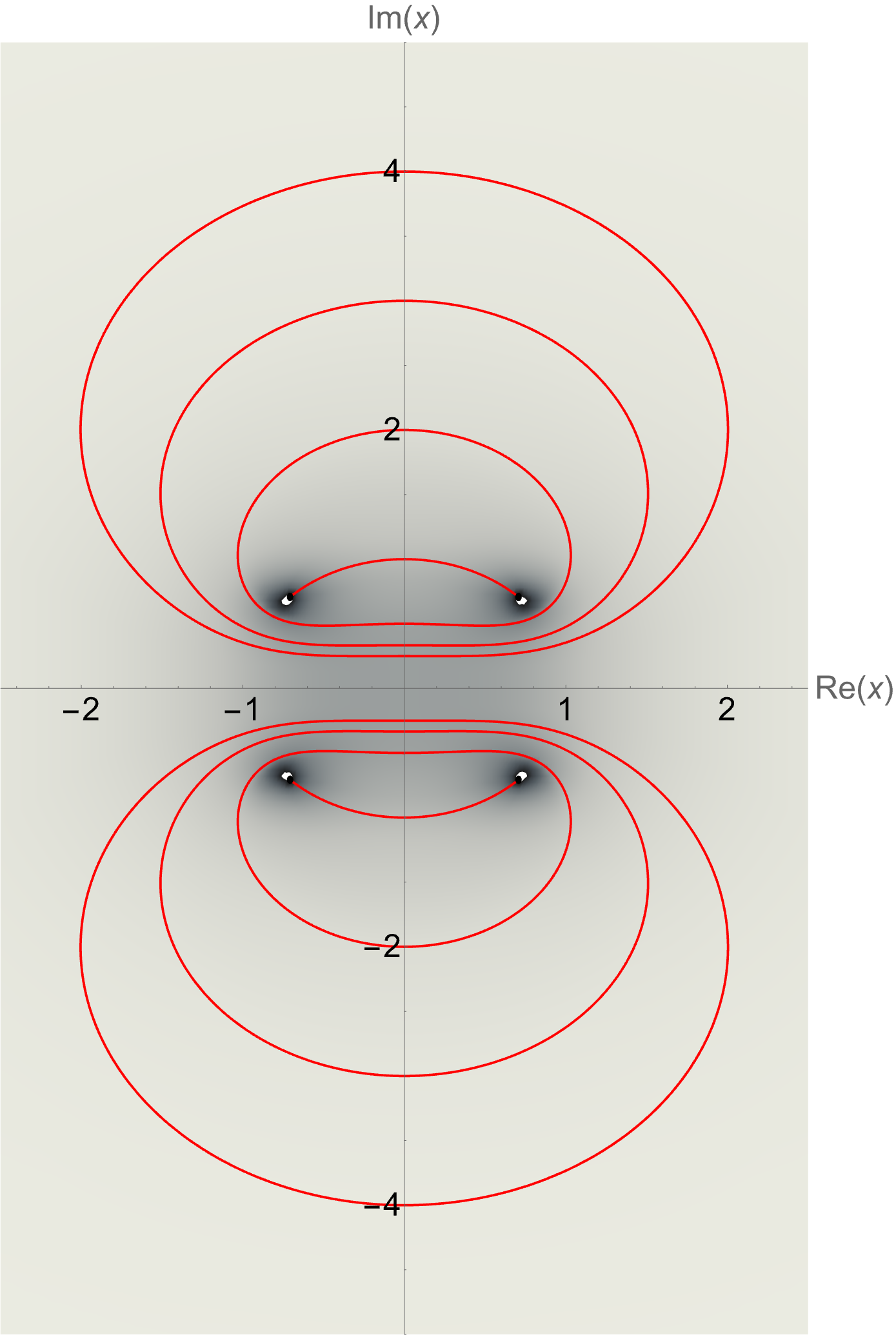}
\caption{Complex classical orbits for the upside-down quartic $-x^4$ potential. All orbits are closed and periodic and all orbits have the same period. As in Fig.~\ref{f17}, particles move slower in areas of heavier shading and the classical turning points are indicated by dots. The set of all orbits exhibits up-down and left-right symmetry and no orbit crosses the real axis. As the orbits get closer to the real axis, they get bigger and approach the shape of a $D$ lying on its side. In the limit as its orbit approaches the real axis, the classical particle zooms out to infinity along a path that is infinitesimally close to the real axis, travels at infinite speed around the rounded part of the $D$, and comes back up the real axis to its starting point, all in finite time. Since the particle moves slowly when it is near the origin and rapidly when it is far from the origin, at any given time the particle is most likely to be found very close to the origin. Thus, the classical particle is in a localized {\it dynamical} bound state centered about the origin.}
\label{f18}
\end{figure}

The situation becomes more complicated when $\vep=2$ because, as we already know from Fig.~\ref{f9}, there are four turning points rather than two (see Fig.~\ref{f18}). There are two turning points in the upper-half $x$-plane and two in the lower-half $x$-plane. The arrangement of these turning points is up-down symmetric because of complex-conjugation symmetry (reality of the Hamiltonian) and also left-right symmetric because of $\cPT$ symmetry. As in Fig.~\ref{f17}, all classical orbits are closed and periodic, and the set of orbits is both left-right and up-down symmetric. The classical orbits in the lower-half $x$-plane correspond to the Stokes sectors in the lower-half plane for the case $\vep=2$ in Fig.~\ref{f9}. Like the classical orbits in Fig.~\ref{f17}, the orbits in Fig.~\ref{f18} all have the same period $P$, which is twice the value of $T$ in (\ref{e4.4}):
\begin{equation}
P=\frac{[\Gamma(1/4)]^2}{4\sqrt{\pi}}\,E^{-1/4}.
\label{e4.5}
\end{equation}
Thus, as in Fig.~\ref{f17}, particles in the outer orbits go faster than particles in the inner orbits.

Figure~\ref{f18} shows that as the orbits increase in size, the lower parts of the upper orbits and the upper parts of the lower orbits become flat and run parallel and very close to the real axis. As an orbit approaches the real axis, a particle that starts at the origin runs off to $\infty$ along a straight line infinitesimally close to the real axis and reaches $\infty$ in finite time $\tfrac{P}{2}$. It then runs at infinite speed in a D-shaped roughly semicircular path from $+\infty$ to $-\infty$ in no time, and finally returns to the origin in time $\tfrac{P}{2}$. The particle does not stay at infinity; rather, it periodically returns to the origin!

How can a particle that travels to $+\infty$ immediately come back from $-\infty$? Isn't $x=+\infty$ infinitely far from $x=-\infty$? The answer is that on the real axis $+\infty$ is indeed infinitely far from $-\infty$, but we are studying classical particle motion in the complex plane. In the complex plane the point at infinity is {\it unique}; $+\infty$ and $-\infty$ are the same point. Thus, the particle does not jump from $+\infty$ to $-\infty$; if it is at $+\infty$, it is {\it already} at $-\infty$.

Most importantly, the particle moves slowly when it is near the origin and rapidly when it is far from the origin. Since the probability of finding the particle in a given segment of its orbit is inversely proportional to the speed of the particle, a particle that begins at the origin will never be found in the complex plane where it is moving infinitely fast, and it is most likely to be found on the real axis near the origin. For this $-x^4$ potential the probability density $\rho(x)$ of finding the particle of energy $E$ on the real axis at any point $x$ is simply [see Chap.~1 in \textcite{pt579}]
\begin{equation}
\rho(x)\propto 1/\sqrt{E+x^4}.
\label{e4.6}
\end{equation}
Since the integral of $\rho(x)$ is finite, the classical particle behaves as if it is in a {\it normalized} dynamical bound state that is localized at the origin. We emphasize that even though the particle is subject to a repulsive potential, it is most likely to be found at the 
origin.\footnote{A classical particle in a right-side-up potential, such as $V(x)=+x^4$, oscillates between the walls of the potential. Such a particle is most likely to be found near the turning points at the {\it edges} of the potential, where it moving slowly.}

\subsection{Quantum particle in a quartic upside-down potential}\label{ss4B}
In the previous subsection we used complex-variable methods to show that a classical particle subject to a $\cPT$-symmetric $-x^4$ potential can be in a dynamically stable bound state that is localized at the origin. However, it is not obvious how a {\it quantum} particle would behave in such a potential. One might try to analyze this problem on the real axis by approximating the $-x^4$ potential with an infinite descending-staircase potential or with a washboard potential \cite{pt684}. However, once again, the simplest way to solve the quantum eigenvalue problem associated with this upside-down quartic potential is to examine it in the complex plane.

The Schr\"odinger equation associated with the Hamiltonian (\ref{e4.1}) is
\begin{equation}
-\psi''(x)-x^4\psi(x)=E\psi(x).
\label{e4.7}
\end{equation}
Following the classical analysis in Subsec.~\ref{ss4A}, we require that the solution $\psi(x)$ to this equation satisfy {\it complex boundary conditions}; namely, that $\psi(x)$ vanishes in the Stokes sectors shown in Fig.~\ref{f9} for $\vep=2$ as $|x|\to\infty$. A numerical calculation shows that the spectrum is real, positive, and discrete (see Fig.~\ref{f4}). The first five eigenvalues are
\begin{eqnarray}
&&E_0=1.477...,~E_1=6.003...,~E_2=11.804...,\nonumber\\
&&\quad E_3=18.459...,~~E_4=25.792...\,.
\label{e4.8}
\end{eqnarray}

This eigenvalue problem cannot be solved in closed form. However, we can perform a leading-order semiclassical WKB calculation using the turning points in the lower-half $x$-plane that are shown in Fig.~\ref{f18}. The standard WKB quantization condition is
$$\int_{x_-}^{x_+}dx\,\sqrt{E+x^4}\sim\big(n+\half\big)\quad(n\to\infty),$$
where in this case the turning points are given by
$$x_-=E^{1/4}e^{-3i\pi/4}~~{\rm and}~~x_+=E^{1/4}e^{-i\pi/4}.$$

To evaluate the WKB integral we integrate from $x_-$ to the origin and then from the origin to $x_+$ along radial lines in the complex-$x$ plane. The resulting integrals are easy to evaluate in terms of Gamma functions:
\begin{equation}
E_n\sim\left[\frac{(6n+3)\sqrt{2\pi}\,\Gamma(3/4)}{2\Gamma(1/4)}\right]^{4/3}\quad(n\to\infty).
\label{e4.9}
\end{equation}
This WKB formula (\ref{e4.9}) gives the following approximations to the first five eigenvalues:
\begin{eqnarray}
&&E_0=1.377...,~E_1=5.956...,~E_2=11.769...,\nonumber\\
&&\quad E_3=18.432...,~~E_4=25.769...\,.
\label{e4.10}
\end{eqnarray}
As we can see, the WKB approximation in (\ref{e4.9}) provides increasingly accurate approximations to the numerical values of the eigenvalues in (\ref{e4.8}) as $n$ increases.

The remarkable feature of the eigenvalues in (\ref{e4.8}), of course, is that these quantum-mechanical energy levels are all {\it real}. Thus, as in the classical case, a quantum particle in this upside-down potential is in a {\it stable bound state}. Even though the potential is upside down, the particle wave function remains localized at the origin; the quantum particle is permanently confined and does not escape to infinity.

\subsection{Elementary proof of spectral reality}\label{ss4C}
We cannot calculate the eigenvalues and eigenfunctions of the Hamiltonians $H=p^2+x^2(ix)^\vep$ $(\vep>0)$ analytically but, as mentioned in Sec.~\ref{s2}, it was proved using sophisticated mathematical techniques that the spectra of these Hamiltonians are all real, positive, and discrete \cite{pt274}. However, the upside-down quartic Hamiltonian (\ref{e4.1}) is special; the unique value $\vep=2$ is the only known case for which one can prove reality, positivity, and discreteness by using nothing more advanced than elementary calculus. An early study of the $\vep=2$ case in \textcite{pt9} was followed by two more papers \cite{pt167,pt169}. We present a short proof of reality below.

We begin with the time-independent Sch\"odinger eigenvalue differential equation associated with $H$ in (\ref{e4.1}),
\begin{eqnarray}
-\frac{\hbar^2}{2m}\psi''(x)-gx^4\psi(x)=E\psi(x),
\label{e4.11}
\end{eqnarray}
where we have inserted the dimensional parameters $\hbar$, $g$, and $m$ for clarity in the subsequent discussion. The boundary conditions on $\psi(x)$, as shown in Fig.~\ref{f9} for the case $\vep=2$, are that $\lim_{|x|\to\infty}\psi(x)=0$ in two Stokes sectors of angular opening $\tfrac{\pi}{3}$ adjacent to and below the positive and negative real-$x$ axis.

For the differential equation (\ref{e4.11}) we impose vanishing boundary conditions on a specially chosen curve in the complex plane that we express in parametric form as
$$x=-2iL\sqrt{1+iy/L}\quad(-\infty<y<+\infty),$$
where $y$ is a real variable. Note that this curve terminates {\it inside the Stokes sectors} at the angles $-\tfrac{\pi}{4}$ and $-\tfrac{3\pi}{4}$.

We use the parametric representation above to replace the independent variable $x$ in (\ref{e4.11}) by $y$ and get the differential equation
\begin{eqnarray}
&&\!\!\!\!\!-\frac{\hbar^2}{2m}\left(1+\frac{iy}{L}\right)\phi''(y)-\frac{i\hbar^2}{4Lm}\phi'(y)\nonumber\\
&&\qquad-16gL^4\left(1+\frac{iy}{L} \right)^2\phi(y)=E\phi(y),
\label{e4.12}
\end{eqnarray}
where $\psi(x)=\phi(y)$. The parameter $L$ has units of length,
$$L=\lambda\left(\frac{\hbar^2}{mg}\right)^{1/6},$$
where $\lambda$ is an arbitrary dimensionless number. The constant $L$ serves to make the subsequent equations dimensionally consistent. (This parameter will drop out at the end of the calculation.)

Next, we take a Fourier transform of (\ref{e4.12}). This transform has the general form $$\tilde f(p)\equiv\int_{-\infty}^{\infty} dy\,e^{-iyp/\hbar}f(y).$$ Since the Fourier transforms of $f'(y)$ and $yf(y)$ are $ip\tilde f(p)/\hbar$ and $i\hbar\tilde f'(p)$, the Fourier transform of (\ref{e4.12}) is
\begin{eqnarray}
&&\!\!\!\!\!\frac{1}{2m}\left(1-\frac{\hbar}{L}\frac{d}{dp}\right)p^2
\tilde\phi(p)+\frac{\hbar}{4Lm}p\tilde\phi(p)\nonumber\\
&&\qquad -16gL^4\left(1-\frac{\hbar}{L}\frac{d}{dp}\right)^2\tilde\phi(p)=
E\tilde\phi(p).
\label{e4.13}
\end{eqnarray}

We simplify (\ref{e4.13}) and get
\begin{eqnarray}
&& \!\!\!\!\!\!\!\!\!\!-16gL^2\hbar^2\tilde\phi''(p)+\left(-\frac{\hbar p^2}
{2mL}+32g L^3\hbar\right)\tilde\phi'(p)\nonumber\\
&&+\left(\frac{p^2}{2m}-\frac{3p\hbar}{4mL}-16gL^4\right)
\tilde\phi(p)=E\tilde\phi(p).
\label{e4.14}
\end{eqnarray}
Note that the variable $p$ in this equation is not a conventional momentum variable. Here, considered as an operator, $p$ represents $-i\hbar\tfrac{d}{dy}$ and not $-i\hbar\frac{d}{dx}$.

The next step is to convert (\ref{e4.14}) to a conventional Schr\"odinger equation by eliminating the one-derivative term. This is done by making the elementary change of dependent variable $$\tilde\phi(p)=e^{Q(p)/2}\Phi(p).$$
Then, to eliminate $\Phi'(p)$ from the resulting equation we require that $Q(p)$ satisfy a first-order differential equation whose solution is $$Q(p)=\frac{2L}{\hbar}p-\frac{1}{96gmL^3\hbar}p^3.$$ The Schr\"odinger equation satisfied by $\Phi(p)$ is then
\begin{eqnarray}
&&\!\!\!\!\!\!-16gL^2\hbar^2\Phi''(p)+\left(-\frac{\hbar p}{4mL}+\frac{p^4}
{256gm^2L^4}\right)\Phi(p)\nonumber\\
&&\qquad =E\Phi(p).\nonumber
\end{eqnarray}

Finally, we perform a scaling transformation of the independent variable,
$$p=zL\sqrt{32mg}.$$
This replaces the $p$ variable, which has units of momentum, by $z$, which is a coordinate variable having units of length. The resulting eigenvalue equation, posed on the {\it real}-$z$ axis, is
\begin{eqnarray}
\!\!\!\!\!\!\!-\frac{\hbar^2}{2m}\Phi''(z)+\left(-\hbar\sqrt{\frac{2g}{m}}\,z+4gz^4\right)\Phi(z)=E\Phi(z).
\label{e4.15}
\end{eqnarray}

This new eigenvalue equation, which no longer contains the length parameter $L$, is the final result of the calculation. The new independent variable $z$ has dimensions of length although it is not a conventional coordinate variable because it is {\it odd under time reversal}.

The eigenvalue equation (\ref{e4.15}) is similar in structure to the original eigenvalue equation (\ref{e4.11}), but there are two significant differences: (i) The quartic term in the potential has {\it changed sign}, so {\it the upside-down quartic potential has become a conventional right-side-up quartic potential}. (ii) There is a new linear potential term proportional to Planck's constant $\hbar$. This linear term is called a {\it quantum anomaly}, which means that it has no classical analog. (The effect of this parity-anomaly term is discussed in Subsec.~\ref{ss4D}.)

The Hamiltonian $\tilde H$ for which (\ref{e4.15}) is the eigenvalue equation is
\begin{equation}
\tilde H=\frac{\tilde p^2}{2m}-\hbar\sqrt{\frac{2g}{m}}\,z+4gz^4.
\label{e4.16}
\end{equation}
This Hamiltonian is quite different from
\begin{equation}
H=\frac{p^2}{2m}-gx^4,
\label{e4.17}
\end{equation}
which is the Hamiltonian $H$ associated with (\ref{e4.11}). At the quantum level the Hamiltonian $\tilde H$ in (\ref{e4.16}) is conventionally Hermitian and the potential is bounded below on the real-$z$ axis; $\tilde H$ is also $\cPT$-symmetric because the sequence of transformations used to obtain $\tilde H$ all respect $\cPT$ symmetry. Note that the potential in (\ref{e4.16}) is $\cPT$-symmetric because while $z$ and $\tilde p$ are canonically conjugate operators satisfying the usual commutation relation $[z,\tilde p]=i$, the new variable $z$ behaves like a {\it momentum} rather than a coordinate variable.

The point of this calculation is simply that while we cannot calculate exactly the eigenvalues of $H$ (with $\cPT$-symmetric boundary conditions imposed in the Stokes sectors shown in Fig.~\ref{f9} for $\vep=2$) or the eigenvalues of $\tilde H$ (with conventional boundary conditions imposed on the real axis), we have shown that these two Hamiltonians are {\it isospectral}; that is, they have exactly the same eigenvalues. Because the potential of $\tilde H$ rises to $+\infty$ as $x\to\pm\infty$, we know that the entire spectrum of $\tilde H$ is real and discrete. This proves that the eigenvalues of the $\cPT$-symmetric Hamiltonian $H$ are also real and discrete.

This elementary calculation easily generalizes if we include a {\it quadratic} term in the $\cPT$-symmetric quartic Hamiltonian:
\begin{equation}
H=p^2+\mu^2x^2-gx^4.
\label{e4.18}
\end{equation}
The corresponding isospectral Hermitian Hamiltonian is just slightly more complicated than that in (\ref{e4.16}):
\begin{equation}
{\tilde H}=p^2-2\hbar\sqrt{g}z+4g\left(z^2-\frac{\mu^2}{4g}\right)^2.
\label{e4.19}
\end{equation}
[This isospectral Hamiltonian is also derived in \textcite{pt9,pt167,pt169}.] However, the calculation in this subsection does not generalize to any other Hamiltonians of the form (\ref{e2.14}); it only works for $\vep=2$. The techniques used in this calculation {\it do} extend to the higher-derivative $\cPT$-symmetric class of Hamiltonians of the form $$H_n=p^n-g(ix)^{n^2}\quad(n=2,\,3,\,4,\,...)$$
but the resulting isospectral Hamiltonians are quite complicated \cite{pt486}.

\subsection{Discussion of the parity anomaly} \label{ss4D}
The equations of motion for the quartic Hamiltonian (\ref{e4.17}) are parity symmetric. The explicit violation of parity symmetry that emerges in the transformed Hamiltonian in (\ref{e4.16}) occurs at large values of $x$ in the $\cPT$-symmetric Hamiltonian $\tilde H$, where the boundary conditions on the eigenfunction $\psi(x)$ in (\ref{e4.11}) are imposed. (These boundary conditions are $\cPT$ symmetric, but they are {\it not} parity symmetric.) Because we have taken a Fourier transform to obtain the Schr\"odinger eigenvalue equation (\ref{e4.15}), the parity-breaking anomaly that we see in the Hamiltonian (\ref{e4.16}) now manifests itself at {\it small} values of $z$ in the Hamiltonian (\ref{e4.16}) and not in the boundary conditions at $z=\pm\infty$.

\begin{figure}
\center
\includegraphics[scale = 0.42]{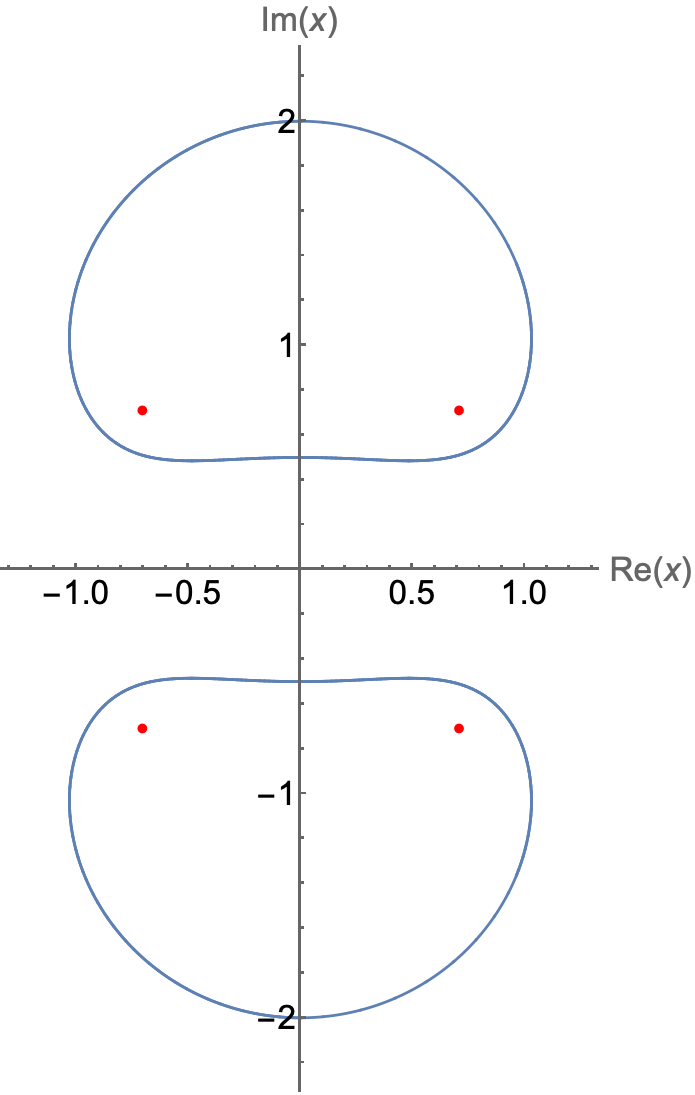}
\caption{Typical complex trajectories for $H$ in \ref{e4.17} with $m=1$, $g=1$, and $E=1$. These trajectories begin at $x=\pm2.0\,i$. The trajectories are similar to the closed trajectories in Fig.~\ref{f18}, but they have a classical period of $2.622060$. The dots indicate the locations of the classical turning points.}
\label{f19}
\end{figure}

The parity-breaking term in $\tilde H$ at $z=0$ is proportional to $\hbar$, so there is no classical analog of this parity-breaking anomaly. Thus, while the two Hamiltonians $H$ and $\tilde H$ have identical quantum-mechanical spectra, these two Hamiltonians determine different classical behaviors, which we demonstrate by comparing the classical trajectories for these two Hamiltonians. First, we plot in Fig.~\ref{f19} a pair of classical trajectories for the $\cPT$-symmetric Hamiltonian $H$ in (\ref{e4.17}) for the case $m=1$, $g=1$, and $E=1$. These trajectories begin at $x=\pm 2.0\,i$ and they resemble the trajectories in Fig.~\ref{f18} although they have a classical period of $P=2.622060$.

\begin{figure}
\center
\includegraphics[scale = 0.48]{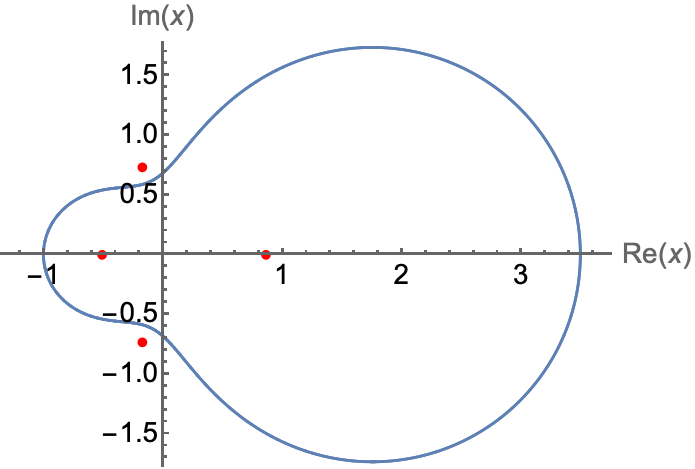}
\caption{Complex classical trajectory for the isospectral
Hamiltonian $\tilde H$ in (\ref{e4.16}) with $m=1$, $g=1$, $\hbar=1$, and $E=1$. The trajectory begins at $x=-1$. The quantum energy levels of this Hamiltonian are the same as those of $H$, but the classical period of the trajectory is $2.495467$, which is shorter than the periods of the trajectories in Fig.~\ref{f19}. The dots indicate the locations of the classical turning points.}
\label{f20}
\end{figure}

Next, using the same values $m=1$, $g=1$, and $E=1$ along with $\hbar=1$, we plot in Fig.~\ref{f20} a typical closed classical trajectory for the isospectral Hermitian Hamiltonian $\tilde H$ in (\ref{e4.16}); this trajectory begins at  $x=-1$. The period of this trajectory is $2.495467$, which is {\it shorter} than the periods of the trajectories in Fig.~\ref{f19}.

\begin{figure}
\center
\includegraphics[scale = 0.47]{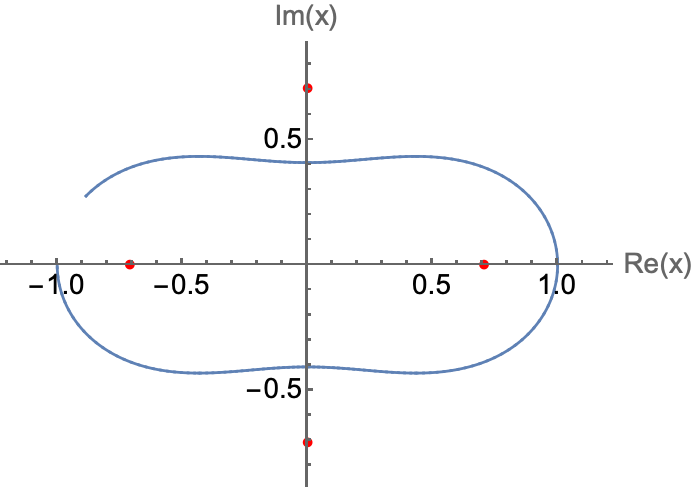}
\caption{Peanut-shaped complex classical trajectory for the Hamiltonian $\tilde H$ in (\ref{e4.16}) with $m=1$, $g=1$, and $E=1$ but with the quantum anomaly removed by setting $\hbar=0$. The period $2.622060$ of this  trajectory is now the same as the period of the trajectories in Fig.~\ref{f19}. However, this trajectory begins at $x=-1$ and is plotted for a time that is equal to the shorter period of the trajectory in Fig.~\ref{f20}. Note that this trajectory does not quite close. This is because while the Hamiltonians $H$ in (\ref{e4.17}) and $\tilde H$ in (\ref{e4.16}) have the same {\it quantum} energy levels, they differ at the classical level because $\tilde H$ has a quantum-anomaly term. As in Figs.~\ref{f19} and \ref{f20}, the dots indicate the locations of the classical turning points.}
\label{f21}
\end{figure}

Finally, we plot in Fig.~\ref{f21} a classical trajectory of $\tilde H$ in (\ref{e4.16}) for the values $m=1$, $g=1$, and $E=1$ except that now we set $\hbar=0$. The period of this trajectory is exactly the same as the period of the classical trajectory of $H$ in Fig.~\ref{f19}. To demonstrate the effect of the quantum anomaly, we plot a trajectory beginning at $x=-1$ for the shorter period $T=2.495467$ from Fig.~\ref{f20}. This trajectory does not have enough time to close.

\subsection{Comparison of $x^4$ and $-x^4$ Potentials}\label{ss4E}
As explained in Sec.~\ref{s2}, to obtain the quartic $\cPT$-symmetric Hamiltonian $H=p^2-x^4$, we smoothly deform in $\vep$ the Hamiltonian $H=p^2+x^2 (ix)^\vep$ from the harmonic-oscillator Hamiltonian $H=p^2+x^2$ at $\vep=0$ to the quartic $\cPT$-symmetric Hamiltonian $H=p^2-x^4$ at $\vep=2$. This exponential analytic continuation in the variable $\vep$ is quite elaborate, and one might wonder if one can obtain this $\cPT$-symmetric Hamiltonian by a simpler procedure such as a complex rotation of a parameter. That is, one might begin with the Hermitian Hamiltonian $H=p^2+x^4$ and insert the {\it multiplicative} parameter $e^{i\theta}$ into the potential term:
\begin{equation}
H_\theta=p^2+e^{i\theta}x^4.
\label{e4.20}
\end{equation}
Now, if $\theta=0$, $H_\theta$ has the right-side-up quartic potential of the Hermitian Hamiltonian, and if we rotate smoothly to $\theta=\pi$, $H_\theta$ has the upside-down potential of the $\cPT$-symmetric Hamiltonian.

If we perform such a rotation, we {\it do} indeed obtain the Hamiltonian operator $H=p^2-x^4$, but this rotation procedure fails to recover the $\cPT$-symmetric properties of the $H=p^2-x^4$ Hamiltonian because {\bf the eigenvalue problems associated with the Hermitian Hamiltonian and the $\cPT$-symmetric Hamiltonian are not analytic continuations of one another.} 

To understand why we cannot rotate smoothly from a conventional Hermitian quartic Hamiltonian to a $\cPT$-symmetric quartic Hamiltonian, let us examine the eigenvalue problem associated with (\ref{e4.20}). Following the procedure explained in Subsec.~\ref{ss3C} for performing a rotation of the harmonic-oscillator eigenvalue problem, we note that when $\theta=0$, the possible large-$|x|$ exponential behaviors of solutions to the eigenvalue differential equation associated with $H_0$ are $\exp\big(\pm\tfrac{1}{3}x^3\big)$. For large-positive $x$ the eigenfunctions decay like $\exp\big(-\tfrac{1}{3}x^3\big)$ and for large-negative $x$ the solutions decay like $\exp\big( \tfrac{1}{3}x^3\big)$. These asymptotic behaviors continue to be valid in Stokes sectors of angular opening $\tfrac{\pi}{3}$ centered about the positive-real-$x$ and negative-real-$x$ axes.

\begin{figure}
\center
\includegraphics[scale = 0.45]{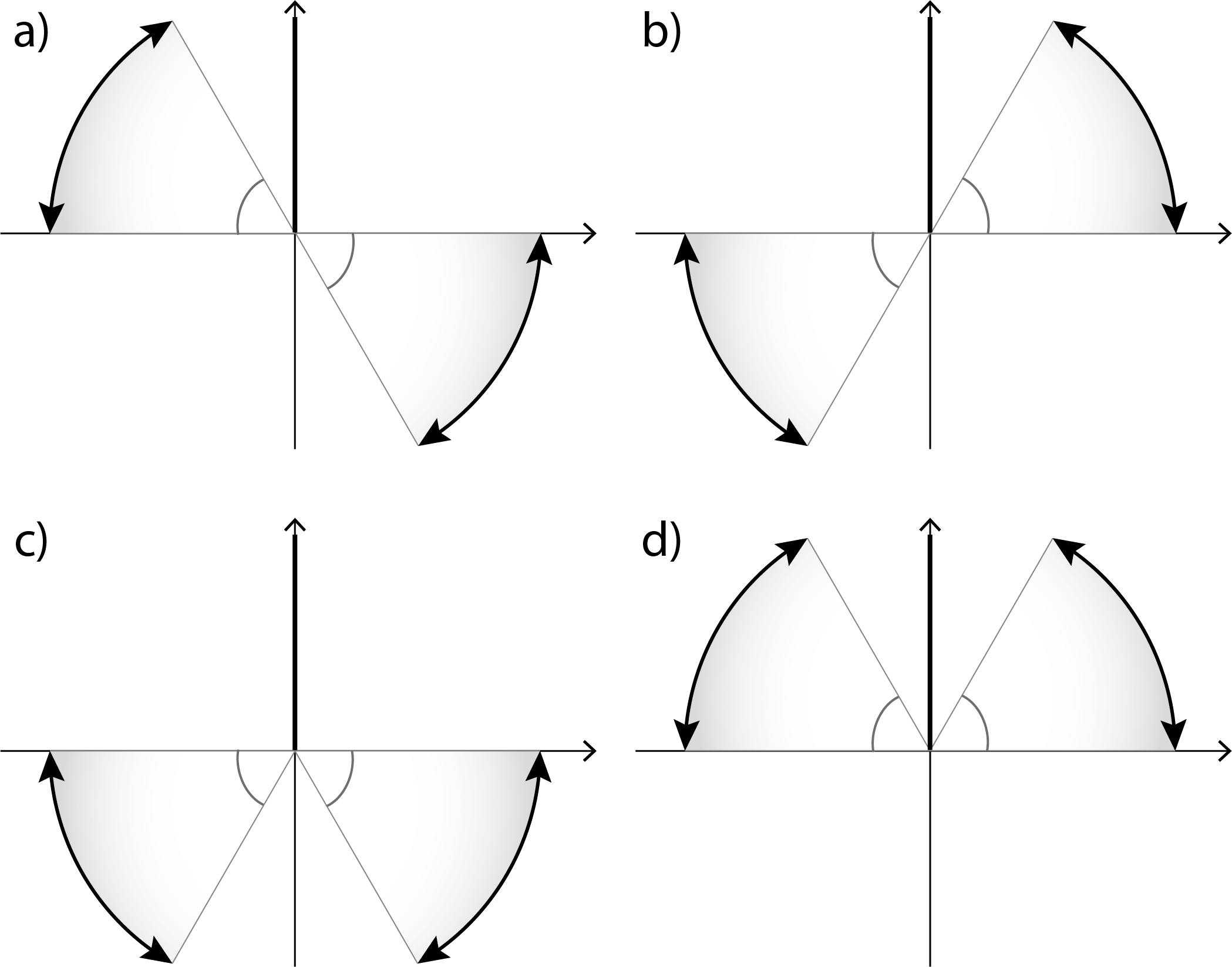}
\caption{Stokes sectors in the complex-$x$ plane in which the eigenfunctions for the upside-down quartic Hamiltonian $H=p^2-x^4$ are required to vanish exponentially. In cases a and b the pair of Stokes sectors is symmetric under parity reflection $\cP$ and the eigenvalues are all complex. However, in cases c and d the pairs of Stokes sectors are $\cPT$ symmetric and the eigenvalues are all real and positive. Thus, there are {\bf two different phases} of the Hamiltonian $H$, and these are distinguished by having different global symmetries, $\cP$ in the case of a and b, and $\cPT$ in the case of c and d.}
\label{f22}
\end{figure}

There are two ways to perform the $\theta$ rotation of $H_\theta$; we can rotate clockwise from $\theta=0$ to $-\pi$ or anticlockwise from $\theta=0$ to $\pi$. If we rotate clockwise, the two sectors in which the eigenfunctions vanish exponentially rotate anticlockwise together (like a propeller) by an angle of $\tfrac{\pi} {6}$. If we rotate anticlockwise, these two sectors rotate together (again like a propeller) by an angle of $\tfrac{\pi}{6}$ but in the clockwise direction. In both cases one sector rotates downward and the other rotates upward; one sector ends up {\it above} the real axis with its lower edge on the real axis and the other ends up {\it below} the real axis with its upper edge on the real axis. These two orientations of the sectors are shown in Fig.~\ref{f22}a and Fig.~\ref{f22}b.

In both orientations (a and b) the original parity symmetry of the Hamiltonian $H_\theta$ at $\theta=0$ is preserved. Under a reflection through the origin the two sectors are interchanged and a point $x$ in one sector goes into the corresponding reflected point $-x$ in the other sector. In both cases the configurations of these sectors are $\cP$ symmetric but {\it not} $\cPT$ symmetric (left-right symmetric).

The energy eigenvalues in the parity-symmetric case are all complex. The $\theta$-rotated Hamiltonian has no bound states; depending on the sign of the imaginary part of the energy eigenvalue, the states either grow or decay exponentially in time. In contrast, the diagrams in Fig.~\ref{f22}c and Fig.~\ref{f22}d {\it are} $\cPT$ symmetric (just like the Stokes sectors in the $\vep=2$ diagram in Fig.~\ref{f9}). In this case the energy eigenvalues are all real and positive.

The Hamiltonians in both the $\cP$-symmetric and the $\cPT$-symmetric cases
are the same, and thus the
time-independent Schr\"odinger equation in both cases is the same. However, the boundary conditions at $|x|= \infty$ are different. These two different models
may be regarded as distinct {\it phases} of the Hamiltonian $H=p^2-x^4$ because they exhibit different global symmetries; namely, $\cP-$reflection and $\cPT$-reflection symmetries. (This distinction resembles the difference in the long-range global symmetries between an ice crystal and liquid water.) 

To better understand what is happening in the $\cP$-symmetric case, let us examine the large-$|x|$ exponential behaviors of the eigenfunctions of the rotated Hamiltonian $H=p^2-x^4$. An exponential WKB approximation to the eigenfunctions shows that on the real-$x$ axis the eigenfunctions are wavelike; they behave for large $|x|$ like $\exp\big(\pm\tfrac{1}{3}ix^3\big)$. Depending on the orientation of the Stokes sectors, waves either {\it approach} the origin symmetrically from the right and left or waves propagate away from the origin symmetrically to the right and left:
\begin{equation}
{\bf\longrightarrow}~~(x=0)~~{\bf\longleftarrow}\qquad{\rm or}\qquad{\bf\longleftarrow}~~(x=0)~{\bf \longrightarrow}.
\label{e4.21}
\end{equation}

If the waves in (\ref{e4.21}) propagate inward, the state is populating in time; if the waves recede from the origin, the state is depopulating in time. In either case, these systems are not in equilibrium. Because the rotated Hamiltonian is real, these systems must be invariant under complex conjugation (up-down reflection), and as we can see, the outgoing and incoming configurations of Stokes sectors become interchanged under this symmetry operation. Correspondingly, the imaginary parts of the eigenvalues change sign under complex conjugation. We conclude that the rotation procedure fails to recover $\cPT$-symmetric boundary conditions.

In contrast, for $\cPT$-symmetric boundary conditions the waves propagate inward on one side of the origin and outward on the other side:
\begin{equation}
{\bf\longrightarrow}~~(x=0)~~{\bf \longrightarrow}\qquad{\rm or}\qquad{\bf\longleftarrow}~~(x=0)~~{\bf\longleftarrow}.
\label{e4.22}
\end{equation}
This means that the $\cPT$-symmetric eigenstates are in dynamic equilibrium; they neither populate nor decay with time. Thus, the energy eigenvalues are real.

Just as there are two $\cP$-symmetric configurations in (\ref{e4.21}) of the Stokes sectors of the rotated Hamiltonian that are interchanged under complex conjugation (up-down reflection), there are also two $\cPT$-symmetric (left-right-symmetric) configurations of the Stokes sectors. In addition to the configuration of Stokes sectors adjacent to and below the real axis ($\vep=2$ diagram in Fig.~\ref{f9}), there is a another possible configuration of the $\cPT$-symmetric (left-right-symmetric) Stokes sectors adjacent to and {\it above} the real axis, as shown in Fig.~\ref{f22}d. We obtain these Stokes sectors by smoothly deforming (analytically continuing) the Hamiltonian $H=p^2+ x^2(-ix)^\vep$ from $\vep=0$ to $\vep=2$. This deformation again breaks parity symmetry but preserves $\cPT$ symmetry. The resulting $\cPT$-symmetric Hamiltonian has the same form as (\ref{e4.1}), but now the eigenfunctions vanish exponentially in a $\cPT$-symmetric pair of Stokes sectors lying {\it above} the real axis. The corresponding eigenvalues are identical to the eigenvalues for the configuration in Fig.~\ref{f9}, third diagram. There is no physical difference between these up-up and down-down configurations because one is obtained from the other by complex conjugation; that is, changing the sign of $i$. (The sign of the complex number $i$ is arbitrary and cannot have any physical consequences.)

The state of dynamic equilibrium exhibited in (\ref{e4.22}) is in perfect analogy with the energy flow in the optical system pictured in Fig.~\ref{f5}; the incoming and outgoing waves to the left and right of the origin in (\ref{e4.22}) show that there is an exact balance of loss and gain. Thus, the system is in dynamic equilibrium. This balance was emphasized and investigated at the quantum level in \cite{pt567}. We can also see this dynamic balance at the classical level in Fig.~\ref{f18}, which shows that classical particles near the real axis are always traveling in the same direction; these particles approach the origin on one side and recede from the origin on the other.

This balanced loss-and-gain behavior has been seen in many optical laser experiments performed on $\cPT$-symmetric systems and is referred to as {\it unidirectional propagation} \cite{pt567,pt142} and {\it reflectionless scattering} \cite{pt203}. The generic phenomenon of reflectionless scattering suggests the possibility of performing new tabletop macroscopic quantum-mechanical experiments in which a one-dimensional particle beam is incident on an upside-down quartic potential (or other upside-down potentials such as $-x^6$ or $-x^8$). In such experiments one measures the amplitude of the reflected wave. If the energy of the incident wave is adjusted so that there is no reflected wave, one has a {\it direct measurement of a $\cPT$-symmetric energy level}. This kind of experiment was originally proposed in \textcite{pt594}; in a recent paper such an experiment was proposed for a cold-atom beam incident on the much wider class of real Hamiltonians of the form $H=p^2-|x|^\vep$ \cite{pt595}. Such Hamiltonians are independently $\cP$-symmetric and $\cT$-symmetric.

\section{Anharmonic oscillators}\label{s5}
Until now we have focused on $\cPT$-symmetric Hamiltonians whose potentials are monomials in $x$ and noninteger powers of $x$. In this section we discuss potentials of quartic anharmonic form $ax^2+bx^4$. The Hermitian anharmonic oscillator, whose Hamiltonian is
\begin{equation}
H=\half p^2+\half m^2x^2+gx^4\quad(g>0),
\label{e5.1}
\end{equation}
has been studied intensively for many decades by mathematical physicists. The time-independent Schr\"odinger differential equation associated with this Hamiltonian is
\begin{equation}
-\half\psi''(x)+\half m^2x^2\psi(x)+gx^4\psi(x)=E\psi(x).
\label{e5.2}
\end{equation}

This differential equation cannot be solved in closed form, but the anharmonic-oscillator model that it represents has been examined in depth because it is an excellent intellectual laboratory in which to analyze perturbation theory, summation theory, Pad\'e theory, dispersion relations, diagrammatic expansions, and variational and semiclassical techniques. This model is especially interesting because it is a quartic quantum field theory in one-dimensional spacetime and does not have the problem of infinities requiring renormalization. Thousands of papers have been written on the properties of the quantum anharmonic oscillator following the early studies \cite{pt13,pt14,pt15, pt17}.

The $\cPT$-symmetric Hamiltonian
\begin{eqnarray}
H=\frac{1}{2m}p^2+\frac{\mu^2}{2}x^2-gx^4 \quad(g>0).
\label{e5.3}
\end{eqnarray}
is obtained formally by changing the sign of $g$ in (\ref{e5.1}). This sign change leads to a significant change in the nature
of the physics depending on whether the unstable boundary conditions in (\ref{e4.21}) or the $\cPT$-symmetric boundary conditions in (\ref{e4.22}) are imposed. 

\begin{figure}
\center
\includegraphics[scale = 0.29]{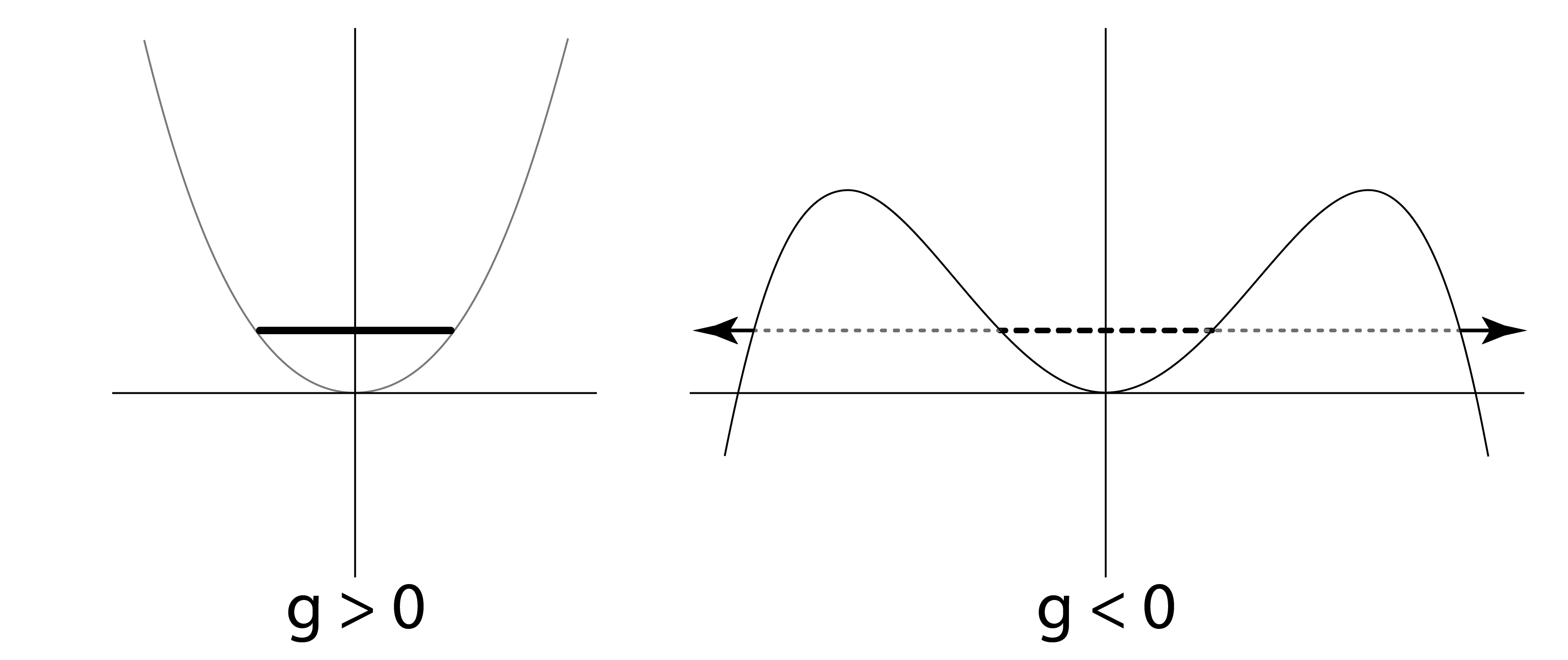}
\caption{Schematic plot of the right-side-up and upside-down potentials for the Hamiltonians (\ref{e5.1}) (left) and (\ref{e5.3}) (right). The usual Hermitian interpretation of these potentials is that if the sign of $g$ is changed from a positive value to a negative value, the stable ground state indicated by a heavy line in the left diagram becomes unstable (dashed line) in the right diagram and it tunnels through the barriers (dotted lines) out to $\infty$. [This instability is also indicated in (\ref{e4.21}).]}
\label{f23}
\end{figure}

\begin{figure}
\center
\includegraphics[scale = 0.29]{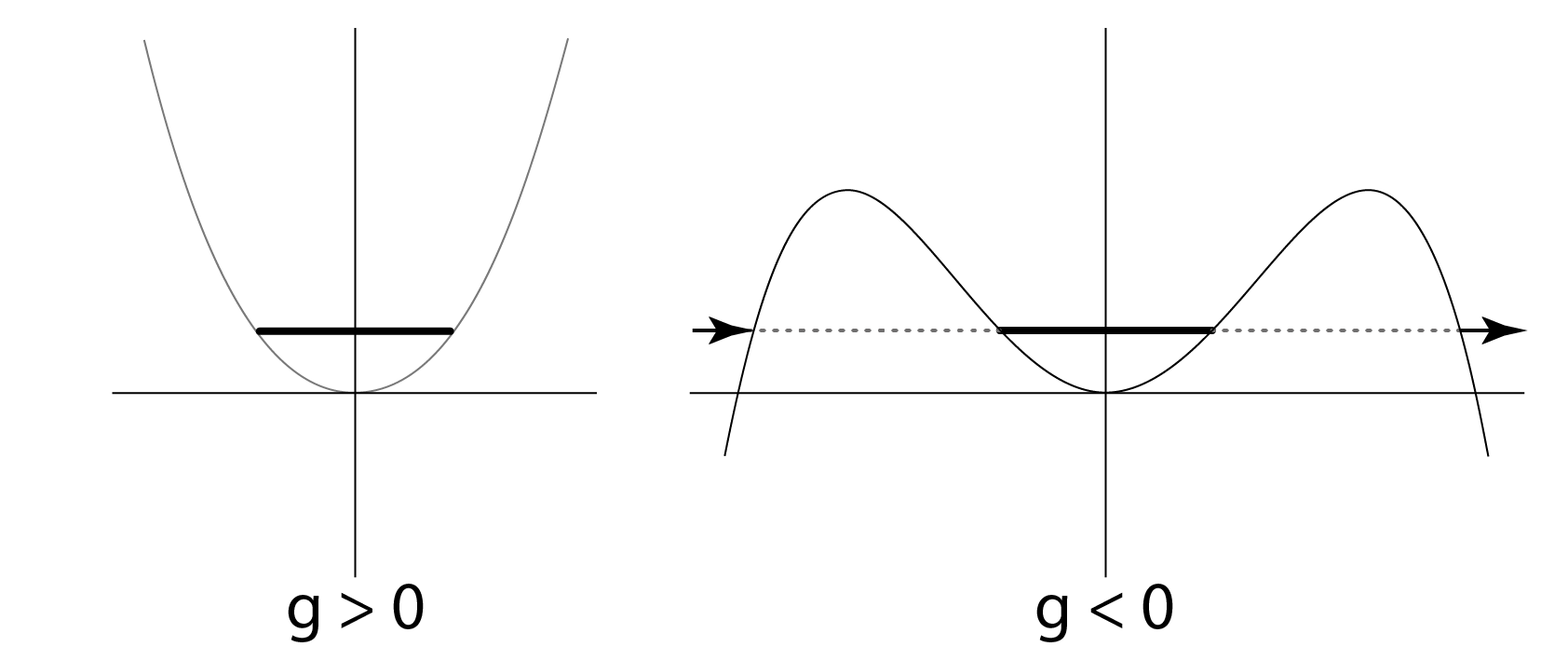}
\caption{Schematic potentials for the Hamiltonians (\ref{e5.1}) (left) and (\ref{e5.3}) (right) plotted as functions of real $x$. The $\cPT$-symmetric boundary conditions shown in the right diagram indicate balanced unidirectional incoming waves on one side of the potential and outgoing waves on the other side. The resulting energy levels of this upside-down potential are all real and the states are all stable. [See also (\ref{e4.22}).]}
\label{f24}
\end{figure}

The anharmonic potentials of $H$ in (\ref{e5.1}) and (\ref{e5.3}) are plotted as functions of real $x$ in Figs.~\ref{f23} and \ref{f24} for $g>0$ (left) and for $g<0$ (right). The left diagrams illustrate the infinitely rising potential, which confines an infinite-real-positive discrete spectrum. The ground-state energy level is also indicated. In the right diagrams the locally rising parabolic potential near $x=0$ is overwhelmed for large $|x|$ by the upside-down quartic potential. The conventional interpretation of this upside-down configuration is that there are no bound states and that the unstable low-energy states tunnel outward to $\infty$, as shown in
Fig.~\ref{f23} (right).

The potential in the Hamiltonian (\ref{e5.3}) is more complicated than the pure quartic potential in (\ref{e4.11}) because (\ref{e5.3}) has a quadratic term, but the same differential-equation analysis used in Subsec.~\ref{ss4C} straightforwardly yields the {\it isospectral} Hermitian Hamiltonian
\begin{eqnarray}
\tilde H=\frac{\tilde p^2}{2m}-\hbar\sqrt{\frac{2g}{m}}\,z+4g\left(z^2-\frac{\mu^2}{8g}\right)^2,
\label{e5.4}
\end{eqnarray}
which has a right-side-up potential. Thus, despite its upside-down potential, the energy levels of $H$ in (\ref{e5.3}) are all discrete, real, and positive. Note also that for the general anharmonic Hamiltonian in (\ref{e5.3}) the form of the parity-anomaly term in (\ref{e5.4}) remains unchanged from that in (\ref{e4.15}).  

As a function of real $x$, the potential in (\ref{e5.3}) is upside down, but the $\cPT$-symmetric interpretation of this Hamiltonian is that there are incoming waves on one side of the potential and outgoing waves on the other side [see (\ref{e4.21}) and (\ref{e4.22})]. Because of these balanced loss-and-gain boundary conditions, the unstable tunneling state in Fig.~\ref{f23} (right) becomes dynamically stable, as indicated by the bold line in Fig.~\ref{f24} (right).

\subsection{Particle-physics analysis of $\cPT$-symmetric bound states} 
\label{ss5A}
In four-dimensional spacetime a $-\phi^4$ quantum field theory is of particular interest because it might be a good model for describing the dynamics of the Higgs sector of the standard model. This is because the $-\phi^4$ theory is asymptotically free and thus is nontrivial \cite{pt586,pt449, pt587}. Furthermore, the one-point Green's function $\langle0|\phi|0\rangle$ is nonvanishing \cite{pt450}.

A four-dimensional quantum field theory is an extremely complicated mathematical construct. However, as stated earlier in this section, the quantum anharmonic oscillator (\ref{e5.3}) is a model quantum field theory for the simple case of one-dimensional spacetime. Such a model describes a universe that has no spatial extent, so nothing moves; all that happens is that this elementary system evolves in time. Nevertheless, this model is instructive because its nontrivial energy spectrum $E_0$, $E_1$, $E_2$, $\dots$ can be interpreted in particle-theory language as the energies of multiparticle states.

We begin by interpreting the quantum-mechanical ground state of (\ref{e5.3}) as the particle-physics {\it vacuum state}. In particle physics the {\it vacuum state} is the state of no particle excitations, and because there is no absolute energy scale, the vacuum state is {\it defined} to have zero energy.

The one-particle state is the lowest-energy excitation above the vacuum state. In particle-physics language this energy is called the {\it mass gap} and it represents the renormalized mass $M$ of a free particle:
$$M\equiv E_1-E_0.$$

The state associated with the second quantum-mechanical energy level $E_2$ is the two-particle state. In particle physics the energy of this state is again measured relative to the vacuum state: $E_2-E_0$. The two-particle state is interpreted as a {\it bound state} if it has negative binding energy $B_2$:
$$B_2=(E_2-E_0)-2M<0.$$
This inequality means that the energy of the two-particle state is {\it less} than the mass $2M$ of two free particles. If $B_2$ is negative, one can think of the two particles as having an {\it attractive force}; energy would be required to decompose the state into two free particles. Since this quantum system has no spatial extent, this force has no spatial dependence. Conversely, if $E_2-E_0$ is greater than $2M$, this force is interpreted to be {\it repulsive}.

The energies of the higher excitations are also measured relative to the vacuum  energy, $E_n-E_0$ ($n=2,\,3,\,4,\,...$), and if the energy $E_n-E_0$ of an $n$-particle state is less than the mass $nM$ of $n$ free particles,
\begin{equation}
B_n\equiv (E_n-E_0)-nM<0,
\label{e5.5}
\end{equation}
we interpret this state as a bound state. Positive energy would be required to decompose this bound state into its $n$ particle constituents. Conversely, if $B_n$ is positive, the state is unbound.

Note that the Hamiltonians whose spectra are shown in Figs.~\ref{f6} and \ref{f7} have no quadratic term, and therefore in particle-theory language these are massless field theories. These spectra become more widely separated with increasing energy and thus the binding energies $B_n$ in (\ref{e5.5}) are all positive. This indicates that these theories have no bound states. 

In contrast, the Hamiltonian (\ref{e5.3}) describes a theory having a more elaborate physical structure. The numerical calculations in \textcite{pt428} show that for small positive $g$ the first few states of $H$ in (\ref{e5.3}) are bound states. That paper found that as $g$ increases, the number of bound states decreases until when $g/\mu^3$ exceeds a critical value of about $0.0465$, there are no bound states at all. Thus, (\ref{e5.3}) describes a system in which there are a finite number of stable particles if $g$ is small.

We have shown that the $\cPT$-symmetric Hamiltonian $H$ in (\ref{e5.3}) and the Hermitian Hamiltonian $\tilde H$ in (\ref{e5.4}) are isospectral.
It is easier to calculate the eigenvalues of a Hermitian than a non-Hermitian Hamiltonian and a straightforward study of $\tilde H$ using the Arnoldi algorithm \cite{pt614} shows that {\it the bound states are a direct consequence of the anomaly term in this Hamiltonian.} To demonstrate this, we insert a multiplicative parameter $\alpha$ that quantifies the strength of the anomaly term in $\tilde H$ and for simplicity we set $m=\mu=\hbar=1$:
\begin{eqnarray}
\tilde H_\alpha=\frac{\tilde p^2}{2}-\alpha\hbar\sqrt{2g}\,z+4g\left(z^2-\frac{1}{8g}
\right)^2.
\label{e5.6}
\end{eqnarray}

\begin{figure}
\center
\includegraphics[scale = 0.23]{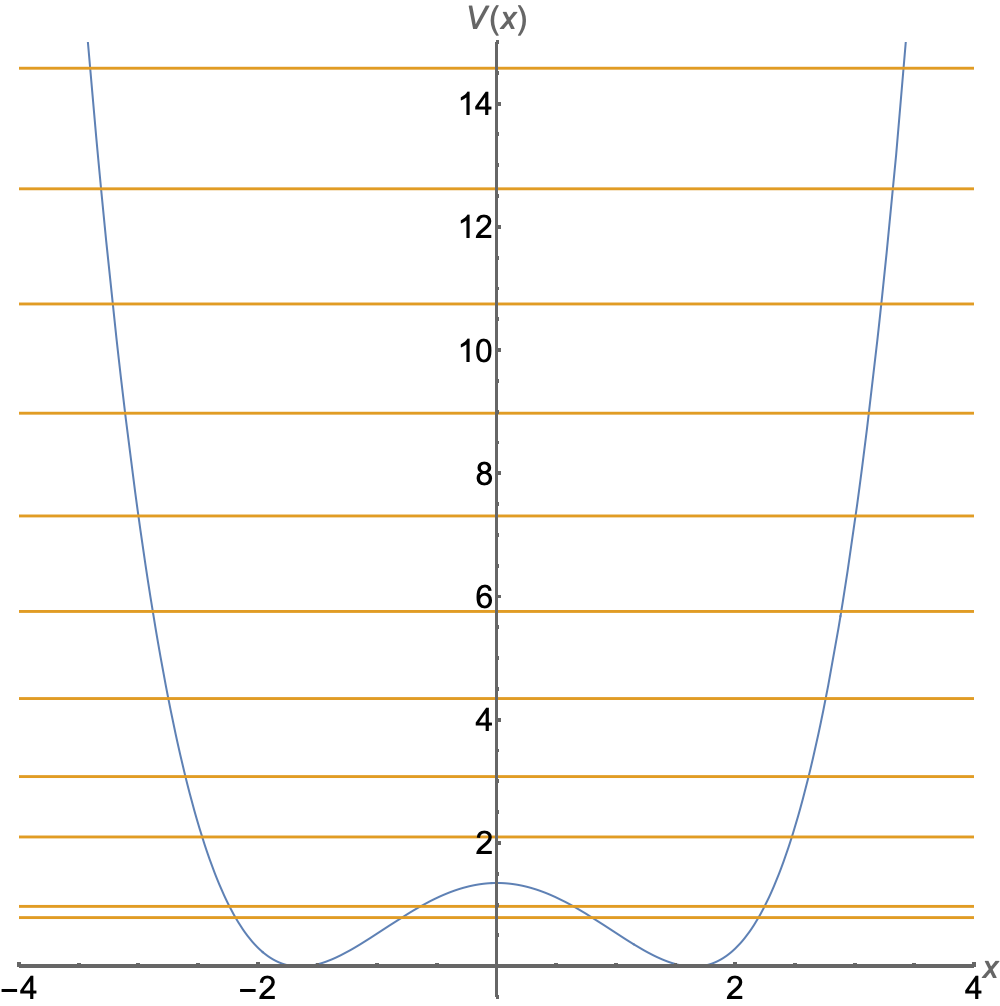}
\caption{Potential of the Hermitian Hamiltonian (\ref{e5.4}) plotted as a function of real $z$ for the case $\alpha=0$ and $g=0.046$ in (\ref{e5.6}). The energy levels are indicated by horizontal lines. Because $\alpha=0$, there is no parity anomaly and the double-well potential is symmetric. Therefore, the mass gap $M$ is very small and consequently there are no bound states at all. This shows that the occurrence of bound states is attributable to the parity anomaly. The eigenvalues shown have the numerical values $0.796906$, $0.978769$, $2.10561$, $3.08558$, $4.35215$, $5.76571$, $7.31558$, $8.98321$, $10.7563$, $12.6253$, $14.5825$.}
\label{f25}
\end{figure}

Setting $\alpha=0$ eliminates the parity-anomaly term in $\tilde H_\alpha$, so the potential becomes a {\it symmetric} double well, as shown in Fig.~\ref{f25}. The mass splitting in a symmetric double well is exponentially small because the splitting is a consequence of semiclassical tunneling between the wells, so the free-particle mass $M$ associated with $H_\alpha$ is small. An elementary WKB calculation shows that for large $n$ the quantum-mechanical energy levels $E_n$ of $\tilde H$ grow like $n^{4/3}$. Thus, $B_n$ in (\ref{e5.5}) is positive for all $n\geq2$ and there cannot be any bound states. Figure~\ref{f25} displays the first eleven quantum-mechanical energies of this system for the case $g=0.046$ and $\alpha=0$ in (\ref{e5.6}).

\begin{figure}
\center
\includegraphics[scale = 0.23]{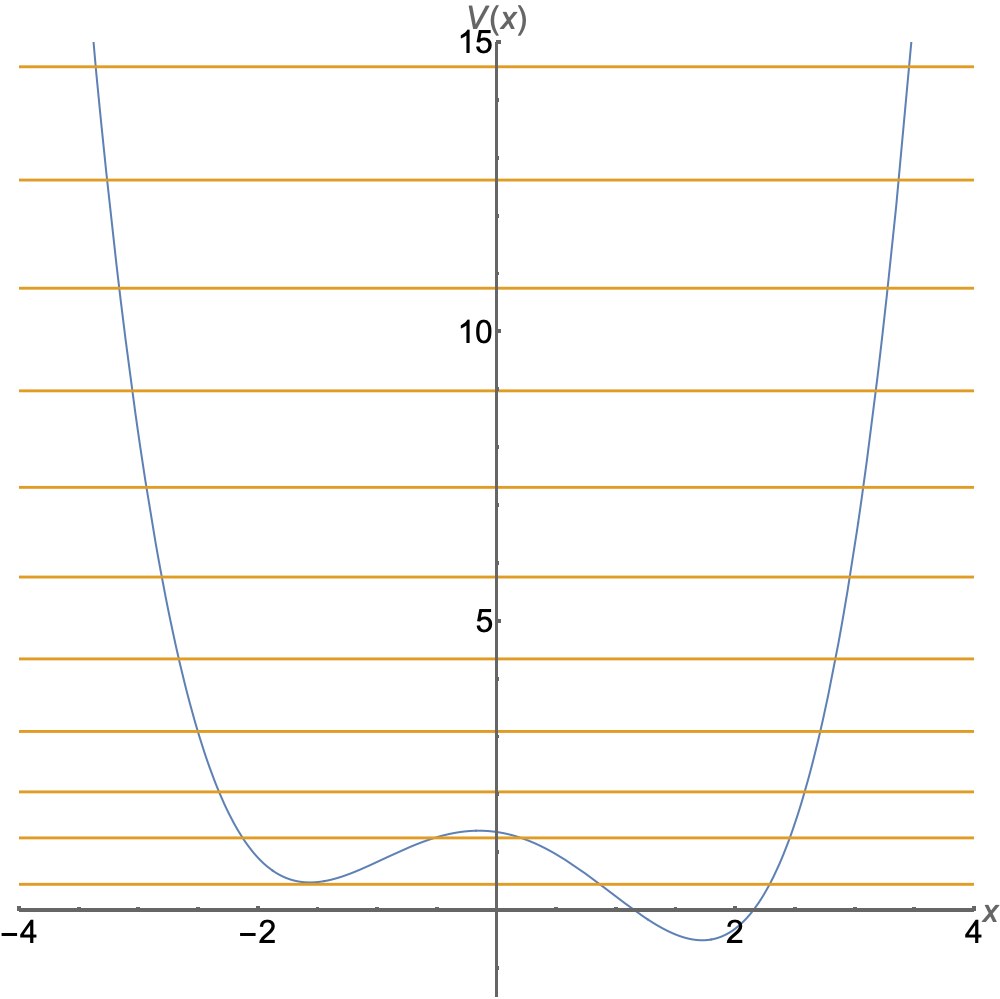}
\caption{Spectrum of the Hamiltonian $\tilde H$ in (\ref{e5.6}) for $g=0.046$, which is slightly smaller than the critical value $g=0.0465$. Unlike Fig.~\ref{f25} where $\alpha=0$, here we have taken $\alpha=1$ so that the parity-breaking anomaly term is present and the potential no longer has a symmetric double well. Now the lowest eleven energy eigenvalues are $0.455521$, $1.25751$, $2.05247$, $3.09657$, $4.35004$, $5.76441$, $7.31388$, $8.98141$, $10.7545$, $12.6235$, $14.5806$. The mass gap $M$ is $1.25751-0.455521=0.80199$. The mass of the two-particle state is $2.05247-0.455521=1.59695$, which is slightly less than $2M$. Therefore, the two-particle state is weakly bound (like a deuterium nucleus). There is only one bound state for this value of $g$.}
\label{f26}
\end{figure}

If $\alpha=1$, the double-well potential becomes {\it asymmetric} so the lowest two states are no longer approximately degenerate (see Fig.~\ref{f26}). As a result, bound states can occur near the bottom of the potential well. However, higher-energy states eventually become unbound because the level spacing increases with increasing $n$. For larger $g$ the number of bound states becomes smaller because the depth of the double well decreases and for sufficiently large $g$ there are no bound states at all. In Fig.~\ref{f26} we display the potential in (\ref{e5.6}) for $\alpha=1$ and $g=0.046$, which is less than the critical value of $g=0.0465$. For this value of $g$ there is just one bound state. 

\begin{figure}
\center
\includegraphics[scale = 0.23]{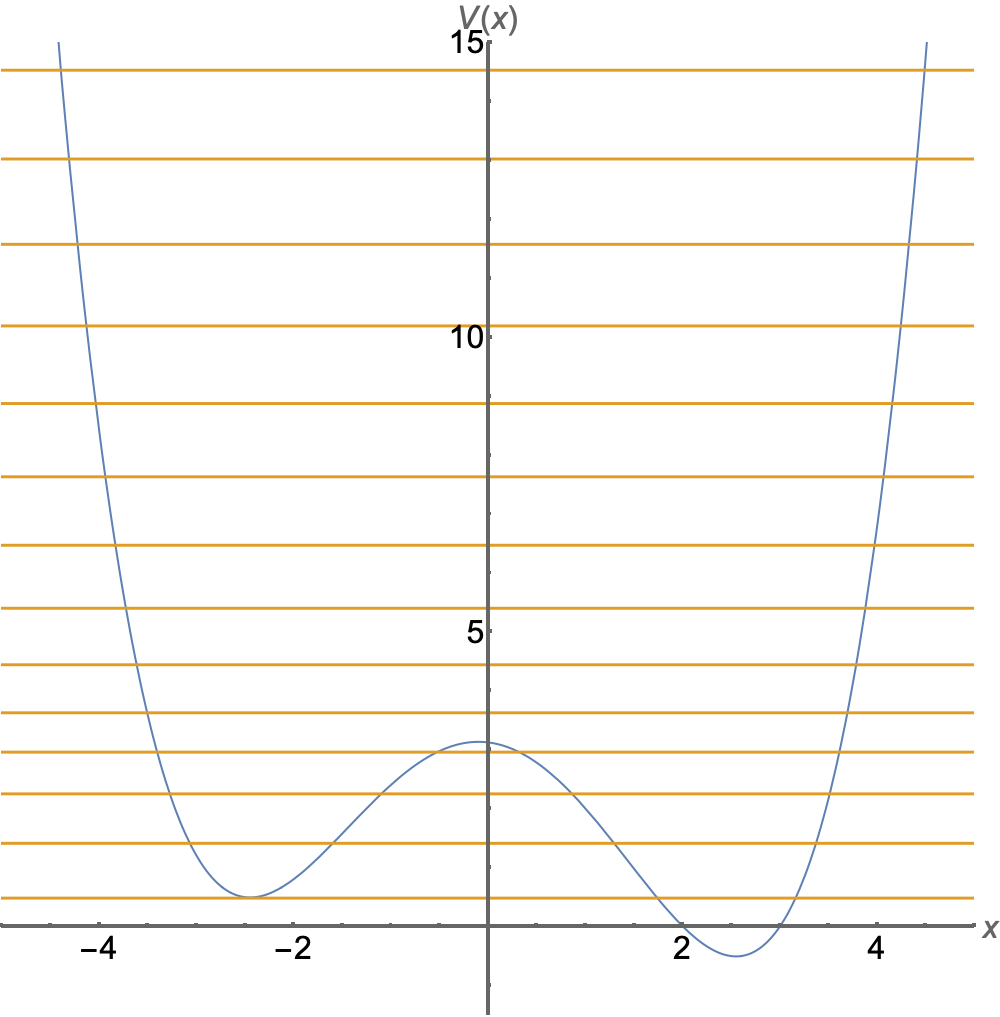}
\caption{Spectrum of the Hamiltonian $\tilde H$ in (\ref{e5.6}) for $g=0.02$ and $\alpha=1$. The first 14 eigenvalues $E_n$ are shown; their numerical values are $0.483726$, $1.41355$, $2.25227$, $2.96041$, $3.62777$, $4.44272$, $5.40171$, $6.47044$, $7.63162$, $8.87451$, $10.1913$, $11.576$, $13.0235$, $14.5299$. From these numbers one can verify that there are seven bound states (see Fig.~\ref{f28}).}
\label{f27}
\end{figure}

\begin{figure}
\center
\includegraphics[scale = 0.23]{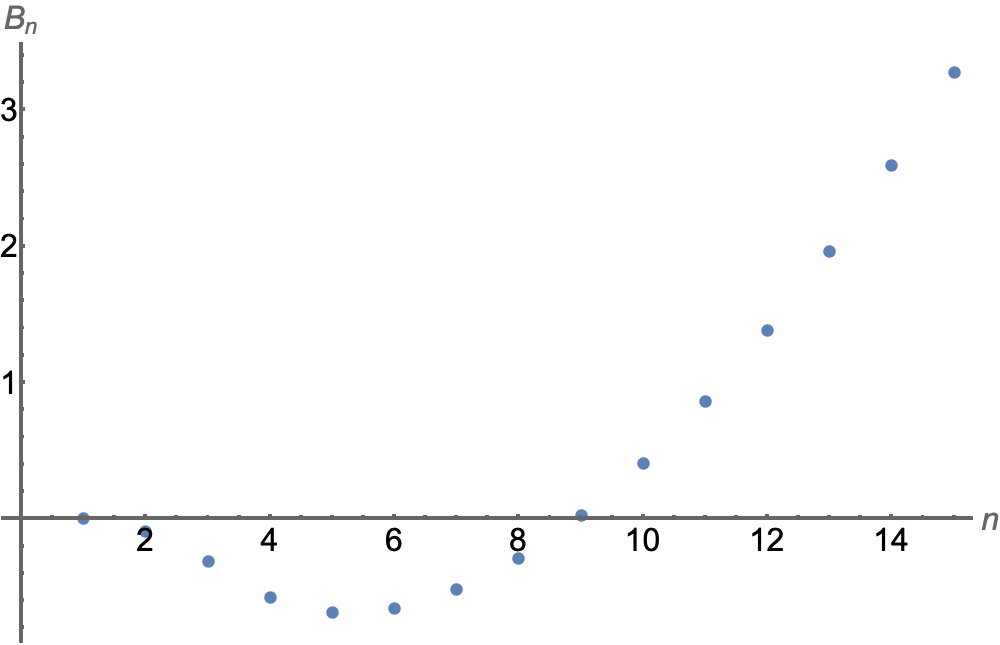}
\caption{Binding energies $B_n=
(E_n-E_0)-nM$ plotted as a function of $n$ for $g =0.008333$ and $\alpha=1$, where $E_n$ are shown in Fig.~\ref{f27}. A negative value of $B_n$ indicates a bound state. There are seven bound states. Note that as $n$ increases, the states become more tightly bound until $n=5$. For larger $n$ the states become less tightly bound and for $n>8$ there are no more bound states.}
\label{f28}
\end{figure}

For smaller values of $g$ the number of bound states increases. In Fig.~\ref{f27} we take $g=0.02$ and $\alpha=1$. In this plot the lowest 14 eigenvalues are shown. There are seven bound states. To see that the spectrum shown in Fig.~\ref{f27} includes seven bound states, we plot in Fig.~\ref{f28} the values of $B_n$ in (\ref{e5.5}) as a function of $n$. Note that the states become more tightly bound as $n$ increases until $n=5$; for higher $n$ the states become more weakly bound, and when $n>8$ there are no more bound states.

\subsection{ Classical symmetric double-well potentials} \label{ss5B}
We have seen in Subsec.~\ref{ss5A} that if we eliminate the parity-breaking anomaly term, which is proportional to $\hbar$, from the quartic Hamiltonian $\tilde H$ in (\ref{e5.4}), we obtain a simpler class of {\it parity-symmetric} anharmonic potentials that no longer exhibit the particle-physics feature of quantum bound states. However, at the {\it classical} level these simple double-well potentials still display a variety of remarkable mathematical behaviors. Specifically, if a classical particle in such a symmetric double-well potential is given a {\it complex} energy, the resulting complex path of the particle may exhibit strange new kinds of resonant tunneling behaviors.

\begin{figure}
\center
\includegraphics[scale = 0.23]{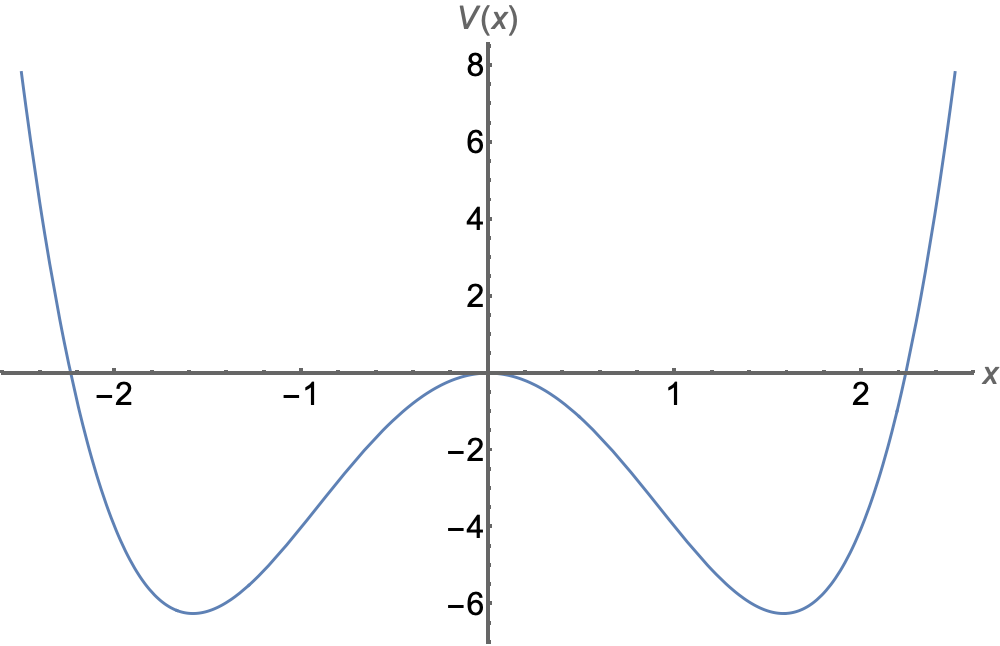}
\caption{Typical example of a symmetric quartic double-well potential. This potential $V(x)=-5x^2 +x^4$, which is associated with the Hamiltonian (\ref{e5.7}), has minima at $x=\pm\sqrt{5/2}$. The depth of the potential is $\tfrac{25}{4}$.}
\label{f29}
\end{figure}

The complex classical tunneling behaviors described in this section appear to be generic, so we restrict our discussion here to just one special case of the Hamiltonian in (\ref{e5.6}) with $\alpha=0$:
\begin{equation}
H=p^2-5x^2+x^4.
\label{e5.7}
\end{equation}
A plot of this parity-symmetric double-well potential for the Hamiltonian (\ref{e5.7}) is given in Fig.~\ref{f29}.

If we put a {\it classical} particle having negative {\it real} energy in one well of the potential $V(x)=-5x^2+x^4$ shown in Fig.~\ref{f29}, the particle will oscillate back and forth between the turning points in that well, but it is classically forbidden to enter the other well. This is true even for complex orbits. To illustrate this, we insert a particle of energy $E=-1$ in various initial positions $x(0)$ in the potential $V(x)$:
\begin{equation}
x(0)=\pm2,~\pm3,~\pm5.
\label{e5.8}
\end{equation}

\begin{figure}
\center
\includegraphics[scale = 0.24]{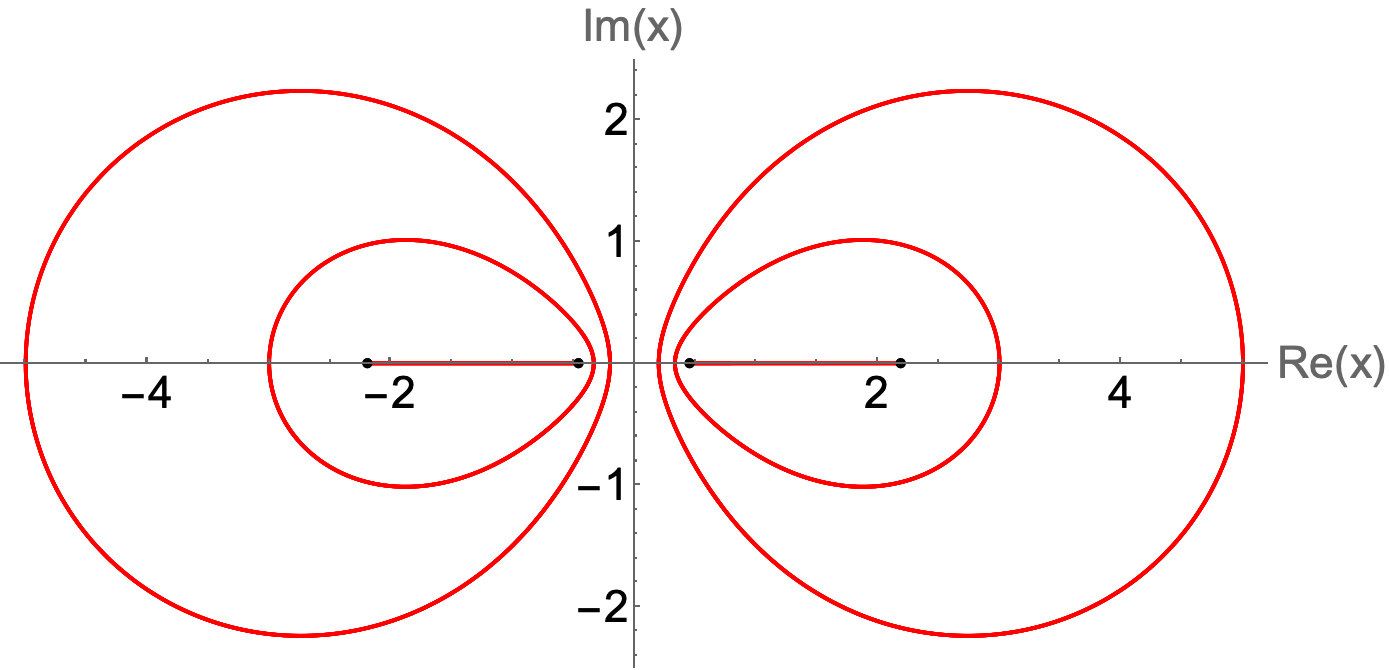}
\caption{Classical trajectories for particles of real energy $E=-1$ in the potential $V(x)=-5x^2+ x^4$, which is plotted in Fig.~\ref{f29}. The four turning points, which are the solutions to the equation $V(x)=-1$, are indicated by dots on the $x$ axis. For the initial condition $x(0)=\pm1$ the particle oscillates between the turning points on the $x$ axis. For the initial conditions $x(0)=\pm3$ and $x(0)=\pm5$ the particle is initially in a classically forbidden region and begins by moving perpendicular to the $x$ axis. The particle then continues in a closed complex orbit that encloses the turning points. In all cases the classical particle can never cross from one well to the other. The set of all orbits on each side of the imaginary axis is space filling and the imaginary axis is a separatrix line.}
\label{f30}
\end{figure}

For the initial condition $\pm2$, the particle remains on the $x$ axis and oscillates between the turning points, as shown in Fig.~\ref{f30}. For the initial condition $\pm3$ or $\pm5$, the particle is initially in a classically forbidden region; in this case the particle travels in a closed {\it complex} orbit that encloses the turning points. In all cases the particle cannot cross from one well to the other. The orbits associated with the left (right) well stay in the left (right) complex-$x$ plane. If the energy of the particle is real, the path of the particle cannot cross the separatrix line ${\rm Re}\,x=0$. Thus, if the energy of the particle is real, tunneling between the left and right wells is strictly forbidden.

\begin{figure}
\center
\includegraphics[scale = 0.24]{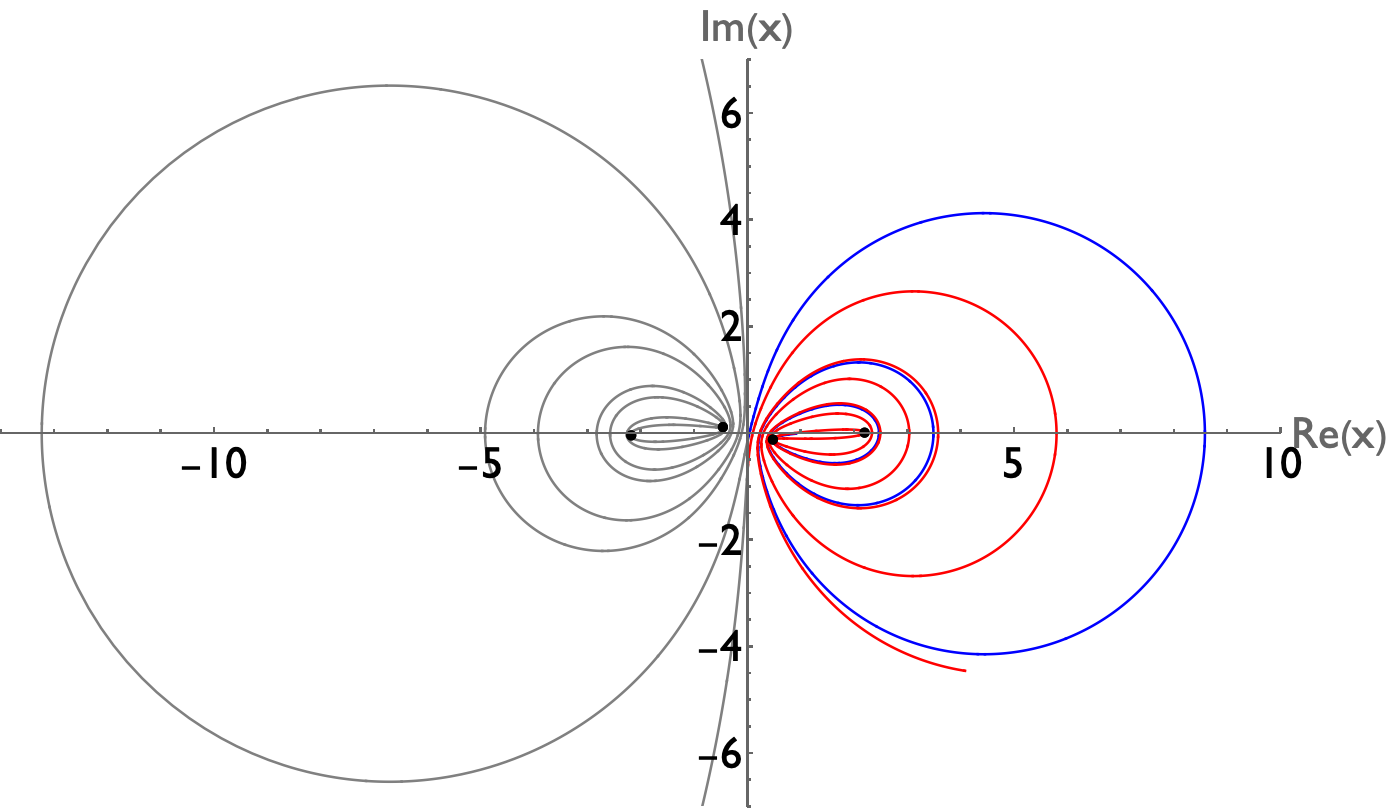}
\caption{Classical analog of quantum tunneling. A particle in the potential $V(x)=-5x^2+x^4$ in Fig.~\ref{f29} is given a complex energy $E=-1+0.5\,i$ and its initial position is $x(0)=0.9$. The particle spirals outward (blue trajectory) and at $t=4.34$ it crosses the imaginary-$x$ axis. (A particle having real energy can never cross the separatrix on the imaginary axis, as Fig.~\ref{f30} shows.) The particle trajectory (now colored gray) spirals inward, crosses the real axis between the left pair of turning points (black dots), and spirals outward. It then crosses the imaginary axis at $t=14.73$ and the trajectory (now colored red) spirals inward, crosses the real axis between the right pair of turning points, and spirals back outward. The plot of the red trajectory terminates at $t=23.56$.}
\label{f31}
\end{figure}

The behavior of the classical orbits shown in Fig.~\ref{f30} is not surprising. However, if the energy of the classical particle is {\it complex}, the classical particle can {\it tunnel} from well to well. To demonstrate this, we take $E=-1+0.5\,i$ and choose the initial condition $x(0)=0.9$. As shown in Fig.~\ref{f31}, the particle spirals outward in the right well and crosses the imaginary axis. It then spirals inward, crosses the real axis between the left pair of turning points, spirals outward and crosses the imaginary axis to the right well, and continues this process. This is a classical analog of quantum tunneling \cite{pt513}. (We have already seen complex classical tunneling in Fig.~\ref{f12}.)

\begin{figure}
\center
\includegraphics[scale = 0.24]{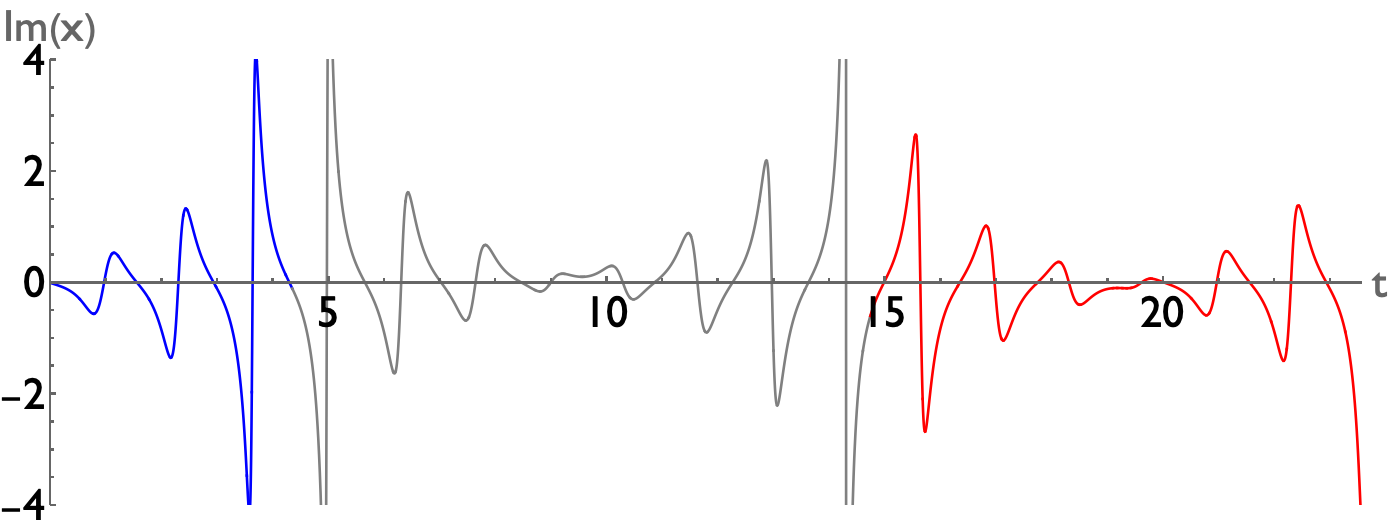}
\caption{Real-time interpretation of Fig.~\ref{f31}. The {\it imaginary part} of the complex trajectory $x(t)$ is plotted as a function of time $t$. As in Fig.~\ref{f31}, the energy of the classical particle is $E=-1+0.5\,i$ and the initial condition is $x(0)=0.9$, and the color coding is unchanged: $0\leq t_{\rm blue}\leq 4.34$; $4.34\leq t_{\rm gray}\leq 14.73$; $14.73\leq t_{\rm red}\leq 23.56$.}
\label{f32}
\end{figure}

To illustrate this classical tunneling process more clearly, we plot in Fig.~\ref{f32} the imaginary part of $x(t)$ as a function of time. Note that the time spent in each well varies somewhat. However, if we average the times spent in each well from the time the particle crosses the imaginary-$x$ axis until it recrosses the imaginary axis and designate this average dwell time as $T$, we find that $T$ is inversely proportional to ${\rm Im}\,E$. {\it This is the complex classical equivalent of the time-energy uncertainty principle} \cite{pt513,pt591}.

For each initial condition $x(0)$ there is a unique
complex-energy particle trajectory in the complex-$x$ plane, and Fig.~\ref{f31} {\it suggests} that a typical complex trajectory is infinitely long, nonintersecting, and space filling. The set of these complex trajectories possesses some unusual topological features. For example, we can plot a classical trajectory for long times and record the points on the imaginary axis where the trajectory crosses this axis. We find that there are intervals containing these points inside of which all trajectories crossing this axis represent particles that are moving in the same direction. In one such case the imaginary axis contains just two such intervals; all trajectories crossing above a point on the imaginary axis represent particles traveling to the right and all trajectories below this point represent particles traveling to the left.

In other cases we find adjacent (arbitrarily close) trajectories crossing the imaginary axis on which particles are moving in {\it opposite directions}. This can happen because the classical equation of motion (Newton's law) is {\it second order} in time, so two trajectories can be arbitrarily close to one another and yet represent particles moving in opposite directions without violating continuity. (This cannot happen in first-order linear dynamical systems.) To understand this nonviolation of continuity, note that the Hamiltonian equation of motion has the form $p^2+ V(x)=E$. Thus, if we are given the complex position $x(t)$ of the particle at time $t$, we know the {\it square} of its velocity $p^2$ at time $t$. Hence, the particle can be moving in one of two directions. The choice of direction is not determined {\it locally} by the equation of motion; rather, it is determined {\it globally} by the initial condition $x(0)$.

It is especially surprising that for  each real value of the energy ${\rm Re}\,E$ there are special unique imaginary values of the energy ${\rm Im}\,E$ for which {\it the classical trajectories are not infinitely long}. These finite-length particle trajectories start at a turning point in one well, unwind, cross the imaginary axis, spiral inward, and {\it terminate} in finite time at a turning point in the other well. These classical trajectories oscillate between pairs of turning points in different wells and they only occur for isolated discrete values of ${\rm Im}\,E$.

\begin{figure}
\center
\includegraphics[scale = 0.24]{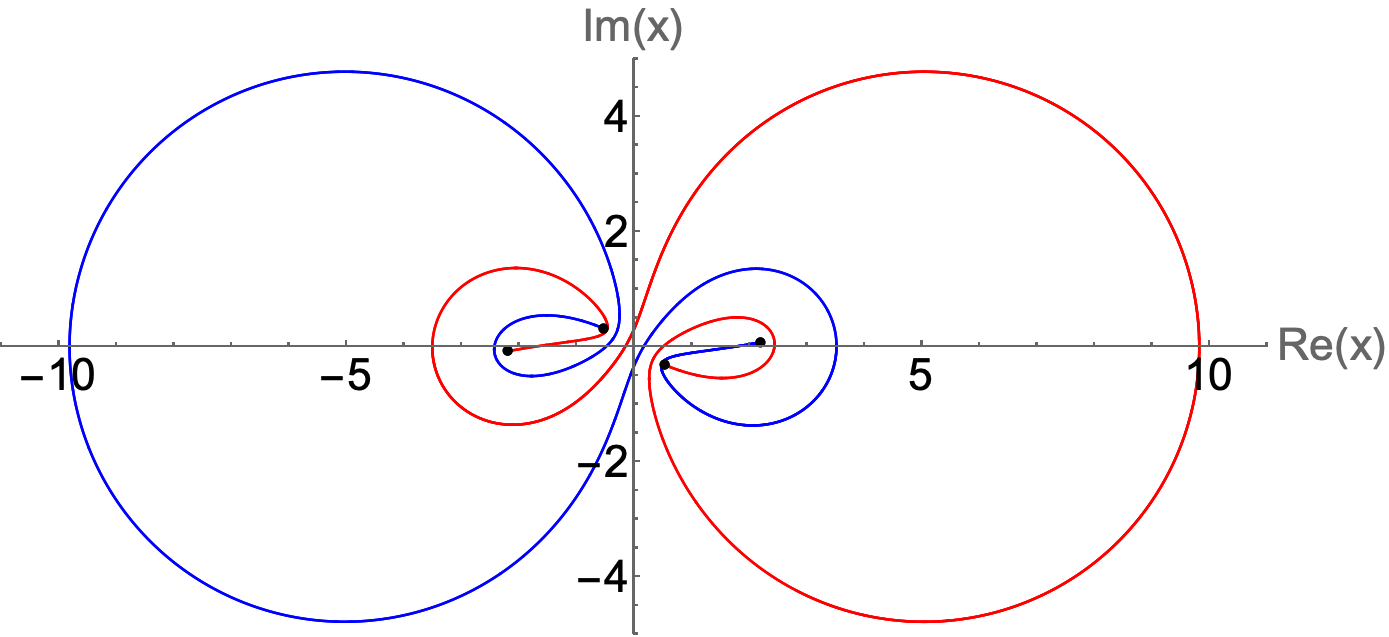}
\caption{Two finite-length classical trajectories that oscillate between a turning point in the left well and a turning point in the right well. This strange resonant behavior is reminiscent of semiclassical quantization. Given a value of the real part of the energy in the well, these finite-length trajectories only occur for isolated discrete values of the imaginary part of the energy. In this picture the classical particle is moving in the potential 
$V(x)=-5x^2+x^4$ shown in Fig.~\ref{f29} and it has complex classical energy $E=-1+ 1.57321\,i$. The trajectories oscillate between turning points at $-0.53377+0.31818\,i$ and 
$-2.19598-0.07734\,i$ in the left well and at $0.53377-0.31818\,i$ and $2.19598+0.07734\,i$ in the right well. Note that this configuration is symmetric if we replace $x$ by $-x$ and interchange the red and blue curves. The oscillation period of each trajectory is $P=8.49314\,$.}
\label{f33}
\end{figure}

Figure~\ref{f33} shows a pair of such paths for the case $E=-1+1.57321i$. The unusual classical trajectories in this figure are complex classical resonant analogs of the semiclassical WKB quantization condition [such as that in (\ref{e3.20})]. If the energy were real, say $E=-1$, the classical particle would be confined to one of the two potential wells in Fig.~\ref{f29}; the classical path would resemble one of the trajectories shown in Fig.~\ref{f30}. Ordinarily, if the energy had a generic nonzero imaginary value $E=-1+ai$, the trajectory would resemble the infinitely long trajectory in Fig.~\ref{f31}. However, for this special nonzero imaginary value of the energy, the particle oscillates from well to well along a {\it finite-length} path in the complex plane like a resonant quantum particle. The period of oscillation of the trajectory in Fig.~\ref{f33} is $P= 8.49314\,$.

\begin{figure}
\center
\includegraphics[scale = 0.24]{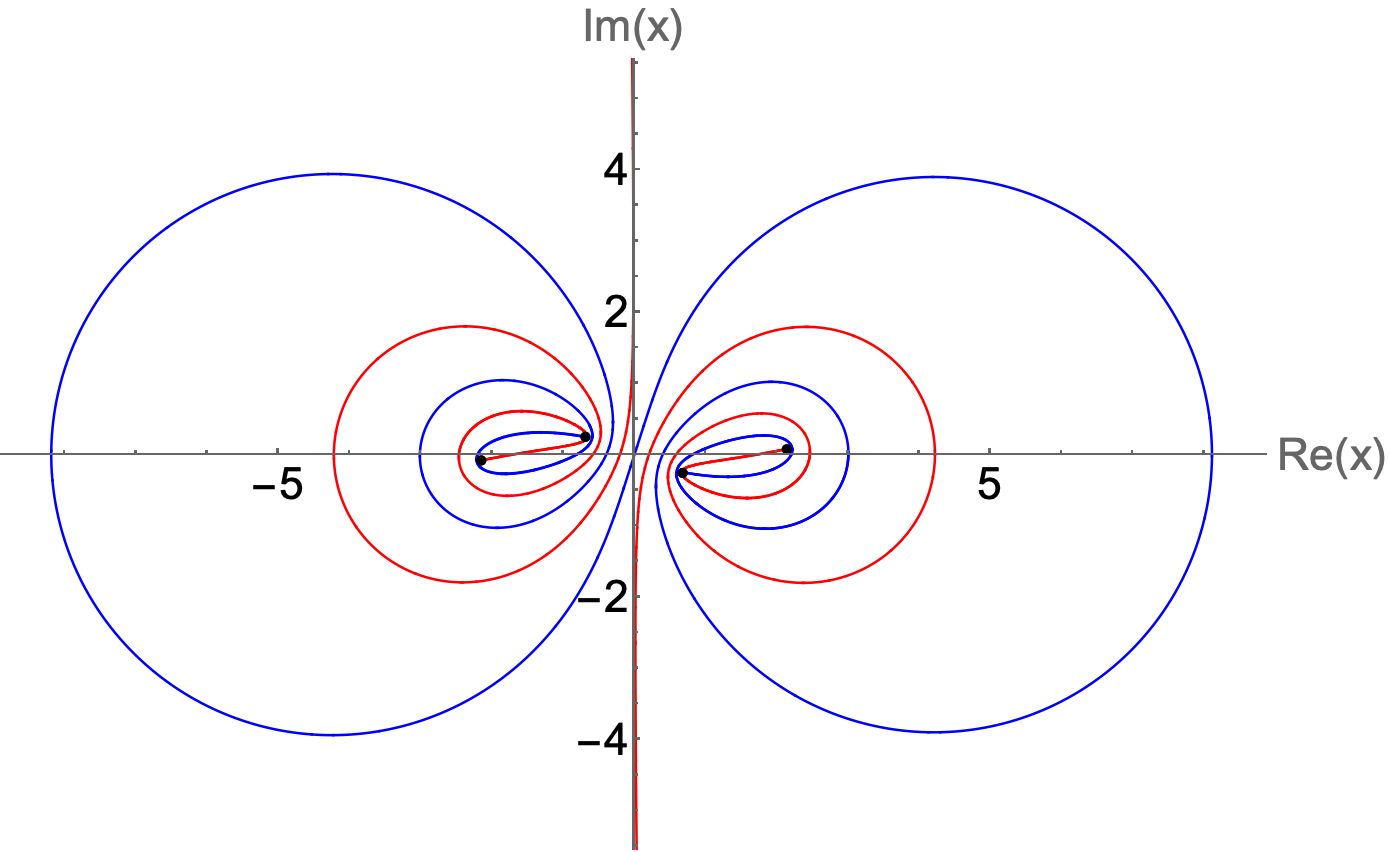}
\caption{Two classical trajectories that oscillate between pairs of turning points, one in the left well and one in the right well of the potential shown in Fig.~\ref{f29}. In this figure the classical energy is $-2+1.42511\,i$, so the real part of the energy is lower in the well than in Fig.~\ref{f33}. The classical paths shown in this figure differ dramatically from those in Fig.~\ref{f33}. While the blue path has finite length, the red path is infinitely long; the red particle zooms off to $\pm i\infty$ and comes back from $\mp i\infty$ in finite time. Unlike the paths in Fig.~\ref{f33}, the red and the blue paths here are {\it independently} symmetric under complex parity: $x\to-x$. The turning points in this figure are located at $-2.1438-0.079533\,i$ and $-0.687039+ 0.248171\,i$ in the left well and at $0.687039-0.248171\,i$ and $2.1438+ 0.079533\,i$ in the right well. Note that the red trajectory manages to go from one well to the other without ever crossing the imaginary axis!}
\label{f34}
\end{figure}

If we take the real part of the energy to be $-2$, the particle is deeper in the potential well in Fig.~\ref{f29} and the resonant complex classical particle takes longer to oscillate from well to well. The precise imaginary part of the energy at which resonant tunneling occurs is ${\rm Im}\, E=1.42511...$ and the oscillation period is $P= 13.9754...\,$. The classical path of this particle is shown in Fig.~\ref{f34}. The period of this motion is longer than the period in Fig.~\ref{f33} because the particle requires two turns in each well before it can escape.

Although the red and blue paths in Fig.~\ref{f34} have the same period, their topology is different: The blue particle trajectory follows a continuous {\it finite-length} path that begins at a turning point in one well, makes two rotations, crosses the imaginary axis {\it at the origin}, makes two more rotations, and terminates at a turning point in the other well. However, the red particle trajectory begins at a turning point, makes two rotations, and {\it runs along the imaginary axis to $\pm i\infty$} in finite time! The particle then comes back along the imaginary axis from $\mp i\infty$, executes two rotations in the other well, and ends at a turning point in the same time as the blue particle. This particle trajectory goes from one well to the other well {\it without crossing the imaginary axis between the wells}! (This path along the imaginary axis is reminiscent of the path out to $\pm \infty$ along the {\it real axis} of the limiting trajectory in Fig.~\ref{f18}.) Each trajectory in Fig.~\ref{f34} is separately invariant under the complex parity reflection $x\to-x$.

\begin{figure*}
\center
\includegraphics[scale = 0.48]{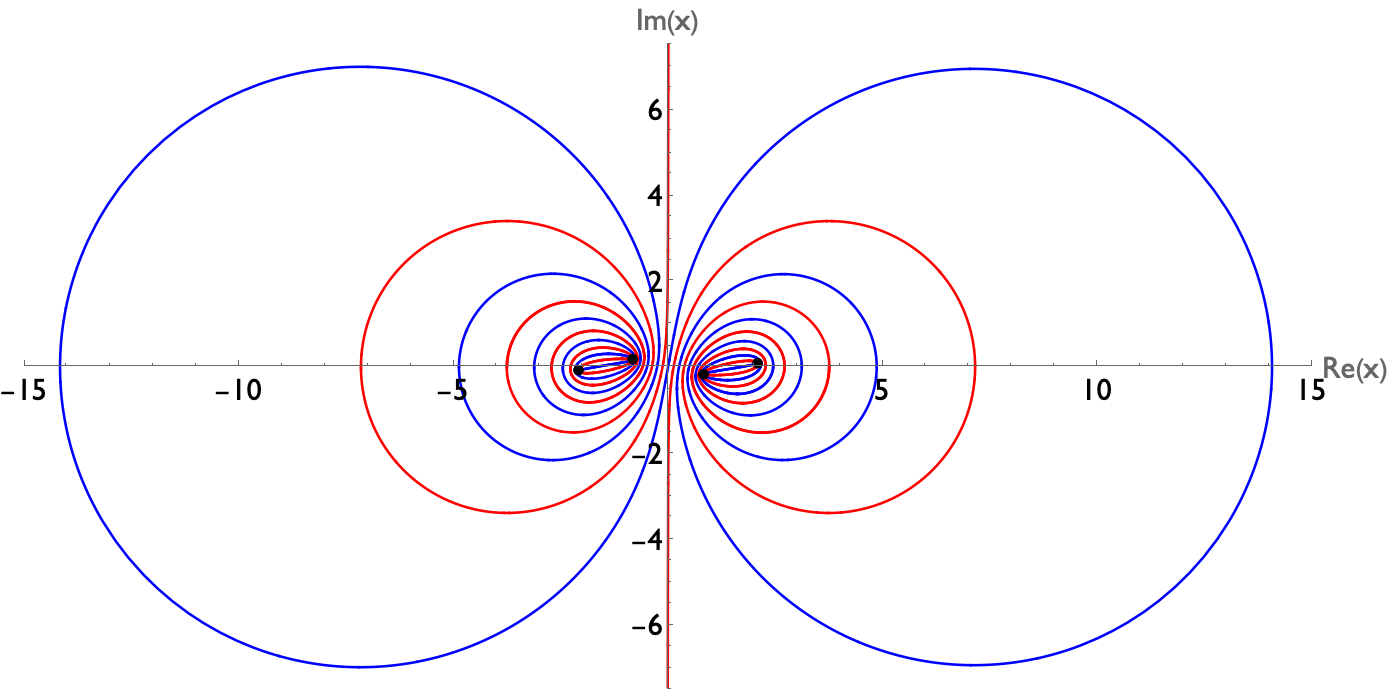}
\caption{Two finite classical trajectories that oscillate between pairs of turning points in the left and right wells in Fig.~\ref{f29}. The energy of the classical particles is $-3+1.13303\,i$, so the real part of the energy is lower down in the well than in Fig.~\ref{f34}. As in Fig.~\ref{f34}, the red path runs up and down the imaginary axis and is infinitely long but the total time of flight is finite. The trajectories terminate at the turning points, which are located at $-2.082-0.0744012\,i$ and $-0.839548-0.184508\,i$ in the left well and at $0.839548 -0.184508\,i$ and $2.082+ 0.0744012\,i$ in the right well. As in Fig.~\ref{f34}, each trajectory is separately symmetric under complex parity $x\to-x$. Again, the red trajectory goes from one well to the other without crossing the imaginary axis.}
\label{f35}
\end{figure*}

If the real part of the energy of the particle is ${\rm Re}\,E=-3$, the particle is even deeper in the well and it takes even longer for the particle to escape from the well; it does so resonantly if the imaginary part of the energy is ${\rm Im}\, E=1.13303...\,$. For this special case Fig.~\ref{f35} shows that the blue trajectory makes {\it five} spiral rotations before it escapes from the well and crosses the imaginary axis into the other well. The red trajectory also makes five rotations before leaving the well and going out to $\pm\infty$, but like the red trajectory in Fig.~\ref{f34} it reaches the well on the other side of the imaginary axis in the same time as the blue trajectory without crossing the imaginary axis. The period of this motion is $P=24.5576...\,$.

Finding the discrete complex {\it classical} energies for this symmetric double-well problem is analogous to finding the discrete {\it quantum} energies in a symmetric potential well. Like the quantum case, the eigenstates in the classical case are finite-period curves connecting the turning points and they are also states of the parity operator. This enforces the observation in \cite{pt513} regarding the similarity between quantum mechanics and complex classical mechanics.

\subsection{ $\cPT$-symmetric quasi-exactly solvable Hamiltonians} \label{ss5C}
In studying $\cPT$-symmetric Hamiltonians one may find new non-Hermitian quantum-mechanical models that exhibit unusual behaviors that were previously discovered for Hermitian Hamiltonians. For example, Hermitian Hamiltonians have been found for which it is possible to calculate exactly some (but not all) of the eigenfunctions and corresponding eigenvalues. Such Hamiltonians are said to be {\it quasi-exactly solvable} (QES). 

Before the introduction of $\cPT$-symmetric Hamiltonians in 1998, it was believed that the lowest-degree one-dimensional QES polynomial potential is sextic \cite{pt157}. This belief was based on the assumption that a Hamiltonian must be Hermitian. However, the quartic $\cPT$-symmetric anharmonic potential
\begin{equation}
V(x)=-x^4+2iax^3+(a^2-2b)x^2+2i(ab-J)x,
\label{e5.9}
\end{equation}
where $a$ and $b$ are continuous real parameters and $J$ is a positive integer, provides a simpler and even more general two-parameter family of QES quartic potentials \cite{pt338}. This potential is an additive generalization of the $-x^4$ potential discussed in Subsec.~\ref{ss4A}. The quasi-exactly solvable spectra of this family of quartic potentials are real, discrete, and bounded below, and the quasi-exact portion of the spectra consists of the lowest $J$ eigenvalues. These eigenvalues are the roots of a $J$th-degree polynomial.

To find the eigenfunctions and eigenvalues of $H=p^2+V(x)$, we substitute
\begin{eqnarray}
\psi(x)=e^{-ix^3/3-ax^2/2-ibx}P_{J-1}(x),
\label{e5.10}
\end{eqnarray}
where
\begin{eqnarray}
P_{J-1}(x)=x^{J-1}+\sum_{k=0}^{J-2}c_k x^k
\label{e5.11}
\end{eqnarray}
is a polynomial of degree $J-1$, into the Schr\"odinger equation
\begin{eqnarray}
&&\!\!\!\!\!\!\!\!\!-\psi''(x)+\big[-x^4+2iax^3+(a^2-2b)x^2\nonumber\\
&&\quad+2i(ab-J)x\big]\psi(x)=E\psi(x)
\label{e5.12}
\end{eqnarray}
associated with the QES potential $V(x)$ in (\ref{e5.9}).

The energy eigenvalues associated with $J=1$, $2$, and $3$ are the zeros of the increasingly higher-order polynomial equations
\begin{eqnarray}
Q_1(E) &=& E -b^2 -a,\nonumber\\
Q_2(E) &=& E^2 -\big(2b^2+4a\big)E+b^4+4ab^2-4b+3a^2,\nonumber\\
Q_3(E) &=& E^3 -\big(3b^2+9a\big)E^2+\big(3b^4+18ab^2-16b\nonumber\\
&&\quad +23a^2\big)E-b^6-9ab^4+16b^3-23a^2b^2\nonumber\\
&&\quad +48ab-15a^3-16.
\label{e5.13}
\end{eqnarray}
The roots of $Q_J(E)$ are the QES portion of the spectrum of $H$. It has been shown numerically that the nontrivial (non-QES) part of the spectrum is entirely real throughout the $(a,b)$ plane \cite{pt338}.

\subsection{Hermitian and $\cPT$-symmetric anharmonic oscillators}\label{ss5D}
As mentioned previously, for decades the quartic anharmonic oscillator has served as a intellectual laboratory for understanding physical mechanisms and developing calculational tools. This subsection summarizes what has been learned from the study of the Hermitian and the $\cPT$-symmetric anharmonic oscillators.

\subsubsection{Hermitian anharmonic oscillator}
The motivation to study the Hermitian anharmonic oscillator can be traced back to the classic \textcite{pt584} paper on the divergence of perturbation theory in quantum electrodynamics (QED). This paper argued that the weak-coupling perturbation expansion in QED, which is a formal series in powers of the fine-structure constant $\alpha\approx\tfrac{1}{137}$, is divergent.

The reasoning in Dyson's paper goes as follows. The classical Coulomb force between a positive and a negative charge is proportional to $\alpha$ and is attractive. But, if the sign of $\alpha$ were changed, this force would become repulsive. A convergent perturbation series in powers of $\alpha$ would define an {\it analytic} (smooth) function of $\alpha$ in the neighborhood of $\alpha =0$. This neighborhood would contain both positive-real and negative-real values of $\alpha$.

The vacuum state (ground state) in QED is stable for positive $\alpha$ and consists of a cloud of virtual electron-positron pairs. However, for negative $\alpha$, like charges would attract and unlike charges would repel, so virtual pairs of electrons and positrons in the vacuum would repel. This suggests that the {\it vacuum state for $\alpha<0$ would be unstable}; it would tunnel to a lower-energy state containing growing heaps of electrons and positrons at opposite ends of the universe. Hence, one would expect there to be an abrupt discontinuity in the physics at $\alpha=0$. This implies that a perturbation series expansion in powers of $\alpha$ would have a zero radius of convergence.

The conclusion of Dyson's novel argument is correct: In general, perturbation series in powers of a coupling constant diverge in quantum mechanics and in quantum field theory. However, the subtle problem with this argument, as the study of the $\cPT$-symmetric anharmonic oscillator shows, is that a repulsive potential can still confine an infinite spectrum of real-energy bound states. In particular, the lowest-energy state (the vacuum state) of a repulsive potential can be stable.

The two anharmonic-oscillator Hamiltonians discussed at the beginning of this section reveal a possible flaw in Dyson's argument. The Hermitian Hamiltonian (\ref{e5.1}) has a positive coupling constant and an attractive potential while the $\cPT$-symmetric Hamiltonian (\ref{e5.3}) has a negative coupling constant and a repulsive potential (on the real axis). Yet the eigenvalues of {\it both} anharmonic-oscillator Hamiltonians are entirely real and positive and the states are stable.

Nevertheless, as Dyson argued, the point $\alpha=0$ in the complex-$\alpha$ plane is indeed a singularity. This singular point is the boundary between two {\it different theories}, one with $\alpha>0$ for which there there is an attractive force between an electron and a positron and the other with $\alpha<0$ for which there is a repulsive force between an electron and a positron. However, the vacuum state in both theories could well be stable.

There is a simple reason why quantum-mechanical and field-theoretic perturbation expansions diverge: The $n$th term in the perturbation expansion is a sum of Feynman diagrams having $n$ vertices. The average size of an $n$-vertex Feynman diagram is typically of order $c^n$, where $c$ is a constant, but the {\it number} of Feynman diagrams grows {\it factorially} for large $n$. This is why perturbation series diverge.

This factorial growth is a direct consequence of graph counting and combinatorics. Following the publication of Dyson's seminal paper, many papers were published in which the perturbation expansions for various field-theoretic models were shown to be divergent \cite{pt588,pt589, pt590,pt598}. These papers established bounds on the size of Feynman diagrams and estimated the growth in the number of such diagrams. Because of this broad interest in the nature of perturbation theory, the Hermitian anharmonic oscillator was identified as an ideal model in which to study perturbation theory in depth.

An analysis of the Hermitian anharmonic oscillator in 1968 showed that the underlying reason for the divergence of perturbation series is the presence of square-root branch-point singularities in the complex coupling-constant plane (the complex-$g$ plane) \cite{pt15,pt14}. Originally, these singularities were discovered in the context of the Hermitian anharmonic oscillator model (\ref{e5.1}), but the appearance of square-root singularities in the complex-coupling constant plane is a generic phenomenon. These singularities, originally called {\it Bender-Wu singularities}, are now usually referred to as {\it exceptional points}. Bender and Wu showed that there are infinite sequences of square-root singularities in the complex-$g$ plane and that these sequences approach the origin $g=0$. Thus, the origin in the coupling-constant plane is a {\it nonisolated} singularity. This explains at a deep level why perturbation series in powers of a coupling constant have a vanishing radius of convergence.

The Bender-Wu singularities provide a clear {\it physical} explanation of the divergence of perturbation series. At each square-root branch-point singularity two different energy levels {\it cross}; that is, two energy levels analytically continue into one another. Thus, in perturbative eigenvalue problems an eigenvalue spectrum is not just a set of unrelated energies $\{E_n(g)\}$. Rather, there is an analytic {\it eigenvalue function} $E(g)$ that is defined on a multisheeted Riemann surface. The $n$th eigenvalue $E_n(g)$ is the value of the complex function $E(g)$ on the $n$th sheet of the Riemann surface. We arrive at an elegant picture of quantization: The quantization of eigenvalues is topological; the $n$th eigenvalue corresponds to the $n$th sheet of a complex-$g$ Riemann surface.

\subsubsection{Elementary model of topological quantization}
The $2\times2$ Hermitian matrix Hamiltonian
\begin{equation}
H=\left(\begin{array}{cc} a & 0 \\ 0 & b \end{array}\right)+g\left(\begin {array}{cc}0 & 1 \\ 1 & 0 \end{array}\right)
\label{e5.14}
\end{equation}
illustrates this topological picture of quantization very clearly. This Hamiltonian describes two states of energy $a$ and $b$ that interact with coupling strength $g$. To calculate the eigenvalues of $H$, we construct the secular equation, which is the quadratic polynomial
\begin{equation}
{\rm det}(H-IE)=E^2-E(a+b)+ab-g^2.
\label{e5.15}
\end{equation}
The roots of this polynomial are the two eigenvalues $E_\pm(g)$ of $H$:
\begin{equation}
E_\pm(g)=\half\left(a+b\pm\sqrt{(a-b)^2+4g^2}\right).
\label{e5.16}
\end{equation}

The eigenvalues $E(g)$ are real and distinct when $g$ is real because $H$ in (\ref{e5.14}) is Hermitian. However, as functions of complex $g$, these eigenvalues become analytic continuations of one another. The function $E(g)$ has two square-root branch-point singularities in the complex-$g$ plane; these branch points lie on the imaginary-$g$ axis at
\begin{equation}
g=\pm\tfrac{1}{2}|a-b|i.
\label{e5.17}
\end{equation}
Thus, the function $E(g)$ is defined on a two-sheeted Riemann surface. These two sheets are joined at the branch cut connecting the two branch points.

This complex picture of quantization is completely general and extends from matrix Hamiltonians to continuum Hamiltonians. For the general perturbed Hamiltonian 
\begin{equation}
H=p^2+V(x)+gW(x),
\label{e5.18}
\end{equation}
if we construct the perturbation series
\begin{equation}
E_N(g)\sim\sum_{n=0}^\infty a_ng^n
\label{e5.19}
\end{equation}
for the $N$th eigenvalue of $H$, the regions of convergence of these series are determined by square-root singularities in the complex-$g$ plane where the $N$th eigenvalue crosses with other eigenvalues \cite{pt615}.

There are rare examples of Hamiltonians for which the singularities are not square-root singularities; there may be cube-root or other higher-order algebraic singularities. [See, for example, the study of the $D$-dimensional square-well in \textcite{pt18}.] However, the generic picture is universal: At branch-point singularities the eigenvalues cross; each eigenvalue is associated with a sheet in a complex-$g$ Riemann surface.

This discussion of perturbation theory suggests a comprehensive research project that has not yet been published. It would be interesting to locate and classify the branch-point singularities in the complex-$g$ plane associated with the quartic $\cPT$-symmetric Hamiltonian (\ref{e5.3}) (as well as for other $\cPT$-symmetric Hamiltonians).

\subsubsection{Large-order behavior of perturbation theory}
The Hermitian anharmonic oscillator is an ideal model to study the {\it large-order behavior} of the coefficients in a perturbation series; that is, the asymptotic growth of the coefficients $a_n$ in (\ref{e5.19}) as $n\to \infty$. Complex-variable theory determines the precise growth of these coefficients.

The eigenvalues of the Hermitian oscillator (\ref{e5.1}) are {\it analytic} functions of the coupling constant $g$ on the principal sheet of the complex-$g$ plane except for a cut on the negative-$g$ axis \cite{pt17, pt616,pt597}. This discovery led to the derivation of a dispersion relation that determines the precise large-order behavior of perturbation coefficients \cite{pt641}. For example, the large-$n$ asymptotic behavior of the perturbation coefficients $a_n$ for the ground-state energy $E_0(g)$ of $H$ in (\ref{e5.1}) is
$$a_n\sim\frac{m}{2}\frac{\sqrt{6}}{\pi^{3/2}}(-1)^{n+1}\Gamma\big(n+\tfrac{1}{2}\big)\left(\frac {24g}{m^3}\right)^n\quad (n\to\infty).$$
The dispersion-relation technique for finding the large-order behavior of perturbation coefficients became known as {\it Bender-Wu theory}. Bender-Wu techniques apply to both $\cPT$-symmetric and Hermitian theories, and they apply in quantum field theory as well as in quantum mechanics \cite{pt601, pt602,pt603,pt604}.

It is necessary to know the large-order behavior of the perturbation coefficients in order to justify using {\it summation techniques}, such as Borel summation and Pad\'e approximation, to {\it sum} divergent perturbation series and obtain arbitrarily accurate results \cite{pt605,pt17, pt607,pt480}. Summation techniques are effective for both Hermitian and $\cPT$-symmetric theories.

\subsubsection{Summation of divergent perturbation series}
An important property of real summable divergent perturbation series is that the coefficients in the series alternate in sign. If this condition holds, Borel and Pad\'e summation typically converge to a real number. If the coefficients in the perturbation series are real but do not alternate in sign, the summation may not converge or may converge to a complex number even if the perturbation series represents a real number.

The {\it cubic} anharmonic-oscillator Hamiltonian
\begin{equation}
H=p^2+x^2+gx^3\quad (g~{\rm real})
\label{e5.20}
\end{equation}
has perturbation expansions that do {\it not} oscillate in sign. For example, the perturbation series about the harmonic-oscillator ground-state energy $E_0(0)=1$ has the form
\begin{equation}
E_0(g)\sim\sum_{n=0}^\infty a_n g^{2n}\quad(g\to0). 
\label{e5.21}
\end{equation}

This series in powers of $g^2$ diverges; the perturbation coefficients $a_n$ are real and grow factorially with $n$. However, the $a_n$ do not alternate in sign with increasing $n$, so this series is not uniquely Borel or Pad\'e summable. The sign of the imaginary part of the Borel sum may be either positive or negative. The series coefficients are real but the series represents a {\it complex} number. This is because the eigenstate of this cubic Hamiltonian either grows or decays in time depending on the boundary conditions on the eigenfunctions.

\subsubsection{Perturbation series for $\cPT$-symmetric oscillators}
If we replace $g$ by $ig$ in the Hamiltonian (\ref{e5.20}) and in the perturbation series (\ref{e5.21}), the perturbation parameter $(ig)^2$ for the cubic $\cPT$-symmetric Hamiltonian
\begin{equation}
H=p^2+x^2+igx^3
\label{e5.22}
\end{equation}
becomes {\it negative}. Now, the perturbation series for this Hamiltonian has real coefficients that {\it alternate} in sign. There is no ambiguity; this series is uniquely Borel and Pad\'e summable, and the Borel and Pad\'e sums of this divergent series converge to the real ground-state energy of this $\cPT$-symmetric Hamiltonian \cite{pt480,pt5,pt605}.

As a second example, let us consider the Hermitian quartic anharmonic oscillator (\ref{e5.1}). The perturbation series for its ground-state energy $E_0(g)$ has the form 
\begin{equation}
\sum_{n=0}^\infty a_n g^n,
\label{e5.23}
\end{equation}
where the perturbation coefficients $a_n$ are real, alternate in sign, and grow factorially with $n$ \cite{pt13,pt14,pt15}. This series is uniquely Borel and Pad\'e summable and these summation methods converge to the exact value of the (real) ground-state energy \cite{pt597}.

The question is, What happens if we replace $g$ by $-g$ in the perturbation series (\ref{e5.23})? The resulting series is not uniquely Borel summable because the coefficients do not alternate in sign. The reason for this is that the nonalternating perturbation series does {\it not} represent the ground-state energy of the $\cPT$-symmetric quartic Hamiltonian (\ref{e5.3}). We cannot simply replace $g$ in the perturbation series for the ground-state energy of the Hermitian Hamiltonian in (\ref{e5.1}) by $-g$ and expect to obtain a perturbation expansion that represents the ground-state energy of the $\cPT$-symmetric Hamiltonian (\ref{e5.3}).

As explained in Subsec.~\ref{ss4E}, making the {\it local} replacement $g\to-g$ leads to one of two possible theories, either the conventional
upside-down-potential theory whose boundary conditions are pictured in (\ref{e4.21}) or the $\cPT$-symmetric theory whose boundary conditions are pictured in (\ref{e4.22}). In the first case, the energy eigenvalues are complex and the sign of the imaginary part of the energy corresponds to whether the boundary conditions represent outgoing or incoming waves. In the second case, the energy is real because we now have a dynamically stable configuration with balanced outgoing and incoming waves (balanced loss and gain).

The perturbation series that correctly represents the energy levels of the $\cPT$-symmetric quartic Hamiltonian (\ref{e5.3}) consists of two parts: First, there is the real nonalternating asymptotic perturbation series obtained by replacing $g$ with $-g$ in (\ref{e5.23}). The terms in this series are obtained by doing a Rayleigh-Schr\"odinger perturbative calculation (or equivalently, by evaluating sums of Feynman diagrams). The Borel sum of this series gives rise to exponentially small imaginary contributions of the form $ie^{-3/g}$. Second, there are nonperturbative contributions coming from saddle points in the complex plane \cite{pt611}. These saddle points contribute exponentially small imaginary terms of the form $ie^{-3/g}$, and these terms {\it exactly cancel the imaginary terms arising from the Borel sum of the nonalternating perturbation series} \cite{pt323}.

Perhaps the most beautiful feature of the $\cPT$-symmetric quantum anharmonic oscillator is this cancellation between perturbative and nonperturbative contributions. This cancellation allows the $\cPT$-symmetric anharmonic oscillator to be a viable quantum-mechanical theory with a {\it positive real energy spectrum}.

\section{This is quantum mechanics (not just complex analysis)}\label{s6}
In the first five sections of this Review we have used complex-variable techniques to study the spectral properties of $\cPT$-symmetric Hamiltonians. We have shown that a $\cPT$-symmetric Hamiltonian can possess an eigenvalue spectrum that is real and bounded below even if the Hamiltonian is non-Hermitian. By itself, this is a remarkable result because the conventional view is that spectral reality and positivity are properties that are reserved for Hermitian Hamiltonians. However, quantum mechanics is more than just the spectral properties of a Hamiltonian. The deeper question is whether a non-Hermitian Hamiltonian can actually define a physical theory of quantum mechanics.

The Hamiltonian for an isolated quantum system must have a real energy spectrum because an energy measurement must give a real number. Furthermore, the energy spectrum must be bounded below to satisfy the physical requirement that an isolated system have a stable ground state. However, the systems discussed in this Review are more general than conventional Hermitian quantum systems because they are coupled to the environment.

Hermiticity is a powerful condition because if $H$ is Hermitian, every eigenvalue of $H$ is real. We proved this in Sec.~\ref{s2} but a quick formal proof goes as follows: Take the Hermitian conjugate of the eigenvalue equation $H|E\rangle=E|E\rangle$ to get $\langle E| H^\dag=\langle E|E^*$ and then multiply this equation on the right by $|E\rangle$. This gives $E\langle E|E\rangle=E^*\langle E|E\rangle$. Therefore, $E=E^*$.

The condition of $\cPT$ symmetry is weaker than the condition of Hermiticity. As shown in Subsec.~\ref{ss2D}, if the Hamiltonian is $\cPT$ symmetric, all we know is that the {\it secular equation} is real.
Ordinarily, systems that are coupled to the environment are not Hermitian because they have loss to the environment (e.g., radioactive decay) or gain from the environment (e.g., scattering). The $\cPT$-symmetric systems discussed here are novel because they are coupled to the environment in such a way that loss to the environment is exactly balanced by gain from the environment. This is illustrated in Figs.~\ref{f5} and \ref{f24} (right diagram). Although one would  expect that the Hamiltonian for a system coupled to the environment would have a complex spectrum, in the unbroken phase a $\cPT$-symmetric system has an entirely real spectrum. 

Moreover, a $\cPT$-symmetric system in the unbroken phase satisfies the additional requisites to be a quantum-mechanical system.
A quantum-mechanical system must have a Hilbert space of state vectors that is endowed with an inner product having a {\it positive norm}. (The norm of a state represents a probability, which must be positive.) Also, the Hamiltonian of the theory must generate {\it unitary time evolution} because probability must be preserved in time. These requirements are met if the Hamiltonian is Hermitian $H^\dagger=H$. 

These requirements are 
also met if the Hamiltonian satisfies the less restrictive condition of unbroken $\cPT$ symmetry. In this section we demonstrate that these requisites are met if the $\cPT$-symmetric Hamiltonian has an unbroken $\cPT$ symmetry and is {\it also} symmetric. (All Hamiltonians discussed in this Review are symmetric.) However, this technical assumption of symmetry is in fact not necessary; these requisites are also met if the $\cPT$-symmetric Hamiltonian is not symmetric, but the argument is more complicated because it requires the use of a biorthogonal basis \cite{pt370,pt371,pt372, pt677}.

\subsection{Finding the inner product} \label{ss6A}
The principal difficulty with showing that a non-Hermitian $\cPT$-symmetric Hamiltonian in a region of unbroken $\cPT$ symmetry defines a theory of quantum mechanics is finding a {\it positive-definite} inner product with respect to which the eigenstates of the Hamiltonian are orthogonal. For the case of a Hermitian Hamiltonian this is not a problem because the inner product is the same for all such Hamiltonians. In coordinate space, the Hermitian inner product is
\begin{equation}
\langle\phi|\psi\rangle=\int dx\,\phi^\dag(x)\psi(x),
\label{e6.1}
\end{equation}
where the adjoint symbol $\dag$ indicates combined matrix transpose and complex conjugation. 

However, for a $\cPT$-symmetric Hamiltonian we do not know in advance the definition of the inner product. Rather, in $\cPT$-symmetric quantum mechanics the Hamiltonian itself determines its own Hilbert space and associated inner product. 

Let us assume that we have found the eigenvalues $E_n$ and eigenfunctions $\psi_n(x)$ of the $\cPT$-symmetric Hamiltonian $H$ by using numerical and/or analytical methods and that the $E_n$ are all real; that is, we are in a region of {\it unbroken $\cPT$ symmetry}. We showed in Subsec.~\ref{ss2B} that this is equivalent to assuming that the eigenfunctions $\psi_n(x)$ of $H$ are also eigenfunctions
of $\cPT$.\footnote{Unlike the case of Hermitian Hamiltonians, the zeros of subsequent eigenfunctions of non-Hermitian $\cPT$-symmetric Hamiltonians do not interlace. These eigenfunctions exhibit the phenomenon of {\it winding}, which is a non-Hermitian analog and generalization of Hermitian interlacing \cite{pt373,pt374, pt375,pt376}.}

Having found the eigenfunctions of $H$, we must find an inner product with respect to which these eigenfunctions are orthogonal. Unlike conventional quantum mechanics where know what the inner product is even before we are given the Hamiltonian, for a $\cPT$-symmetric Hamiltonian we begin by making a plausible guess for what the inner product might be. We argue by analogy: For a Hermitian Hamiltonian, where $H=H^\dag$, the inner product is given in (\ref{e6.1}), so for a $\cPT$-symmetric Hamiltonian, where $H=H^\cPT$, we replace the symbol $\dag$ by $\cPT$ and conjecture that the appropriate inner product is
\begin{eqnarray}
(\psi,\phi)^\cPT &\equiv&\int_C dx\,[\psi(x)]^{\cPT}\phi(x)\nonumber\\ &=&\int_C dx\,[\psi(-x)]^*\phi(x),
\label{e6.2}
\end{eqnarray}
where $C$ is an integration contour that terminates in the Stokes sectors in which we impose the boundary conditions on the eigenvalue equation associated with the Hamiltonian. We refer to the inner product (\ref{e6.2}) as the $\cPT$ {\it inner product}. With this conjecture for the $\cPT$ inner product we can immediately use integration by parts to verify that eigenfunctions of $H$ associated with different eigenvalues are orthogonal.

Although the $\cPT$ inner product establishes orthogonality, it is not yet acceptable because there is no guarantee that the norm of every eigenfunction is positive; we seek a positive-definite inner product. [Nonpositive inner products, which are associated with Krein spaces, are still of interest in mathematical physics. See Chap.~8 of \textcite{pt579}.] Therefore, we must improve the definition of the inner product in (\ref{e6.2}). To do so we construct a new operator called $\cC$ that ensures that the norms of all eigenfunctions are strictly positive.

\subsection{The $\cC$ operator and the $\cCPT$ inner product}\label{ss6B}
If the parameters of a $\cPT$-symmetric Hamiltonian $H$ are chosen so that all of the eigenvalues are real (the $\cPT$ symmetry is unbroken), there exists a new linear operator $\cC$ that commutes with both $H$ and $\cPT$. The existence of an operator that commutes with $H$ reveals that in the unbroken $\cPT$-symmetric region the Hamiltonian has a new and unexpected symmetry. (We use the symbol $\cC$ to represent this symmetry operator because the properties of $\cC$ are similar to those of the charge-conjugation operator in particle physics.) The $\cC$ operator is the missing component of the conjectured $\cPT$ inner product in (\ref{e6.2}).

The appropriate adjoint for the definition of the inner product combines $\cC$ conjugation and $\cPT$ conjugation:
\begin{equation}
\langle\psi|\phi\rangle^{\cCPT}\equiv\int_C dx\,\psi^{\cCPT}(x)\phi(x),
\label{e6.3}
\end{equation}
where
\begin{equation}
\psi^\cCPT(x)=\int_C dy\,{\cC}(x,y)\psi^*(-y).
\label{e6.4}
\end{equation}
The inner product (\ref{e6.3}) is called the $\cCPT$ {\it inner product}. This inner product fulfills the requirements that a quantum theory have a Hilbert space with a {\it positive-definite} norm and be a unitary theory. 

We will construct the $\cC$ operator as a sum over the eigenfunctions of the Hamiltonian $H$, but we must first normalize these eigenfunctions. We have already shown in (\ref{e2.9}) that the eigenfunctions $\psi_n(x)$ of $H$ are also eigenfunctions of the $\cPT$ operator with eigenvalue $\lambda$. Furthermore, (\ref{e2.10}) shows that $\lambda$ is a pure phase: $\lambda=e^{i\alpha}$, where $\lambda$ and $\alpha$ depend on $n$.

We define $\cPT$-{\it normalized} eigenfunctions $\phi_n(x)$ as
\begin{equation}
\phi_n(x)\equiv e^{-i\alpha/2}\psi_n(x).
\label{e6.5}
\end{equation}
With this definition $\phi_n(x)$ is still an eigenfunction of $H$ and by construction it is also an eigenfunction of $\cPT$ with eigenvalue $1$. One can also show both numerically and analytically that for $H$ in (\ref{e2.14}) the algebraic sign of the $\cPT$ norm in (\ref{e6.2}) of $\phi_n(x)$ is $(-1)^n$ for all $n$ and for all values of $\vep>0$ \cite{pt90,pt178}. Thus, we multiply each eigenfunction of $H$ by a real number so that their $\cPT$ norms are now exactly $(-1)^n$:
\begin{eqnarray}
\int_C dx\,[\phi_n(x)]^{\cPT}\phi_n(x)
&=&\int_C dx\,[\phi_n(-x)]^*\phi_n(x)\nonumber\\
&=& (-1)^n,
\label{e6.6}
\end{eqnarray}
where the contour $C$ lies in the Stokes sectors in Fig.~\ref{f9}.

In terms of these $\cPT$-normalized eigenfunctions there is a simple but unusual statement of completeness:
\begin{equation}
\sum_{n=0}^\infty(-1)^n\phi_n(x)\phi_n(y)=\delta(x-y).
\label{e6.7}
\end{equation}
This statement of completeness has been verified analytically and numerically to great precision for all $\vep>0$ \cite{pt377,pt378} and a proof is given in \textcite{pt370}. Using (\ref{e6.6}) one can verify that the left side of (\ref{e6.7}) satisfies the delta-function integral identity $\int dy\,\delta(x-y)\delta(y-z)=\delta(x-z)$.

To illustrate this normalization procedure we consider the elementary case of the harmonic-oscillator eigenfunctions. For the Hamiltonian $H=p^2+x^2$, which is both Hermitian and $\cPT$ symmetric, the eigenfunctions are Gaussians multiplied by Hermite polynomials:
\begin{eqnarray}
\psi_0(x)&=&\exp\left(-\half x^2\right),\nonumber\\
\psi_1(x)&=&x\exp\left(-\half x^2\right),\nonumber\\
\psi_2(x)&=&(2x^2-1)\exp\left(-\half x^2\right),\nonumber\\
\psi_3(x)&=&(2x^3-3x)\exp\left(-\half x^2 \right).\nonumber
\end{eqnarray}
We normalize these eigenfunctions so that they are also eigenfunctions of the
$\cPT$ operator with eigenvalue $1$:
\begin{eqnarray}
\phi_0(x)&=&a_0\exp\left(-\half x^2\right),\nonumber\\
\phi_1(x)&=&a_1ix\exp\left(-\half x^2\right),\nonumber\\
\phi_2(x)&=&a_2(2x^2-1)\exp\left(-\half x^2\right),\nonumber\\
\phi_3(x)&=&a_3i(2x^3-3x)\exp\left(-\half x^2\right).\nonumber
\end{eqnarray}
The real numbers $a_n$ are chosen so that the integral in (\ref{e6.6}) evaluates to $(-1)^n$ for all $n$. If we then substitute the eigenfunctions $\phi_n(x)$ into (\ref{e6.7}) and perform the summation, we obtain the delta function in (\ref{e6.7}).

Finally, we construct the $\cC$ operator. Recall from (\ref{e6.6}) that half of the energy eigenstates of $H$ have positive $\cPT$ norm and half have negative $\cPT$ norm. This condition is similar to what Dirac encountered in formulating the spinor wave equation for fermionic quantum theory \cite{pt379}. To solve this problem Dirac interpreted the negative-norm states as antiparticles and constructed a new inner product that changed the negative signs to positive signs. We follow Dirac's construction: Note that if the Hamiltonian $H$ has an unbroken $\cPT$ symmetry, there exists an additional symmetry of $H$ represented by a linear operator that we call $\cC$. This operator arises because there are equal numbers of positive-$\cPT$-norm and negative-$\cPT$-norm states all with real energy eigenvalues. In coordinate space the operator $\cC$ that embodies this symmetry is a sum over the $\cPT$-normalized eigenfunctions of the $\cPT$-symmetric Hamiltonian:
\begin{equation}
\cC(x,y)\equiv\sum_{n=0}^\infty\phi_n(x)\phi_n(y).
\label{e6.8}
\end{equation}

This equation looks like the unusual completeness condition in (\ref{e6.7}) except that {\it the factor of $(-1)^n$ is absent}. We can use (\ref{e6.6}) and (\ref{e6.7}) to verify that the square of $\cC$ is unity: $\cC^2={\bf 1}$. In coordinate space this equation reads
\begin{equation}
\int_C dy\,\cC(x,y)\cC(y,z)=\delta(x-z).
\label{e6.9}
\end{equation}

The eigenvalues of $\cC$ are $\pm1$, so $\cC$ is a reflection operator. Also, it is easy to verify that $[\cC,H]=0$, and because $\cC$ is linear it follows that the eigenstates of $H$ have definite $\cC$ values:
\begin{eqnarray}
\cC\phi_n(x)&=&\int_C dy\,\cC(x,y)\phi_n(y)\nonumber\\
&=&\sum_{m=0}^\infty\phi_m(x)\int_C dy\,\phi_m(y)\phi_n(y)\nonumber\\
&=&(-1)^n\phi_n(x).
\label{e6.10}
\end{eqnarray}
As a reflection operator, the $\cC$ operator resembles the parity operator in quantum mechanics or the charge-conjugation operator in quantum field theory.

The precise meaning of $\cC$ is that it represents the measurement of the sign of the $\cPT$ norm in (\ref{e6.6}). The operators $\cP$ and $\cC$ are both reflection operators and they are distinct square roots of the unity operator $\delta(x-y)$. That is, $\cP^2=\cC^2=1$. However, $\cP\neq \cC$. The parity operator in coordinate space is real because $\cP(x,y)=\delta(x+y)$, but the $\cC(x,y)$ operator is complex because it is a sum of products of complex functions. The operators $\cP$ and $\cC$ do not commute, but $\cC$ {\it does} commute with $\cPT$.

Let us now examine the properties of the $\cCPT$ inner product defined in (\ref{e6.3}). Like the $\cPT$ inner product in (\ref{e6.2}), this new inner product is phase independent. Also, because the time-evolution operator is $e^{-iHt}$ (as in ordinary quantum mechanics) and because $H$ commutes with $\cPT$ and with $\cCPT$, both the $\cPT$ inner product and the $\cCPT$ inner product remain time independent as the states evolve. However, the advantage of the $\cCPT$ inner product is that, unlike the $\cPT$ inner product, the $\cCPT$ inner product is {\it positive definite}. This is because $\cC$ contributes a factor of $-1$ when it acts on states with negative $\cPT$ norm. In terms of the $\cCPT$ conjugate, the completeness condition in coordinate space reads
\begin{equation}
\sum_{n=0}^\infty\phi_n(x)[\cCPT\phi_n(y)]=\delta(x-y).
\label{e6.11}
\end{equation}

We can also construct representations of other linear operators in terms of the eigenstates $\phi_n(x)$: The coordinate-space representation of the parity operator is
\begin{equation}
\cP(x,y)=\delta(x+y)=\sum_{n=0}^\infty(-1)^n\phi_n(x)\phi_n(-y),\nonumber
\end{equation}
and the coordinate-space representations of the Hamiltonian and of the Green's function (the inverse of the Hamiltonian operator) are
\begin{eqnarray}
H(x,y)&=&\sum_{n=0}^\infty(-1)^nE_n\phi_n(x)\phi_n(y),\nonumber\\
G(x,y)&=&\sum_{n=0}^\infty(-1)^n\frac{1}{E_n}\phi_n(x)\phi_n(y).\nonumber
\end{eqnarray}

From the equation for $G(x,y)$ above, we can verify that this Green's function is the {\it functional inverse} of $H$: $\int dy\,H(x,y)G(y,z)=\delta(x-z)$. For example, for the $\cPT$-symmetric Hamiltonian (\ref{e2.14}) the Green's function $G(x,y)$ satisfies the differential equation
$$\left[-\frac{d^2}{dx^2}+x^2(ix)^\vep\right]G(x,y)=\delta(x-y).$$
We solve this differential equation in terms of associated Bessel functions in the two regions $x>y$ and $x<y$ and then patch the solutions together at $x=y$ to obtain a closed-form coordinate-space expression for $G(x,y)$ \cite{pt377,pt378}. We then integrate $G(x,x)$ with respect to $x$ to obtain an exact formula for the spectral $\zeta$ function for all values of $\vep>0$:
\begin{eqnarray}
\zeta(\vep)=\sum_{n=0}^\infty\frac{1}{E_n(\vep)}&=&\left[1+\frac{\cos\left(\frac{3\vep\pi}{2\vep+8}\right)
\sin\left(\frac{\pi}{4+\vep}\right)}{\cos\left(\frac{\vep\pi}{4+2\vep}\right)
\sin\left(\frac{3\pi}{4+\vep}\right)}\right]\nonumber\\
&&\quad\times\frac{\Gamma\left(\frac{1}{4+\vep}\right)\Gamma\left(\frac{2}{4+\vep}\right)\Gamma\left(\frac{\vep}{4+\vep}\right)}{(4+\vep)^\frac{4+2\vep}{4+\vep}\Gamma\left(\frac{1+\vep}{4+\vep}\right)\Gamma\left(\frac{2+\vep}{4+\vep}\right)}.\nonumber
\end{eqnarray}
This formula has been verified numerically to over 20 decimal places \cite{pt378}.

\subsection{Discussion of the $\cCPT$-inner product}\label{ss6C}
For a quantum theory defined by a Hermitian Hamiltonian, the Hilbert space of physical states is specified even before the Hamiltonian $H$ is known. The inner product for this Hilbert space is defined with respect to Dirac Hermitian conjugation (combined complex conjugation and matrix transposition). Once a Hermitian Hamiltonian has been selected, the standard procedure is to find the eigenvectors and eigenvalues of this Hamiltonian.

In contrast, for a $\cPT$-symmetric quantum theory, the inner product is determined by the Hamiltonian itself, and thus is determined {\it dynamically}. One must first solve for the eigenvalues and eigenstates of $H$ and then show that we are in an unbroken $\cPT$-symmetric region before we can know what the Hilbert space and the associated inner product of the theory are. This is because the $\cC$ operator is constructed from the eigenstates of the Hamiltonian, and the $\cCPT$ inner product is determined by these eigenstates. The Hilbert space consists of all complex linear combinations of the eigenstates of $H$.

The $\cC$ operator does not exist as a distinct entity in conventional Hermitian quantum mechanics. For example, if we let the parameter $\vep$ in (\ref{e2.14}) tend to 0, the $\cC$ operator in this limit becomes identical to $\cP$ and the $\cCPT$ operator becomes $\cT$, which performs complex conjugation. Hence, as $\vep$ tends to 0, the inner product defined with respect to $\cCPT$ conjugation reduces to the inner product of conventional quantum mechanics and (\ref{e6.7}) reduces to the usual Hermitian statement of completeness for a symmetric Hamiltonian:
$$\sum_n\phi_n(x)\phi_n^*(y)=\delta(x-y).$$

The $\cCPT$ inner product in (\ref{e6.3}) is independent of the choice of integration contour $C$ as long as $C$ terminates inside the Stokes sectors associated with the boundary conditions for the eigenvalue problem. [In ordinary quantum mechanics the positive-definite inner product has the form $\int dx\,f^*(x)g(x)$, where the integral is taken along the real axis. The path of integration cannot be deformed into the complex plane because the integrand is not analytic.] The path integrals representing both the $\cPT$ inner product and the $\cCPT$ inner product have the advantage of analyticity and path independence. The $\cPT$ inner product is not positive-definite, but it is satisfying that one can still construct a positive-definite metric by using $\cCPT$ conjugation without disturbing the path independence of the inner-product integral.

The operator $e^{-iHt}$ performs time evolution regardless of whether the theory is determined by a $\cPT$-symmetric Hamiltonian or a conventional Hermitian Hamiltonian. To show that time evolution in a $\cPT$-symmetric quantum system is unitary, we must demonstrate that the norm of the state does not change as the state vector evolves in time: Let $\psi_0(x)$ be a vector at $t=0$ in the Hilbert space spanned by the energy eigenstates. It evolves into the state $\psi_t(x)$ at time $t$ according to $$\psi_t(x)=e^{-iHt}\psi_0(x).$$  With respect to the $\cCPT$ inner product the norm of $\psi_t(x)$ does not  change in time because $H$ commutes with $\cCPT$.

\subsection{Example:  $\cPT$-symmetric $2\times2$ matrix
Hamiltonian}\label{ss6D} 
We illustrate the formal procedures for calculating the $\cC$ operator and the inner product by using the $\cPT$-symmetric matrix Hamiltonian in (\ref{e2.20}), which we rewrite in polar form as
\begin{equation}
H=\left(\begin{array}{cc} re^{i\theta} & g \cr g & re^{-i\theta}\end{array}\right).
\label{e6.12}
\end{equation}
The three parameters $r$, $s$, and $\theta$ are real. This symmetric Hamiltonian is not Hermitian. However, as shown in Subsec.~\ref{ss2E}, it is $\cPT$ symmetric, where the parity operator is given in (\ref{e2.21}) and the $\cT$ operator performs complex conjugation.

Following the procedures outlined above, we first calculate the eigenvalues and eigenfunctions of the Hamiltonian (\ref{e6.12}). From (\ref{e2.23}) we obtain
\begin{equation}
E_\pm=r\cos\theta\pm\sqrt{g^2-r^2\sin^2\theta}.
\label{e6.13}
\end{equation}
As we saw for $H$ in (\ref{e2.20}), there are two parametric regions for $H$ in (\ref{e6.12}), one for which the square root is real and the other for which it is imaginary. In the region of broken $\cPT$ symmetry $g^2<r^2\sin^2\theta$, the eigenvalues form a complex-conjugate pair. In the region of unbroken $\cPT$ symmetry $g^2\geq r^2\sin^2\theta$, the eigenvalues are real and the simultaneous eigenstates of the operators $H$ and $\cPT$ are
\begin{eqnarray}
|E_+\rangle&=&\frac{1}{\sqrt{2\cos\alpha}}\left(\begin{array}{c} e^{i\alpha/2}\cr e^{-i\alpha/2}\end{array}\right),\nonumber\\
|E_-\rangle&=&\frac{i}{\sqrt{2\cos\alpha}}\left(\begin{array}{c}e^{-i\alpha/2}\cr-e^{i\alpha/2}\end{array}\right),
\label{e6.14}
\end{eqnarray}
where $\sin\alpha=\tfrac{r}{g}\,\sin\theta$.

Next we calculate the $\cPT$ inner product:
\begin{equation}
(E_{\pm},E_{\pm})^\cPT=\pm1\quad{\rm and}\quad (E_{\pm},E_{\mp})=0,
\label{e6.15}
\end{equation}
where $(u,v)^\cPT=(\cPT u)\cdot v$. As expected from (\ref{e6.6}), the $\cPT$ inner product has a metric of signature $(+,-)$. Also, note that if the condition $g^2>r^2\sin^2\theta$ for unbroken $\cPT$ symmetry is violated, the states (\ref{e6.14}) are {\it not} eigenstates of $\cPT$ because $\alpha$ is imaginary. Also, in the region of broken $\cPT$ symmetry, the $\cPT$ norms of the energy eigenstates vanish.

We then use (\ref{e6.8}) to construct the $\cC$ operator:
\begin{equation}
\cC=\frac{1}{\cos\alpha}\left(\begin{array}{cc} i\sin\alpha & 1 \cr 1 & -i\sin
\alpha\end{array}\right).
\label{e6.16}
\end{equation}
A brief calculation verifies that, as predicted, the $\cC$ operator commutes with $H$ and satisfies $\cC^2=1$. Also, as expected, the eigenvectors of $H$ are simultaneously eigenvectors of $\cC$ with eigenvalues $+1$ and $-1$:
\begin{equation}
\cC|E_{\pm}\rangle=\pm|E_{\pm}\rangle.
\label{e6.17}
\end{equation}
Again, as expected, the eigenvalues of $\cC$ are the signs of the $\cPT$ norms of the corresponding eigenstates.

Finally, we construct the $\cCPT$ inner product
$$\langle u|v\rangle^\cCPT=(\cCPT u)\cdot v$$
by using the operator $\cC$. This inner product is positive definite because $\langle E_{\pm}|E_{\pm}\rangle=1$. The two-dimensional Hilbert space spanned by $|E_\pm\rangle$ and with inner product $\langle\cdot|\cdot\rangle$ has the signature $(+,+)$. This completes the construction of the Hilbert space.

It is instructive to demonstrate that the $\cCPT$ norm of {\it any} vector in the Hilbert space (that is, any linear combination of the eigenvectors) is positive. For the vector $\psi=\left(\begin{array}{c} a \\ b \end{array}\right)$, where $a$ and $b$ are arbitrary complex numbers, we calculate that
\begin{eqnarray}
&&\cT\psi=\left(\begin{array}{c}a^*\cr b^*\end{array}\right),\qquad
\cPT\psi=\left(\begin{array}{c} b^*\cr a^*\end{array}\right),\nonumber\\
&&\cC\cPT\psi=\frac{1}{\cos\alpha}\,\left(\begin{array}{c}a^*+ib^*\sin\alpha
\cr b^*-ia^*\sin\alpha\end{array}\right).
\label{e6.18}
\end{eqnarray}
Hence,
\begin{eqnarray}
\langle\psi|\psi\rangle^\cCPT&=&(\cC\cPT\psi)\cdot\psi\nonumber\\
&=&\frac{1}{\cos\alpha}\big[a^*a+b^*b+i\big(b^*b-a^*a\big)\sin\alpha\big].
\nonumber
\end{eqnarray}
Thus, if we take $a=x+iy$ and $b=u+iv$ with $x$, $y$, $u$, and $v$ real, then
\begin{eqnarray}
\langle\psi|\psi\rangle^\cCPT&=&\frac{1}{\cos\alpha}\big[x^2+v^2+y^2+u^2
\nonumber\\
&&\quad+2(xv-yu)\sin\alpha\big].
\label{e6.19}
\end{eqnarray}
It is easy to prove that this $\cCPT$ inner product is positive and that it vanishes only if $x=y=u=v=0$.

We can now formulate the statement of completeness for the $\cPT$-symmetric Hamiltonian (\ref{e6.12}). Following the notation used in conventional quantum mechanics, we denote the $\cCPT$-conjugate of $|u\rangle$ as $\langle u|$; that is, the $\cCPT$ adjoint operator converts the {\it ket} vector into the {\it bra} vector. The completeness condition then has the form
\begin{equation}
|E_+\rangle\langle E_+|+|E_-\rangle\langle E_-|=\left(\begin{array}{cc} 1 & 0
\cr 0 & 1\end{array}\right).
\label{e6.20}
\end{equation}
Furthermore, using the $\cCPT$ conjugate $\langle E_\pm|$, we can represent the $\cC$ operator for $H$ in (\ref{e6.12}) as
\begin{equation}
\cC=|E_+\rangle\langle E_+|-|E_-\rangle\langle E_-|.
\label{e6.21}
\end{equation}
The change in sign between (\ref{e6.20}) and (\ref{e6.21}) is the two-dimensional analog of the sign change in the formulas for completeness in (\ref{e6.7}) and for the $\cC$ operator in (\ref{e6.8}).

Note that as $\theta$ approaches $0$, the Hamiltonian (\ref{e6.12}) becomes Hermitian and the $\cC$ operator reduces to the parity operator $\cP$. This illustrates that for a symmetric Hermitian matrix, $\cCPT$ invariance reduces to the conventional condition of Hermiticity: $H=H^*$.

\subsection{ Calculating the $\cC$ Operator} \label{ss6E}
The difference between $\cPT$-symmetric quantum mechanics and Hermitian quantum mechanics is the existence of a $\cC$ operator. Only a non-Hermitian $\cPT$-symmetric Hamiltonian possesses a $\cC$ operator that is distinct from the parity operator $\cP$. However, although the $\cC$ operator is represented as the formal infinite series in (\ref{e6.8}), it is not easy to calculate $\cC$ by summing this series directly. In Subsec.~\ref{ss2D} we evaluated the sum for the case of a $2\times2$ matrix, but evaluating the sum for a continuum theory would require that we find analytic expressions for all of the eigenfunctions $\phi_n(x)$ of $H$. Moreover, such a procedure would be hopeless in quantum field theory because in field theory there is no simple analog of the Schr\"odinger eigenvalue equation.

The first calculation of the $\cC$ operator was perturbative \cite{pt402}. This paper considers the cubic $\cPT$-symmetric Hamiltonian
\begin{equation}
H=\tfrac{1}{2}p^2+\tfrac{1}{2}x^2+i\vep x^3
\label{e6.22}
\end{equation}
and treats $\vep$ as a small real parameter. When $\vep=0$, the Hamiltonian $H$ reduces to the harmonic oscillator Hamiltonian whose eigenfunctions are Hermite functions. Each eigenfunction of $H$ is then expressed as a perturbation series in powers of $\vep$ and the first few terms of these perturbation series are calculated. The perturbation series are substituted into (\ref{e6.8}), and the summation over the eigenfunctions is done for the first few powers of $\vep$. The result is a formal perturbation expansion of the $\cC$ operator in powers of $\vep$.

This calculation is complicated but instructive because it shows that the result for $\cC$ simplifies dramatically if the $\cC$ operator is written as the exponential of an operator $Q$ (which in quantum mechanics is a real function of the operators $x$ and $p$) multiplying the parity operator $\cP$:
\begin{equation}
\cC=e^{Q(x,p)}\cP.
\label{e6.23}
\end{equation}
Furthermore, it shows that $\lim_{ \vep\to0}Q=0$, which verifies that $\cC$ approaches $\cP$ in the limit as the Hamiltonian becomes Hermitian.

We now exploit the representation (\ref{e6.23}) for the $\cC$ operator for two elementary cases. First, we consider the complex shifted harmonic oscillator $H=\half p^2+\half x^2+i\vep x$ in (\ref{e2.6}). As shown in Subsec.~\ref{ss2A}, this Hamiltonian has an unbroken $\cPT$ symmetry for all real $\vep$ and its eigenvalues $E_n=n+\half+\half\vep^2$ are all real. The {\it exact} formula for $\cC$ for this Hamiltonian is $\cC=e^Q\cP$, where
\begin{equation}
Q=-\vep p.
\label{e6.24}
\end{equation}
In the limit $\vep\to0$ the shifted Hamiltonian (\ref{e2.6}) becomes Hermitian and $\cC$ reduces to the parity operator $\cP$.

Next, we reconsider the non-Hermitian $2\times2$ matrix Hamiltonian (\ref{e6.12}). We rewrite the $\cC$ operator in (\ref{e6.16}) for this Hamiltonian as $\cC=e^Q\cP$, where
\begin{equation}
Q=\half\sigma_2\log\left(\frac{1-\sin\alpha}{1+\sin\alpha}\right)
\label{e6.25}
\end{equation}
and $\sigma_2$ is the Pauli matrix
\begin{equation}
\sigma_2=\left(\begin{array}{cc}0 & -i\cr i & 0\end{array}\right).
\label{e6.26}
\end{equation}
Once again, in the limit $\theta\to0$, the Hamiltonian (\ref{e6.12}) becomes Hermitian and the $\cC$ operator reduces to the parity operator $\cP$.

\subsection{
Algebraic calculation of the $\cC$ operator} \label{ss6F}
Calculating the $\cC$ operator is difficult and only rarely has it been done exactly. However, in this subsection we show that there is a simple-looking set of three {\it algebraic} equations that serve as a general starting point for the calculation. These algebraic equations avoid the formidable task of evaluating the sum over eigenstates in (\ref{e6.8}). Furthermore, the algebraic technique generalizes from quantum mechanics, where the eigenstates are known at least in principle, to quantum field theory.

For any $\cPT$-symmetric Hamiltonian $H$ the $\cC$ operator has three fundamental algebraic properties. First, $\cC$ commutes with the spacetime reflection operator $\cPT$:
\begin{eqnarray}
[\cC,\cPT]=0,
\label{e6.27}
\end{eqnarray}
although $\cC$ does not commute with $\cP$ or $\cT$ separately. Second, $\cC$ is a reflection operator, so 
\begin{eqnarray}
\cC^2=1.
\label{e6.28}
\end{eqnarray}
These first two conditions are {\it kinematic} in character and are valid for any $\cPT$-symmetric Hamiltonian in the unbroken $\cPT$-symmetric region. The third condition is that $\cC$ commutes with the Hamiltonian
\begin{eqnarray}
[\cC,H]=0,
\label{e6.29}
\end{eqnarray}
so $\cC$ is time independent. This third condition is {\it dynamical} in character because it depends on the Hamiltonian. To summarize, $\cC$ is a time-independent $\cPT$-symmetric reflection operator.

To solve (\ref{e6.27})-(\ref{e6.29}) for a quantum-mechanical system defined by a Hamiltonian of the form $H=H(x,p)$, we substitute the operator representation for $\cC$ in (\ref{e6.23}) into (\ref{e6.27})--(\ref{e6.29}) and try to solve the resulting equations for $Q(x,p)$. Substituting (\ref{e6.23}) into (\ref{e6.27}) gives
$$e^{Q(x,p)}=\cPT e^{Q(x,p)}\cPT=e^{Q(-x,p)}.$$
This implies that $Q(x,p)$ is an {\it even} function of $x$. Next, we substitute (\ref{e6.23}) into (\ref{e6.28}):
$$e^{Q(x,p)}\cP e^{Q(x,p)}\cP=e^{Q(x,p)}e^{Q(-x,-p)}=1.$$
This equation implies that $Q(x,p)=-Q(-x,-p)$, and because $Q(x,p)$ is an even function of $x$, it must be an {\it odd} function of $p$. These two properties of $Q$ are general and hold for any $\cPT$-symmetric Hamiltonian..

The third condition (\ref{e6.29}) imposes the dynamics specified by the Hamiltonian that defines the quantum-mechanical theory. Substituting $\cC=e^{Q(x,p)}\cP$ into (\ref{e6.29}), we obtain an exact equation satisfied by $Q$:
\begin{equation}
e^{Q(x,p)}[\cP,H]+[e^{Q(x,p)},H]\cP=0.
\label{e6.30}
\end{equation}

Finding an exact solution to (\ref{e6.30}) is difficult, but perturbative methods may be used to solve for $Q$.
[Nonperturbative techniques, such as semiclassical approximations, may also be helpful \cite{pt403}.] To illustrate the perturbative approach, we examine (\ref{e6.30}) for the case of the cubic Hamiltonian $H$ in (\ref{e6.22}). We write $H=H_0+\vep H_1$, where $H_0=\half p^2+\half x^2$ is the harmonic-oscillator Hamiltonian, which commutes with $\cP$, and $H_1=ix^3$, which anticommutes with $\cP$. Then, (\ref{e6.30}) simplifies to
\begin{equation}
2\vep e^{Q(x,p)}H_1=[e^{Q(x,p)},H].
\label{e6.31}
\end{equation}

Next, we expand $Q(x,p)$ as a perturbation series in {\it odd} powers of $\vep$,
\begin{eqnarray}
Q(x,p)=\vep Q_1(x,p)+\vep^3Q_3(x,p)+\cdots,
\label{e6.32}
\end{eqnarray}
and substitute this expansion into the exponential $e^{Q(x,p)}$. This leads to a sequence of equations that can be solved systematically for the operator functions $Q_n(x,p)$ $(n=1,3,5,...)$ subject to the symmetry constraints that ensure that the conditions (\ref{e6.27}) and (\ref{e6.28}) are met. The first two of these equations are
\begin{eqnarray}
\big[H_0,Q_1\big] &=& -2H_1,\nonumber\\
\big[H_0,Q_3\big] &=&-\tfrac{1}{6}\big[Q_1,\big[Q_1,H_1\big]\big].
\label{e6.33}
\end{eqnarray}

To solve these equations we substitute a polynomial with arbitrary coefficients for $Q_n$ and then calculate the coefficients. For example, to solve the first equation in (\ref{e6.33}), we substitute for $Q_1$ the most general Hermitian {\it cubic} polynomial that is {\it even} in $x$ and {\it odd} in $p$:
$$Q_1(x,p)=Ap^3+Bxpx,$$
where $A$ and $B$ are numerical coefficients. The operator equation for $Q_1$ is satisfied if $A=-\tfrac{4}{3}$ and $B=-2$. It is tedious but routine to continue this process to obtain higher-order perturbative approximations to $Q$.

\subsection{ Isospectral Hermitian and $\cPT$-symmetric Hamiltonians} \label{ss6G}
In Sec.~\ref{s5} we constructed a quartic Hermitian Hamiltonian whose eigenvalues are the same as the eigenvalues of a quartic $\cPT$-symmetric Hamiltonian. The transformation from the $\cPT$-symmetric to the Hermitian Hamiltonian in that case was simple and explicit, but to date no other nontrivial examples of exact isometric pairs of $\cPT$-symmetric and Hermitian Hamiltonians of the form $H=p^2+V(x)$ have been found. However, it is possible to construct a {\it formal} (nonexplicit) similarity mapping from a $\cPT$-symmetric Hamiltonian $H$ to an equivalent Hermitian Hamiltonian $h$:
\begin{equation}
h=e^{-Q/2}He^{Q/2}.
\label{e6.34}
\end{equation}
This mapping makes use of the {\it square root} of the positive operator $e^Q$ \cite{pt140, pt181,pt389,pt385, pt386}.

The new Hamiltonian $h$ is equivalent to $H$ in the sense that it is isospectral to (has the same eigenvalues as) $H$. To explain the similarity transformation (\ref{e6.34}) we begin by recalling from (\ref{e6.23}) that the $\cC$ operator has the general form $\cC=e^Q\cP$, where $Q=Q(x,p)$ is a real function of the dynamical operator variables $x$ and $p$ of the quantum theory. We then multiply $\cC$ on the right by the parity operator $\cP$ to obtain $e^Q$:
\begin{equation}
e^Q=\cC\cP.
\label{e6.35}
\end{equation}
This shows that $\cC\cP$ is a positive and invertible operator.

To show that $h$ in (\ref{e6.34}) is Hermitian, we take the Hermitian conjugate of (\ref{e6.34}),
\begin{equation}
h^\dag=e^{Q/2}H^\dag e^{-Q/2},
\label{e6.36}
\end{equation}
and rewrite this formula as $h^\dag=e^{-Q/2}e^Q H^\dag e^{-Q}e^{Q/2}$. We then
use (\ref{e6.35}) to replace $e^Q$ by $\cC\cP$ and $e^{-Q}$ by $\cP\cC$:
\begin{equation}
h^\dag=e^{-Q/2}\cC\cP H^\dag\cP\cC e^{Q/2}.
\label{e6.37}
\end{equation}
Next, we use (\ref{e6.12}) to replace $H^\dag$ by $\cP H\cP$. This gives $$h^\dag=e^{-Q/2}\cC\cP\cP H\cP\cP\cC e^{Q/2}=e^{-Q/2}\cC H\cC e^{Q/2}.$$ Finally, recalling that $\cC$ commutes with $H$ [see (\ref{e6.29})] and that the square of $\cC$ is unity [see (\ref{e6.28})], we reduce the right side of (\ref{e6.37}) to the right side of (\ref{e6.34}). This verifies that $h$ is Dirac Hermitian.

We have shown that for a non-Hermitian $\cPT$-symmetric Hamiltonian $H$ whose $\cPT$ symmetry is unbroken, it is possible, at least in principle, to use (\ref{e6.34}) to construct a Hermitian Hamiltonian $h$ that is isospectral to $H$. Assuming that $H$ has an {\it unbroken} $\cPT$ symmetry is crucial because this allows us to construct $\cC$, which in turn allows us to construct the similarity operator $e^{Q/2}$.

This similarity transformation raises a crucial question that has not yet been answered: Are $\cPT$-symmetric Hamiltonians physically new and distinct from ordinary Hermitian Hamiltonians or do they describe exactly the same physical processes that ordinary Hermitian Hamiltonians describe? That is, is $\cPT$-symmetric quantum mechanics physically new and different from conventional Hermitian quantum mechanics or are $\cPT$-symmetric Hamiltonians just complicated similarity transformations of conventional Hermitian Hamiltonians?

One possible response to this question is calculational in character. Equation (\ref{e6.34}) is a {\it formal} similarity transformation that maps a $\cPT$-symmetric Hamiltonian $H$ having unbroken $\cPT$ symmetry to a Hermitian Hamiltonian $h$ that is isospectral with $H$. However, except in rare cases, actually constructing this transformation is hopelessly difficult. [See, for example, the discussion of the $\cC$ operator for a $\cPT$-symmetric square well \cite{pt579,pt160, pt327,pt382, pt383}.] Of course, perturbative calculations can be performed, but these are also complicated and not likely to give transparent analytical insight. A $\cPT$-symmetric Hamiltonian $H$ may have a simple local structure that makes it easy to use in performing calculations, but the isospectral Hermitian $h$ is typically {\it nonlocal} because its interaction term has arbitrarily high powers of the variables $x$ and $p$ \cite{pt409}. Thus, it is not just difficult to calculate using the Hermitian Hamiltonian $h$; it is virtually impossible because the usual regulation schemes, especially in quantum field theory, are inadequate.

Even for the one nontrivial example in Sec.~\ref{s5} for which there is actually a closed-form quartic potential for both $H$ and $h$, it is not even possible to construct the $\cC$ operator in simple closed form. This is because the mapping from $H$ to $h$ involves a Fourier transform and thus the mapping is nonlocal (see Subsec.~\ref{ss4C}). 

Moreover, while the $\cPT$-symmetric $H$ and the Hermitian $h$ are isospectral, we saw in Subsecs.~\ref{ss4C} and \ref{ss4D} that there are profound differences between these Hamiltonians at the {\it classical} level; $h$ has a quantum anomaly, that is, $h$ depends explicitly on Planck's constant $\hbar$. We conclude that although the formal mapping from $H$ to $h$ is theoretically interesting, it does not have any practical calculational value.

A second response to the question of whether $\cPT$-symmetric Hamiltonians are physically new is of greater significance because it has far-reaching theoretical implications. The mapping from $H$ to $h$ in (\ref{e6.34}) is a {\it similarity} transformation and not a {\it unitary} transformation. A unitary transformation is {\it bounded}, but the similarity transformation operator $e^Q$ is unbounded. Thus, the Hamiltonians $H$ and $h$ are {\it formally} isospectral, but the mapping between the Hamiltonians is unlikely to be one-to-one because it may not map all the vectors in the Hilbert space associated with $h$ onto all the vectors in the Hilbert space associated with $H$. 

Of course, for a finite-dimensional matrix Hamiltonian like that in (\ref{e6.12}), the operator $e^Q$ is bounded [see (\ref{e6.16}) and (\ref{e6.25})]. However, for continuum Hamiltonians such as (\ref{e2.14}) the mapping $e^Q$ is unbounded. Thus, there may be experimentally measurable physical differences between $H$ and $h$, perhaps some sort of Bohm-Aharanov-like effect. This issue has theoretical importance because it relates to whether $\cPT$-symmetric quantum mechanics is physically different from ordinary quantum mechanics. An entire chapter of \textcite{pt579},
Chap.~8, is devoted to a discussion of this question.

Finally, we emphasize that the formal mapping from a $\cPT$-symmetric Hamiltonian $H$ to a Hermitian Hamiltonian $h$ is not possible unless $H$ has an unbroken $\cPT$ symmetry. The mapping from $H$ to $h$ cannot be analytically continued from the unbroken to the broken region. For example, the entire analysis in Subsec.~\ref{ss6D} fails if $g^2<r^2\sin^2(\theta)$. If $H$ has a broken $\cPT$ symmetry, there are usually a finite number of real eigenvalues and an infinite number of complex eigenvalues, as in Fig.~\ref{f3}. In fact, this is what we see in the real world; there are a small number of stable fundamental particles and the rest are unstable \cite{pt174}.

\subsection{Observables}\label{ss6H}
In Hermitian quantum mechanics a linear operator $A$ must be Hermitian to be an observable: $A=A^\dag$. This condition ensures that the expectation value of $A$ in a state is real. Furthermore, operators in the Heisenberg picture evolve in time according to $A(t)=e^{iHt}A(0) e^{-iHt}$, so the Hermiticity condition is maintained in time.

In $\cPT$-symmetric quantum mechanics the condition for an operator $A$ to be an observable is
\begin{equation}
A^{\rm T}=\cCPT A\,\cCPT,
\label{e6.38}
\end{equation}
where $A^{\rm T}$ is the {\it matrix transpose} of $A$ \cite{pt90,pt178}. If (\ref{e6.38}) holds at $t=0$, then it continues to hold for $t>0$ because we have assumed that the Hamiltonian $H$ is symmetric ($H=H^{\rm T}$). This condition also guarantees that the expectation value of $A$ in any state is real. As stated above, the symmetry requirement (\ref{e6.38}) for $A$ to be an observable involves matrix transposition. Thus, (\ref{e6.38}) is more restrictive than necessary and it has been generalized in \cite{pt380,pt391,pt392}. 
Again, if the matrix-transpose-symmetry condition on the Hamiltonian is removed, we must introduce a biorthogonal basis \cite{pt370,pt371,pt372}.

The operator $\cC$ itself satisfies (\ref{e6.38}), so it is an observable. The Hamiltonian is also an observable. However, as explained in Sec.~\ref{ss3F}, the $x$ operator in $\cPT$-symmetric quantum theory is not an observable and the $p$ operator is also not an observable. Indeed, as noted earlier, the expectation value of $x$ in the ground state is a negative imaginary number. Thus, $\cPT$-symmetric quantum mechanics is somewhat similar to fermionic quantum field theories, in which the fermion field is complex and does not have a classical limit.

To see why the expectation value of the $x$ operator in $\cPT$-symmetric quantum mechanics is a negative imaginary number, we examine the classical trajectories shown in Fig.~\ref{f9}. All these classical trajectories have left-right ($\cPT$) symmetry, but not up-down symmetry. Furthermore, the classical paths spend more time in the lower-half complex-$x$ plane. Thus, the average classical position must be a negative imaginary number. Just as the classical particle moves about in the complex plane, the quantum probability current flows about in the complex plane \cite{pt491,pt525}. One plausible interpretation of $\cPT$-symmetric quantum mechanics is that it describes extended rather than pointlike objects, where these objects extend in the imaginary as well as in the real direction.

\section{Opportunities for Research}\label{s7}
The study of $\cPT$ symmetry is a rich source of theoretical and experimental research problems. This section gives a small and limited summary of some of the research accomplishments so far. We are confident that this area of physics will continue to provide a vast range of new and exciting research challenges.

Early theoretical work on $\cPT$ symmetry used complex-variable methods to elucidate the properties of quantum-mechanical models, to formulate proofs of spectral reality, and to examine the mathematical structure of the space of states. Early work not discussed in this Review includes studies of $\cPT$-symmetric periodic potentials \cite{pt282,pt681}.
Also not discussed here is the work on classical nonlinear $\cPT$-symmetric wave equations, which were shown to exhibit a wide range of unexpected properties \cite{pt45,pt46,pt64,pt66,pt67,pt644,pt645,pt646, pt647,pt648}. Remarkable connections between the instablilites of nonlinear partial differential equations, such as the Painlev\'e transcendents, and $\cPT$-symmetric Schr\"odinger equations were discovered \cite{pt47,pt59,pt48,pt531,pt532}. 

Early experimental work focused on observing the transition between regions of broken and unbroken $\cPT$ symmetry in {\it classical} systems having balanced loss and gain. Experiments using lasers and wave guides observed the phenomenon of unidirectional transmission \cite{pt193,pt544}. These experiments suggest the possibility of developing $\cPT$-symmetric optical components to manipulate and control optical beams \cite{pt184,pt702}. The $\cPT$ transition was also observed in studies of superconducting wires \cite{pt691, pt692,pt693} and in $\cPT$-symmetric electronic circuits \cite{pt488}. It was found that electronic $\cPT$-symmetric systems can be used to optimize electrical energy transfer \cite{pt485}; this may have important commercial applications. 

Additionally, the $\cPT$ transition was seen in mechanical \cite{pt487}, acoustic \cite{pt242,pt243, pt100,pt549,pt686}, and optomechanical systems \cite{pt699,pt700,pt701}, and also in optical and microwave cavities \cite{pt465,pt499,pt664, pt683}. Fluid-flow instabilities, such as the Kelvin-Helmholtz instability, also have a $\cPT$-symmetric interpretation \cite{pt617,pt618,pt619}. 

These early experiments led to studies of photonic crystals \cite{pt543,pt547,pt548,pt552,pt665}, multidimensional classical systems \cite{pt237,pt238,pt239}, plasmonics \cite{pt667}, and also recent work on $\cPT$-symmetric
condensed-matter systems \cite{pt629, pt630,pt631, pt632,pt633}.

There have been many analytical studies of $\cPT$-symmetric quantum field theory and particle physics. The renormalized Lee model and the Pais-Uhlenbeck model were shown to be $\cPT$ symmetric \cite{pt436,pt530,pt634}. Scalar $\cPT$-symmetric
quantum field theories were explored \cite{pt703,pt704}, more complicated interacting theories were studied \cite{pt267,pt455,pt622,pt643,pt651,pt652,pt653,pt654,pt655,pt656,pt657,pt658,pt659,pt660,pt661}, and $\cPT$-symmetric effects in QCD were examined \cite{pt695,pt696, pt697,pt698}. Connections between the massless $\cPT$-symmetric Dirac equation and neutrino oscillations were found \cite{pt563,pt564,pt572, pt659}.

In Hermitian quantum systems exceptional points (square-root singularities) are of theoretical interest but are not experimentally accessible. However, experimental studies of exceptional points are possible in $\cPT$-symmetric systems \cite{pt663,pt668,pt669, pt670,pt671, pt672,pt673, pt674,pt675,pt141}. Near exceptional points one can make extremely accurate physical measurements. This phenomenon is known as {\it enhanced sensitivity} \cite{pt509,pt511,pt592,pt706}. Enhanced-sensitivity techniques are currently being implemented for delicate measurements of gravitational waves \cite{pt705}.

Cold-atom-beam experiments are currently being designed that will make direct measurements of $\cPT$-symmetric quantum energy levels in upside-down potentials \cite{pt567,pt594,pt595}. Also, discrete statistical-mechanical models are being studied and noteworthy topological effects have been found \cite{pt592}.

{\it Broken}-$\cPT$-symmetric phases typically have a small number of real eigenvalues and many complex eigenvalues \cite{pt292}. This mirrors what is seen in nature, where there are a few stable particles and many unstable particles \cite{pt174}. This warrants further investigation.

Another possible research problem concerns the connection between different Hermitian and $\cPT$-symmetric phases of a Hamiltonian. For example, in conventional QED the interaction term $eA^\mu J_\mu$ is both $\cP$ and $\cT$ symmetric. Without changing the classical field equations, we might interpret the $A^\mu$ field as a pseudovector instead of a vector. The new theory is parity violating. If we then replace $e$ by $ie$, in the resulting non-Hermitian $\cPT$-symmetric phase, like charges attract. This theory behaves very much like a theory of magnetic charge \cite{pt453,pt454}.

Similarly, one might obtain a non-Hermitian large-distance alternative phase of a quantum gravitational theory in which masses repel instead of attracting. Perhaps such an investigation might explain dark energy and the accelerating expansion of the universe \cite{pt639,pt642,pt666,pt688,pt490}.

\acknowledgments
We thank our many colleagues and collaborators for their insightful and informative discussions over the years. In particular, we thank Michael V. Berry, Stefan Boettcher, Vincenzo Branchina, Dorje C. Brody, Junhua Chen, Demetrios N. Christodoulides,
Patrick Dorey, Andreas Fring, Mariagiovanna Gianfreda, Uwe G\"unther, Naomichi Hatano, Darryl D. Holm, Hugh F. Jones, Sandra P. Klevansky, Philip D. Mannheim, Nikolaos E. Mavromatos, Peter N. Meisinger, Kimball A. Milton, Ali Mostafazadeh, Michael C. Ogilvie, Raymond J. Rivers, Sarben Sarkar, Micheline Soley, A. Douglas Stone, Qing-hai Wang, and Miloslav Znojil. We are especially indebted to Jessica Bender for her tireless, persistent, and careful editing and proofreading. CMB thanks the Alexander von Humboldt and Simons Foundations, and the UK Engineering and Physical Sciences Research Council for financial support.

\appendix
\section{Increasing role of complex variables in physics}\label{sa}
The theme of this Review is that although we can only observe and measure real quantities, the dynamical behavior of a quantum-mechanical system is determined in part by what happens in the complex plane. For example, in particle physics a resonance in a scattering amplitude is associated with a Breit-Wigner pole in the complex-energy plane. In quantum mechanics when a particle tunnels, it vanishes on one side of a potential barrier and reappears on the other side; the motion of the particle as it passes through the barrier can be expressed as a sum over {\it complex} semiclassical trajectories.

As mentioned in Sec.~\ref{s1}, complex numbers did not play a role in physical theories until the advent of quantum mechanics. This Appendix emphasizes the importance of complex numbers in the formulation of modern physical theories as well as in mathematics, and in particular in $\cPT$-symmetric quantum mechanics.

The complex number system has been studied for centuries. In the sixteenth century G.~Cardano found that complex numbers were required to represent the roots of some polynomial equations and that the real number system alone was not sufficient. (Cardano was particularly interested in cubic equations.) In the ensuing years it became increasingly clear to mathematicians that the real number system was incomplete and that the real number line naturally extends to the complex plane. 

There is a subtle difference between extending the real line to the real plane and extending the real line to the {\it complex} plane. A point in the real plane represents a pair of unrelated real numbers $x$ and $y$, whereas a point in the complex plane is a {\it single} complex number $z=x+
iy$. In complex analysis one studies functions of {\it one} argument, namely, the complex variable $z$, and not functions of two independent real arguments $x$ and $y$.

Real analysis is empowered by embedding the real line in the complex plane; new techniques become available that can be used to solve difficult problems, and some features of real functions can be better understood. For example, complex analysis easily explains why the Taylor series
$\sum_{n=0}^\infty (-x^2)^n$,
which represents the real rational function $f(x)=\frac{1}{1+x^2}$, ceases to converge when $|x|\geq1$ even though $f(x)$ is smooth and infinitely differentiable for all real $x$. (This is due to singularities in the complex-$x$ plane at $\pm i$.) This example shows that what happens in the complex plane strongly affects what happens on the real axis. In modern physics we find that while the complex world is invisible to inhabitants of the real world, what happens in the complex plane has a powerful influence on what we observe in the real world.

There are countless examples of the power of complex-variable theory in explaining the behavior of real mathematics. For example, just a few lines of complex analysis establishes the fundamental theorem of algebra. Moreover, complex analysis explains the nature of real multivalued functions such as $\sqrt{x}$ and $\log(x)$ by defining them on Riemann surfaces. (Without resorting to a Riemann surface one cannot explain how the square-root function can be double-valued and still be a function.)

Complex-variable techniques are often needed to explain the behavior of {\it real} integrals. For example, Laplace's method (a standard tool of real analysis) fails to determine the asymptotic behavior of the real integral 
$$I(x)=\int_0^1 dt\,e^{-4xt^2}\cos\big(5xt-xt^3\big)$$
for large positive $x$. However, in just a few lines the complex method of steepest descents readily establishes that the asymptotic behavior of $I(x)$ is given by
$$I(x)\sim\sqrt {\pi/x}\,e^{-2x} \quad(x\to+\infty)$$ \cite{pt16}.
Even though the integral $I(x)$ is a {\it real} function of $x$, real analysis alone is 
insufficient to derive this asymptotic approximation. 

Like the case of the Taylor series for the function $1/(x^2+1)$, the asymptotic behavior of $I(x)$ is determined by a {\it saddle point} that lies off the real axis and in the complex plane at $t=i$. If we insist on using Laplace's method to obtain the asymptotic behavior, we run into trouble: Using real analysis, we would argue that for large positive $x$ the contribution to the integral comes from a localized range of $t$ around $t=0$ because the integrand contains a Gaussian. This statement is correct, but it suggests (wrongly) that for large $x$ we can replace the cosine term in the integrand by 1 (its value at $t=0$) and evaluate the resulting Gaussian integral. If we do so, we obtain the {\it incorrect} result
$$I(x)\sim\half\sqrt{\pi/x}\quad({\rm wrong!}).$$
If we try to improve this argument by approximating the cosine term by $\cos(5xt)$ instead of 1 (neglecting $t^3$ compared with $t$) and then evaluate the resulting Gaussian integral, we obtain a different but still incorrect result:
$$I(x)\sim\half\sqrt{\pi/x}\exp(-25x/16)\quad({\rm wrong!}).$$

Why is complex analysis so powerful? One can only approach a point on the real line from the left and from the right. In contrast, one can approach a point in the complex domain from infinitely many directions. Because limits are required to construct a quantity such as the derivative of a function, the properties of the limiting quantity are more constrained in complex-variable theory than they are in real-variable theory.

For the case of a derivative, this constraint takes the form of the Cauchy-Riemann equations and leads to the property of {\it analyticity}. In real-variable theory if the derivative $f'(x)$ exists in an open region of the real-$x$ axis, we cannot know whether higher derivatives of $f(x)$ exist in that region. However, in complex-variable theory if $f'(z)$ exists in an open region $R$ of the complex-$z$ plane, then $f(z)$ is analytic in $R$ and all higher derivatives of $f(z)$ exist in $R$ as well. Furthermore, the Taylor series for $f(z)$ about $z_0\in R$ converges inside the largest circle about $z_0$ that fits inside $R$. Analyticity enables us to use path independence and residue theory (for which there is no real-variable analog) to evaluate difficult real integrals and to perform complex transforms. Analyticity enables us to derive dispersion relations, which are immensely powerful tools in optics and scattering theory.

Another consequence of the complex limiting process is that, unlike the case of real-variable theory where there are two different infinite numbers, $+\infty$ and $-\infty$, in complex-variable theory there is a {\it unique} point at infinity \cite{pt583}. The point $\infty$ in the complex plane is defined as $\lim_{z\to0}\,1/z$, where $z$ approaches 0 {\it along any path} in the complex-$z$ plane. This is crucial in understanding how a classical particle at $x=0$ in a $\cPT$-symmetric $-x^4$ potential can drift off to $x=+\infty$ and then come back from $x=-\infty$ (see Fig.~\ref{f18}) or how a particle in a $-5x^2+x^4$ potential can drift off to $x=i\infty$ and come back from $x=-i\infty$ (see Figs.~\ref{f34} and \ref{f35}). Of course, a particle that begins at $x=0$ in a $-x^4$ potential may slide down to $x=+ \infty$ in finite time and simply {\it stay there} and never come back. Such a trajectory is not forbidden, but complex-variable theory {\it allows} such a particle to exhibit an alternative behavior by identifying the points $+\infty$ and $-\infty$ at opposite ends of the real axis.

Despite the analytical power of complex-variable theory, which was understood and used by theoretical physicists for centuries, complex variables were not incorporated into the many physical theories that were developed in the nineteenth century. Complex analysis provided powerful tools for solving differential equations, evaluating integrals, or deriving asymptotic approximations, but the number $i$ did not appear in the physical theories of classical mechanics, fluid dynamics, classical gravity, or electrodynamics. The number $i$ did not even appear in the formulation of special and general relativity in the early twentieth century. 

Virtually all of the fundamental physical phenomena that were studied before the discovery of quantum mechanics exhibit three kinds of characteristic behaviors, propagation, equilibrium, and diffusion. These three behaviors can be modeled by three classical linear partial differential equations. Propagation is modeled by the hyperbolic wave equation $u_{tt}= u_{xx}$, equilibrium is modeled by the elliptic equation $u_{xx}+u_{yy}=0$ (Laplace's equation), and diffusion is modeled by the parabolic equation $u_t= u_{xx}$. In none of these equations does the number $i$ appear explicitly and the only solutions that are of physical interest are real. (Interestingly, the real solutions to the two-dimensional Laplace's equation are the real or the imaginary parts of analytic functions.)

It was not until the theory of nonrelativistic quantum mechanics was formulated in the 1920s that the complex number $i$ first appeared in a fundamental way in the Schr\"odinger equation $i\psi_t=-\psi_{xx}+V(x)\psi$. Among the early successes of nonrelativistic quantum mechanics was the development of BCS theory in 1958; the complex number $i$ is essential in formulating this theory. Analyticity of scattering amplitudes is the key assumption in S-matrix theory \cite{pt612}, which had its heyday in the 1950s, and this led to studies of complex angular momentum \cite{pt613,pt626}. The Dirac equation and Fermi statistics rely explicitly on the complex number $i$.

The central mathematical content of quantum theory is that while a physical probabilistic prediction is a real number, one obtains this real number from a complex amplitude, which is determined by solving the complex Schr\"odinger equation or by evaluating a complex path integral. The principal point of this Review is that if we extend and deform real potentials into the complex domain and interpret these new potentials using complex-variable techniques, we obtain new, viable, and physically-allowed $\cPT$-symmetric quantum theories that can be studied both theoretically and experimentally.

\bibliography{pt}

\begin{thebibliography}{272}%
\makeatletter
\providecommand \@ifxundefined [1]{%
 \@ifx{#1\undefined}
}%
\providecommand \@ifnum [1]{%
 \ifnum #1\expandafter \@firstoftwo
 \else \expandafter \@secondoftwo
 \fi
}%
\providecommand \@ifx [1]{%
 \ifx #1\expandafter \@firstoftwo
 \else \expandafter \@secondoftwo
 \fi
}%
\providecommand \natexlab [1]{#1}%
\providecommand \enquote  [1]{``#1''}%
\providecommand \bibnamefont  [1]{#1}%
\providecommand \bibfnamefont [1]{#1}%
\providecommand \citenamefont [1]{#1}%
\providecommand \href@noop [0]{\@secondoftwo}%
\providecommand \href [0]{\begingroup \@sanitize@url \@href}%
\providecommand \@href[1]{\@@startlink{#1}\@@href}%
\providecommand \@@href[1]{\endgroup#1\@@endlink}%
\providecommand \@sanitize@url [0]{\catcode `\\12\catcode `\$12\catcode
  `\&12\catcode `\#12\catcode `\^12\catcode `\_12\catcode `\%12\relax}%
\providecommand \@@startlink[1]{}%
\providecommand \@@endlink[0]{}%
\providecommand \url  [0]{\begingroup\@sanitize@url \@url }%
\providecommand \@url [1]{\endgroup\@href {#1}{\urlprefix }}%
\providecommand \urlprefix  [0]{URL }%
\providecommand \Eprint [0]{\href }%
\providecommand \doibase [0]{https://doi.org/}%
\providecommand \selectlanguage [0]{\@gobble}%
\providecommand \bibinfo  [0]{\@secondoftwo}%
\providecommand \bibfield  [0]{\@secondoftwo}%
\providecommand \translation [1]{[#1]}%
\providecommand \BibitemOpen [0]{}%
\providecommand \bibitemStop [0]{}%
\providecommand \bibitemNoStop [0]{.\EOS\space}%
\providecommand \EOS [0]{\spacefactor3000\relax}%
\providecommand \BibitemShut  [1]{\csname bibitem#1\endcsname}%
\let\auto@bib@innerbib\@empty
\bibitem [{\citenamefont {Abbasi}\ \emph {et~al.}(2022)\citenamefont {Abbasi},
  \citenamefont {Chen}, \citenamefont {Naghiloo}, \citenamefont {Joglekar},\
  and\ \citenamefont {Murch}}]{pt632}%
  \BibitemOpen
  \bibfield  {author} {\bibinfo {author} {\bibnamefont {Abbasi}, \bibfnamefont
  {Maryam}}, \bibinfo {author} {\bibfnamefont {Weijian}\ \bibnamefont {Chen}},
  \bibinfo {author} {\bibfnamefont {Mahdi}\ \bibnamefont {Naghiloo}}, \bibinfo
  {author} {\bibfnamefont {Yogesh~N}\ \bibnamefont {Joglekar}}, and\ \bibinfo
  {author} {\bibfnamefont {Kater~W}\ \bibnamefont {Murch}}} (\bibinfo {year}
  {2022}),\ \bibfield  {title} {\enquote {\bibinfo {title} {Topological quantum
  state control through exceptional-point proximity},}\ }\href
  {https://doi.org/10.1103/PhysRevLett.128.160401} {\bibfield  {journal}
  {\bibinfo  {journal} {Physical Review Letters}\ }\textbf {\bibinfo {volume}
  {128}},\ 10.1103/PhysRevLett.128.160401}\BibitemShut {NoStop}%
\bibitem [{\citenamefont {Ablowitz}(2023)}]{pt645}%
  \BibitemOpen
  \bibfield  {author} {\bibinfo {author} {\bibnamefont {Ablowitz},
  \bibfnamefont {Mark~J}}} (\bibinfo {year} {2023}),\ \bibfield  {title}
  {\enquote {\bibinfo {title} {Nonlinear waves and the inverse scattering
  transform},}\ }\href {https://doi.org/10.1016/j.ijleo.2023.170710} {\bibfield
   {journal} {\bibinfo  {journal} {Optik}\ }\textbf {\bibinfo {volume} {278}},\
  10.1016/j.ijleo.2023.170710}\BibitemShut {NoStop}%
\bibitem [{\citenamefont {Ablowitz}\ \emph {et~al.}(2020)\citenamefont
  {Ablowitz}, \citenamefont {Luo},\ and\ \citenamefont {Musslimani}}]{pt644}%
  \BibitemOpen
  \bibfield  {author} {\bibinfo {author} {\bibnamefont {Ablowitz},
  \bibfnamefont {Mark~J}}, \bibinfo {author} {\bibfnamefont {Xu-Dan}\
  \bibnamefont {Luo}}, and\ \bibinfo {author} {\bibfnamefont {Ziad~H}\
  \bibnamefont {Musslimani}}} (\bibinfo {year} {2020}),\ \bibfield  {title}
  {\enquote {\bibinfo {title} {Discrete nonlocal nonlinear schr{\"o}dinger
  systems: Integrability, inverse scattering and solitons},}\ }\href
  {https://doi.org/10.1088/1361-6544/ab74ae} {\bibfield  {journal} {\bibinfo
  {journal} {Nonlinearity}\ }\textbf {\bibinfo {volume} {33}},\
  10.1088/1361-6544/ab74ae}\BibitemShut {NoStop}%
\bibitem [{\citenamefont {Ablowitz}\ and\ \citenamefont
  {Musslimani}(2013)}]{pt647}%
  \BibitemOpen
  \bibfield  {author} {\bibinfo {author} {\bibnamefont {Ablowitz},
  \bibfnamefont {Mark~J}}, and\ \bibinfo {author} {\bibfnamefont {Ziad~H}\
  \bibnamefont {Musslimani}}} (\bibinfo {year} {2013}),\ \bibfield  {title}
  {\enquote {\bibinfo {title} {Integrable nonlocal nonlinear schr{\"o}dinger
  equation},}\ }\href {https://doi.org/10.1103/physrevlett.110.064105}
  {\bibfield  {journal} {\bibinfo  {journal} {Physical Review Letters}\
  }\textbf {\bibinfo {volume} {110}},\
  10.1103/physrevlett.110.064105}\BibitemShut {NoStop}%
\bibitem [{\citenamefont {Ablowitz}\ and\ \citenamefont
  {Musslimani}(2014)}]{pt646}%
  \BibitemOpen
  \bibfield  {author} {\bibinfo {author} {\bibnamefont {Ablowitz},
  \bibfnamefont {Mark~J}}, and\ \bibinfo {author} {\bibfnamefont {Ziad~H}\
  \bibnamefont {Musslimani}}} (\bibinfo {year} {2014}),\ \bibfield  {title}
  {\enquote {\bibinfo {title} {Integrable discrete pt symmetric model},}\
  }\href {https://doi.org/10.1103/physreve.90.032912} {\bibfield  {journal}
  {\bibinfo  {journal} {Physical Review E}\ }\textbf {\bibinfo {volume} {90}},\
  10.1103/physreve.90.032912}\BibitemShut {NoStop}%
\bibitem [{\citenamefont {Ablowitz}\ and\ \citenamefont
  {Musslimani}(2016)}]{pt648}%
  \BibitemOpen
  \bibfield  {author} {\bibinfo {author} {\bibnamefont {Ablowitz},
  \bibfnamefont {Mark~J}}, and\ \bibinfo {author} {\bibfnamefont {Ziad~H}\
  \bibnamefont {Musslimani}}} (\bibinfo {year} {2016}),\ \bibfield  {title}
  {\enquote {\bibinfo {title} {Inverse scattering transform for the integrable
  nonlocal nonlinear schr\"{o}dinger equation},}\ }\href
  {https://doi.org/10.1088/0951-7715/29/3/915} {\bibfield  {journal} {\bibinfo
  {journal} {Nonlinearity}\ }\textbf {\bibinfo {volume} {29}},\
  10.1088/0951-7715/29/3/915}\BibitemShut {NoStop}%
\bibitem [{\citenamefont {Ahmed}\ \emph {et~al.}(2005)\citenamefont {Ahmed},
  \citenamefont {Bender},\ and\ \citenamefont {Berry}}]{pt567}%
  \BibitemOpen
  \bibfield  {author} {\bibinfo {author} {\bibnamefont {Ahmed}, \bibfnamefont
  {Zafar}}, \bibinfo {author} {\bibfnamefont {Carl~M}\ \bibnamefont {Bender}},
  and\ \bibinfo {author} {\bibfnamefont {M~V}\ \bibnamefont {Berry}}} (\bibinfo
  {year} {2005}),\ \bibfield  {title} {\enquote {\bibinfo {title}
  {Reflectionless potentials and $\mathcal{PT}$ symmetry},}\ }\href
  {https://doi.org/10.1088/0305-4470/38/39/L01} {\bibfield  {journal} {\bibinfo
   {journal} {Journal of Physics A: Mathematical and General}\ }\textbf
  {\bibinfo {volume} {38}},\ \bibinfo {pages} {L627}}\BibitemShut {NoStop}%
\bibitem [{\citenamefont {Alaeian}\ \emph {et~al.}(2016)\citenamefont
  {Alaeian}, \citenamefont {Baum}, \citenamefont {Jankovic}, \citenamefont
  {Lawrence},\ and\ \citenamefont {Dionne}}]{pt559}%
  \BibitemOpen
  \bibfield  {author} {\bibinfo {author} {\bibnamefont {Alaeian}, \bibfnamefont
  {Hadiseh}}, \bibinfo {author} {\bibfnamefont {Brian}\ \bibnamefont {Baum}},
  \bibinfo {author} {\bibfnamefont {Vladan}\ \bibnamefont {Jankovic}}, \bibinfo
  {author} {\bibfnamefont {Mark}\ \bibnamefont {Lawrence}}, and\ \bibinfo
  {author} {\bibfnamefont {Jennifer~A}\ \bibnamefont {Dionne}}} (\bibinfo
  {year} {2016}),\ \bibfield  {title} {\enquote {\bibinfo {title} {Towards
  nanoscale multiplexing with parity-time-symmetric plasmonic coaxial
  waveguides},}\ }\href {https://doi.org/10.1103/PhysRevB.93.205439} {\bibfield
   {journal} {\bibinfo  {journal} {Physical Review B}\ }\textbf {\bibinfo
  {volume} {93}},\ \bibinfo {pages} {205439}}\BibitemShut {NoStop}%
\bibitem [{\citenamefont {Alaeian}\ and\ \citenamefont {Dionne}(2014)}]{pt560}%
  \BibitemOpen
  \bibfield  {author} {\bibinfo {author} {\bibnamefont {Alaeian}, \bibfnamefont
  {Hadiseh}}, and\ \bibinfo {author} {\bibfnamefont {Jennifer~A}\ \bibnamefont
  {Dionne}}} (\bibinfo {year} {2014}),\ \bibfield  {title} {\enquote {\bibinfo
  {title} {Parity-time-symmetric plasmonic metamaterials},}\ }\href
  {https://doi.org/10.1103/PhysRevA.89.033829} {\bibfield  {journal} {\bibinfo
  {journal} {Physical Review A}\ }\textbf {\bibinfo {volume} {89}},\ \bibinfo
  {pages} {033829}}\BibitemShut {NoStop}%
\bibitem [{\citenamefont {Alexandre}\ \emph
  {et~al.}(2020{\natexlab{a}})\citenamefont {Alexandre}, \citenamefont
  {Ellis},\ and\ \citenamefont {Millington}}]{pt655}%
  \BibitemOpen
  \bibfield  {author} {\bibinfo {author} {\bibnamefont {Alexandre},
  \bibfnamefont {Jean}}, \bibinfo {author} {\bibfnamefont {John}\ \bibnamefont
  {Ellis}}, and\ \bibinfo {author} {\bibfnamefont {Peter}\ \bibnamefont
  {Millington}}} (\bibinfo {year} {2020}{\natexlab{a}}),\ \bibfield  {title}
  {\enquote {\bibinfo {title} {Discrete spacetime symmetries and particle
  mixing in non-hermitian scalar quantum field theories},}\ }\href
  {https://doi.org/10.1103/physrevd.102.125030} {\bibfield  {journal} {\bibinfo
   {journal} {Physical Review D}\ }\textbf {\bibinfo {volume} {102}},\
  10.1103/physrevd.102.125030}\BibitemShut {NoStop}%
\bibitem [{\citenamefont {Alexandre}\ \emph
  {et~al.}(2020{\natexlab{b}})\citenamefont {Alexandre}, \citenamefont
  {Ellis},\ and\ \citenamefont {Millington}}]{pt656}%
  \BibitemOpen
  \bibfield  {author} {\bibinfo {author} {\bibnamefont {Alexandre},
  \bibfnamefont {Jean}}, \bibinfo {author} {\bibfnamefont {John}\ \bibnamefont
  {Ellis}}, and\ \bibinfo {author} {\bibfnamefont {Peter}\ \bibnamefont
  {Millington}}} (\bibinfo {year} {2020}{\natexlab{b}}),\ \bibfield  {title}
  {\enquote {\bibinfo {title} {Pt-symmetric non-hermitian quantum field
  theories with supersymmetry},}\ }\href
  {https://doi.org/10.1103/physrevd.101.085015} {\bibfield  {journal} {\bibinfo
   {journal} {Physical Review D}\ }\textbf {\bibinfo {volume} {101}},\
  10.1103/physrevd.101.085015}\BibitemShut {NoStop}%
\bibitem [{\citenamefont {Alexandre}\ \emph {et~al.}(2022)\citenamefont
  {Alexandre}, \citenamefont {Ellis},\ and\ \citenamefont
  {Millington}}]{pt657}%
  \BibitemOpen
  \bibfield  {author} {\bibinfo {author} {\bibnamefont {Alexandre},
  \bibfnamefont {Jean}}, \bibinfo {author} {\bibfnamefont {John}\ \bibnamefont
  {Ellis}}, and\ \bibinfo {author} {\bibfnamefont {Peter}\ \bibnamefont
  {Millington}}} (\bibinfo {year} {2022}),\ \bibfield  {title} {\enquote
  {\bibinfo {title} {Discrete spacetime symmetries, second quantization, and
  inner products in a non-hermitian dirac fermionic field theory},}\ }\href
  {https://doi.org/10.1103/physrevd.106.065003} {\bibfield  {journal} {\bibinfo
   {journal} {Physical Review D}\ }\textbf {\bibinfo {volume} {106}},\
  10.1103/physrevd.106.065003}\BibitemShut {NoStop}%
\bibitem [{\citenamefont {Alexandre}\ \emph {et~al.}(2018)\citenamefont
  {Alexandre}, \citenamefont {Ellis}, \citenamefont {Millington},\ and\
  \citenamefont {Seynaeve}}]{pt652}%
  \BibitemOpen
  \bibfield  {author} {\bibinfo {author} {\bibnamefont {Alexandre},
  \bibfnamefont {Jean}}, \bibinfo {author} {\bibfnamefont {John}\ \bibnamefont
  {Ellis}}, \bibinfo {author} {\bibfnamefont {Peter}\ \bibnamefont
  {Millington}}, and\ \bibinfo {author} {\bibfnamefont {Dries}\ \bibnamefont
  {Seynaeve}}} (\bibinfo {year} {2018}),\ \bibfield  {title} {\enquote
  {\bibinfo {title} {Spontaneous symmetry breaking and the goldstone theorem in
  non-hermitian field theories},}\ }\href
  {https://doi.org/10.1103/physrevd.98.045001} {\bibfield  {journal} {\bibinfo
  {journal} {Physical Review D}\ }\textbf {\bibinfo {volume} {98}},\
  10.1103/physrevd.98.045001}\BibitemShut {NoStop}%
\bibitem [{\citenamefont {Alexandre}\ \emph {et~al.}(2019)\citenamefont
  {Alexandre}, \citenamefont {Ellis}, \citenamefont {Millington},\ and\
  \citenamefont {Seynaeve}}]{pt653}%
  \BibitemOpen
  \bibfield  {author} {\bibinfo {author} {\bibnamefont {Alexandre},
  \bibfnamefont {Jean}}, \bibinfo {author} {\bibfnamefont {John}\ \bibnamefont
  {Ellis}}, \bibinfo {author} {\bibfnamefont {Peter}\ \bibnamefont
  {Millington}}, and\ \bibinfo {author} {\bibfnamefont {Dries}\ \bibnamefont
  {Seynaeve}}} (\bibinfo {year} {2019}),\ \bibfield  {title} {\enquote
  {\bibinfo {title} {Gauge invariance and the englert-brout-higgs mechanism in
  non-hermitian field theories},}\ }\href
  {https://doi.org/10.1103/physrevd.99.075024} {\bibfield  {journal} {\bibinfo
  {journal} {Physical Review D}\ }\textbf {\bibinfo {volume} {99}},\
  10.1103/physrevd.99.075024}\BibitemShut {NoStop}%
\bibitem [{\citenamefont {Alexandre}\ \emph
  {et~al.}(2020{\natexlab{c}})\citenamefont {Alexandre}, \citenamefont {Ellis},
  \citenamefont {Millington},\ and\ \citenamefont {Seynaeve}}]{pt654}%
  \BibitemOpen
  \bibfield  {author} {\bibinfo {author} {\bibnamefont {Alexandre},
  \bibfnamefont {Jean}}, \bibinfo {author} {\bibfnamefont {John}\ \bibnamefont
  {Ellis}}, \bibinfo {author} {\bibfnamefont {Peter}\ \bibnamefont
  {Millington}}, and\ \bibinfo {author} {\bibfnamefont {Dries}\ \bibnamefont
  {Seynaeve}}} (\bibinfo {year} {2020}{\natexlab{c}}),\ \bibfield  {title}
  {\enquote {\bibinfo {title} {Spontaneously breaking non-abelian gauge
  symmetry in non-hermitian field theories},}\ }\href
  {https://doi.org/10.1103/physrevd.101.035008} {\bibfield  {journal} {\bibinfo
   {journal} {Physical Review D}\ }\textbf {\bibinfo {volume} {101}},\
  10.1103/physrevd.101.035008}\BibitemShut {NoStop}%
\bibitem [{\citenamefont {Alexandre}\ \emph
  {et~al.}(2020{\natexlab{d}})\citenamefont {Alexandre}, \citenamefont
  {Mavromatos},\ and\ \citenamefont {Soto}}]{pt661}%
  \BibitemOpen
  \bibfield  {author} {\bibinfo {author} {\bibnamefont {Alexandre},
  \bibfnamefont {Jean}}, \bibinfo {author} {\bibfnamefont {Nick~E}\
  \bibnamefont {Mavromatos}}, and\ \bibinfo {author} {\bibfnamefont {Alex}\
  \bibnamefont {Soto}}} (\bibinfo {year} {2020}{\natexlab{d}}),\ \bibfield
  {title} {\enquote {\bibinfo {title} {Dynamical majorana neutrino masses and
  axions i},}\ }\href {https://doi.org/10.1016/j.nuclphysb.2020.115212}
  {\bibfield  {journal} {\bibinfo  {journal} {Nuclear Physics B}\ }\textbf
  {\bibinfo {volume} {961}},\ 10.1016/j.nuclphysb.2020.115212}\BibitemShut
  {NoStop}%
\bibitem [{\citenamefont {Ambichl}\ \emph {et~al.}(2013)\citenamefont
  {Ambichl}, \citenamefont {Makris}, \citenamefont {Ge}, \citenamefont {Chong},
  \citenamefont {Stone},\ and\ \citenamefont {Rotter}}]{pt689}%
  \BibitemOpen
  \bibfield  {author} {\bibinfo {author} {\bibnamefont {Ambichl}, \bibfnamefont
  {Philipp}}, \bibinfo {author} {\bibfnamefont {Konstantinos~G}\ \bibnamefont
  {Makris}}, \bibinfo {author} {\bibfnamefont {Li}~\bibnamefont {Ge}}, \bibinfo
  {author} {\bibfnamefont {Yidong}\ \bibnamefont {Chong}}, \bibinfo {author}
  {\bibfnamefont {A~Douglas}\ \bibnamefont {Stone}}, and\ \bibinfo {author}
  {\bibfnamefont {Stefan}\ \bibnamefont {Rotter}}} (\bibinfo {year} {2013}),\
  \bibfield  {title} {\enquote {\bibinfo {title} {Breaking of pt symmetry in
  bounded and unbounded scattering systems},}\ }\href
  {https://doi.org/10.1103/physrevx.3.041030} {\bibfield  {journal} {\bibinfo
  {journal} {Physical Review X}\ }\textbf {\bibinfo {volume} {3}},\
  10.1103/physrevx.3.041030}\BibitemShut {NoStop}%
\bibitem [{\citenamefont {Anderson}\ \emph {et~al.}(2011)\citenamefont
  {Anderson}, \citenamefont {Bender},\ and\ \citenamefont {Morone}}]{pt23}%
  \BibitemOpen
  \bibfield  {author} {\bibinfo {author} {\bibnamefont {Anderson},
  \bibfnamefont {Alexander~G}}, \bibinfo {author} {\bibfnamefont {Carl~M}\
  \bibnamefont {Bender}}, and\ \bibinfo {author} {\bibfnamefont {Uriel~I}\
  \bibnamefont {Morone}}} (\bibinfo {year} {2011}),\ \bibfield  {title}
  {\enquote {\bibinfo {title} {Periodic orbits for classical particles having
  complex energy},}\ }\href {https://doi.org/10.1016/j.physleta.2011.07.051}
  {\bibfield  {journal} {\bibinfo  {journal} {Physics Letters A}\ }\textbf
  {\bibinfo {volume} {375}}~(\bibinfo {number} {39}),\ \bibinfo {pages}
  {3399--3404}}\BibitemShut {NoStop}%
\bibitem [{\citenamefont {Arnoldi}(1951)}]{pt614}%
  \BibitemOpen
  \bibfield  {author} {\bibinfo {author} {\bibnamefont {Arnoldi}, \bibfnamefont
  {W~E}}} (\bibinfo {year} {1951}),\ \bibfield  {title} {\enquote {\bibinfo
  {title} {The principle of minimized iterations in the solution of the matrix
  eigenvalue problem},}\ }\href {https://doi.org/10.1090/qam/42792} {\bibfield
  {journal} {\bibinfo  {journal} {Quarterly of Applied Mathematics}\ }\textbf
  {\bibinfo {volume} {9}},\ 10.1090/qam/42792}\BibitemShut {NoStop}%
\bibitem [{\citenamefont {Arpornthip}\ and\ \citenamefont
  {Bender}(2009)}]{pt516}%
  \BibitemOpen
  \bibfield  {author} {\bibinfo {author} {\bibnamefont {Arpornthip},
  \bibfnamefont {Tanwa}}, and\ \bibinfo {author} {\bibfnamefont {Carl~M}\
  \bibnamefont {Bender}}} (\bibinfo {year} {2009}),\ \bibfield  {title}
  {\enquote {\bibinfo {title} {Conduction bands in classical periodic
  potentials},}\ }\href {https://doi.org/10.1007/s12043-009-0117-5} {\bibfield
  {journal} {\bibinfo  {journal} {Pramana}\ }\textbf {\bibinfo {volume} {73}},\
  \bibinfo {pages} {259--267}}\BibitemShut {NoStop}%
\bibitem [{\citenamefont {Ashida}\ \emph {et~al.}(2017)\citenamefont {Ashida},
  \citenamefont {Furukawa},\ and\ \citenamefont {Ueda}}]{pt555}%
  \BibitemOpen
  \bibfield  {author} {\bibinfo {author} {\bibnamefont {Ashida}, \bibfnamefont
  {Yuto}}, \bibinfo {author} {\bibfnamefont {Shunsuke}\ \bibnamefont
  {Furukawa}}, and\ \bibinfo {author} {\bibfnamefont {Masahito}\ \bibnamefont
  {Ueda}}} (\bibinfo {year} {2017}),\ \bibfield  {title} {\enquote {\bibinfo
  {title} {Parity-time-symmetric quantum critical phenomena},}\ }\href
  {https://doi.org/10.1038/ncomms15791} {\bibfield  {journal} {\bibinfo
  {journal} {Nature Communications}\ }\textbf {\bibinfo {volume} {8}},\
  \bibinfo {pages} {15791}}\BibitemShut {NoStop}%
\bibitem [{\citenamefont {Assawaworrarit}\ \emph {et~al.}(2017)\citenamefont
  {Assawaworrarit}, \citenamefont {Yu},\ and\ \citenamefont {Fan}}]{pt485}%
  \BibitemOpen
  \bibfield  {author} {\bibinfo {author} {\bibnamefont {Assawaworrarit},
  \bibfnamefont {Sid}}, \bibinfo {author} {\bibfnamefont {Xiaofang}\
  \bibnamefont {Yu}}, and\ \bibinfo {author} {\bibfnamefont {Shanhui}\
  \bibnamefont {Fan}}} (\bibinfo {year} {2017}),\ \bibfield  {title} {\enquote
  {\bibinfo {title} {Robust wireless power transfer using a nonlinear
  parity-time-symmetric circuit},}\ }\href
  {https://doi.org/10.1038/nature22404} {\bibfield  {journal} {\bibinfo
  {journal} {Nature}\ }\textbf {\bibinfo {volume} {546}},\
  10.1038/nature22404}\BibitemShut {NoStop}%
\bibitem [{\citenamefont {Assis}\ and\ \citenamefont {Fring}(2009)}]{pt45}%
  \BibitemOpen
  \bibfield  {author} {\bibinfo {author} {\bibnamefont {Assis}, \bibfnamefont
  {P~E~G}}, and\ \bibinfo {author} {\bibfnamefont {A}~\bibnamefont {Fring}}}
  (\bibinfo {year} {2009}),\ \bibfield  {title} {\enquote {\bibinfo {title}
  {Integrable models from $\cal{PT}$-symmetric deformations},}\ }\href
  {https://doi.org/10.1088/1751-8113/42/10/105206} {\bibfield  {journal}
  {\bibinfo  {journal} {Journal of Physics A: Mathematical and Theoretical}\
  }\textbf {\bibinfo {volume} {42}},\ \bibinfo {pages} {105206}}\BibitemShut
  {NoStop}%
\bibitem [{\citenamefont {Assis}\ and\ \citenamefont {Fring}(2010)}]{pt46}%
  \BibitemOpen
  \bibfield  {author} {\bibinfo {author} {\bibnamefont {Assis}, \bibfnamefont
  {P~E~G}}, and\ \bibinfo {author} {\bibfnamefont {A}~\bibnamefont {Fring}}}
  (\bibinfo {year} {2010}),\ \bibfield  {title} {\enquote {\bibinfo {title}
  {Compactons versus solitons},}\ }\href
  {https://doi.org/10.1007/s12043-010-0078-8} {\bibfield  {journal} {\bibinfo
  {journal} {Pramana-Journal of Physics}\ }\textbf {\bibinfo {volume} {74}},\
  \bibinfo {pages} {857--865}}\BibitemShut {NoStop}%
\bibitem [{\citenamefont {Aur\'{e}gan}\ and\ \citenamefont
  {Pagneux}(2017)}]{pt500}%
  \BibitemOpen
  \bibfield  {author} {\bibinfo {author} {\bibnamefont {Aur\'{e}gan},
  \bibfnamefont {Yves}}, and\ \bibinfo {author} {\bibfnamefont {Vincent}\
  \bibnamefont {Pagneux}}} (\bibinfo {year} {2017}),\ \bibfield  {title}
  {\enquote {\bibinfo {title} {$\mathcal{PT}$-symmetric scattering in flow duct
  acoustics},}\ }\href {https://doi.org/10.1103/physrevlett.118.174301}
  {\bibfield  {journal} {\bibinfo  {journal} {Physical Review Letters}\
  }\textbf {\bibinfo {volume} {118}},\
  10.1103/physrevlett.118.174301}\BibitemShut {NoStop}%
\bibitem [{\citenamefont {Bagchi}\ \emph {et~al.}(2002)\citenamefont {Bagchi},
  \citenamefont {Mallik},\ and\ \citenamefont {Quesne}}]{pt382}%
  \BibitemOpen
  \bibfield  {author} {\bibinfo {author} {\bibnamefont {Bagchi}, \bibfnamefont
  {B}}, \bibinfo {author} {\bibfnamefont {S}~\bibnamefont {Mallik}}, and\
  \bibinfo {author} {\bibfnamefont {C}~\bibnamefont {Quesne}}} (\bibinfo {year}
  {2002}),\ \bibfield  {title} {\enquote {\bibinfo {title}
  {$\mathcal{PT}$-symmetric square well and the associated {SUSY}
  hierarchies},}\ }\href {https://doi.org/10.1142/S0217732302008009} {\bibfield
   {journal} {\bibinfo  {journal} {Modern Physics Letters A}\ }\textbf
  {\bibinfo {volume} {17}},\ 10.1142/S0217732302008009}\BibitemShut {NoStop}%
\bibitem [{\citenamefont {Baranov}\ \emph {et~al.}(2017)\citenamefont
  {Baranov}, \citenamefont {Krasnok}, \citenamefont {Shegai}, \citenamefont
  {Al\`{u}},\ and\ \citenamefont {Chong}}]{pt542}%
  \BibitemOpen
  \bibfield  {author} {\bibinfo {author} {\bibnamefont {Baranov}, \bibfnamefont
  {Denis~G}}, \bibinfo {author} {\bibfnamefont {Alex}\ \bibnamefont {Krasnok}},
  \bibinfo {author} {\bibfnamefont {Timur}\ \bibnamefont {Shegai}}, \bibinfo
  {author} {\bibfnamefont {Andrea}\ \bibnamefont {Al\`{u}}}, and\ \bibinfo
  {author} {\bibfnamefont {Yidong}\ \bibnamefont {Chong}}} (\bibinfo {year}
  {2017}),\ \bibfield  {title} {\enquote {\bibinfo {title} {Coherent perfect
  absorbers: linear control of light with light},}\ }\href
  {https://doi.org/10.1038/natrevmats.2017.64} {\bibfield  {journal} {\bibinfo
  {journal} {Nature Reviews: Materials}\ }\textbf {\bibinfo {volume} {2}},\
  \bibinfo {pages} {17064}}\BibitemShut {NoStop}%
\bibitem [{\citenamefont {Barashenkov}\ \emph {et~al.}(2013)\citenamefont
  {Barashenkov}, \citenamefont {Jackson},\ and\ \citenamefont {Flach}}]{pt222}%
  \BibitemOpen
  \bibfield  {author} {\bibinfo {author} {\bibnamefont {Barashenkov},
  \bibfnamefont {I~V}}, \bibinfo {author} {\bibfnamefont {G~S}\ \bibnamefont
  {Jackson}}, and\ \bibinfo {author} {\bibfnamefont {S}~\bibnamefont {Flach}}}
  (\bibinfo {year} {2013}),\ \bibfield  {title} {\enquote {\bibinfo {title}
  {Blow-up regimes in the $\cal{PT}$-symmetric coupler and the actively coupled
  dimer},}\ }\href {https://doi.org/10.1103/PhysRevA.88.053817} {\bibfield
  {journal} {\bibinfo  {journal} {Physical Review A}\ }\textbf {\bibinfo
  {volume} {88}},\ \bibinfo {pages} {053817}}\BibitemShut {NoStop}%
\bibitem [{\citenamefont {Bender}\ and\ \citenamefont
  {Boettcher}(1998{\natexlab{a}})}]{pt338}%
  \BibitemOpen
  \bibfield  {author} {\bibinfo {author} {\bibnamefont {Bender}, \bibfnamefont
  {C~M}}, and\ \bibinfo {author} {\bibfnamefont {S}~\bibnamefont {Boettcher}}}
  (\bibinfo {year} {1998}{\natexlab{a}}),\ \bibfield  {title} {\enquote
  {\bibinfo {title} {Quasi-exactly solvable quartic potential},}\ }\href
  {https://doi.org/10.1088/0305-4470/31/14/001} {\bibfield  {journal} {\bibinfo
   {journal} {Journal of Physics A: Mathematical and General}\ }\textbf
  {\bibinfo {volume} {31}},\ \bibinfo {pages} {L273--L277}}\BibitemShut
  {NoStop}%
\bibitem [{\citenamefont {Bender}\ \emph
  {et~al.}(2007{\natexlab{a}})\citenamefont {Bender}, \citenamefont {Brody},
  \citenamefont {Chen},\ and\ \citenamefont {Furlan}}]{pt66}%
  \BibitemOpen
  \bibfield  {author} {\bibinfo {author} {\bibnamefont {Bender}, \bibfnamefont
  {C~M}}, \bibinfo {author} {\bibfnamefont {D~C}\ \bibnamefont {Brody}},
  \bibinfo {author} {\bibfnamefont {J-H}\ \bibnamefont {Chen}}, and\ \bibinfo
  {author} {\bibfnamefont {E}~\bibnamefont {Furlan}}} (\bibinfo {year}
  {2007}{\natexlab{a}}),\ \bibfield  {title} {\enquote {\bibinfo {title}
  {$\cal{PT}$-symmetric extension of the {K}orteweg-de {V}ries equation},}\
  }\href {https://doi.org/10.1088/1751-8113/40/5/F02} {\bibfield  {journal}
  {\bibinfo  {journal} {Journal of Physics A: Mathematical and Theoretical}\
  }\textbf {\bibinfo {volume} {40}}~(\bibinfo {number} {5}),\ \bibinfo {pages}
  {F153--F160}}\BibitemShut {NoStop}%
\bibitem [{\citenamefont {Bender}\ \emph
  {et~al.}(2009{\natexlab{a}})\citenamefont {Bender}, \citenamefont {Cooper},
  \citenamefont {Khare}, \citenamefont {Mihaila},\ and\ \citenamefont
  {Saxena}}]{pt67}%
  \BibitemOpen
  \bibfield  {author} {\bibinfo {author} {\bibnamefont {Bender}, \bibfnamefont
  {C~M}}, \bibinfo {author} {\bibfnamefont {F}~\bibnamefont {Cooper}}, \bibinfo
  {author} {\bibfnamefont {Avinash}\ \bibnamefont {Khare}}, \bibinfo {author}
  {\bibfnamefont {B}~\bibnamefont {Mihaila}}, and\ \bibinfo {author}
  {\bibfnamefont {A}~\bibnamefont {Saxena}}} (\bibinfo {year}
  {2009}{\natexlab{a}}),\ \bibfield  {title} {\enquote {\bibinfo {title}
  {Compactons in $\cal{PT}$-symmetric generalized {K}orteweg-de {V}ries
  equations},}\ }\href {https://doi.org/10.1007/s12043-009-0129-1} {\bibfield
  {journal} {\bibinfo  {journal} {Pramana-Journal of Physics}\ }\textbf
  {\bibinfo {volume} {73}}~(\bibinfo {number} {2}),\ \bibinfo {pages}
  {375--385}}\BibitemShut {NoStop}%
\bibitem [{\citenamefont {Bender}\ and\ \citenamefont {Feinberg}(2008)}]{pt64}%
  \BibitemOpen
  \bibfield  {author} {\bibinfo {author} {\bibnamefont {Bender}, \bibfnamefont
  {C~M}}, and\ \bibinfo {author} {\bibfnamefont {J}~\bibnamefont {Feinberg}}}
  (\bibinfo {year} {2008}),\ \bibfield  {title} {\enquote {\bibinfo {title}
  {Does the complex deformation of the {R}iemann equation exhibit shocks?}}\
  }\href {https://doi.org/10.1088/1751-8113/41/24/244004} {\bibfield  {journal}
  {\bibinfo  {journal} {Journal of Physics A: Mathematical and Theoretical}\
  }\textbf {\bibinfo {volume} {41}}~(\bibinfo {number} {24}),\ \bibinfo {pages}
  {244004}}\BibitemShut {NoStop}%
\bibitem [{\citenamefont {Bender}(2005)}]{pt175}%
  \BibitemOpen
  \bibfield  {author} {\bibinfo {author} {\bibnamefont {Bender}, \bibfnamefont
  {Carl~M}}} (\bibinfo {year} {2005}),\ \bibfield  {title} {\enquote {\bibinfo
  {title} {Introduction to $\cal{PT}$-symmetric quantum theory},}\ }\href
  {https://doi.org/10.1080/00107500072632} {\bibfield  {journal} {\bibinfo
  {journal} {Contemporary Physics}\ }\textbf {\bibinfo {volume} {46}},\
  \bibinfo {pages} {277--292}}\BibitemShut {NoStop}%
\bibitem [{\citenamefont {Bender}\ \emph
  {et~al.}(2013{\natexlab{a}})\citenamefont {Bender}, \citenamefont {Berntson},
  \citenamefont {Parker},\ and\ \citenamefont {Samuel}}]{pt487}%
  \BibitemOpen
  \bibfield  {author} {\bibinfo {author} {\bibnamefont {Bender}, \bibfnamefont
  {Carl~M}}, \bibinfo {author} {\bibfnamefont {Bjorn~K}\ \bibnamefont
  {Berntson}}, \bibinfo {author} {\bibfnamefont {David}\ \bibnamefont
  {Parker}}, and\ \bibinfo {author} {\bibfnamefont {E}~\bibnamefont {Samuel}}}
  (\bibinfo {year} {2013}{\natexlab{a}}),\ \bibfield  {title} {\enquote
  {\bibinfo {title} {Observation of $\mathcal{PT}$ phase transition in a simple
  mechanical system},}\ }\href {https://doi.org/10.1119/1.4789549} {\bibfield
  {journal} {\bibinfo  {journal} {American Journal of Physics}\ }\textbf
  {\bibinfo {volume} {81}},\ 10.1119/1.4789549}\BibitemShut {NoStop}%
\bibitem [{\citenamefont {Bender}\ and\ \citenamefont
  {Boettcher}(1998{\natexlab{b}})}]{pt1}%
  \BibitemOpen
  \bibfield  {author} {\bibinfo {author} {\bibnamefont {Bender}, \bibfnamefont
  {Carl~M}}, and\ \bibinfo {author} {\bibfnamefont {Stefan}\ \bibnamefont
  {Boettcher}}} (\bibinfo {year} {1998}{\natexlab{b}}),\ \bibfield  {title}
  {\enquote {\bibinfo {title} {Real spectra in non-{H}ermitian {H}amiltonians
  having $\mathcal{PT}$ symmetry},}\ }\href
  {https://doi.org/10.1103/PhysRevLett.80.5243} {\bibfield  {journal} {\bibinfo
   {journal} {Physical Review Letters}\ }\textbf {\bibinfo {volume} {80}},\
  \bibinfo {pages} {5243--5246}}\BibitemShut {NoStop}%
\bibitem [{\citenamefont {Bender}\ \emph
  {et~al.}(1999{\natexlab{a}})\citenamefont {Bender}, \citenamefont
  {Boettcher}, \citenamefont {Jones},\ and\ \citenamefont {Savage}}]{pt327}%
  \BibitemOpen
  \bibfield  {author} {\bibinfo {author} {\bibnamefont {Bender}, \bibfnamefont
  {Carl~M}}, \bibinfo {author} {\bibfnamefont {Stefan}\ \bibnamefont
  {Boettcher}}, \bibinfo {author} {\bibfnamefont {H~F}\ \bibnamefont {Jones}},
  and\ \bibinfo {author} {\bibfnamefont {Van~M}\ \bibnamefont {Savage}}}
  (\bibinfo {year} {1999}{\natexlab{a}}),\ \bibfield  {title} {\enquote
  {\bibinfo {title} {Complex square well -- a new exactly solvable quantum
  mechanical model},}\ }\href {https://doi.org/10.1088/0305-4470/32/39/305}
  {\bibfield  {journal} {\bibinfo  {journal} {Journal of Physics A:
  Mathematical and General}\ }\textbf {\bibinfo {volume} {32}},\ \bibinfo
  {pages} {6771--6781}}\BibitemShut {NoStop}%
\bibitem [{\citenamefont {Bender}\ \emph
  {et~al.}(2001{\natexlab{a}})\citenamefont {Bender}, \citenamefont
  {Boettcher}, \citenamefont {Jones}, \citenamefont {Meisinger},\ and\
  \citenamefont {{\d S}im{\d s}ek}}]{pt428}%
  \BibitemOpen
  \bibfield  {author} {\bibinfo {author} {\bibnamefont {Bender}, \bibfnamefont
  {Carl~M}}, \bibinfo {author} {\bibfnamefont {Stefan}\ \bibnamefont
  {Boettcher}}, \bibinfo {author} {\bibfnamefont {Hugh~F}\ \bibnamefont
  {Jones}}, \bibinfo {author} {\bibfnamefont {Peter~N}\ \bibnamefont
  {Meisinger}}, and\ \bibinfo {author} {\bibfnamefont {Mehmet}\ \bibnamefont
  {{\d S}im{\d s}ek}}} (\bibinfo {year} {2001}{\natexlab{a}}),\ \bibfield
  {title} {\enquote {\bibinfo {title} {Bound states of non-{H}ermitian quantum
  field theories},}\ }\href {https://doi.org/10.1016/S0375-9601(01)00745-9}
  {\bibfield  {journal} {\bibinfo  {journal} {Physics Letters A}\ }\textbf
  {\bibinfo {volume} {291}},\ 10.1016/S0375-9601(01)00745-9}\BibitemShut
  {NoStop}%
\bibitem [{\citenamefont {Bender}\ \emph {et~al.}(1992)\citenamefont {Bender},
  \citenamefont {Boettcher},\ and\ \citenamefont {Lipatov}}]{pt18}%
  \BibitemOpen
  \bibfield  {author} {\bibinfo {author} {\bibnamefont {Bender}, \bibfnamefont
  {Carl~M}}, \bibinfo {author} {\bibfnamefont {Stefan}\ \bibnamefont
  {Boettcher}}, and\ \bibinfo {author} {\bibfnamefont {Lev}\ \bibnamefont
  {Lipatov}}} (\bibinfo {year} {1992}),\ \bibfield  {title} {\enquote {\bibinfo
  {title} {Almost zero-dimensional quantum field theories},}\ }\href
  {https://doi.org/10.1103/PhysRevD.46.5557} {\bibfield  {journal} {\bibinfo
  {journal} {Physical Review D}\ }\textbf {\bibinfo {volume} {46}},\ \bibinfo
  {pages} {5557--5573}}\BibitemShut {NoStop}%
\bibitem [{\citenamefont {Bender}\ \emph
  {et~al.}(1999{\natexlab{b}})\citenamefont {Bender}, \citenamefont
  {Boettcher},\ and\ \citenamefont {Meisinger}}]{pt173}%
  \BibitemOpen
  \bibfield  {author} {\bibinfo {author} {\bibnamefont {Bender}, \bibfnamefont
  {Carl~M}}, \bibinfo {author} {\bibfnamefont {Stefan}\ \bibnamefont
  {Boettcher}}, and\ \bibinfo {author} {\bibfnamefont {Peter~N}\ \bibnamefont
  {Meisinger}}} (\bibinfo {year} {1999}{\natexlab{b}}),\ \bibfield  {title}
  {\enquote {\bibinfo {title} {$\cal{PT}$-symmetric quantum mechanics},}\
  }\href {https://doi.org/10.1063/1.532860} {\bibfield  {journal} {\bibinfo
  {journal} {Journal of Mathematical Physics}\ }\textbf {\bibinfo {volume}
  {40}},\ \bibinfo {pages} {2201--2229}}\BibitemShut {NoStop}%
\bibitem [{\citenamefont {Bender}\ \emph
  {et~al.}(2000{\natexlab{a}})\citenamefont {Bender}, \citenamefont
  {Boettcher},\ and\ \citenamefont {Savage}}]{pt373}%
  \BibitemOpen
  \bibfield  {author} {\bibinfo {author} {\bibnamefont {Bender}, \bibfnamefont
  {Carl~M}}, \bibinfo {author} {\bibfnamefont {Stefan}\ \bibnamefont
  {Boettcher}}, and\ \bibinfo {author} {\bibfnamefont {Van~M}\ \bibnamefont
  {Savage}}} (\bibinfo {year} {2000}{\natexlab{a}}),\ \bibfield  {title}
  {\enquote {\bibinfo {title} {Conjecture on the interlacing of zeros in
  complex {S}turm-{L}iouville problems},}\ }\href
  {https://doi.org/10.1063/1.1288247} {\bibfield  {journal} {\bibinfo
  {journal} {Journal of Mathematical Physics}\ }\textbf {\bibinfo {volume}
  {41}},\ 10.1063/1.1288247}\BibitemShut {NoStop}%
\bibitem [{\citenamefont {Bender}\ \emph
  {et~al.}(2013{\natexlab{b}})\citenamefont {Bender}, \citenamefont
  {Branchina},\ and\ \citenamefont {Messina}}]{pt611}%
  \BibitemOpen
  \bibfield  {author} {\bibinfo {author} {\bibnamefont {Bender}, \bibfnamefont
  {Carl~M}}, \bibinfo {author} {\bibfnamefont {V}~\bibnamefont {Branchina}},
  and\ \bibinfo {author} {\bibfnamefont {Emanuele}\ \bibnamefont {Messina}}}
  (\bibinfo {year} {2013}{\natexlab{b}}),\ \bibfield  {title} {\enquote
  {\bibinfo {title} {Critical behavior of the $\mathcal{PT}$-symmetric
  $i\phi^3$ quantum field theory},}\ }\href
  {https://doi.org/10.1103/PhysRevD.87.085029} {\bibfield  {journal} {\bibinfo
  {journal} {Physical Review D}\ }\textbf {\bibinfo {volume} {87}},\
  10.1103/PhysRevD.87.085029}\BibitemShut {NoStop}%
\bibitem [{\citenamefont {Bender}\ \emph
  {et~al.}(2005{\natexlab{a}})\citenamefont {Bender}, \citenamefont {Brandt},
  \citenamefont {Chen},\ and\ \citenamefont {Wang}}]{pt436}%
  \BibitemOpen
  \bibfield  {author} {\bibinfo {author} {\bibnamefont {Bender}, \bibfnamefont
  {Carl~M}}, \bibinfo {author} {\bibfnamefont {S~F}\ \bibnamefont {Brandt}},
  \bibinfo {author} {\bibfnamefont {J~H}\ \bibnamefont {Chen}}, and\ \bibinfo
  {author} {\bibfnamefont {Qinghai}\ \bibnamefont {Wang}}} (\bibinfo {year}
  {2005}{\natexlab{a}}),\ \bibfield  {title} {\enquote {\bibinfo {title} {Ghost
  busting: $\mathcal{PT}$-symmetric interpretation of the {L}ee model},}\
  }\href {https://doi.org/10.1103/PhysRevD.71.025014} {\bibfield  {journal}
  {\bibinfo  {journal} {Physical Review D}\ }\textbf {\bibinfo {volume} {71}},\
  10.1103/PhysRevD.71.025014}\BibitemShut {NoStop}%
\bibitem [{\citenamefont {Bender}\ \emph
  {et~al.}(2006{\natexlab{a}})\citenamefont {Bender}, \citenamefont {Brody},
  \citenamefont {Chen}, \citenamefont {Jones}, \citenamefont {Milton},\ and\
  \citenamefont {Ogilvie}}]{pt169}%
  \BibitemOpen
  \bibfield  {author} {\bibinfo {author} {\bibnamefont {Bender}, \bibfnamefont
  {Carl~M}}, \bibinfo {author} {\bibfnamefont {Dorje~C}\ \bibnamefont {Brody}},
  \bibinfo {author} {\bibfnamefont {Jun-Hua}\ \bibnamefont {Chen}}, \bibinfo
  {author} {\bibfnamefont {Hugh~F}\ \bibnamefont {Jones}}, \bibinfo {author}
  {\bibfnamefont {Kimball~A}\ \bibnamefont {Milton}}, and\ \bibinfo {author}
  {\bibfnamefont {Michael~C.}\ \bibnamefont {Ogilvie}}} (\bibinfo {year}
  {2006}{\natexlab{a}}),\ \bibfield  {title} {\enquote {\bibinfo {title}
  {Equivalence of a complex $\cal{PT}$-symmetric quartic {H}amiltonian and a
  {H}ermitian quartic {H}amiltonian with an anomaly},}\ }\href
  {https://doi.org/10.1103/PhysRevD.74.025016} {\bibfield  {journal} {\bibinfo
  {journal} {Physical Review D}\ }\textbf {\bibinfo {volume} {74}},\ \bibinfo
  {pages} {025016}}\BibitemShut {NoStop}%
\bibitem [{\citenamefont {Bender}\ \emph {et~al.}(2008)\citenamefont {Bender},
  \citenamefont {Brody},\ and\ \citenamefont {Hook}}]{pt513}%
  \BibitemOpen
  \bibfield  {author} {\bibinfo {author} {\bibnamefont {Bender}, \bibfnamefont
  {Carl~M}}, \bibinfo {author} {\bibfnamefont {Dorje~C}\ \bibnamefont {Brody}},
  and\ \bibinfo {author} {\bibfnamefont {Daniel~W}\ \bibnamefont {Hook}}}
  (\bibinfo {year} {2008}),\ \bibfield  {title} {\enquote {\bibinfo {title}
  {Quantum effects in classical systems having complex energy},}\ }\href
  {https://doi.org/10.1088/1751-8113/41/35/352003} {\bibfield  {journal}
  {\bibinfo  {journal} {Journal of Physics A: Mathematical and Theoretical}\
  }\textbf {\bibinfo {volume} {41}},\ \bibinfo {pages} {352003}}\BibitemShut
  {NoStop}%
\bibitem [{\citenamefont {Bender}\ \emph {et~al.}(2002)\citenamefont {Bender},
  \citenamefont {Brody},\ and\ \citenamefont {Jones}}]{pt90}%
  \BibitemOpen
  \bibfield  {author} {\bibinfo {author} {\bibnamefont {Bender}, \bibfnamefont
  {Carl~M}}, \bibinfo {author} {\bibfnamefont {Dorje~C}\ \bibnamefont {Brody}},
  and\ \bibinfo {author} {\bibfnamefont {Hugh~F}\ \bibnamefont {Jones}}}
  (\bibinfo {year} {2002}),\ \bibfield  {title} {\enquote {\bibinfo {title}
  {Complex extension of quantum mechanics},}\ }\href
  {https://doi.org/10.1103/PhysRevLett.89.270401} {\bibfield  {journal}
  {\bibinfo  {journal} {Physical Review Letters}\ }\textbf {\bibinfo {volume}
  {89}},\ \bibinfo {pages} {270401}}\BibitemShut {NoStop}%
\bibitem [{\citenamefont {Bender}\ \emph {et~al.}(2004)\citenamefont {Bender},
  \citenamefont {Brody},\ and\ \citenamefont {Jones}}]{pt178}%
  \BibitemOpen
  \bibfield  {author} {\bibinfo {author} {\bibnamefont {Bender}, \bibfnamefont
  {Carl~M}}, \bibinfo {author} {\bibfnamefont {Dorje~C}\ \bibnamefont {Brody}},
  and\ \bibinfo {author} {\bibfnamefont {Hugh~F}\ \bibnamefont {Jones}}}
  (\bibinfo {year} {2004}),\ \bibfield  {title} {\enquote {\bibinfo {title}
  {Complex extension of quantum mechanics ({E}rratum)},}\ }\href
  {https://doi.org/10.1103/PhysRevLett.92.119902} {\bibfield  {journal}
  {\bibinfo  {journal} {Physical Review Letters}\ }\textbf {\bibinfo {volume}
  {92}},\ \bibinfo {pages} {119902(E)}}\BibitemShut {NoStop}%
\bibitem [{\citenamefont {Bender}\ \emph
  {et~al.}(2005{\natexlab{b}})\citenamefont {Bender}, \citenamefont
  {Cavero-Palaez}, \citenamefont {Milton},\ and\ \citenamefont
  {Shajesh}}]{pt454}%
  \BibitemOpen
  \bibfield  {author} {\bibinfo {author} {\bibnamefont {Bender}, \bibfnamefont
  {Carl~M}}, \bibinfo {author} {\bibfnamefont {Ines}\ \bibnamefont
  {Cavero-Palaez}}, \bibinfo {author} {\bibfnamefont {Kimball~A}\ \bibnamefont
  {Milton}}, and\ \bibinfo {author} {\bibfnamefont {K~V}\ \bibnamefont
  {Shajesh}}} (\bibinfo {year} {2005}{\natexlab{b}}),\ \bibfield  {title}
  {\enquote {\bibinfo {title} {$\mathcal{PT}$-symmetric quantum
  electrodynamics},}\ }\href {https://doi.org/10.1016/j.physletb.2005.03.032}
  {\bibfield  {journal} {\bibinfo  {journal} {Physics Letters B}\ }\textbf
  {\bibinfo {volume} {613}},\ 10.1016/j.physletb.2005.03.032}\BibitemShut
  {NoStop}%
\bibitem [{\citenamefont {Bender}\ \emph
  {et~al.}(2006{\natexlab{b}})\citenamefont {Bender}, \citenamefont {Chen},
  \citenamefont {Darg},\ and\ \citenamefont {Milton}}]{pt431}%
  \BibitemOpen
  \bibfield  {author} {\bibinfo {author} {\bibnamefont {Bender}, \bibfnamefont
  {Carl~M}}, \bibinfo {author} {\bibfnamefont {Jun-Hua}\ \bibnamefont {Chen}},
  \bibinfo {author} {\bibfnamefont {Daniel~W}\ \bibnamefont {Darg}}, and\
  \bibinfo {author} {\bibfnamefont {Kimball~A}\ \bibnamefont {Milton}}}
  (\bibinfo {year} {2006}{\natexlab{b}}),\ \bibfield  {title} {\enquote
  {\bibinfo {title} {Classical trajectories for complex {H}amiltonians},}\
  }\href {https://doi.org/10.1088/0305-4470/39/16/009} {\bibfield  {journal}
  {\bibinfo  {journal} {Journal of Physics A: Mathematical and General}\
  }\textbf {\bibinfo {volume} {39}},\ 10.1088/0305-4470/39/16/009}\BibitemShut
  {NoStop}%
\bibitem [{\citenamefont {Bender}\ \emph
  {et~al.}(2006{\natexlab{c}})\citenamefont {Bender}, \citenamefont {Chen},\
  and\ \citenamefont {Milton}}]{pt409}%
  \BibitemOpen
  \bibfield  {author} {\bibinfo {author} {\bibnamefont {Bender}, \bibfnamefont
  {Carl~M}}, \bibinfo {author} {\bibfnamefont {Jun-Hua}\ \bibnamefont {Chen}},
  and\ \bibinfo {author} {\bibfnamefont {Kimball~A}\ \bibnamefont {Milton}}}
  (\bibinfo {year} {2006}{\natexlab{c}}),\ \bibfield  {title} {\enquote
  {\bibinfo {title} {$\mathcal{PT}$-symmetric versus {H}ermitian formulations
  of quantum mechanics},}\ }\href {https://doi.org/10.1088/0305-4470/39/7/010}
  {\bibfield  {journal} {\bibinfo  {journal} {Journal of Physics A:
  Mathematical and General}\ }\textbf {\bibinfo {volume} {39}},\
  10.1088/0305-4470/39/7/010}\BibitemShut {NoStop}%
\bibitem [{\citenamefont {Bender}\ and\ \citenamefont {Darg}(2007)}]{pt434}%
  \BibitemOpen
  \bibfield  {author} {\bibinfo {author} {\bibnamefont {Bender}, \bibfnamefont
  {Carl~M}}, and\ \bibinfo {author} {\bibfnamefont {Daniel~W}\ \bibnamefont
  {Darg}}} (\bibinfo {year} {2007}),\ \bibfield  {title} {\enquote {\bibinfo
  {title} {Spontaneous breaking of classical $\mathcal{PT}$ symmetry},}\ }\href
  {https://doi.org/10.1063/1.2720279} {\bibfield  {journal} {\bibinfo
  {journal} {Journal of Mathematical Physics}\ }\textbf {\bibinfo {volume}
  {48}},\ 10.1063/1.2720279}\BibitemShut {NoStop}%
\bibitem [{\citenamefont {Bender}\ \emph {et~al.}(2019)\citenamefont {Bender},
  \citenamefont {Dorey}, \citenamefont {Dunning}, \citenamefont {Fring},
  \citenamefont {Hook}, \citenamefont {Jones}, \citenamefont {Kuzhel},
  \citenamefont {L\'evai},\ and\ \citenamefont {Tateo}}]{pt579}%
  \BibitemOpen
  \bibfield  {author} {\bibinfo {author} {\bibnamefont {Bender}, \bibfnamefont
  {Carl~M}}, \bibinfo {author} {\bibfnamefont {Patrick~E}\ \bibnamefont
  {Dorey}}, \bibinfo {author} {\bibfnamefont {Clare}\ \bibnamefont {Dunning}},
  \bibinfo {author} {\bibfnamefont {Andreas}\ \bibnamefont {Fring}}, \bibinfo
  {author} {\bibfnamefont {Daniel~W}\ \bibnamefont {Hook}}, \bibinfo {author}
  {\bibfnamefont {Hugh~F}\ \bibnamefont {Jones}}, \bibinfo {author}
  {\bibfnamefont {Sergii}\ \bibnamefont {Kuzhel}}, \bibinfo {author}
  {\bibfnamefont {G\'eza}\ \bibnamefont {L\'evai}}, and\ \bibinfo {author}
  {\bibfnamefont {Roberto}\ \bibnamefont {Tateo}}} (\bibinfo {year} {2019}),\
  \href@noop {} {\emph {\bibinfo {title} {PT Symmetry in Quantum and Classical
  Physics}}}\ (\bibinfo  {publisher} {World Scientific},\ \bibinfo {address}
  {Singapore})\BibitemShut {NoStop}%
\bibitem [{\citenamefont {Bender}\ and\ \citenamefont {Dunne}(1996)}]{pt323}%
  \BibitemOpen
  \bibfield  {author} {\bibinfo {author} {\bibnamefont {Bender}, \bibfnamefont
  {Carl~M}}, and\ \bibinfo {author} {\bibfnamefont {Gerald~V}\ \bibnamefont
  {Dunne}}} (\bibinfo {year} {1996}),\ \bibfield  {title} {\enquote {\bibinfo
  {title} {Quasiexactly solvable systems and orthogonal polynomials},}\ }\href
  {https://doi.org/10.1063/1.531373} {\bibfield  {journal} {\bibinfo  {journal}
  {Journal of Mathematical Physics}\ }\textbf {\bibinfo {volume} {37}},\
  \bibinfo {pages} {6--11}}\BibitemShut {NoStop}%
\bibitem [{\citenamefont {Bender}\ and\ \citenamefont {Dunne}(1999)}]{pt480}%
  \BibitemOpen
  \bibfield  {author} {\bibinfo {author} {\bibnamefont {Bender}, \bibfnamefont
  {Carl~M}}, and\ \bibinfo {author} {\bibfnamefont {Gerald~V}\ \bibnamefont
  {Dunne}}} (\bibinfo {year} {1999}),\ \bibfield  {title} {\enquote {\bibinfo
  {title} {Large-order perturbation theory for a non-{H}ermitian
  $\mathcal{PT}$-symmetric {H}amiltonian},}\ }\href
  {https://doi.org/10.1063/1.532991} {\bibfield  {journal} {\bibinfo  {journal}
  {Journal of Mathematical Physics}\ }\textbf {\bibinfo {volume} {40}},\
  10.1063/1.532991}\BibitemShut {NoStop}%
\bibitem [{\citenamefont {Bender}\ \emph
  {et~al.}(1999{\natexlab{c}})\citenamefont {Bender}, \citenamefont {Dunne},\
  and\ \citenamefont {Meisinger}}]{pt282}%
  \BibitemOpen
  \bibfield  {author} {\bibinfo {author} {\bibnamefont {Bender}, \bibfnamefont
  {Carl~M}}, \bibinfo {author} {\bibfnamefont {Gerald~V}\ \bibnamefont
  {Dunne}}, and\ \bibinfo {author} {\bibfnamefont {Peter~N}\ \bibnamefont
  {Meisinger}}} (\bibinfo {year} {1999}{\natexlab{c}}),\ \bibfield  {title}
  {\enquote {\bibinfo {title} {Complex periodic potentials with real band
  spectra},}\ }\href {https://doi.org/10.1016/S0375-9601(98)00960-8} {\bibfield
   {journal} {\bibinfo  {journal} {Physics Letters A}\ }\textbf {\bibinfo
  {volume} {252}},\ \bibinfo {pages} {272--276}}\BibitemShut {NoStop}%
\bibitem [{\citenamefont {Bender}\ \emph
  {et~al.}(2009{\natexlab{b}})\citenamefont {Bender}, \citenamefont {Feinberg},
  \citenamefont {Hook},\ and\ \citenamefont {Weir}}]{pt650}%
  \BibitemOpen
  \bibfield  {author} {\bibinfo {author} {\bibnamefont {Bender}, \bibfnamefont
  {Carl~M}}, \bibinfo {author} {\bibfnamefont {Joshua}\ \bibnamefont
  {Feinberg}}, \bibinfo {author} {\bibfnamefont {Daniel~W}\ \bibnamefont
  {Hook}}, and\ \bibinfo {author} {\bibfnamefont {David~J}\ \bibnamefont
  {Weir}}} (\bibinfo {year} {2009}{\natexlab{b}}),\ \bibfield  {title}
  {\enquote {\bibinfo {title} {Chaotic systems in complex phase space},}\
  }\href {https://doi.org/10.1007/s12043-009-0099-3} {\bibfield  {journal}
  {\bibinfo  {journal} {Pramana}\ }\textbf {\bibinfo {volume} {73}},\
  10.1007/s12043-009-0099-3}\BibitemShut {NoStop}%
\bibitem [{\citenamefont {Bender}\ \emph
  {et~al.}(2014{\natexlab{a}})\citenamefont {Bender}, \citenamefont {Fring},\
  and\ \citenamefont {Komijani}}]{pt531}%
  \BibitemOpen
  \bibfield  {author} {\bibinfo {author} {\bibnamefont {Bender}, \bibfnamefont
  {Carl~M}}, \bibinfo {author} {\bibfnamefont {Andreas}\ \bibnamefont {Fring}},
  and\ \bibinfo {author} {\bibfnamefont {Javad}\ \bibnamefont {Komijani}}}
  (\bibinfo {year} {2014}{\natexlab{a}}),\ \bibfield  {title} {\enquote
  {\bibinfo {title} {Nonlinear eigenvalue problems},}\ }\href
  {https://doi.org/10.1088/1751-8113/47/23/235204} {\bibfield  {journal}
  {\bibinfo  {journal} {Journal of Physics A: Mathematical and Theoretical}\
  }\textbf {\bibinfo {volume} {47}},\ \bibinfo {pages} {235204}}\BibitemShut
  {NoStop}%
\bibitem [{\citenamefont {Bender}\ and\ \citenamefont
  {Gianfreda}(2018)}]{pt594}%
  \BibitemOpen
  \bibfield  {author} {\bibinfo {author} {\bibnamefont {Bender}, \bibfnamefont
  {Carl~M}}, and\ \bibinfo {author} {\bibfnamefont {Mariagiovanna}\
  \bibnamefont {Gianfreda}}} (\bibinfo {year} {2018}),\ \bibfield  {title}
  {\enquote {\bibinfo {title} {Scattering off $\mathcal{PT}$-symmetric
  upside-down potentials},}\ }\href
  {https://doi.org/10.1103/PhysRevA.98.052118} {\bibfield  {journal} {\bibinfo
  {journal} {Physical Review A}\ }\textbf {\bibinfo {volume} {98}},\
  10.1103/PhysRevA.98.052118}\BibitemShut {NoStop}%
\bibitem [{\citenamefont {Bender}\ \emph {et~al.}(1974)\citenamefont {Bender},
  \citenamefont {Happ},\ and\ \citenamefont {Svetitsky}}]{pt615}%
  \BibitemOpen
  \bibfield  {author} {\bibinfo {author} {\bibnamefont {Bender}, \bibfnamefont
  {Carl~M}}, \bibinfo {author} {\bibfnamefont {Henry~J}\ \bibnamefont {Happ}},
  and\ \bibinfo {author} {\bibfnamefont {Benjamin}\ \bibnamefont {Svetitsky}}}
  (\bibinfo {year} {1974}),\ \bibfield  {title} {\enquote {\bibinfo {title}
  {Numerical study of energy-level crossing},}\ }\href
  {https://doi.org/10.1103/PhysRevD.9.2324} {\bibfield  {journal} {\bibinfo
  {journal} {Physical Review D}\ }\textbf {\bibinfo {volume} {9}},\
  10.1103/PhysRevD.9.2324}\BibitemShut {NoStop}%
\bibitem [{\citenamefont {Bender}\ \emph {et~al.}(2017)\citenamefont {Bender},
  \citenamefont {Hassanpour}, \citenamefont {Hook}, \citenamefont {Klevansky},
  \citenamefont {S{\"u}nderhauf},\ and\ \citenamefont {Wen}}]{pt292}%
  \BibitemOpen
  \bibfield  {author} {\bibinfo {author} {\bibnamefont {Bender}, \bibfnamefont
  {Carl~M}}, \bibinfo {author} {\bibfnamefont {Nima}\ \bibnamefont
  {Hassanpour}}, \bibinfo {author} {\bibfnamefont {Daniel~W}\ \bibnamefont
  {Hook}}, \bibinfo {author} {\bibfnamefont {S~P}\ \bibnamefont {Klevansky}},
  \bibinfo {author} {\bibfnamefont {Christoph}\ \bibnamefont {S{\"u}nderhauf}},
  and\ \bibinfo {author} {\bibfnamefont {Zichao}\ \bibnamefont {Wen}}}
  (\bibinfo {year} {2017}),\ \bibfield  {title} {\enquote {\bibinfo {title}
  {{Behavior of eigenvalues in a region of broken-$\cal{PT}$ symmetry}},}\
  }\href {https://doi.org/10.1103/PhysRevA.95.052113} {\bibfield  {journal}
  {\bibinfo  {journal} {Physical Review A}\ }\textbf {\bibinfo {volume} {95}},\
  \bibinfo {pages} {052113}}\BibitemShut {NoStop}%
\bibitem [{\citenamefont {Bender}\ \emph
  {et~al.}(2007{\natexlab{b}})\citenamefont {Bender}, \citenamefont {Holm},\
  and\ \citenamefont {Hook}}]{pt22}%
  \BibitemOpen
  \bibfield  {author} {\bibinfo {author} {\bibnamefont {Bender}, \bibfnamefont
  {Carl~M}}, \bibinfo {author} {\bibfnamefont {Darryl~D}\ \bibnamefont {Holm}},
  and\ \bibinfo {author} {\bibfnamefont {Daniel~W}\ \bibnamefont {Hook}}}
  (\bibinfo {year} {2007}{\natexlab{b}}),\ \bibfield  {title} {\enquote
  {\bibinfo {title} {Complexified dynamical systems},}\ }\href
  {https://doi.org/10.1088/1751-8113/40/32/F02} {\bibfield  {journal} {\bibinfo
   {journal} {Journal of Physics A: Mathematical and Theoretical}\ }\textbf
  {\bibinfo {volume} {40}}~(\bibinfo {number} {32}),\ \bibinfo {pages}
  {F793--F804}}\BibitemShut {NoStop}%
\bibitem [{\citenamefont {Bender}\ \emph
  {et~al.}(2007{\natexlab{c}})\citenamefont {Bender}, \citenamefont {Holm},\
  and\ \citenamefont {Hook}}]{pt432}%
  \BibitemOpen
  \bibfield  {author} {\bibinfo {author} {\bibnamefont {Bender}, \bibfnamefont
  {Carl~M}}, \bibinfo {author} {\bibfnamefont {Daryl~D}\ \bibnamefont {Holm}},
  and\ \bibinfo {author} {\bibfnamefont {Daniel~W}\ \bibnamefont {Hook}}}
  (\bibinfo {year} {2007}{\natexlab{c}}),\ \bibfield  {title} {\enquote
  {\bibinfo {title} {Complex trajectories of a simple pendulum},}\ }\href
  {https://doi.org/10.1088/1751-8113/40/3/F01} {\bibfield  {journal} {\bibinfo
  {journal} {Journal of Physics A: Mathematical and Theoretical}\ }\textbf
  {\bibinfo {volume} {40}},\ 10.1088/1751-8113/40/3/F01}\BibitemShut {NoStop}%
\bibitem [{\citenamefont {Bender}\ and\ \citenamefont {Hook}(2008)}]{pt486}%
  \BibitemOpen
  \bibfield  {author} {\bibinfo {author} {\bibnamefont {Bender}, \bibfnamefont
  {Carl~M}}, and\ \bibinfo {author} {\bibfnamefont {Daniel~W}\ \bibnamefont
  {Hook}}} (\bibinfo {year} {2008}),\ \bibfield  {title} {\enquote {\bibinfo
  {title} {Exact isospectral pairs of $\mathcal{PT}$ symmetric
  {H}amiltonians},}\ }\href {https://doi.org/10.1088/1751-8113/41/24/244005}
  {\bibfield  {journal} {\bibinfo  {journal} {Journal of Physics A:
  Mathematical and Theoretical}\ }\textbf {\bibinfo {volume} {41}},\
  10.1088/1751-8113/41/24/244005}\BibitemShut {NoStop}%
\bibitem [{\citenamefont {Bender}\ and\ \citenamefont {Hook}(2011)}]{pt591}%
  \BibitemOpen
  \bibfield  {author} {\bibinfo {author} {\bibnamefont {Bender}, \bibfnamefont
  {Carl~M}}, and\ \bibinfo {author} {\bibfnamefont {Daniel~W}\ \bibnamefont
  {Hook}}} (\bibinfo {year} {2011}),\ \bibfield  {title} {\enquote {\bibinfo
  {title} {Quantum tunneling as a classical anomaly},}\ }\href
  {https://doi.org/10.1088/1751-8113/44/37/372001} {\bibfield  {journal}
  {\bibinfo  {journal} {Journal of Physics A: Mathematical and Theoretical}\
  }\textbf {\bibinfo {volume} {44}},\
  10.1088/1751-8113/44/37/372001}\BibitemShut {NoStop}%
\bibitem [{\citenamefont {Bender}\ and\ \citenamefont {Hook}(2012)}]{pt170}%
  \BibitemOpen
  \bibfield  {author} {\bibinfo {author} {\bibnamefont {Bender}, \bibfnamefont
  {Carl~M}}, and\ \bibinfo {author} {\bibfnamefont {Daniel~W}\ \bibnamefont
  {Hook}}} (\bibinfo {year} {2012}),\ \bibfield  {title} {\enquote {\bibinfo
  {title} {Universal spectral behavior of $x^2(ix)^\varepsilon$ potentials},}\
  }\href {https://doi.org/10.1103/PhysRevA.86.022113} {\bibfield  {journal}
  {\bibinfo  {journal} {Physical Review A}\ }\textbf {\bibinfo {volume} {86}},\
  \bibinfo {pages} {022113}}\BibitemShut {NoStop}%
\bibitem [{\citenamefont {Bender}\ and\ \citenamefont {Hook}(2014)}]{pt593}%
  \BibitemOpen
  \bibfield  {author} {\bibinfo {author} {\bibnamefont {Bender}, \bibfnamefont
  {Carl~M}}, and\ \bibinfo {author} {\bibfnamefont {Daniel~W}\ \bibnamefont
  {Hook}}} (\bibinfo {year} {2014}),\ \bibfield  {title} {\enquote {\bibinfo
  {title} {Complex classical motion in potentials with poles and turning
  points},}\ }\href {https://doi.org/10.1111/sapm.12059} {\bibfield  {journal}
  {\bibinfo  {journal} {Studies in Applied Mathematics}\ }\textbf {\bibinfo
  {volume} {133}},\ 10.1111/sapm.12059}\BibitemShut {NoStop}%
\bibitem [{\citenamefont {Bender}\ \emph {et~al.}(2012)\citenamefont {Bender},
  \citenamefont {Hook},\ and\ \citenamefont {Klevansky}}]{pt489}%
  \BibitemOpen
  \bibfield  {author} {\bibinfo {author} {\bibnamefont {Bender}, \bibfnamefont
  {Carl~M}}, \bibinfo {author} {\bibfnamefont {Daniel~W}\ \bibnamefont {Hook}},
  and\ \bibinfo {author} {\bibfnamefont {S~P}\ \bibnamefont {Klevansky}}}
  (\bibinfo {year} {2012}),\ \bibfield  {title} {\enquote {\bibinfo {title}
  {Negative-energy $\mathcal{PT}$-symmetric {H}amiltonian},}\ }\href
  {https://doi.org/10.1088/1751-8113/45/44/444003} {\bibfield  {journal}
  {\bibinfo  {journal} {Journal of Physics A: Mathematical and Theoretical}\
  }\textbf {\bibinfo {volume} {45}},\
  10.1088/1751-8113/45/44/444003}\BibitemShut {NoStop}%
\bibitem [{\citenamefont {Bender}\ \emph
  {et~al.}(2010{\natexlab{a}})\citenamefont {Bender}, \citenamefont {Hook},\
  and\ \citenamefont {Kooner}}]{pt514}%
  \BibitemOpen
  \bibfield  {author} {\bibinfo {author} {\bibnamefont {Bender}, \bibfnamefont
  {Carl~M}}, \bibinfo {author} {\bibfnamefont {Daniel~W}\ \bibnamefont {Hook}},
  and\ \bibinfo {author} {\bibfnamefont {Karta~Singh}\ \bibnamefont {Kooner}}}
  (\bibinfo {year} {2010}{\natexlab{a}}),\ \bibfield  {title} {\enquote
  {\bibinfo {title} {Classical particle in a complex elliptic potential},}\
  }\href {https://doi.org/10.1088/1751-8113/43/16/165201} {\bibfield  {journal}
  {\bibinfo  {journal} {Journal of Physics A: Mathematical and Theoretical}\
  }\textbf {\bibinfo {volume} {43}},\ \bibinfo {pages} {3165201}}\BibitemShut
  {NoStop}%
\bibitem [{\citenamefont {Bender}\ \emph
  {et~al.}(2014{\natexlab{b}})\citenamefont {Bender}, \citenamefont {Hook},
  \citenamefont {Mavromatos},\ and\ \citenamefont {Sarkar}}]{pt490}%
  \BibitemOpen
  \bibfield  {author} {\bibinfo {author} {\bibnamefont {Bender}, \bibfnamefont
  {Carl~M}}, \bibinfo {author} {\bibfnamefont {Daniel~W}\ \bibnamefont {Hook}},
  \bibinfo {author} {\bibfnamefont {Nick~E}\ \bibnamefont {Mavromatos}}, and\
  \bibinfo {author} {\bibfnamefont {Sarben}\ \bibnamefont {Sarkar}}} (\bibinfo
  {year} {2014}{\natexlab{b}}),\ \bibfield  {title} {\enquote {\bibinfo {title}
  {Infinite class of $\mathcal{PT}$-symmetric theories from one timelike
  {L}iouville {L}agrangian},}\ }\href
  {https://doi.org/10.1103/physrevlett.113.231605} {\bibfield  {journal}
  {\bibinfo  {journal} {Physical Review Letters}\ }\textbf {\bibinfo {volume}
  {113}},\ 10.1103/physrevlett.113.231605}\BibitemShut {NoStop}%
\bibitem [{\citenamefont {Bender}\ \emph {et~al.}(2016)\citenamefont {Bender},
  \citenamefont {Hook}, \citenamefont {Mavromatos},\ and\ \citenamefont
  {Sarkar}}]{pt174}%
  \BibitemOpen
  \bibfield  {author} {\bibinfo {author} {\bibnamefont {Bender}, \bibfnamefont
  {Carl~M}}, \bibinfo {author} {\bibfnamefont {Daniel~W}\ \bibnamefont {Hook}},
  \bibinfo {author} {\bibfnamefont {Nick~E}\ \bibnamefont {Mavromatos}}, and\
  \bibinfo {author} {\bibfnamefont {Sarben}\ \bibnamefont {Sarkar}}} (\bibinfo
  {year} {2016}),\ \bibfield  {title} {\enquote {\bibinfo {title}
  {$\cal{PT}$-symmetric interpretation of unstable effective potentials},}\
  }\href {https://doi.org/10.1088/1751-8113/49/45/45LT01} {\bibfield  {journal}
  {\bibinfo  {journal} {Journal of Physics A: Mathematical and Theoretical}\
  }\textbf {\bibinfo {volume} {49}},\ \bibinfo {pages} {45LT01}}\BibitemShut
  {NoStop}%
\bibitem [{\citenamefont {Bender}\ \emph
  {et~al.}(2010{\natexlab{b}})\citenamefont {Bender}, \citenamefont {Hook},
  \citenamefont {Meisinger},\ and\ \citenamefont {Wang}}]{pt525}%
  \BibitemOpen
  \bibfield  {author} {\bibinfo {author} {\bibnamefont {Bender}, \bibfnamefont
  {Carl~M}}, \bibinfo {author} {\bibfnamefont {Daniel~W}\ \bibnamefont {Hook}},
  \bibinfo {author} {\bibfnamefont {Peter~N}\ \bibnamefont {Meisinger}}, and\
  \bibinfo {author} {\bibfnamefont {Qinghai}\ \bibnamefont {Wang}}} (\bibinfo
  {year} {2010}{\natexlab{b}}),\ \bibfield  {title} {\enquote {\bibinfo {title}
  {Complex correspondence principle},}\ }\href
  {https://doi.org/10.1103/physrevlett.104.061601} {\bibfield  {journal}
  {\bibinfo  {journal} {Physical Review Letters}\ }\textbf {\bibinfo {volume}
  {104}},\ \bibinfo {pages} {061601}}\BibitemShut {NoStop}%
\bibitem [{\citenamefont {Bender}\ \emph
  {et~al.}(2010{\natexlab{c}})\citenamefont {Bender}, \citenamefont {Hook},
  \citenamefont {Meisinger},\ and\ \citenamefont {Wang}}]{pt491}%
  \BibitemOpen
  \bibfield  {author} {\bibinfo {author} {\bibnamefont {Bender}, \bibfnamefont
  {Carl~M}}, \bibinfo {author} {\bibfnamefont {Daniel~W}\ \bibnamefont {Hook}},
  \bibinfo {author} {\bibfnamefont {Peter~N}\ \bibnamefont {Meisinger}}, and\
  \bibinfo {author} {\bibfnamefont {Qinghai}\ \bibnamefont {Wang}}} (\bibinfo
  {year} {2010}{\natexlab{c}}),\ \bibfield  {title} {\enquote {\bibinfo {title}
  {Probability density in the complex plane},}\ }\href
  {https://doi.org/10.1016/j.aop.2010.02.011} {\bibfield  {journal} {\bibinfo
  {journal} {Annals of Physics}\ }\textbf {\bibinfo {volume} {325}},\
  10.1016/j.aop.2010.02.011}\BibitemShut {NoStop}%
\bibitem [{\citenamefont {Bender}\ and\ \citenamefont {Jones}(2004)}]{pt403}%
  \BibitemOpen
  \bibfield  {author} {\bibinfo {author} {\bibnamefont {Bender}, \bibfnamefont
  {Carl~M}}, and\ \bibinfo {author} {\bibfnamefont {Hugh~F}\ \bibnamefont
  {Jones}}} (\bibinfo {year} {2004}),\ \bibfield  {title} {\enquote {\bibinfo
  {title} {Semiclassical calculation of the $\mathcal{C}$ operator in
  $\mathcal{PT}$-symmetric quantum mechanics},}\ }\href
  {https://doi.org/10.1016/j.physleta.2004.05.063} {\bibfield  {journal}
  {\bibinfo  {journal} {Physics Letters A}\ }\textbf {\bibinfo {volume}
  {328}},\ 10.1016/j.physleta.2004.05.063}\BibitemShut {NoStop}%
\bibitem [{\citenamefont {Bender}\ and\ \citenamefont {Jones}(2011)}]{pt587}%
  \BibitemOpen
  \bibfield  {author} {\bibinfo {author} {\bibnamefont {Bender}, \bibfnamefont
  {Carl~M}}, and\ \bibinfo {author} {\bibfnamefont {Hugh~F}\ \bibnamefont
  {Jones}}} (\bibinfo {year} {2011}),\ \bibfield  {title} {\enquote {\bibinfo
  {title} {Bound states of $\mathcal{PT}$-symmetric separable potentials},}\
  }\href {https://doi.org/10.1103/PhysRevA.84.032103} {\bibfield  {journal}
  {\bibinfo  {journal} {Physical Review A}\ }\textbf {\bibinfo {volume} {84}},\
  10.1103/PhysRevA.84.032103}\BibitemShut {NoStop}%
\bibitem [{\citenamefont {Bender}\ \emph
  {et~al.}(2005{\natexlab{c}})\citenamefont {Bender}, \citenamefont {Jones},\
  and\ \citenamefont {Rivers}}]{pt455}%
  \BibitemOpen
  \bibfield  {author} {\bibinfo {author} {\bibnamefont {Bender}, \bibfnamefont
  {Carl~M}}, \bibinfo {author} {\bibfnamefont {Hugh~F}\ \bibnamefont {Jones}},
  and\ \bibinfo {author} {\bibfnamefont {Ray~J}\ \bibnamefont {Rivers}}}
  (\bibinfo {year} {2005}{\natexlab{c}}),\ \bibfield  {title} {\enquote
  {\bibinfo {title} {Dual $\mathcal{PT}$-symmetric quantum field theories},}\
  }\href {https://doi.org/10.1016/j.physletb.2005.08.087} {\bibfield  {journal}
  {\bibinfo  {journal} {Physics Letters B}\ }\textbf {\bibinfo {volume}
  {625}},\ 10.1016/j.physletb.2005.08.087}\BibitemShut {NoStop}%
\bibitem [{\citenamefont {Bender}\ \emph
  {et~al.}(2023{\natexlab{a}})\citenamefont {Bender}, \citenamefont
  {Karapoulitidis},\ and\ \citenamefont {Klevansky}}]{pt658}%
  \BibitemOpen
  \bibfield  {author} {\bibinfo {author} {\bibnamefont {Bender}, \bibfnamefont
  {Carl~M}}, \bibinfo {author} {\bibfnamefont {C}~\bibnamefont
  {Karapoulitidis}}, and\ \bibinfo {author} {\bibfnamefont {S~P}\ \bibnamefont
  {Klevansky}}} (\bibinfo {year} {2023}{\natexlab{a}}),\ \bibfield  {title}
  {\enquote {\bibinfo {title} {Dyson-schwinger equations in zero dimensions and
  polynomial approximations},}\ }\href
  {https://doi.org/10.1103/physrevd.108.056002} {\bibfield  {journal} {\bibinfo
   {journal} {Physical Review D}\ }\textbf {\bibinfo {volume} {108}},\
  10.1103/physrevd.108.056002}\BibitemShut {NoStop}%
\bibitem [{\citenamefont {Bender}\ \emph {et~al.}(2021)\citenamefont {Bender},
  \citenamefont {Karapoulitidis},\ and\ \citenamefont {Klevansky}}]{pt704}%
  \BibitemOpen
  \bibfield  {author} {\bibinfo {author} {\bibnamefont {Bender}, \bibfnamefont
  {Carl~M}}, \bibinfo {author} {\bibfnamefont {Christos}\ \bibnamefont
  {Karapoulitidis}}, and\ \bibinfo {author} {\bibfnamefont {S~P}\ \bibnamefont
  {Klevansky}}} (\bibinfo {year} {2021}),\ \bibfield  {title} {\enquote
  {\bibinfo {title} {Underdetermined dyson-schwinger equations},}\ }\href
  {https://doi.org/10.1103/physrevlett.130.101602} {\bibfield  {journal}
  {\bibinfo  {journal} {Physical Review Letters}\ }\textbf {\bibinfo {volume}
  {130}},\ 10.1103/physrevlett.130.101602}\BibitemShut {NoStop}%
\bibitem [{\citenamefont {Bender}\ and\ \citenamefont
  {Klevansky}(2009)}]{pt267}%
  \BibitemOpen
  \bibfield  {author} {\bibinfo {author} {\bibnamefont {Bender}, \bibfnamefont
  {Carl~M}}, and\ \bibinfo {author} {\bibfnamefont {S~P}\ \bibnamefont
  {Klevansky}}} (\bibinfo {year} {2009}),\ \bibfield  {title} {\enquote
  {\bibinfo {title} {Nonunique $\cal{C}$ operator in $\cal{PT}$ quantum
  mechanics},}\ }\href {https://doi.org/10.1016/j.physleta.2009.05.066}
  {\bibfield  {journal} {\bibinfo  {journal} {Physics Letters A}\ }\textbf
  {\bibinfo {volume} {373}},\ \bibinfo {pages} {2670--2674}}\BibitemShut
  {NoStop}%
\bibitem [{\citenamefont {Bender}\ and\ \citenamefont
  {Klevansky}(2010)}]{pt369}%
  \BibitemOpen
  \bibfield  {author} {\bibinfo {author} {\bibnamefont {Bender}, \bibfnamefont
  {Carl~M}}, and\ \bibinfo {author} {\bibfnamefont {S~P}\ \bibnamefont
  {Klevansky}}} (\bibinfo {year} {2010}),\ \bibfield  {title} {\enquote
  {\bibinfo {title} {Families of particles with different masses in
  $\mathcal{PT}$-symmetric quantum field theory},}\ }\href
  {https://doi.org/10.1103/PhysRevLett.105.031601} {\bibfield  {journal}
  {\bibinfo  {journal} {Physical Review Letters}\ }\textbf {\bibinfo {volume}
  {105}},\ 10.1103/PhysRevLett.105.031601}\BibitemShut {NoStop}%
\bibitem [{\citenamefont {Bender}\ and\ \citenamefont
  {Komijani}(2015)}]{pt532}%
  \BibitemOpen
  \bibfield  {author} {\bibinfo {author} {\bibnamefont {Bender}, \bibfnamefont
  {Carl~M}}, and\ \bibinfo {author} {\bibfnamefont {Javad}\ \bibnamefont
  {Komijani}}} (\bibinfo {year} {2015}),\ \bibfield  {title} {\enquote
  {\bibinfo {title} {Painlev\'{e} transcendents and $\mathcal{PT}$-symmetric
  hamiltonians},}\ }\href {https://doi.org/10.1088/1751-8113/48/47/475202}
  {\bibfield  {journal} {\bibinfo  {journal} {Journal of Physics A:
  Mathematical and Theoretical}\ }\textbf {\bibinfo {volume} {48}},\ \bibinfo
  {pages} {475202}}\BibitemShut {NoStop}%
\bibitem [{\citenamefont {Bender}\ and\ \citenamefont
  {Mannheim}(2008{\natexlab{a}})}]{pt634}%
  \BibitemOpen
  \bibfield  {author} {\bibinfo {author} {\bibnamefont {Bender}, \bibfnamefont
  {Carl~M}}, and\ \bibinfo {author} {\bibfnamefont {Philip~D}\ \bibnamefont
  {Mannheim}}} (\bibinfo {year} {2008}{\natexlab{a}}),\ \bibfield  {title}
  {\enquote {\bibinfo {title} {Giving up the ghost},}\ }\href
  {https://doi.org/10.1088/1751-8113/41/30/304018} {\bibfield  {journal}
  {\bibinfo  {journal} {Journal of Physics A: Mathematical and Theoretical}\
  }\textbf {\bibinfo {volume} {41}},\
  10.1088/1751-8113/41/30/304018}\BibitemShut {NoStop}%
\bibitem [{\citenamefont {Bender}\ and\ \citenamefont
  {Mannheim}(2008{\natexlab{b}})}]{pt530}%
  \BibitemOpen
  \bibfield  {author} {\bibinfo {author} {\bibnamefont {Bender}, \bibfnamefont
  {Carl~M}}, and\ \bibinfo {author} {\bibfnamefont {Philip~D}\ \bibnamefont
  {Mannheim}}} (\bibinfo {year} {2008}{\natexlab{b}}),\ \bibfield  {title}
  {\enquote {\bibinfo {title} {No-ghost theorem for the fourth-order derivative
  pais-uhlenbeck oscillator model},}\ }\href
  {https://doi.org/10.1103/physrevlett.100.110402} {\bibfield  {journal}
  {\bibinfo  {journal} {Physical Review Letters}\ }\textbf {\bibinfo {volume}
  {100}},\ \bibinfo {pages} {110402}}\BibitemShut {NoStop}%
\bibitem [{\citenamefont {Bender}\ \emph
  {et~al.}(2023{\natexlab{b}})\citenamefont {Bender}, \citenamefont
  {Mavromatos},\ and\ \citenamefont {Sarkar}}]{pt688}%
  \BibitemOpen
  \bibfield  {author} {\bibinfo {author} {\bibnamefont {Bender}, \bibfnamefont
  {Carl~M}}, \bibinfo {author} {\bibfnamefont {Nick~E}\ \bibnamefont
  {Mavromatos}}, and\ \bibinfo {author} {\bibfnamefont {Sarben}\ \bibnamefont
  {Sarkar}}} (\bibinfo {year} {2023}{\natexlab{b}}),\ \bibfield  {title}
  {\enquote {\bibinfo {title} {Non-hermitian chern-simons interactions and the
  repulsive nature of gravity at large scales},}\ }\href@noop {} {\bibinfo
  {journal} {Submitted}\ }\BibitemShut {NoStop}%
\bibitem [{\citenamefont {Bender}\ \emph {et~al.}(2003)\citenamefont {Bender},
  \citenamefont {Meisinger},\ and\ \citenamefont {Wang}}]{pt402}%
  \BibitemOpen
\bibfield  {journal} {  }\bibfield  {author} {\bibinfo {author} {\bibnamefont
  {Bender}, \bibfnamefont {Carl~M}}, \bibinfo {author} {\bibfnamefont
  {Peter~N}\ \bibnamefont {Meisinger}}, and\ \bibinfo {author} {\bibfnamefont
  {Qinghai}\ \bibnamefont {Wang}}} (\bibinfo {year} {2003}),\ \bibfield
  {title} {\enquote {\bibinfo {title} {Calculation of the hidden symmetry
  operator in $\mathcal{PT}$-symmetric quantum mechanics},}\ }\href
  {https://doi.org/10.1088/0305-4470/36/7/312} {\bibfield  {journal} {\bibinfo
  {journal} {Journal of Physics A: Mathematical and General}\ }\textbf
  {\bibinfo {volume} {36}},\ 10.1088/0305-4470/36/7/312}\BibitemShut {NoStop}%
\bibitem [{\citenamefont {Bender}\ \emph
  {et~al.}(2001{\natexlab{b}})\citenamefont {Bender}, \citenamefont
  {Meisinger},\ and\ \citenamefont {Yang}}]{pt450}%
  \BibitemOpen
  \bibfield  {author} {\bibinfo {author} {\bibnamefont {Bender}, \bibfnamefont
  {Carl~M}}, \bibinfo {author} {\bibfnamefont {Peter~N}\ \bibnamefont
  {Meisinger}}, and\ \bibinfo {author} {\bibfnamefont {Haitang}\ \bibnamefont
  {Yang}}} (\bibinfo {year} {2001}{\natexlab{b}}),\ \bibfield  {title}
  {\enquote {\bibinfo {title} {Calculation of the one-point {G}reen's function
  for a $-g\phi^4$ quantum field theory},}\ }\href
  {https://doi.org/10.1103/PhysRevD.63.045001} {\bibfield  {journal} {\bibinfo
  {journal} {Physical Review D}\ }\textbf {\bibinfo {volume} {63}},\
  10.1103/PhysRevD.63.045001}\BibitemShut {NoStop}%
\bibitem [{\citenamefont {Bender}\ and\ \citenamefont {Milton}(1999)}]{pt453}%
  \BibitemOpen
  \bibfield  {author} {\bibinfo {author} {\bibnamefont {Bender}, \bibfnamefont
  {Carl~M}}, and\ \bibinfo {author} {\bibfnamefont {Kimball~A}\ \bibnamefont
  {Milton}}} (\bibinfo {year} {1999}),\ \bibfield  {title} {\enquote {\bibinfo
  {title} {A nonunitary version of massless quantum electrodynamics possessing
  a critical point},}\ }\href {https://doi.org/10.1088/0305-4470/32/7/001}
  {\bibfield  {journal} {\bibinfo  {journal} {Journal of Physics A:
  Mathematical and General}\ }\textbf {\bibinfo {volume} {32}},\
  10.1088/0305-4470/32/7/001}\BibitemShut {NoStop}%
\bibitem [{\citenamefont {Bender}\ \emph
  {et~al.}(2000{\natexlab{b}})\citenamefont {Bender}, \citenamefont {Milton},\
  and\ \citenamefont {Savage}}]{pt449}%
  \BibitemOpen
  \bibfield  {author} {\bibinfo {author} {\bibnamefont {Bender}, \bibfnamefont
  {Carl~M}}, \bibinfo {author} {\bibfnamefont {Kimball~A}\ \bibnamefont
  {Milton}}, and\ \bibinfo {author} {\bibfnamefont {Van~M}\ \bibnamefont
  {Savage}}} (\bibinfo {year} {2000}{\natexlab{b}}),\ \bibfield  {title}
  {\enquote {\bibinfo {title} {Solution of {S}chwinger-{D}yson equations for
  $\mathcal{PT}$-symmetric quantum field theory},}\ }\href
  {https://doi.org/10.1103/PhysRevD.62.085001} {\bibfield  {journal} {\bibinfo
  {journal} {Physical Review D}\ }\textbf {\bibinfo {volume} {62}},\
  10.1103/PhysRevD.62.085001}\BibitemShut {NoStop}%
\bibitem [{\citenamefont {Bender}\ and\ \citenamefont {Orszag}(1999)}]{pt16}%
  \BibitemOpen
  \bibfield  {author} {\bibinfo {author} {\bibnamefont {Bender}, \bibfnamefont
  {Carl~M}}, and\ \bibinfo {author} {\bibfnamefont {Steven~A}\ \bibnamefont
  {Orszag}}} (\bibinfo {year} {1999}),\ \href@noop {} {\emph {\bibinfo {title}
  {Advanced Mathematical Methods for Scientists and Engineers I}}}\ (\bibinfo
  {publisher} {Springer},\ \bibinfo {address} {New York})\BibitemShut {NoStop}%
\bibitem [{\citenamefont {Bender}\ and\ \citenamefont {Wang}(2001)}]{pt378}%
  \BibitemOpen
  \bibfield  {author} {\bibinfo {author} {\bibnamefont {Bender}, \bibfnamefont
  {Carl~M}}, and\ \bibinfo {author} {\bibfnamefont {Qinghai}\ \bibnamefont
  {Wang}}} (\bibinfo {year} {2001}),\ \bibfield  {title} {\enquote {\bibinfo
  {title} {Comment on a recent paper by {M}ezincescu},}\ }\href
  {https://doi.org/10.1088/0305-4470/34/15/401} {\bibfield  {journal} {\bibinfo
   {journal} {Journal of Physics A: Mathematical and General}\ }\textbf
  {\bibinfo {volume} {33}},\ 10.1088/0305-4470/34/15/401}\BibitemShut {NoStop}%
\bibitem [{\citenamefont {Bender}\ and\ \citenamefont {Weniger}(2001)}]{pt607}%
  \BibitemOpen
  \bibfield  {author} {\bibinfo {author} {\bibnamefont {Bender}, \bibfnamefont
  {Carl~M}}, and\ \bibinfo {author} {\bibfnamefont {Ernst~Joachim}\
  \bibnamefont {Weniger}}} (\bibinfo {year} {2001}),\ \bibfield  {title}
  {\enquote {\bibinfo {title} {Numerical evidence that the perturbation
  expansion for a non-hermitian $\mathcal{PT}$-symmetric hamiltonian is
  stieltjes},}\ }\href {https://doi.org/10.1063/1.1362287} {\bibfield
  {journal} {\bibinfo  {journal} {Journal of Mathematical Physics}\ }\textbf
  {\bibinfo {volume} {42}},\ 10.1063/1.1362287}\BibitemShut {NoStop}%
\bibitem [{\citenamefont {Bender}\ and\ \citenamefont {Wu}(1968)}]{pt15}%
  \BibitemOpen
  \bibfield  {author} {\bibinfo {author} {\bibnamefont {Bender}, \bibfnamefont
  {Carl~M}}, and\ \bibinfo {author} {\bibfnamefont {Tai~Tsun}\ \bibnamefont
  {Wu}}} (\bibinfo {year} {1968}),\ \bibfield  {title} {\enquote {\bibinfo
  {title} {Analytic structure of energy levels in a field-theory model},}\
  }\href {https://doi.org/10.1103/PhysRevLett.21.406} {\bibfield  {journal}
  {\bibinfo  {journal} {Physical Review Letters}\ }\textbf {\bibinfo {volume}
  {21}},\ \bibinfo {pages} {406--409}}\BibitemShut {NoStop}%
\bibitem [{\citenamefont {Bender}\ and\ \citenamefont {Wu}(1969)}]{pt14}%
  \BibitemOpen
  \bibfield  {author} {\bibinfo {author} {\bibnamefont {Bender}, \bibfnamefont
  {Carl~M}}, and\ \bibinfo {author} {\bibfnamefont {Tai~Tsun}\ \bibnamefont
  {Wu}}} (\bibinfo {year} {1969}),\ \bibfield  {title} {\enquote {\bibinfo
  {title} {Anharmonic oscillator},}\ }\href
  {https://doi.org/10.1103/PhysRev.184.1231} {\bibfield  {journal} {\bibinfo
  {journal} {Physical Review}\ }\textbf {\bibinfo {volume} {184}},\ \bibinfo
  {pages} {1231--1260}}\BibitemShut {NoStop}%
\bibitem [{\citenamefont {Bender}\ and\ \citenamefont {Wu}(1971)}]{pt641}%
  \BibitemOpen
  \bibfield  {author} {\bibinfo {author} {\bibnamefont {Bender}, \bibfnamefont
  {Carl~M}}, and\ \bibinfo {author} {\bibfnamefont {Tai~Tsun}\ \bibnamefont
  {Wu}}} (\bibinfo {year} {1971}),\ \bibfield  {title} {\enquote {\bibinfo
  {title} {Large-order behavior of perturbation theory},}\ }\href
  {https://doi.org/10.1103/PhysRevLett.27.461} {\bibfield  {journal} {\bibinfo
  {journal} {Physical Review Letters}\ }\textbf {\bibinfo {volume} {27}},\
  10.1103/PhysRevLett.27.461}\BibitemShut {NoStop}%
\bibitem [{\citenamefont {Bender}\ and\ \citenamefont {Wu}(1973)}]{pt13}%
  \BibitemOpen
  \bibfield  {author} {\bibinfo {author} {\bibnamefont {Bender}, \bibfnamefont
  {Carl~M}}, and\ \bibinfo {author} {\bibfnamefont {Tai~Tsun}\ \bibnamefont
  {Wu}}} (\bibinfo {year} {1973}),\ \bibfield  {title} {\enquote {\bibinfo
  {title} {Anharmonic oscillator. {II}. {A} study of perturbation theory in
  large order},}\ }\href {https://doi.org/10.1103/PhysRevD.7.1620} {\bibfield
  {journal} {\bibinfo  {journal} {Physical Review D}\ }\textbf {\bibinfo
  {volume} {7}},\ \bibinfo {pages} {1620--1636}}\BibitemShut {NoStop}%
\bibitem [{\citenamefont {Berry}(2020)}]{pt662}%
  \BibitemOpen
  \bibfield  {author} {\bibinfo {author} {\bibnamefont {Berry}, \bibfnamefont
  {M~V}}} (\bibinfo {year} {2020}),\ \bibfield  {title} {\enquote {\bibinfo
  {title} {Classical and quantum complex hamiltonian curl forces},}\ }\href
  {https://doi.org/10.1088/1751-8121/abad77} {\bibfield  {journal} {\bibinfo
  {journal} {Journal of Physics A: Mathematical and Theoretical}\ }\textbf
  {\bibinfo {volume} {53}},\ 10.1088/1751-8121/abad77}\BibitemShut {NoStop}%
\bibitem [{\citenamefont {Bittner}\ \emph {et~al.}(2012)\citenamefont
  {Bittner}, \citenamefont {Dietz}, \citenamefont {G{\"u}nther}, \citenamefont
  {Harney}, \citenamefont {Miski-Oglu}, \citenamefont {Richter},\ and\
  \citenamefont {Sch{\"a}fer}}]{pt465}%
  \BibitemOpen
  \bibfield  {author} {\bibinfo {author} {\bibnamefont {Bittner}, \bibfnamefont
  {S}}, \bibinfo {author} {\bibfnamefont {B}~\bibnamefont {Dietz}}, \bibinfo
  {author} {\bibfnamefont {U}~\bibnamefont {G{\"u}nther}}, \bibinfo {author}
  {\bibfnamefont {H~L}\ \bibnamefont {Harney}}, \bibinfo {author}
  {\bibfnamefont {M}~\bibnamefont {Miski-Oglu}}, \bibinfo {author}
  {\bibfnamefont {A}~\bibnamefont {Richter}}, and\ \bibinfo {author}
  {\bibfnamefont {F}~\bibnamefont {Sch{\"a}fer}}} (\bibinfo {year} {2012}),\
  \bibfield  {title} {\enquote {\bibinfo {title} {$\mathcal{PT}$ symmetry and
  spontaneous symmetry breaking in a microwave billiard},}\ }\href
  {https://doi.org/10.1103/PhysRevLett.108.024101} {\bibfield  {journal}
  {\bibinfo  {journal} {Physical Review Letters}\ }\textbf {\bibinfo {volume}
  {108}},\ 10.1103/PhysRevLett.108.024101}\BibitemShut {NoStop}%
\bibitem [{\citenamefont {Br\'ezin}\ \emph {et~al.}(1977)\citenamefont
  {Br\'ezin}, \citenamefont {Guillou},\ and\ \citenamefont
  {Zinn-Justin}}]{pt602}%
  \BibitemOpen
  \bibfield  {author} {\bibinfo {author} {\bibnamefont {Br\'ezin},
  \bibfnamefont {E}}, \bibinfo {author} {\bibfnamefont {J~C~Le}\ \bibnamefont
  {Guillou}}, and\ \bibinfo {author} {\bibfnamefont {J}~\bibnamefont
  {Zinn-Justin}}} (\bibinfo {year} {1977}),\ \bibfield  {title} {\enquote
  {\bibinfo {title} {Perturbation theory at large order. i. the $\phi^{2N}$
  interaction},}\ }\href {https://doi.org/10.1103/PhysRevD.15.15446} {\bibfield
   {journal} {\bibinfo  {journal} {Physics Review D}\ }\textbf {\bibinfo
  {volume} {15}},\ 10.1103/PhysRevD.15.15446}\BibitemShut {NoStop}%
\bibitem [{\citenamefont {Brody}(2014)}]{pt372}%
  \BibitemOpen
  \bibfield  {author} {\bibinfo {author} {\bibnamefont {Brody}, \bibfnamefont
  {D~C}}} (\bibinfo {year} {2014}),\ \bibfield  {title} {\enquote {\bibinfo
  {title} {Biorthogonal quantum mechanics},}\ }\href
  {https://doi.org/10.1088/1751-8113/47/3/035305} {\bibfield  {journal}
  {\bibinfo  {journal} {Journal of Physics A: Mathematical and Theoretical}\
  }\textbf {\bibinfo {volume} {47}},\
  10.1088/1751-8113/47/3/035305}\BibitemShut {NoStop}%
\bibitem [{\citenamefont {Brody}\ and\ \citenamefont {Graefe}(2011)}]{pt678}%
  \BibitemOpen
  \bibfield  {author} {\bibinfo {author} {\bibnamefont {Brody}, \bibfnamefont
  {Dorje~C}}, and\ \bibinfo {author} {\bibfnamefont {Eva-Maria}\ \bibnamefont
  {Graefe}}} (\bibinfo {year} {2011}),\ \bibfield  {title} {\enquote {\bibinfo
  {title} {On complexified mechanics and coquaternions},}\ }\href
  {https://doi.org/10.1088/1751-8113/44/7/072001} {\bibfield  {journal}
  {\bibinfo  {journal} {Journal of Physics A: Mathematical and Theoretical}\
  }\textbf {\bibinfo {volume} {44}},\
  10.1088/1751-8113/44/7/072001}\BibitemShut {NoStop}%
\bibitem [{\citenamefont {Buslaev}\ and\ \citenamefont {Grecchi}(1993)}]{pt9}%
  \BibitemOpen
  \bibfield  {author} {\bibinfo {author} {\bibnamefont {Buslaev}, \bibfnamefont
  {Vladimir}}, and\ \bibinfo {author} {\bibfnamefont {Vincenzo}\ \bibnamefont
  {Grecchi}}} (\bibinfo {year} {1993}),\ \bibfield  {title} {\enquote {\bibinfo
  {title} {Equivalence of unstable anharmonic oscillators and double wells},}\
  }\href {https://doi.org/10.1088/0305-4470/26/20/035} {\bibfield  {journal}
  {\bibinfo  {journal} {Journal of Physics A: Mathematical and General}\
  }\textbf {\bibinfo {volume} {26}}~(\bibinfo {number} {20}),\ \bibinfo {pages}
  {5541--5549}}\BibitemShut {NoStop}%
\bibitem [{\citenamefont {Caliceti}\ \emph {et~al.}(2006)\citenamefont
  {Caliceti}, \citenamefont {Cannata},\ and\ \citenamefont {Graffi}}]{pt605}%
  \BibitemOpen
  \bibfield  {author} {\bibinfo {author} {\bibnamefont {Caliceti},
  \bibfnamefont {E}}, \bibinfo {author} {\bibfnamefont {F}~\bibnamefont
  {Cannata}}, and\ \bibinfo {author} {\bibfnamefont {S}~\bibnamefont {Graffi}}}
  (\bibinfo {year} {2006}),\ \bibfield  {title} {\enquote {\bibinfo {title}
  {Perturbation theory of $\mathcal{PT}$-symmetric hamiltonians},}\ }\href
  {https://doi.org/10.1088/0305-4470/39/32/S06} {\bibfield  {journal} {\bibinfo
   {journal} {Journal of Physics A: Mathematical and General}\ }\textbf
  {\bibinfo {volume} {39}},\ 10.1088/0305-4470/39/32/S06}\BibitemShut {NoStop}%
\bibitem [{\citenamefont {Caliceti}\ \emph {et~al.}(1980)\citenamefont
  {Caliceti}, \citenamefont {Graffi},\ and\ \citenamefont {Maioli}}]{pt5}%
  \BibitemOpen
  \bibfield  {author} {\bibinfo {author} {\bibnamefont {Caliceti},
  \bibfnamefont {E}}, \bibinfo {author} {\bibfnamefont {S}~\bibnamefont
  {Graffi}}, and\ \bibinfo {author} {\bibfnamefont {M}~\bibnamefont {Maioli}}}
  (\bibinfo {year} {1980}),\ \bibfield  {title} {\enquote {\bibinfo {title}
  {Perturbation theory of odd anharmonic oscillators},}\ }\href
  {https://doi.org/10.1007/BF01962591} {\bibfield  {journal} {\bibinfo
  {journal} {Communications in Mathematical Physics}\ }\textbf {\bibinfo
  {volume} {75}}~(\bibinfo {number} {1}),\ \bibinfo {pages}
  {51--66}}\BibitemShut {NoStop}%
\bibitem [{\citenamefont {Cartarius}\ and\ \citenamefont
  {Wunner}(2022)}]{pt668}%
  \BibitemOpen
  \bibfield  {author} {\bibinfo {author} {\bibnamefont {Cartarius},
  \bibfnamefont {Holger}}, and\ \bibinfo {author} {\bibfnamefont {G{\"u}nter}\
  \bibnamefont {Wunner}}} (\bibinfo {year} {2022}),\ \bibfield  {title}
  {\enquote {\bibinfo {title} {Model of a pt-symmetric bose-einstein condensate
  in a $\delta$-function double-well potential},}\ }\href
  {https://doi.org/10.1103/physreva.86.013612} {\bibfield  {journal} {\bibinfo
  {journal} {Physical Review A}\ }\textbf {\bibinfo {volume} {86}},\
  10.1103/physreva.86.013612}\BibitemShut {NoStop}%
\bibitem [{\citenamefont {Cavaglia}\ \emph {et~al.}(2011)\citenamefont
  {Cavaglia}, \citenamefont {Fring},\ and\ \citenamefont {Bagchi}}]{pt48}%
  \BibitemOpen
  \bibfield  {author} {\bibinfo {author} {\bibnamefont {Cavaglia},
  \bibfnamefont {A}}, \bibinfo {author} {\bibfnamefont {A}~\bibnamefont
  {Fring}}, and\ \bibinfo {author} {\bibfnamefont {B}~\bibnamefont {Bagchi}}}
  (\bibinfo {year} {2011}),\ \bibfield  {title} {\enquote {\bibinfo {title}
  {$\cal{PT}$-symmetry breaking in complex nonlinear wave equations and their
  deformations},}\ }\href {https://doi.org/10.1088/1751-8113/44/32/325201}
  {\bibfield  {journal} {\bibinfo  {journal} {Journal of Physics A:
  Mathematical and Theoretical}\ }\textbf {\bibinfo {volume} {44}},\ \bibinfo
  {pages} {325201}}\BibitemShut {NoStop}%
\bibitem [{\citenamefont {Cerjan}\ \emph {et~al.}(2016)\citenamefont {Cerjan},
  \citenamefont {Raman},\ and\ \citenamefont {Fan}}]{pt552}%
  \BibitemOpen
  \bibfield  {author} {\bibinfo {author} {\bibnamefont {Cerjan}, \bibfnamefont
  {Alexander}}, \bibinfo {author} {\bibfnamefont {Aaswath}\ \bibnamefont
  {Raman}}, and\ \bibinfo {author} {\bibfnamefont {Shanhui}\ \bibnamefont
  {Fan}}} (\bibinfo {year} {2016}),\ \bibfield  {title} {\enquote {\bibinfo
  {title} {Exceptional contours and band structure design in parity-time
  symmetric photonic crystals},}\ }\href
  {https://doi.org/10.1103/physrevlett.116.203902} {\bibfield  {journal}
  {\bibinfo  {journal} {Physical Review Letters}\ }\textbf {\bibinfo {volume}
  {116}},\ \bibinfo {pages} {203902}}\BibitemShut {NoStop}%
\bibitem [{\citenamefont {Chen}\ and\ \citenamefont {Jung}(2016)}]{pt592}%
  \BibitemOpen
  \bibfield  {author} {\bibinfo {author} {\bibnamefont {Chen}, \bibfnamefont
  {Pai-Yen}}, and\ \bibinfo {author} {\bibfnamefont {Jeil}\ \bibnamefont
  {Jung}}} (\bibinfo {year} {2016}),\ \bibfield  {title} {\enquote {\bibinfo
  {title} {$\mathcal{PT}$ symmetry and singularity-enhanced sensing based on
  photoexcited graphene metasurfaces},}\ }\href
  {https://doi.org/10.1103/PhysRevApplied.5.064018} {\bibfield  {journal}
  {\bibinfo  {journal} {Physical Review Applied}\ }\textbf {\bibinfo {volume}
  {5}},\ 10.1103/PhysRevApplied.5.064018}\BibitemShut {NoStop}%
\bibitem [{\citenamefont {Chen}\ \emph {et~al.}(2022)\citenamefont {Chen},
  \citenamefont {Abbasi}, \citenamefont {Ha}, \citenamefont {Erdamar},
  \citenamefont {Joglekar},\ and\ \citenamefont {Murch}}]{pt631}%
  \BibitemOpen
  \bibfield  {author} {\bibinfo {author} {\bibnamefont {Chen}, \bibfnamefont
  {Weijian}}, \bibinfo {author} {\bibfnamefont {Maryam}\ \bibnamefont
  {Abbasi}}, \bibinfo {author} {\bibfnamefont {Byung}\ \bibnamefont {Ha}},
  \bibinfo {author} {\bibfnamefont {Serra}\ \bibnamefont {Erdamar}}, \bibinfo
  {author} {\bibfnamefont {Yogesh~N}\ \bibnamefont {Joglekar}}, and\ \bibinfo
  {author} {\bibfnamefont {Kater~W}\ \bibnamefont {Murch}}} (\bibinfo {year}
  {2022}),\ \bibfield  {title} {\enquote {\bibinfo {title} {Decoherence-induced
  exceptional points in a dissipative superconducting qubit},}\ }\href
  {https://doi.org/10.1103/PhysRevLett.128.110402} {\bibfield  {journal}
  {\bibinfo  {journal} {Physical Review Letters}\ }\textbf {\bibinfo {volume}
  {128}},\ 10.1103/PhysRevLett.128.110402}\BibitemShut {NoStop}%
\bibitem [{\citenamefont {Chen}\ \emph {et~al.}(2021)\citenamefont {Chen},
  \citenamefont {Abbasi}, \citenamefont {Joglekar},\ and\ \citenamefont
  {Murch}}]{pt630}%
  \BibitemOpen
  \bibfield  {author} {\bibinfo {author} {\bibnamefont {Chen}, \bibfnamefont
  {Weijian}}, \bibinfo {author} {\bibfnamefont {Maryam}\ \bibnamefont
  {Abbasi}}, \bibinfo {author} {\bibfnamefont {Yogesh~N}\ \bibnamefont
  {Joglekar}}, and\ \bibinfo {author} {\bibfnamefont {Kater~W}\ \bibnamefont
  {Murch}}} (\bibinfo {year} {2021}),\ \bibfield  {title} {\enquote {\bibinfo
  {title} {Quantum jumps in the non-{H}ermitian dynamics of a superconducting
  qubit},}\ }\href {https://doi.org/10.1103/PhysRevLett.127.140504} {\bibfield
  {journal} {\bibinfo  {journal} {Physical Review Letters}\ }\textbf {\bibinfo
  {volume} {127}},\ 10.1103/PhysRevLett.127.140504}\BibitemShut {NoStop}%
\bibitem [{\citenamefont {Chen}\ \emph {et~al.}(2017)\citenamefont {Chen},
  \citenamefont {\c{S}ahin Kaya~\"{O}zdemir}, \citenamefont {Zhao},
  \citenamefont {Wiersig},\ and\ \citenamefont {Yang}}]{pt553}%
  \BibitemOpen
  \bibfield  {author} {\bibinfo {author} {\bibnamefont {Chen}, \bibfnamefont
  {Weijian}}, \bibinfo {author} {\bibnamefont {\c{S}ahin Kaya~\"{O}zdemir}},
  \bibinfo {author} {\bibfnamefont {Guangming}\ \bibnamefont {Zhao}}, \bibinfo
  {author} {\bibfnamefont {Jan}\ \bibnamefont {Wiersig}}, and\ \bibinfo
  {author} {\bibfnamefont {Lan}\ \bibnamefont {Yang}}} (\bibinfo {year}
  {2017}),\ \bibfield  {title} {\enquote {\bibinfo {title} {Exceptional points
  enhance sensing in an optical microcavity},}\ }\href
  {https://doi.org/10.1038/nature23281} {\bibfield  {journal} {\bibinfo
  {journal} {Nature}\ }\textbf {\bibinfo {volume} {548}},\ \bibinfo {pages}
  {192--196}}\BibitemShut {NoStop}%
\bibitem [{\citenamefont {Cheng}\ \emph {et~al.}(2018)\citenamefont {Cheng},
  \citenamefont {Cirillo}, \citenamefont {Salina},\ and\ \citenamefont
  {Grønbech-Jensen}}]{pt684}%
  \BibitemOpen
  \bibfield  {author} {\bibinfo {author} {\bibnamefont {Cheng}, \bibfnamefont
  {Chungho}}, \bibinfo {author} {\bibfnamefont {Matteo}\ \bibnamefont
  {Cirillo}}, \bibinfo {author} {\bibfnamefont {Gaetano}\ \bibnamefont
  {Salina}}, and\ \bibinfo {author} {\bibfnamefont {Niels}\ \bibnamefont
  {Grønbech-Jensen}}} (\bibinfo {year} {2018}),\ \bibfield  {title} {\enquote
  {\bibinfo {title} {Nonequilibrium transient phenomena in the washboard
  potential},}\ }\href {https://doi.org/10.1103/physreve.98.012140} {\bibfield
  {journal} {\bibinfo  {journal} {Physical Review E}\ }\textbf {\bibinfo
  {volume} {98}},\ 10.1103/physreve.98.012140}\BibitemShut {NoStop}%
\bibitem [{\citenamefont {Chew}\ \emph {et~al.}(1962)\citenamefont {Chew},
  \citenamefont {Frautschi},\ and\ \citenamefont {Mandelstam}}]{pt612}%
  \BibitemOpen
  \bibfield  {author} {\bibinfo {author} {\bibnamefont {Chew}, \bibfnamefont
  {Geoffrey~F}}, \bibinfo {author} {\bibfnamefont {Steven~C}\ \bibnamefont
  {Frautschi}}, and\ \bibinfo {author} {\bibfnamefont {Stanley}\ \bibnamefont
  {Mandelstam}}} (\bibinfo {year} {1962}),\ \bibfield  {title} {\enquote
  {\bibinfo {title} {Regge poles in $\pi-\pi$ scattering},}\ }\href
  {https://doi.org/10.1103/PhysRev.126.1202} {\bibfield  {journal} {\bibinfo
  {journal} {Physical Review}\ }\textbf {\bibinfo {volume} {126}},\
  10.1103/PhysRev.126.1202}\BibitemShut {NoStop}%
\bibitem [{\citenamefont {Chong}\ \emph {et~al.}(2010)\citenamefont {Chong},
  \citenamefont {Ge}, \citenamefont {Cao},\ and\ \citenamefont
  {Stone}}]{pt203}%
  \BibitemOpen
  \bibfield  {author} {\bibinfo {author} {\bibnamefont {Chong}, \bibfnamefont
  {Y~D}}, \bibinfo {author} {\bibfnamefont {Li}~\bibnamefont {Ge}}, \bibinfo
  {author} {\bibfnamefont {Hui}\ \bibnamefont {Cao}}, and\ \bibinfo {author}
  {\bibfnamefont {A~D}\ \bibnamefont {Stone}}} (\bibinfo {year} {2010}),\
  \bibfield  {title} {\enquote {\bibinfo {title} {Coherent perfect absorbers:
  Time-reversed lasers},}\ }\href
  {https://doi.org/10.1103/PhysRevLett.105.053901} {\bibfield  {journal}
  {\bibinfo  {journal} {Physical Review Letters}\ }\textbf {\bibinfo {volume}
  {105}},\ \bibinfo {pages} {053901}}\BibitemShut {NoStop}%
\bibitem [{\citenamefont {Chong}\ \emph {et~al.}(2011)\citenamefont {Chong},
  \citenamefont {Ge},\ and\ \citenamefont {Stone}}]{pt544}%
  \BibitemOpen
  \bibfield  {author} {\bibinfo {author} {\bibnamefont {Chong}, \bibfnamefont
  {Y~D}}, \bibinfo {author} {\bibfnamefont {Li}~\bibnamefont {Ge}}, and\
  \bibinfo {author} {\bibfnamefont {A~Douglas}\ \bibnamefont {Stone}}}
  (\bibinfo {year} {2011}),\ \bibfield  {title} {\enquote {\bibinfo {title}
  {$\mathcal{PT}$-symmetry breaking and laser-absorber modes in optical
  scattering systems},}\ }\href
  {https://doi.org/10.1103/physrevlett.106.093902} {\bibfield  {journal}
  {\bibinfo  {journal} {Physical Review Letters}\ }\textbf {\bibinfo {volume}
  {106}},\ \bibinfo {pages} {093902}}\BibitemShut {NoStop}%
\bibitem [{\citenamefont {Christenson}\ \emph {et~al.}(1964)\citenamefont
  {Christenson}, \citenamefont {Cronin}, \citenamefont {Fitch},\ and\
  \citenamefont {Turlay}}]{pt582}%
  \BibitemOpen
  \bibfield  {author} {\bibinfo {author} {\bibnamefont {Christenson},
  \bibfnamefont {J~H}}, \bibinfo {author} {\bibfnamefont {J~W}\ \bibnamefont
  {Cronin}}, \bibinfo {author} {\bibfnamefont {V~L}\ \bibnamefont {Fitch}},
  and\ \bibinfo {author} {\bibfnamefont {R}~\bibnamefont {Turlay}}} (\bibinfo
  {year} {1964}),\ \bibfield  {title} {\enquote {\bibinfo {title} {Evidence for
  the $2\pi$ decay of the $k_2^0$ meson},}\ }\href
  {https://doi.org/10.1103/PhysRevLett.13.138} {\bibfield  {journal} {\bibinfo
  {journal} {Physical Review Letters}\ }\textbf {\bibinfo {volume} {13}},\
  10.1103/PhysRevLett.13.138}\BibitemShut {NoStop}%
\bibitem [{\citenamefont {Chtchelkatchev}\ \emph {et~al.}(2012)\citenamefont
  {Chtchelkatchev}, \citenamefont {Golubov}, \citenamefont {Baturina},\ and\
  \citenamefont {Vinokur}}]{pt691}%
  \BibitemOpen
  \bibfield  {author} {\bibinfo {author} {\bibnamefont {Chtchelkatchev},
  \bibfnamefont {N~M}}, \bibinfo {author} {\bibfnamefont {A~A}\ \bibnamefont
  {Golubov}}, \bibinfo {author} {\bibfnamefont {T~I}\ \bibnamefont {Baturina}},
  and\ \bibinfo {author} {\bibfnamefont {V~M}\ \bibnamefont {Vinokur}}}
  (\bibinfo {year} {2012}),\ \bibfield  {title} {\enquote {\bibinfo {title}
  {Stimulation of the fluctuation superconductivity by pt symmetry},}\ }\href
  {https://doi.org/10.1103/physrevlett.109.150405} {\bibfield  {journal}
  {\bibinfo  {journal} {Physical Review Letters}\ }\textbf {\bibinfo {volume}
  {109}},\ 10.1103/physrevlett.109.150405}\BibitemShut {NoStop}%
\bibitem [{\citenamefont {Copson}(1960)}]{pt583}%
  \BibitemOpen
  \bibfield  {author} {\bibinfo {author} {\bibnamefont {Copson}, \bibfnamefont
  {Edward~Thomas}}} (\bibinfo {year} {1960}),\ \href@noop {} {\emph {\bibinfo
  {title} {An Introduction to the theory of functions of a complex variable}}}\
  (\bibinfo  {publisher} {Clarendon Press},\ \bibinfo {address}
  {Oxford})\BibitemShut {NoStop}%
\bibitem [{\citenamefont {Cummer}\ \emph {et~al.}(2016)\citenamefont {Cummer},
  \citenamefont {Christensen},\ and\ \citenamefont {Al\`{u}}}]{pt549}%
  \BibitemOpen
  \bibfield  {author} {\bibinfo {author} {\bibnamefont {Cummer}, \bibfnamefont
  {Steven~A}}, \bibinfo {author} {\bibfnamefont {Johan}\ \bibnamefont
  {Christensen}}, and\ \bibinfo {author} {\bibfnamefont {Andrea}\ \bibnamefont
  {Al\`{u}}}} (\bibinfo {year} {2016}),\ \bibfield  {title} {\enquote {\bibinfo
  {title} {Controlling sound with acoustic metamaterials},}\ }\href
  {https://doi.org/10.1038/natrevmats.2016.1} {\bibfield  {journal} {\bibinfo
  {journal} {Nature Reviews: Materials}\ }\textbf {\bibinfo {volume} {1}},\
  \bibinfo {pages} {16001}}\BibitemShut {NoStop}%
\bibitem [{\citenamefont {Curtright}\ and\ \citenamefont
  {Fairlie}(2008)}]{pt65}%
  \BibitemOpen
  \bibfield  {author} {\bibinfo {author} {\bibnamefont {Curtright},
  \bibfnamefont {T~L}}, and\ \bibinfo {author} {\bibfnamefont {D~B}\
  \bibnamefont {Fairlie}}} (\bibinfo {year} {2008}),\ \bibfield  {title}
  {\enquote {\bibinfo {title} {Euler incognito},}\ }\href
  {https://doi.org/10.1088/1751-8113/41/24/244009} {\bibfield  {journal}
  {\bibinfo  {journal} {Journal of Physics A: Mathematical and Theoretical}\
  }\textbf {\bibinfo {volume} {41}}~(\bibinfo {number} {24}),\ \bibinfo {pages}
  {244009}}\BibitemShut {NoStop}%
\bibitem [{\citenamefont {Curtright}\ and\ \citenamefont
  {Mezincescu}(2007)}]{pt371}%
  \BibitemOpen
  \bibfield  {author} {\bibinfo {author} {\bibnamefont {Curtright},
  \bibfnamefont {Thomas}}, and\ \bibinfo {author} {\bibfnamefont {Luca}\
  \bibnamefont {Mezincescu}}} (\bibinfo {year} {2007}),\ \bibfield  {title}
  {\enquote {\bibinfo {title} {Biorthogonal quantum systems},}\ }\href
  {https://doi.org/10.1063/1.2196243} {\bibfield  {journal} {\bibinfo
  {journal} {Journal of Mathematical Physics}\ }\textbf {\bibinfo {volume}
  {48}},\ 10.1063/1.2196243}\BibitemShut {NoStop}%
\bibitem [{\citenamefont {Dirac}(1942)}]{pt379}%
  \BibitemOpen
  \bibfield  {author} {\bibinfo {author} {\bibnamefont {Dirac}, \bibfnamefont
  {Paul A~M}}} (\bibinfo {year} {1942}),\ \bibfield  {title} {\enquote
  {\bibinfo {title} {{B}akerian {L}ecture. {T}he physical interpretation of
  quantum mechanics},}\ }\href {https://doi.org/10.1098/rspa.1942.0023}
  {\bibfield  {journal} {\bibinfo  {journal} {Proceedings of the Royal Society
  A}\ }\textbf {\bibinfo {volume} {180}},\ 10.1098/rspa.1942.0023}\BibitemShut
  {NoStop}%
\bibitem [{\citenamefont {Dorey}\ \emph {et~al.}(2008)\citenamefont {Dorey},
  \citenamefont {Dunning}, \citenamefont {Gliozzi},\ and\ \citenamefont
  {Tateo}}]{pt295}%
  \BibitemOpen
  \bibfield  {author} {\bibinfo {author} {\bibnamefont {Dorey}, \bibfnamefont
  {Patrick}}, \bibinfo {author} {\bibfnamefont {Clare}\ \bibnamefont
  {Dunning}}, \bibinfo {author} {\bibfnamefont {Ferdinando}\ \bibnamefont
  {Gliozzi}}, and\ \bibinfo {author} {\bibfnamefont {Roberto}\ \bibnamefont
  {Tateo}}} (\bibinfo {year} {2008}),\ \bibfield  {title} {\enquote {\bibinfo
  {title} {{On the {ODE/IM} correspondence for minimal models}},}\ }\href
  {https://doi.org/10.1088/1751-8113/41/13/132001} {\bibfield  {journal}
  {\bibinfo  {journal} {Journal of Physics A: Mathematical and Theoretical}\
  }\textbf {\bibinfo {volume} {41}},\ \bibinfo {pages} {132001}}\BibitemShut
  {NoStop}%
\bibitem [{\citenamefont {Dorey}\ \emph {et~al.}(2009)\citenamefont {Dorey},
  \citenamefont {Dunning}, \citenamefont {Lishman},\ and\ \citenamefont
  {Tateo}}]{pt312}%
  \BibitemOpen
  \bibfield  {author} {\bibinfo {author} {\bibnamefont {Dorey}, \bibfnamefont
  {Patrick}}, \bibinfo {author} {\bibfnamefont {Clare}\ \bibnamefont
  {Dunning}}, \bibinfo {author} {\bibfnamefont {Anna}\ \bibnamefont {Lishman}},
  and\ \bibinfo {author} {\bibfnamefont {Roberto}\ \bibnamefont {Tateo}}}
  (\bibinfo {year} {2009}),\ \bibfield  {title} {\enquote {\bibinfo {title}
  {$\cal{PT}$ symmetry breaking and exceptional points for a class of
  inhomogeneous complex potentials},}\ }\href
  {https://doi.org/10.1088/1751-8113/42/46/465302} {\bibfield  {journal}
  {\bibinfo  {journal} {Journal of Physics A: Mathematical and Theoretical}\
  }\textbf {\bibinfo {volume} {42}},\ \bibinfo {pages} {465302}}\BibitemShut
  {NoStop}%
\bibitem [{\citenamefont {Dorey}\ \emph {et~al.}(2001)\citenamefont {Dorey},
  \citenamefont {Dunning},\ and\ \citenamefont {Tateo}}]{pt274}%
  \BibitemOpen
  \bibfield  {author} {\bibinfo {author} {\bibnamefont {Dorey}, \bibfnamefont
  {Patrick}}, \bibinfo {author} {\bibfnamefont {Clare}\ \bibnamefont
  {Dunning}}, and\ \bibinfo {author} {\bibfnamefont {Roberto}\ \bibnamefont
  {Tateo}}} (\bibinfo {year} {2001}),\ \bibfield  {title} {\enquote {\bibinfo
  {title} {Spectral equivalence, {B}ethe ansatz, and reality properties in
  $\cal{PT}$-symmetric quantum mechanics},}\ }\href
  {https://doi.org/10.1088/0305-4470/34/28/305} {\bibfield  {journal} {\bibinfo
   {journal} {Journal of Physics A: Mathematical and General}\ }\textbf
  {\bibinfo {volume} {34}},\ \bibinfo {pages} {5679--5704}}\BibitemShut
  {NoStop}%
\bibitem [{\citenamefont {Dorey}\ \emph {et~al.}(2007)\citenamefont {Dorey},
  \citenamefont {Dunning},\ and\ \citenamefont {Tateo}}]{pt294}%
  \BibitemOpen
  \bibfield  {author} {\bibinfo {author} {\bibnamefont {Dorey}, \bibfnamefont
  {Patrick}}, \bibinfo {author} {\bibfnamefont {Clare}\ \bibnamefont
  {Dunning}}, and\ \bibinfo {author} {\bibfnamefont {Roberto}\ \bibnamefont
  {Tateo}}} (\bibinfo {year} {2007}),\ \bibfield  {title} {\enquote {\bibinfo
  {title} {{The {ODE/IM} correspondence}},}\ }\href
  {https://doi.org/10.1088/1751-8113/40/32/R01} {\bibfield  {journal} {\bibinfo
   {journal} {Journal of Physics A: Mathematical and Theoretical}\ }\textbf
  {\bibinfo {volume} {40}},\ \bibinfo {pages} {R205--R283}}\BibitemShut
  {NoStop}%
\bibitem [{\citenamefont {Dorey}\ \emph {et~al.}(2012)\citenamefont {Dorey},
  \citenamefont {Dunning},\ and\ \citenamefont {Tateo}}]{pt297}%
  \BibitemOpen
  \bibfield  {author} {\bibinfo {author} {\bibnamefont {Dorey}, \bibfnamefont
  {Patrick}}, \bibinfo {author} {\bibfnamefont {Clare}\ \bibnamefont
  {Dunning}}, and\ \bibinfo {author} {\bibfnamefont {Roberto}\ \bibnamefont
  {Tateo}}} (\bibinfo {year} {2012}),\ \bibfield  {title} {\enquote {\bibinfo
  {title} {{Quasi-exact solvability, resonances and trivial monodromy in
  ordinary differential equations}},}\ }\href
  {https://doi.org/10.1088/1751-8113/45/44/444013} {\bibfield  {journal}
  {\bibinfo  {journal} {Journal of Physics A: Mathematical and Theoretical}\
  }\textbf {\bibinfo {volume} {45}},\ \bibinfo {pages} {444013}}\BibitemShut
  {NoStop}%
\bibitem [{\citenamefont {Dorey}\ and\ \citenamefont {Tateo}(1999)}]{pt298}%
  \BibitemOpen
  \bibfield  {author} {\bibinfo {author} {\bibnamefont {Dorey}, \bibfnamefont
  {Patrick}}, and\ \bibinfo {author} {\bibfnamefont {Roberto}\ \bibnamefont
  {Tateo}}} (\bibinfo {year} {1999}),\ \bibfield  {title} {\enquote {\bibinfo
  {title} {{On the relation between {S}tokes multipliers and the {T-Q} systems
  of conformal field theory}},}\ }\href
  {https://doi.org/10.1016/S0550-3213(99)00609-4} {\bibfield  {journal}
  {\bibinfo  {journal} {Nuclear Physics B}\ }\textbf {\bibinfo {volume}
  {563}},\ \bibinfo {pages} {573--602}},\ \bibinfo {note} {[Erratum: Nucl.
  Phys.B603,581(2001)]}\BibitemShut {NoStop}%
\bibitem [{\citenamefont {Dutt}\ \emph {et~al.}(1995)\citenamefont {Dutt},
  \citenamefont {Khare},\ and\ \citenamefont {Varshni}}]{pt100}%
  \BibitemOpen
  \bibfield  {author} {\bibinfo {author} {\bibnamefont {Dutt}, \bibfnamefont
  {Ranabir}}, \bibinfo {author} {\bibfnamefont {Avinash}\ \bibnamefont
  {Khare}}, and\ \bibinfo {author} {\bibfnamefont {Y~P}\ \bibnamefont
  {Varshni}}} (\bibinfo {year} {1995}),\ \bibfield  {title} {\enquote {\bibinfo
  {title} {New class of conditionally exactly solvable potentials in quantum
  mechanics},}\ }\href {https://doi.org/10.1088/0305-4470/28/3/008} {\bibfield
  {journal} {\bibinfo  {journal} {Journal of Physics A: Mathematical and
  General}\ }\textbf {\bibinfo {volume} {28}},\ \bibinfo {pages}
  {L107--L113}}\BibitemShut {NoStop}%
\bibitem [{\citenamefont {Dyson}(1952)}]{pt584}%
  \BibitemOpen
  \bibfield  {author} {\bibinfo {author} {\bibnamefont {Dyson}, \bibfnamefont
  {F~J}}} (\bibinfo {year} {1952}),\ \bibfield  {title} {\enquote {\bibinfo
  {title} {Divergence of perturbation theory in quantum electrodynamics},}\
  }\href {https://doi.org/10.1103/PhysRev.85.631} {\bibfield  {journal}
  {\bibinfo  {journal} {Physical Review}\ }\textbf {\bibinfo {volume} {85}},\
  10.1103/PhysRev.85.631}\BibitemShut {NoStop}%
\bibitem [{\citenamefont {El-Ganainy}\ \emph {et~al.}(2007)\citenamefont
  {El-Ganainy}, \citenamefont {Makris}, \citenamefont {Christodoulides},\ and\
  \citenamefont {Musslimani}}]{pt182}%
  \BibitemOpen
  \bibfield  {author} {\bibinfo {author} {\bibnamefont {El-Ganainy},
  \bibfnamefont {R}}, \bibinfo {author} {\bibfnamefont {K~G}\ \bibnamefont
  {Makris}}, \bibinfo {author} {\bibfnamefont {D~N}\ \bibnamefont
  {Christodoulides}}, and\ \bibinfo {author} {\bibfnamefont {Z~H}\ \bibnamefont
  {Musslimani}}} (\bibinfo {year} {2007}),\ \bibfield  {title} {\enquote
  {\bibinfo {title} {Theory of coupled optical $\cal{PT}$-symmetric
  structures},}\ }\href {https://doi.org/10.1364/OL.32.002632} {\bibfield
  {journal} {\bibinfo  {journal} {Optics Letters}\ }\textbf {\bibinfo {volume}
  {32}},\ \bibinfo {pages} {2632--2634}}\BibitemShut {NoStop}%
\bibitem [{\citenamefont {El-Ganainy}\ \emph
  {et~al.}(2018{\natexlab{a}})\citenamefont {El-Ganainy}, \citenamefont
  {Makris}, \citenamefont {Khajavikhan}, \citenamefont {Musslimani},
  \citenamefont {Rotter},\ and\ \citenamefont {Christodoulides}}]{pt557}%
  \BibitemOpen
  \bibfield  {author} {\bibinfo {author} {\bibnamefont {El-Ganainy},
  \bibfnamefont {Ramy}}, \bibinfo {author} {\bibfnamefont {Konstantinos~G}\
  \bibnamefont {Makris}}, \bibinfo {author} {\bibfnamefont {Mercedeh}\
  \bibnamefont {Khajavikhan}}, \bibinfo {author} {\bibfnamefont {Ziad~H}\
  \bibnamefont {Musslimani}}, \bibinfo {author} {\bibfnamefont {Stefan}\
  \bibnamefont {Rotter}}, and\ \bibinfo {author} {\bibfnamefont {Demetrios~N}\
  \bibnamefont {Christodoulides}}} (\bibinfo {year} {2018}{\natexlab{a}}),\
  \bibfield  {title} {\enquote {\bibinfo {title} {Non-{H}ermitian physics and
  $\mathcal{PT}$ symmetry},}\ }\href {https://doi.org/10.1038/nphys4323}
  {\bibfield  {journal} {\bibinfo  {journal} {Nature Physics}\ }\textbf
  {\bibinfo {volume} {14}},\ \bibinfo {pages} {11--19}}\BibitemShut {NoStop}%
\bibitem [{\citenamefont {El-Ganainy}\ \emph
  {et~al.}(2018{\natexlab{b}})\citenamefont {El-Ganainy}, \citenamefont
  {Makris}, \citenamefont {Khajavikhan}, \citenamefont {Musslimani},
  \citenamefont {Rotter},\ and\ \citenamefont {Christodoulides}}]{pt683}%
  \BibitemOpen
  \bibfield  {author} {\bibinfo {author} {\bibnamefont {El-Ganainy},
  \bibfnamefont {Ramy}}, \bibinfo {author} {\bibfnamefont {Konstantinos~G}\
  \bibnamefont {Makris}}, \bibinfo {author} {\bibfnamefont {Mercedeh}\
  \bibnamefont {Khajavikhan}}, \bibinfo {author} {\bibfnamefont {Ziad~H}\
  \bibnamefont {Musslimani}}, \bibinfo {author} {\bibfnamefont {Stefan}\
  \bibnamefont {Rotter}}, and\ \bibinfo {author} {\bibfnamefont {Demetrios~N}\
  \bibnamefont {Christodoulides}}} (\bibinfo {year} {2018}{\natexlab{b}}),\
  \bibfield  {title} {\enquote {\bibinfo {title} {Non-hermitian physics and pt
  symmetry},}\ }\href {https://doi.org/10.1038/nphys4323} {\bibfield  {journal}
  {\bibinfo  {journal} {Nature Physics}\ }\textbf {\bibinfo {volume} {14}},\
  10.1038/nphys4323}\BibitemShut {NoStop}%
\bibitem [{\citenamefont {Eleuch}\ and\ \citenamefont
  {Rotter}(2016{\natexlab{a}})}]{pt674}%
  \BibitemOpen
  \bibfield  {author} {\bibinfo {author} {\bibnamefont {Eleuch}, \bibfnamefont
  {Hichem}}, and\ \bibinfo {author} {\bibfnamefont {Ingrid}\ \bibnamefont
  {Rotter}}} (\bibinfo {year} {2016}{\natexlab{a}}),\ \bibfield  {title}
  {\enquote {\bibinfo {title} {Clustering of exceptional points and dynamical
  phase transitions},}\ }\href {https://doi.org/10.1103/physreva.93.042116}
  {\bibfield  {journal} {\bibinfo  {journal} {Physical Review A}\ }\textbf
  {\bibinfo {volume} {93}},\ 10.1103/physreva.93.042116}\BibitemShut {NoStop}%
\bibitem [{\citenamefont {Eleuch}\ and\ \citenamefont
  {Rotter}(2016{\natexlab{b}})}]{pt675}%
  \BibitemOpen
  \bibfield  {author} {\bibinfo {author} {\bibnamefont {Eleuch}, \bibfnamefont
  {Hichem}}, and\ \bibinfo {author} {\bibfnamefont {Ingrid}\ \bibnamefont
  {Rotter}}} (\bibinfo {year} {2016}{\natexlab{b}}),\ \bibfield  {title}
  {\enquote {\bibinfo {title} {Clustering of exceptional points and dynamical
  phase transitions},}\ }\href {https://doi.org/10.1103/physreva.93.042116}
  {\bibfield  {journal} {\bibinfo  {journal} {Physical Review A}\ }\textbf
  {\bibinfo {volume} {93}},\ 10.1103/physreva.93.042116}\BibitemShut {NoStop}%
\bibitem [{\citenamefont {Eremenko}\ \emph {et~al.}(2008)\citenamefont
  {Eremenko}, \citenamefont {Gabrielov},\ and\ \citenamefont
  {Shapiro}}]{pt374}%
  \BibitemOpen
  \bibfield  {author} {\bibinfo {author} {\bibnamefont {Eremenko},
  \bibfnamefont {Alexandre}}, \bibinfo {author} {\bibfnamefont {Andrei}\
  \bibnamefont {Gabrielov}}, and\ \bibinfo {author} {\bibfnamefont {Boris}\
  \bibnamefont {Shapiro}}} (\bibinfo {year} {2008}),\ \bibfield  {title}
  {\enquote {\bibinfo {title} {High energy eigenfunctions of one-dimensional
  {S}chr{\"o}dinger operators with polynomial potentials},}\ }\href
  {https://doi.org/10.1007/BF03321702} {\bibfield  {journal} {\bibinfo
  {journal} {Computational Methods and Function Theory}\ }\textbf {\bibinfo
  {volume} {8}},\ 10.1007/BF03321702}\BibitemShut {NoStop}%
\bibitem [{\citenamefont {Felski}\ \emph {et~al.}(2021)\citenamefont {Felski},
  \citenamefont {Bender}, \citenamefont {Klevansky},\ and\ \citenamefont
  {Sarkar}}]{pt703}%
  \BibitemOpen
  \bibfield  {author} {\bibinfo {author} {\bibnamefont {Felski}, \bibfnamefont
  {Alexander}}, \bibinfo {author} {\bibfnamefont {Carl~M}\ \bibnamefont
  {Bender}}, \bibinfo {author} {\bibfnamefont {S~P}\ \bibnamefont {Klevansky}},
  and\ \bibinfo {author} {\bibfnamefont {Sarben}\ \bibnamefont {Sarkar}}}
  (\bibinfo {year} {2021}),\ \bibfield  {title} {\enquote {\bibinfo {title}
  {Towards perturbative renormalization of $\phi^2(i\phi)^\varepsilon$ quantum
  field theory},}\ }\href {https://doi.org/10.1103/physrevd.104.085011}
  {\bibfield  {journal} {\bibinfo  {journal} {Physical Review D}\ }\textbf
  {\bibinfo {volume} {104}},\ 10.1103/physrevd.104.085011}\BibitemShut
  {NoStop}%
\bibitem [{\citenamefont {Feng}\ \emph {et~al.}(2017)\citenamefont {Feng},
  \citenamefont {El-Ganainy},\ and\ \citenamefont {Ge}}]{pt556}%
  \BibitemOpen
  \bibfield  {author} {\bibinfo {author} {\bibnamefont {Feng}, \bibfnamefont
  {Liang}}, \bibinfo {author} {\bibfnamefont {Ramy}\ \bibnamefont
  {El-Ganainy}}, and\ \bibinfo {author} {\bibfnamefont {Li}~\bibnamefont {Ge}}}
  (\bibinfo {year} {2017}),\ \bibfield  {title} {\enquote {\bibinfo {title}
  {Non-{H}ermitian photonics based on parity-time symmetry},}\ }\href
  {https://doi.org/10.1038/s41566-017-0031-1} {\bibfield  {journal} {\bibinfo
  {journal} {Nature Photonics}\ }\textbf {\bibinfo {volume} {11}},\ \bibinfo
  {pages} {752--762}}\BibitemShut {NoStop}%
\bibitem [{\citenamefont {Feng}\ \emph {et~al.}(2014)\citenamefont {Feng},
  \citenamefont {Wong}, \citenamefont {Ma}, \citenamefont {Wang},\ and\
  \citenamefont {Zhang}}]{pt205}%
  \BibitemOpen
  \bibfield  {author} {\bibinfo {author} {\bibnamefont {Feng}, \bibfnamefont
  {Liang}}, \bibinfo {author} {\bibfnamefont {Zi~Jing}\ \bibnamefont {Wong}},
  \bibinfo {author} {\bibfnamefont {Ren-Min}\ \bibnamefont {Ma}}, \bibinfo
  {author} {\bibfnamefont {Yuan}\ \bibnamefont {Wang}}, and\ \bibinfo {author}
  {\bibfnamefont {Xiang}\ \bibnamefont {Zhang}}} (\bibinfo {year} {2014}),\
  \bibfield  {title} {\enquote {\bibinfo {title} {Single-mode laser by
  parity-time symmetry breaking},}\ }\href
  {https://doi.org/10.1126/science.1258479} {\bibfield  {journal} {\bibinfo
  {journal} {Science}\ }\textbf {\bibinfo {volume} {346}},\ \bibinfo {pages}
  {972--975}}\BibitemShut {NoStop}%
\bibitem [{\citenamefont {Fleury}\ \emph {et~al.}(2016)\citenamefont {Fleury},
  \citenamefont {Khanikaev},\ and\ \citenamefont {Al\`{u}}}]{pt551}%
  \BibitemOpen
  \bibfield  {author} {\bibinfo {author} {\bibnamefont {Fleury}, \bibfnamefont
  {Romain}}, \bibinfo {author} {\bibfnamefont {Alexander~B}\ \bibnamefont
  {Khanikaev}}, and\ \bibinfo {author} {\bibfnamefont {Andrea}\ \bibnamefont
  {Al\`{u}}}} (\bibinfo {year} {2016}),\ \bibfield  {title} {\enquote {\bibinfo
  {title} {Floquet topological insulators for sound},}\ }\href
  {https://doi.org/10.1038/ncomms11744} {\bibfield  {journal} {\bibinfo
  {journal} {Nature Communications}\ }\textbf {\bibinfo {volume} {7}},\
  \bibinfo {pages} {11744}}\BibitemShut {NoStop}%
\bibitem [{\citenamefont {Fleury}\ \emph {et~al.}(2015)\citenamefont {Fleury},
  \citenamefont {Monticone},\ and\ \citenamefont {Al\'{u}}}]{pt238}%
  \BibitemOpen
  \bibfield  {author} {\bibinfo {author} {\bibnamefont {Fleury}, \bibfnamefont
  {Romain}}, \bibinfo {author} {\bibfnamefont {Francesco}\ \bibnamefont
  {Monticone}}, and\ \bibinfo {author} {\bibfnamefont {Andrea}\ \bibnamefont
  {Al\'{u}}}} (\bibinfo {year} {2015}),\ \bibfield  {title} {\enquote {\bibinfo
  {title} {Invisibility and cloaking: Origins, present, and future
  perspectives},}\ }\href {https://doi.org/10.1103/PhysRevApplied.4.037001}
  {\bibfield  {journal} {\bibinfo  {journal} {Physical Review Applied}\
  }\textbf {\bibinfo {volume} {4}},\ \bibinfo {pages} {037001}}\BibitemShut
  {NoStop}%
\bibitem [{\citenamefont {Fleury}\ \emph
  {et~al.}(2014{\natexlab{a}})\citenamefont {Fleury}, \citenamefont {Sounas},\
  and\ \citenamefont {Al{\'u}}}]{pt243}%
  \BibitemOpen
  \bibfield  {author} {\bibinfo {author} {\bibnamefont {Fleury}, \bibfnamefont
  {Romain}}, \bibinfo {author} {\bibfnamefont {Dimitrios}\ \bibnamefont
  {Sounas}}, and\ \bibinfo {author} {\bibfnamefont {Andrea}\ \bibnamefont
  {Al{\'u}}}} (\bibinfo {year} {2014}{\natexlab{a}}),\ \bibfield  {title}
  {\enquote {\bibinfo {title} {An invisible acoustic sensor based on
  parity-time symmetry},}\ }\href {https://doi.org/10.1038/ncomms6905}
  {\bibfield  {journal} {\bibinfo  {journal} {Nature Communications}\ }\textbf
  {\bibinfo {volume} {6}},\ \bibinfo {pages} {5905}}\BibitemShut {NoStop}%
\bibitem [{\citenamefont {Fleury}\ \emph
  {et~al.}(2014{\natexlab{b}})\citenamefont {Fleury}, \citenamefont {Sounas},\
  and\ \citenamefont {Al\'{u}}}]{pt239}%
  \BibitemOpen
  \bibfield  {author} {\bibinfo {author} {\bibnamefont {Fleury}, \bibfnamefont
  {Romain}}, \bibinfo {author} {\bibfnamefont {Dimitrios~L}\ \bibnamefont
  {Sounas}}, and\ \bibinfo {author} {\bibfnamefont {Andrea}\ \bibnamefont
  {Al\'{u}}}} (\bibinfo {year} {2014}{\natexlab{b}}),\ \bibfield  {title}
  {\enquote {\bibinfo {title} {Negative refraction and planar focusing based on
  parity-time symmetric metasurfaces},}\ }\href
  {https://doi.org/10.1103/PhysRevLett.113.023903} {\bibfield  {journal}
  {\bibinfo  {journal} {Physical Review Letters}\ }\textbf {\bibinfo {volume}
  {113}},\ \bibinfo {pages} {023903}}\BibitemShut {NoStop}%
\bibitem [{\citenamefont {Fring}(2007)}]{pt47}%
  \BibitemOpen
  \bibfield  {author} {\bibinfo {author} {\bibnamefont {Fring}, \bibfnamefont
  {A}}} (\bibinfo {year} {2007}),\ \bibfield  {title} {\enquote {\bibinfo
  {title} {$\cal{PT}$-symmetric deformations of the {K}orteweg-de {V}ries
  equation},}\ }\href {https://doi.org/10.1088/1751-8113/40/15/012} {\bibfield
  {journal} {\bibinfo  {journal} {Journal of Physics A: Mathematical and
  Theoretical}\ }\textbf {\bibinfo {volume} {40}},\ \bibinfo {pages}
  {4215--4224}}\BibitemShut {NoStop}%
\bibitem [{\citenamefont {Fring}(2013)}]{pt59}%
  \BibitemOpen
  \bibfield  {author} {\bibinfo {author} {\bibnamefont {Fring}, \bibfnamefont
  {A}}} (\bibinfo {year} {2013}),\ \bibfield  {title} {\enquote {\bibinfo
  {title} {$\cal{PT}$-symmetric deformations of integrable models},}\ }\href
  {https://doi.org/10.1098/rsta.2012.0046} {\bibfield  {journal} {\bibinfo
  {journal} {Philosophical Transactions of the Royal Society of London A:
  Mathematical, Physical and Engineering Sciences}\ }\textbf {\bibinfo {volume}
  {371}}~(\bibinfo {number} {1989}),\ \bibinfo {pages} {20120046}}\BibitemShut
  {NoStop}%
\bibitem [{\citenamefont {Fring}\ and\ \citenamefont {Znojil}(2008)}]{pt51}%
  \BibitemOpen
  \bibfield  {author} {\bibinfo {author} {\bibnamefont {Fring}, \bibfnamefont
  {A}}, and\ \bibinfo {author} {\bibfnamefont {M}~\bibnamefont {Znojil}}}
  (\bibinfo {year} {2008}),\ \bibfield  {title} {\enquote {\bibinfo {title}
  {$\cal{PT}$-symmetric deformations of {C}alogero models},}\ }\href
  {https://doi.org/10.1088/1751-8113/41/19/194010} {\bibfield  {journal}
  {\bibinfo  {journal} {Journal of Physics A: Mathematical and Theoretical}\
  }\textbf {\bibinfo {volume} {41}},\ \bibinfo {pages} {194010}}\BibitemShut
  {NoStop}%
\bibitem [{\citenamefont {Fring}\ and\ \citenamefont {Taira}(2020)}]{pt651}%
  \BibitemOpen
  \bibfield  {author} {\bibinfo {author} {\bibnamefont {Fring}, \bibfnamefont
  {Andreas}}, and\ \bibinfo {author} {\bibfnamefont {Takanobu}\ \bibnamefont
  {Taira}}} (\bibinfo {year} {2020}),\ \bibfield  {title} {\enquote {\bibinfo
  {title} {'t hooft-polyakov monopoles in non-hermitian quantum field
  theory},}\ }\href {https://doi.org/10.1016/j.physletb.2020.135583} {\bibfield
   {journal} {\bibinfo  {journal} {Physics Letters B}\ }\textbf {\bibinfo
  {volume} {807}},\ 10.1016/j.physletb.2020.135583}\BibitemShut {NoStop}%
\bibitem [{\citenamefont {Fu}\ and\ \citenamefont {Qin}(2020)}]{pt618}%
  \BibitemOpen
  \bibfield  {author} {\bibinfo {author} {\bibnamefont {Fu}, \bibfnamefont
  {Yichen}}, and\ \bibinfo {author} {\bibfnamefont {Hong}\ \bibnamefont {Qin}}}
  (\bibinfo {year} {2020}),\ \bibfield  {title} {\enquote {\bibinfo {title}
  {The physics of spontaneous parity-time symmetry breaking in the
  kelvin-helmholtz instability},}\ }\href
  {https://doi.org/10.1088/1367-2630/aba38f} {\bibfield  {journal} {\bibinfo
  {journal} {New Journal of Physics}\ }\textbf {\bibinfo {volume} {22}},\
  10.1088/1367-2630/aba38f}\BibitemShut {NoStop}%
\bibitem [{\citenamefont {Galda}\ and\ \citenamefont {Vinokur}(2016)}]{pt692}%
  \BibitemOpen
  \bibfield  {author} {\bibinfo {author} {\bibnamefont {Galda}, \bibfnamefont
  {Alexey}}, and\ \bibinfo {author} {\bibfnamefont {Valerii~M}\ \bibnamefont
  {Vinokur}}} (\bibinfo {year} {2016}),\ \bibfield  {title} {\enquote {\bibinfo
  {title} {Parity-time symmetry breaking in magnetic systems},}\ }\href
  {https://doi.org/10.1103/physrevb.94.020408} {\bibfield  {journal} {\bibinfo
  {journal} {Physical Review B}\ }\textbf {\bibinfo {volume} {94}},\
  10.1103/physrevb.94.020408}\BibitemShut {NoStop}%
\bibitem [{\citenamefont {Gelfand}\ \emph {et~al.}(1963)\citenamefont
  {Gelfand}, \citenamefont {Minlos},\ and\ \citenamefont {Shapiro}}]{pt599}%
  \BibitemOpen
  \bibfield  {author} {\bibinfo {author} {\bibnamefont {Gelfand}, \bibfnamefont
  {I~M}}, \bibinfo {author} {\bibfnamefont {R~A}\ \bibnamefont {Minlos}}, and\
  \bibinfo {author} {\bibfnamefont {Z~Ya}\ \bibnamefont {Shapiro}}} (\bibinfo
  {year} {1963}),\ \href@noop {} {\emph {\bibinfo {title} {Representations of
  the Rotation and Lorentz Groups and Their Applications}}}\ (\bibinfo
  {publisher} {Pergamon Press},\ \bibinfo {address} {Oxford})\BibitemShut
  {NoStop}%
\bibitem [{\citenamefont {Gelfand}\ and\ \citenamefont
  {Naimark}(1962)}]{pt600}%
  \BibitemOpen
  \bibfield  {author} {\bibinfo {author} {\bibnamefont {Gelfand}, \bibfnamefont
  {I~M}}, and\ \bibinfo {author} {\bibfnamefont {M~A}\ \bibnamefont {Naimark}}}
  (\bibinfo {year} {1962}),\ \href@noop {} {\emph {\bibinfo {title} {Lie
  Groups}}}\ (\bibinfo  {publisher} {American Mathematical Society},\ \bibinfo
  {address} {Rhode Island})\BibitemShut {NoStop}%
\bibitem [{\citenamefont {Glimm}\ and\ \citenamefont {Jaffe}(1968)}]{pt598}%
  \BibitemOpen
  \bibfield  {author} {\bibinfo {author} {\bibnamefont {Glimm}, \bibfnamefont
  {James}}, and\ \bibinfo {author} {\bibfnamefont {Arthur}\ \bibnamefont
  {Jaffe}}} (\bibinfo {year} {1968}),\ \bibfield  {title} {\enquote {\bibinfo
  {title} {A $\lambda \phi^4$ quantum field theory without cutoffs. i},}\
  }\href {https://doi.org/10.1103/PhysRev.176.1945} {\bibfield  {journal}
  {\bibinfo  {journal} {Physical Review}\ }\textbf {\bibinfo {volume} {176}},\
  10.1103/PhysRev.176.1945}\BibitemShut {NoStop}%
\bibitem [{\citenamefont {Goldzak}\ \emph {et~al.}(2018)\citenamefont
  {Goldzak}, \citenamefont {Mailybaev},\ and\ \citenamefont
  {Moiseyev}}]{pt568}%
  \BibitemOpen
  \bibfield  {author} {\bibinfo {author} {\bibnamefont {Goldzak}, \bibfnamefont
  {Tamar}}, \bibinfo {author} {\bibfnamefont {Alexei~A}\ \bibnamefont
  {Mailybaev}}, and\ \bibinfo {author} {\bibfnamefont {Nimrod}\ \bibnamefont
  {Moiseyev}}} (\bibinfo {year} {2018}),\ \bibfield  {title} {\enquote
  {\bibinfo {title} {Light stops at exceptional points},}\ }\href
  {https://doi.org/10.1103/PhysRevLett.120.013901} {\bibfield  {journal}
  {\bibinfo  {journal} {Physical Review Letters}\ }\textbf {\bibinfo {volume}
  {120}},\ \bibinfo {pages} {013901}}\BibitemShut {NoStop}%
\bibitem [{\citenamefont {Graefe}\ and\ \citenamefont {Jones}(2011)}]{pt681}%
  \BibitemOpen
  \bibfield  {author} {\bibinfo {author} {\bibnamefont {Graefe}, \bibfnamefont
  {Eva-Maria}}, and\ \bibinfo {author} {\bibfnamefont {H~F}\ \bibnamefont
  {Jones}}} (\bibinfo {year} {2011}),\ \bibfield  {title} {\enquote {\bibinfo
  {title} {Pt-symmetric sinusoidal optical lattices at the symmetry-breaking
  threshold},}\ }\href {https://doi.org/10.1103/physreva.84.013818} {\bibfield
  {journal} {\bibinfo  {journal} {Physical Review A}\ }\textbf {\bibinfo
  {volume} {84}},\ 10.1103/physreva.84.013818}\BibitemShut {NoStop}%
\bibitem [{\citenamefont {G{\"u}nther}\ \emph {et~al.}(2007)\citenamefont
  {G{\"u}nther}, \citenamefont {Rotter},\ and\ \citenamefont
  {Samsonov}}]{pt673}%
  \BibitemOpen
  \bibfield  {author} {\bibinfo {author} {\bibnamefont {G{\"u}nther},
  \bibfnamefont {Uwe}}, \bibinfo {author} {\bibfnamefont {Ingrid}\ \bibnamefont
  {Rotter}}, and\ \bibinfo {author} {\bibfnamefont {Boris~F}\ \bibnamefont
  {Samsonov}}} (\bibinfo {year} {2007}),\ \bibfield  {title} {\enquote
  {\bibinfo {title} {Projective hilbert space structures at exceptional
  points},}\ }\href {https://doi.org/10.1088/1751-8113/40/30/014} {\bibfield
  {journal} {\bibinfo  {journal} {Journal of Physics A: Mathematical and
  Theoretical}\ }\textbf {\bibinfo {volume} {40}},\
  10.1088/1751-8113/40/30/014}\BibitemShut {NoStop}%
\bibitem [{\citenamefont {Guo}\ \emph {et~al.}(2009)\citenamefont {Guo},
  \citenamefont {Salamo}, \citenamefont {Duchesne}, \citenamefont {Morandotti},
  \citenamefont {Volatier-Ravat}, \citenamefont {Aimez}, \citenamefont
  {Siviloglou},\ and\ \citenamefont {Christodoulides}}]{pt186}%
  \BibitemOpen
  \bibfield  {author} {\bibinfo {author} {\bibnamefont {Guo}, \bibfnamefont
  {A}}, \bibinfo {author} {\bibfnamefont {G~J}\ \bibnamefont {Salamo}},
  \bibinfo {author} {\bibfnamefont {D}~\bibnamefont {Duchesne}}, \bibinfo
  {author} {\bibfnamefont {R}~\bibnamefont {Morandotti}}, \bibinfo {author}
  {\bibfnamefont {M}~\bibnamefont {Volatier-Ravat}}, \bibinfo {author}
  {\bibfnamefont {V}~\bibnamefont {Aimez}}, \bibinfo {author} {\bibfnamefont
  {G~A}\ \bibnamefont {Siviloglou}}, and\ \bibinfo {author} {\bibfnamefont
  {D~N}\ \bibnamefont {Christodoulides}}} (\bibinfo {year} {2009}),\ \bibfield
  {title} {\enquote {\bibinfo {title} {Observation of $\cal{PT}$-symmetry
  breaking in complex optical potentials},}\ }\href
  {https://doi.org/10.1103/PhysRevLett.103.093902} {\bibfield  {journal}
  {\bibinfo  {journal} {Physical Review Letters}\ }\textbf {\bibinfo {volume}
  {103}},\ \bibinfo {pages} {093902}}\BibitemShut {NoStop}%
\bibitem [{\citenamefont {Heiss}(2012)}]{pt671}%
  \BibitemOpen
  \bibfield  {author} {\bibinfo {author} {\bibnamefont {Heiss}, \bibfnamefont
  {W~D}}} (\bibinfo {year} {2012}),\ \bibfield  {title} {\enquote {\bibinfo
  {title} {The physics of exceptional points},}\ }\href
  {https://doi.org/10.1088/1751-8113/45/44/444016} {\bibfield  {journal}
  {\bibinfo  {journal} {Journal of Physics A: Mathematical and Theoretical}\
  }\textbf {\bibinfo {volume} {45}},\
  10.1088/1751-8113/45/44/444016}\BibitemShut {NoStop}%
\bibitem [{\citenamefont {Heiss}\ \emph {et~al.}(2013)\citenamefont {Heiss},
  \citenamefont {Cartarius}, \citenamefont {Wunner},\ and\ \citenamefont
  {Main}}]{pt670}%
  \BibitemOpen
  \bibfield  {author} {\bibinfo {author} {\bibnamefont {Heiss}, \bibfnamefont
  {W~D}}, \bibinfo {author} {\bibfnamefont {H}~\bibnamefont {Cartarius}},
  \bibinfo {author} {\bibfnamefont {G}~\bibnamefont {Wunner}}, and\ \bibinfo
  {author} {\bibfnamefont {J}~\bibnamefont {Main}}} (\bibinfo {year} {2013}),\
  \bibfield  {title} {\enquote {\bibinfo {title} {Spectral singularities in
  $pt$-symmetric bose-einstein condensates},}\ }\href
  {https://doi.org/10.1088/1751-8113/46/27/275307} {\bibfield  {journal}
  {\bibinfo  {journal} {Journal of Physics A: Mathematical and Theoretical}\
  }\textbf {\bibinfo {volume} {46}},\
  10.1088/1751-8113/46/27/275307}\BibitemShut {NoStop}%
\bibitem [{\citenamefont {Hezaro}(2008)}]{pt375}%
  \BibitemOpen
  \bibfield  {author} {\bibinfo {author} {\bibnamefont {Hezaro}, \bibfnamefont
  {Hamid}}} (\bibinfo {year} {2008}),\ \bibfield  {title} {\enquote {\bibinfo
  {title} {Complex zeros of eigenfunctions of 1{D} {S}chr{\"o}dinger
  operators},}\ }\href {https://doi.org/10.1007/10.1093/imrn/rnm148} {\bibfield
   {journal} {\bibinfo  {journal} {International Mathematics Research Notices}\
  }\textbf {\bibinfo {volume} {2008}},\
  10.1007/10.1093/imrn/rnm148}\BibitemShut {NoStop}%
\bibitem [{\citenamefont {Hodaei}\ \emph
  {et~al.}(2017{\natexlab{a}})\citenamefont {Hodaei}, \citenamefont {Hassan},
  \citenamefont {Garcia-Gracia}, \citenamefont {Hayenga}, \citenamefont
  {Christodoulides},\ and\ \citenamefont {Khajavikhan}}]{pt511}%
  \BibitemOpen
  \bibfield  {author} {\bibinfo {author} {\bibnamefont {Hodaei}, \bibfnamefont
  {H}}, \bibinfo {author} {\bibfnamefont {A~U}\ \bibnamefont {Hassan}},
  \bibinfo {author} {\bibfnamefont {H}~\bibnamefont {Garcia-Gracia}}, \bibinfo
  {author} {\bibfnamefont {W~E}\ \bibnamefont {Hayenga}}, \bibinfo {author}
  {\bibfnamefont {D~N}\ \bibnamefont {Christodoulides}}, and\ \bibinfo {author}
  {\bibfnamefont {M}~\bibnamefont {Khajavikhan}}} (\bibinfo {year}
  {2017}{\natexlab{a}}),\ \bibfield  {title} {\enquote {\bibinfo {title}
  {Enhanced sensitivity in $\mathcal{PT}$-symmetric coupled resonators},}\ }in\
  \href {https://doi.org/10.1117/12.2250047} {\emph {\bibinfo {booktitle}
  {Proceedings of {SPIE}: {L}aser {R}esonators, {M}icroresonators, and {B}eam
  {C}ontrol {XIX}}}},\ Vol.\ \bibinfo {volume} {10090},\ \bibinfo {editor}
  {edited by\ \bibinfo {editor} {\bibfnamefont {Alexis~V}\ \bibnamefont
  {Kudryashov}}, \bibinfo {editor} {\bibfnamefont {Alan~H}\ \bibnamefont
  {Paxton}}, \ and\ \bibinfo {editor} {\bibfnamefont {Vladimir~S}\ \bibnamefont
  {Ilchenko}}},\ p.\ \bibinfo {pages} {6 pages}\BibitemShut {NoStop}%
\bibitem [{\citenamefont {Hodaei}\ \emph
  {et~al.}(2017{\natexlab{b}})\citenamefont {Hodaei}, \citenamefont {Hassan},
  \citenamefont {Wittek}, \citenamefont {Garcia-Gracia}, \citenamefont
  {El-Ganainy}, \citenamefont {Christodoulides},\ and\ \citenamefont
  {Khajavikhan}}]{pt706}%
  \BibitemOpen
  \bibfield  {author} {\bibinfo {author} {\bibnamefont {Hodaei}, \bibfnamefont
  {Hossein}}, \bibinfo {author} {\bibfnamefont {Absar~U}\ \bibnamefont
  {Hassan}}, \bibinfo {author} {\bibfnamefont {Steffen}\ \bibnamefont
  {Wittek}}, \bibinfo {author} {\bibfnamefont {Hipolito}\ \bibnamefont
  {Garcia-Gracia}}, \bibinfo {author} {\bibfnamefont {Ramy}\ \bibnamefont
  {El-Ganainy}}, \bibinfo {author} {\bibfnamefont {Demetrios~N}\ \bibnamefont
  {Christodoulides}}, and\ \bibinfo {author} {\bibfnamefont {Mercedeh}\
  \bibnamefont {Khajavikhan}}} (\bibinfo {year} {2017}{\natexlab{b}}),\
  \bibfield  {title} {\enquote {\bibinfo {title} {Enhanced sensitivity at
  higher-order exceptional points},}\ }\href
  {https://doi.org/10.1038/nature23280} {\bibfield  {journal} {\bibinfo
  {journal} {Nature}\ }\textbf {\bibinfo {volume} {548}},\
  10.1038/nature23280}\BibitemShut {NoStop}%
\bibitem [{\citenamefont {Hodaei}\ \emph
  {et~al.}(2014{\natexlab{a}})\citenamefont {Hodaei}, \citenamefont {Miri},
  \citenamefont {Heinrich}, \citenamefont {Christodoulides},\ and\
  \citenamefont {Khajavikhan}}]{pt209}%
  \BibitemOpen
  \bibfield  {author} {\bibinfo {author} {\bibnamefont {Hodaei}, \bibfnamefont
  {Hossein}}, \bibinfo {author} {\bibfnamefont {Mohammad-Ali}\ \bibnamefont
  {Miri}}, \bibinfo {author} {\bibfnamefont {Matthias}\ \bibnamefont
  {Heinrich}}, \bibinfo {author} {\bibfnamefont {Demetrios~N}\ \bibnamefont
  {Christodoulides}}, and\ \bibinfo {author} {\bibfnamefont {Mercedeh}\
  \bibnamefont {Khajavikhan}}} (\bibinfo {year} {2014}{\natexlab{a}}),\
  \bibfield  {title} {\enquote {\bibinfo {title} {Parity-time-symmetric
  microring lasers},}\ }\href {https://doi.org/10.1126/science.1258480}
  {\bibfield  {journal} {\bibinfo  {journal} {Science}\ }\textbf {\bibinfo
  {volume} {346}},\ \bibinfo {pages} {975--978}}\BibitemShut {NoStop}%
\bibitem [{\citenamefont {Hodaei}\ \emph
  {et~al.}(2014{\natexlab{b}})\citenamefont {Hodaei}, \citenamefont {Miri},
  \citenamefont {Heinrich}, \citenamefont {Chistodoulides},\ and\ \citenamefont
  {Khajavikhan}}]{pt210}%
  \BibitemOpen
  \bibfield  {author} {\bibinfo {author} {\bibnamefont {Hodaei}, \bibfnamefont
  {Hossein}}, \bibinfo {author} {\bibfnamefont {Mohammad-Ali}\ \bibnamefont
  {Miri}}, \bibinfo {author} {\bibfnamefont {Mattias}\ \bibnamefont
  {Heinrich}}, \bibinfo {author} {\bibfnamefont {Demetrios~N}\ \bibnamefont
  {Chistodoulides}}, and\ \bibinfo {author} {\bibfnamefont {Mercedeh}\
  \bibnamefont {Khajavikhan}}} (\bibinfo {year} {2014}{\natexlab{b}}),\
  \bibfield  {title} {\enquote {\bibinfo {title} {$\mathcal{PT}$-symmetric
  microring lasers: {S}elf-adapting broadband mode-selective resonators},}\
  }\href {http://arxiv.org/abs/1405.2103} {\bibinfo  {journal} {ArXiv}\
  }\BibitemShut {NoStop}%
\bibitem [{\citenamefont {Hook}\ \emph {et~al.}(2018)\citenamefont {Hook},
  \citenamefont {Porter},\ and\ \citenamefont {Herzog}}]{pt578}%
  \BibitemOpen
\bibfield  {journal} {  }\bibfield  {author} {\bibinfo {author} {\bibnamefont
  {Hook}, \bibfnamefont {Daniel~W}}, \bibinfo {author} {\bibfnamefont
  {Simon~J}\ \bibnamefont {Porter}}, and\ \bibinfo {author} {\bibfnamefont
  {Christian}\ \bibnamefont {Herzog}}} (\bibinfo {year} {2018}),\ \bibfield
  {title} {\enquote {\bibinfo {title} {Dimensions: Building context for search
  and evaluation},}\ }\href {https://doi.org/10.3389/frma.2018.00023}
  {\bibfield  {journal} {\bibinfo  {journal} {Frontiers in Resesearch Metrics
  and Analytics}\ }\textbf {\bibinfo {volume} {3}},\
  10.3389/frma.2018.00023}\BibitemShut {NoStop}%
\bibitem [{\citenamefont {Hu}\ \emph {et~al.}(2011)\citenamefont {Hu},
  \citenamefont {Ma}, \citenamefont {Lu}, \citenamefont {Yang}, \citenamefont
  {Zheng},\ and\ \citenamefont {Hu}}]{pt229}%
  \BibitemOpen
  \bibfield  {author} {\bibinfo {author} {\bibnamefont {Hu}, \bibfnamefont
  {Sumei}}, \bibinfo {author} {\bibfnamefont {Xuekai}\ \bibnamefont {Ma}},
  \bibinfo {author} {\bibfnamefont {Daquan}\ \bibnamefont {Lu}}, \bibinfo
  {author} {\bibfnamefont {Zhenjun}\ \bibnamefont {Yang}}, \bibinfo {author}
  {\bibfnamefont {Yizhou}\ \bibnamefont {Zheng}}, and\ \bibinfo {author}
  {\bibfnamefont {Wei}\ \bibnamefont {Hu}}} (\bibinfo {year} {2011}),\
  \bibfield  {title} {\enquote {\bibinfo {title} {Solitons supported by complex
  $\cal{PT}$-symmetric gaussian potentials},}\ }\href
  {https://doi.org/10.1103/PhysRevA.84.043818} {\bibfield  {journal} {\bibinfo
  {journal} {Physical Review A}\ }\textbf {\bibinfo {volume} {84}},\ \bibinfo
  {pages} {043818}}\BibitemShut {NoStop}%
\bibitem [{\citenamefont {Hurst}(1952)}]{pt588}%
  \BibitemOpen
  \bibfield  {author} {\bibinfo {author} {\bibnamefont {Hurst}, \bibfnamefont
  {C~A}}} (\bibinfo {year} {1952}),\ \bibfield  {title} {\enquote {\bibinfo
  {title} {An example of a divergent perturbation expansion in field theory},}\
  }\href {https://doi.org/10.1017/S0305004100076416} {\bibfield  {journal}
  {\bibinfo  {journal} {Mathematical Proceedings of the Cambridge Philosophical
  Society}\ }\textbf {\bibinfo {volume} {48}},\
  10.1017/S0305004100076416}\BibitemShut {NoStop}%
\bibitem [{\citenamefont {Itzykson}\ \emph {et~al.}(1977)\citenamefont
  {Itzykson}, \citenamefont {Parisi},\ and\ \citenamefont {Zuber}}]{pt601}%
  \BibitemOpen
  \bibfield  {author} {\bibinfo {author} {\bibnamefont {Itzykson},
  \bibfnamefont {C}}, \bibinfo {author} {\bibfnamefont {G}~\bibnamefont
  {Parisi}}, and\ \bibinfo {author} {\bibfnamefont {J-B}\ \bibnamefont
  {Zuber}}} (\bibinfo {year} {1977}),\ \bibfield  {title} {\enquote {\bibinfo
  {title} {Asymptotic estimates in quantum electrodynamics},}\ }\href
  {https://doi.org/10.1103/PhysRevD.16.996} {\bibfield  {journal} {\bibinfo
  {journal} {Physics Review D}\ }\textbf {\bibinfo {volume} {16}},\
  10.1103/PhysRevD.16.996}\BibitemShut {NoStop}%
\bibitem [{\citenamefont {Jahromi}\ \emph {et~al.}(2017)\citenamefont
  {Jahromi}, \citenamefont {Hassan}, \citenamefont {Christodoulides},\ and\
  \citenamefont {Abouraddy}}]{pt554}%
  \BibitemOpen
  \bibfield  {author} {\bibinfo {author} {\bibnamefont {Jahromi}, \bibfnamefont
  {Ali~K}}, \bibinfo {author} {\bibfnamefont {Absar~U}\ \bibnamefont {Hassan}},
  \bibinfo {author} {\bibfnamefont {Demetrios~N}\ \bibnamefont
  {Christodoulides}}, and\ \bibinfo {author} {\bibfnamefont {Ayman~F}\
  \bibnamefont {Abouraddy}}} (\bibinfo {year} {2017}),\ \bibfield  {title}
  {\enquote {\bibinfo {title} {Statistical parity-time-symmetric lasing in an
  optical fibre network},}\ }\href {https://doi.org/10.1038/s41467-017-00958-x}
  {\bibfield  {journal} {\bibinfo  {journal} {Nature Communications}\ }\textbf
  {\bibinfo {volume} {8}},\ \bibinfo {pages} {1359}}\BibitemShut {NoStop}%
\bibitem [{\citenamefont {Jing}\ \emph {et~al.}(2015)\citenamefont {Jing},
  \citenamefont {\c{S}ahin K~\"{O}zdemir}, \citenamefont {Geng}, \citenamefont
  {Zhang}, \citenamefont {L\"u}, \citenamefont {Peng}, \citenamefont {Yang},\
  and\ \citenamefont {Nori}}]{pt701}%
  \BibitemOpen
  \bibfield  {author} {\bibinfo {author} {\bibnamefont {Jing}, \bibfnamefont
  {H}}, \bibinfo {author} {\bibnamefont {\c{S}ahin K~\"{O}zdemir}}, \bibinfo
  {author} {\bibfnamefont {Z}~\bibnamefont {Geng}}, \bibinfo {author}
  {\bibfnamefont {Jing}\ \bibnamefont {Zhang}}, \bibinfo {author}
  {\bibfnamefont {Xin-You}\ \bibnamefont {L\"u}}, \bibinfo {author}
  {\bibfnamefont {Bo}~\bibnamefont {Peng}}, \bibinfo {author} {\bibfnamefont
  {Lan}\ \bibnamefont {Yang}}, and\ \bibinfo {author} {\bibfnamefont {Franco}\
  \bibnamefont {Nori}}} (\bibinfo {year} {2015}),\ \bibfield  {title} {\enquote
  {\bibinfo {title} {Optomechanically-induced transparency in
  parity-time-symmetric microresonators},}\ }\href
  {https://doi.org/10.1038/srep09663} {\bibfield  {journal} {\bibinfo
  {journal} {Scientific Reports}\ }\textbf {\bibinfo {volume} {5}},\
  10.1038/srep09663}\BibitemShut {NoStop}%
\bibitem [{\citenamefont {Jing}\ \emph {et~al.}(2017)\citenamefont {Jing},
  \citenamefont {\c{S} K~\"{O}zdemir}, \citenamefont {L\"{u}},\ and\
  \citenamefont {Nori}}]{pt700}%
  \BibitemOpen
  \bibfield  {author} {\bibinfo {author} {\bibnamefont {Jing}, \bibfnamefont
  {H}}, \bibinfo {author} {\bibnamefont {\c{S} K~\"{O}zdemir}}, \bibinfo
  {author} {\bibfnamefont {H}~\bibnamefont {L\"{u}}}, and\ \bibinfo {author}
  {\bibfnamefont {Franco}\ \bibnamefont {Nori}}} (\bibinfo {year} {2017}),\
  \bibfield  {title} {\enquote {\bibinfo {title} {High-order exceptional points
  in optomechanics},}\ }\href {https://doi.org/10.1038/s41598-017-03546-7}
  {\bibfield  {journal} {\bibinfo  {journal} {Scientific Reports}\ }\textbf
  {\bibinfo {volume} {7}},\ 10.1038/s41598-017-03546-7}\BibitemShut {NoStop}%
\bibitem [{\citenamefont {Jones}(2005)}]{pt392}%
  \BibitemOpen
  \bibfield  {author} {\bibinfo {author} {\bibnamefont {Jones}, \bibfnamefont
  {H~F}}} (\bibinfo {year} {2005}),\ \bibfield  {title} {\enquote {\bibinfo
  {title} {On pseudo-{H}ermitian {H}amiltonians and their {H}ermitian
  counterparts},}\ }\href {https://doi.org/10.1088/0305-4470/38/8/010}
  {\bibfield  {journal} {\bibinfo  {journal} {Journal of Physics A:
  Mathematical and General}\ }\textbf {\bibinfo {volume} {38}},\
  10.1088/0305-4470/38/8/010}\BibitemShut {NoStop}%
\bibitem [{\citenamefont {Jones}\ and\ \citenamefont {Mateo}(2006)}]{pt167}%
  \BibitemOpen
  \bibfield  {author} {\bibinfo {author} {\bibnamefont {Jones}, \bibfnamefont
  {Hugh~F}}, and\ \bibinfo {author} {\bibfnamefont {J}~\bibnamefont {Mateo}}}
  (\bibinfo {year} {2006}),\ \bibfield  {title} {\enquote {\bibinfo {title}
  {Equivalent {H}ermitian {H}amiltonian for the non-{H}ermitian $-x^4$
  potential},}\ }\href {https://doi.org/10.1103/PhysRevD.73.085002} {\bibfield
  {journal} {\bibinfo  {journal} {Physical Review D}\ }\textbf {\bibinfo
  {volume} {73}},\ \bibinfo {pages} {085002}}\BibitemShut {NoStop}%
\bibitem [{\citenamefont {Jones-Smith}\ and\ \citenamefont
  {Mathur}(2014)}]{pt563}%
  \BibitemOpen
  \bibfield  {author} {\bibinfo {author} {\bibnamefont {Jones-Smith},
  \bibfnamefont {Katherine}}, and\ \bibinfo {author} {\bibfnamefont {Harsh}\
  \bibnamefont {Mathur}}} (\bibinfo {year} {2014}),\ \bibfield  {title}
  {\enquote {\bibinfo {title} {Relativistic non-{H}ermitian quantum
  mechanics},}\ }\href {https://doi.org/10.1103/physrevd.89.125014} {\bibfield
  {journal} {\bibinfo  {journal} {Physical Review D}\ }\textbf {\bibinfo
  {volume} {89}},\ \bibinfo {pages} {125014}}\BibitemShut {NoStop}%
\bibitem [{\citenamefont {Jones-Smith}\ and\ \citenamefont
  {Mathur}(2016)}]{pt564}%
  \BibitemOpen
  \bibfield  {author} {\bibinfo {author} {\bibnamefont {Jones-Smith},
  \bibfnamefont {Katherine}}, and\ \bibinfo {author} {\bibfnamefont {Harsh}\
  \bibnamefont {Mathur}}} (\bibinfo {year} {2016}),\ \bibfield  {title}
  {\enquote {\bibinfo {title} {Non-{H}ermitian neutrino oscillations in matter
  with $\mathcal{PT}$ symmetric {H}amiltonians},}\ }\href
  {https://doi.org/10.1209/0295-5075/113/61001} {\bibfield  {journal} {\bibinfo
   {journal} {Europhysics Letters}\ }\textbf {\bibinfo {volume} {113}},\
  \bibinfo {pages} {61001}}\BibitemShut {NoStop}%
\bibitem [{\citenamefont {Kim}\ \emph {et~al.}(2016)\citenamefont {Kim},
  \citenamefont {Hwang}, \citenamefont {Kim}, \citenamefont {Choi},
  \citenamefont {No},\ and\ \citenamefont {Park}}]{pt545}%
  \BibitemOpen
  \bibfield  {author} {\bibinfo {author} {\bibnamefont {Kim}, \bibfnamefont
  {Kyoung-Ho}}, \bibinfo {author} {\bibfnamefont {Min-Soo}\ \bibnamefont
  {Hwang}}, \bibinfo {author} {\bibfnamefont {Ha-Reem}\ \bibnamefont {Kim}},
  \bibinfo {author} {\bibfnamefont {Jae-Hyuck}\ \bibnamefont {Choi}}, \bibinfo
  {author} {\bibfnamefont {You-Shin}\ \bibnamefont {No}}, and\ \bibinfo
  {author} {\bibfnamefont {Hong-Gyu}\ \bibnamefont {Park}}} (\bibinfo {year}
  {2016}),\ \bibfield  {title} {\enquote {\bibinfo {title} {Direct observation
  of exceptional points in coupled photonic-crystal lasers with asymmetric
  optical gains},}\ }\href {https://doi.org/10.1038/ncomms13893} {\bibfield
  {journal} {\bibinfo  {journal} {Nature Communications}\ }\textbf {\bibinfo
  {volume} {7}},\ \bibinfo {pages} {13893}}\BibitemShut {NoStop}%
\bibitem [{\citenamefont {Kottos}(2010)}]{pt702}%
  \BibitemOpen
  \bibfield  {author} {\bibinfo {author} {\bibnamefont {Kottos}, \bibfnamefont
  {Tsampikos}}} (\bibinfo {year} {2010}),\ \bibfield  {title} {\enquote
  {\bibinfo {title} {Broken symmetry makes light work},}\ }\href
  {https://doi.org/10.1038/nphys1612} {\bibfield  {journal} {\bibinfo
  {journal} {Nature Physics}\ }\textbf {\bibinfo {volume} {6}},\
  10.1038/nphys1612}\BibitemShut {NoStop}%
\bibitem [{\citenamefont {Krej{\v{c}}i{\v{r}}{\'{i}}k}\ \emph
  {et~al.}(2006)\citenamefont {Krej{\v{c}}i{\v{r}}{\'{i}}k}, \citenamefont
  {B{\'{i}}la},\ and\ \citenamefont {Znojil}}]{pt677}%
  \BibitemOpen
  \bibfield  {author} {\bibinfo {author} {\bibnamefont
  {Krej{\v{c}}i{\v{r}}{\'{i}}k}, \bibfnamefont {D}}, \bibinfo {author}
  {\bibfnamefont {H}~\bibnamefont {B{\'{i}}la}}, and\ \bibinfo {author}
  {\bibfnamefont {M}~\bibnamefont {Znojil}}} (\bibinfo {year} {2006}),\
  \bibfield  {title} {\enquote {\bibinfo {title} {Closed formula for the metric
  in the hilbert space of a $\mathcal{PT}$-symmetric model},}\ }\href
  {https://doi.org/10.1088/0305-4470/39/32/s15} {\bibfield  {journal} {\bibinfo
   {journal} {Journal of Physics A: Mathematical and Theoretical}\ }\textbf
  {\bibinfo {volume} {39}},\ 10.1088/0305-4470/39/32/s15}\BibitemShut {NoStop}%
\bibitem [{\citenamefont {Lee}\ and\ \citenamefont {Yang}(1956)}]{pt580}%
  \BibitemOpen
  \bibfield  {author} {\bibinfo {author} {\bibnamefont {Lee}, \bibfnamefont
  {T~D}}, and\ \bibinfo {author} {\bibfnamefont {C~N}\ \bibnamefont {Yang}}}
  (\bibinfo {year} {1956}),\ \bibfield  {title} {\enquote {\bibinfo {title}
  {Question of parity conservation in weak interactions},}\ }\href
  {https://doi.org/10.1103/PhysRev.104.254} {\bibfield  {journal} {\bibinfo
  {journal} {Physical Review}\ }\textbf {\bibinfo {volume} {104}},\
  10.1103/PhysRev.104.254}\BibitemShut {NoStop}%
\bibitem [{\citenamefont {L\'evai}\ and\ \citenamefont {Znojil}(2000)}]{pt128}%
  \BibitemOpen
  \bibfield  {author} {\bibinfo {author} {\bibnamefont {L\'evai}, \bibfnamefont
  {G\'eza}}, and\ \bibinfo {author} {\bibfnamefont {Miloslav}\ \bibnamefont
  {Znojil}}} (\bibinfo {year} {2000}),\ \bibfield  {title} {\enquote {\bibinfo
  {title} {Systematic search for $\cal{PT}$-symmetric potentials with real
  energy spectra},}\ }\href {https://doi.org/10.1088/0305-4470/33/40/313}
  {\bibfield  {journal} {\bibinfo  {journal} {Journal of Physics A:
  Mathematical and General}\ }\textbf {\bibinfo {volume} {33}},\ \bibinfo
  {pages} {7165--7180}}\BibitemShut {NoStop}%
\bibitem [{\citenamefont {L\'evai}\ and\ \citenamefont {Znojil}(2001)}]{pt129}%
  \BibitemOpen
  \bibfield  {author} {\bibinfo {author} {\bibnamefont {L\'evai}, \bibfnamefont
  {G\'eza}}, and\ \bibinfo {author} {\bibfnamefont {Miloslav}\ \bibnamefont
  {Znojil}}} (\bibinfo {year} {2001}),\ \bibfield  {title} {\enquote {\bibinfo
  {title} {Conditions for complex spectra in a class of $\cal{PT}$-symmetric
  potentials},}\ }\href {https://doi.org/10.1142/S0217732301005321} {\bibfield
  {journal} {\bibinfo  {journal} {Modern Physics Letters A}\ }\textbf {\bibinfo
  {volume} {16}},\ \bibinfo {pages} {1973--1981}}\BibitemShut {NoStop}%
\bibitem [{\citenamefont {Li}\ \emph {et~al.}(2019)\citenamefont {Li},
  \citenamefont {Harter}, \citenamefont {Liu}, \citenamefont {de~Melo},
  \citenamefont {Joglekar},\ and\ \citenamefont {Luo}}]{pt609}%
  \BibitemOpen
  \bibfield  {author} {\bibinfo {author} {\bibnamefont {Li}, \bibfnamefont
  {Jiaming}}, \bibinfo {author} {\bibfnamefont {Andrew~K}\ \bibnamefont
  {Harter}}, \bibinfo {author} {\bibfnamefont {Ji}~\bibnamefont {Liu}},
  \bibinfo {author} {\bibfnamefont {Leonardo}\ \bibnamefont {de~Melo}},
  \bibinfo {author} {\bibfnamefont {Yogesh~N}\ \bibnamefont {Joglekar}}, and\
  \bibinfo {author} {\bibfnamefont {Le}~\bibnamefont {Luo}}} (\bibinfo {year}
  {2019}),\ \bibfield  {title} {\enquote {\bibinfo {title} {Observation of
  parity-time symmetry breaking transitions in a dissipative floquet system of
  ultracold atoms},}\ }\href {https://doi.org/10.1038/s41467-019-08596-1}
  {\bibfield  {journal} {\bibinfo  {journal} {Nature Communications}\ }\textbf
  {\bibinfo {volume} {10}},\ 10.1038/s41467-019-08596-1}\BibitemShut {NoStop}%
\bibitem [{\citenamefont {Li}\ \emph {et~al.}(2023{\natexlab{a}})\citenamefont
  {Li}, \citenamefont {Wang}, \citenamefont {Luo}, \citenamefont {Vemuri},\
  and\ \citenamefont {Joglekar}}]{pt687}%
  \BibitemOpen
  \bibfield  {author} {\bibinfo {author} {\bibnamefont {Li}, \bibfnamefont
  {Jiaming}}, \bibinfo {author} {\bibfnamefont {Tishuo}\ \bibnamefont {Wang}},
  \bibinfo {author} {\bibfnamefont {Le}~\bibnamefont {Luo}}, \bibinfo {author}
  {\bibfnamefont {Sreya}\ \bibnamefont {Vemuri}}, and\ \bibinfo {author}
  {\bibfnamefont {Yogesh~N}\ \bibnamefont {Joglekar}}} (\bibinfo {year}
  {2023}{\natexlab{a}}),\ \bibfield  {title} {\enquote {\bibinfo {title}
  {Unification of quantum zeno-anti zeno effects and parity-time symmetry
  breaking transitions},}\ }\href
  {https://doi.org/10.1103/physrevresearch.5.023204} {\bibfield  {journal}
  {\bibinfo  {journal} {Physical Review Research}\ }\textbf {\bibinfo {volume}
  {5}},\ 10.1103/physrevresearch.5.023204}\BibitemShut {NoStop}%
\bibitem [{\citenamefont {Li}\ and\ \citenamefont {Kevrekidis}(2011)}]{pt223}%
  \BibitemOpen
  \bibfield  {author} {\bibinfo {author} {\bibnamefont {Li}, \bibfnamefont
  {K}}, and\ \bibinfo {author} {\bibfnamefont {P~G}\ \bibnamefont
  {Kevrekidis}}} (\bibinfo {year} {2011}),\ \bibfield  {title} {\enquote
  {\bibinfo {title} {$\cal{PT}$-symmetric oligomers: Analytical solutions,
  linear stability, and nonlinear dynamics},}\ }\href
  {https://doi.org/10.1103/PhysRevE.83.066608} {\bibfield  {journal} {\bibinfo
  {journal} {Physical Review E}\ }\textbf {\bibinfo {volume} {83}},\ \bibinfo
  {pages} {066608}}\BibitemShut {NoStop}%
\bibitem [{\citenamefont {Li}\ \emph {et~al.}(2013)\citenamefont {Li},
  \citenamefont {Kevrekidis}, \citenamefont {Frantzeskakis}, \citenamefont
  {R\"{u}ter},\ and\ \citenamefont {Kip}}]{pt224}%
  \BibitemOpen
  \bibfield  {author} {\bibinfo {author} {\bibnamefont {Li}, \bibfnamefont
  {K}}, \bibinfo {author} {\bibfnamefont {P~G}\ \bibnamefont {Kevrekidis}},
  \bibinfo {author} {\bibfnamefont {D~J}\ \bibnamefont {Frantzeskakis}},
  \bibinfo {author} {\bibfnamefont {C~E}\ \bibnamefont {R\"{u}ter}}, and\
  \bibinfo {author} {\bibfnamefont {D}~\bibnamefont {Kip}}} (\bibinfo {year}
  {2013}),\ \bibfield  {title} {\enquote {\bibinfo {title} {Revisiting the
  $\cal{PT}$-symmetric trimer: bifurcations, ghost states and associated
  dynamics},}\ }\href {https://doi.org/10.1088/1751-8113/46/37/375304}
  {\bibfield  {journal} {\bibinfo  {journal} {Journal of Physics A:
  Mathematical and Theoretical}\ }\textbf {\bibinfo {volume} {46}},\ \bibinfo
  {pages} {375304}}\BibitemShut {NoStop}%
\bibitem [{\citenamefont {Li}\ \emph {et~al.}(2012)\citenamefont {Li},
  \citenamefont {Kevrekidis}, \citenamefont {Malomed},\ and\ \citenamefont
  {G\"{u}nther}}]{pt225}%
  \BibitemOpen
  \bibfield  {author} {\bibinfo {author} {\bibnamefont {Li}, \bibfnamefont
  {Kai}}, \bibinfo {author} {\bibfnamefont {P~G}\ \bibnamefont {Kevrekidis}},
  \bibinfo {author} {\bibfnamefont {Boris~A}\ \bibnamefont {Malomed}}, and\
  \bibinfo {author} {\bibfnamefont {Uwe}\ \bibnamefont {G\"{u}nther}}}
  (\bibinfo {year} {2012}),\ \bibfield  {title} {\enquote {\bibinfo {title}
  {Nonlinear $\cal{PT}$-symmetric plaquettes},}\ }\href
  {https://doi.org/10.1088/1751-8113/45/44/444021} {\bibfield  {journal}
  {\bibinfo  {journal} {Journal of Physics A: Mathematical and Theoretical}\
  }\textbf {\bibinfo {volume} {45}},\ \bibinfo {pages} {444021}}\BibitemShut
  {NoStop}%
\bibitem [{\citenamefont {Li}\ \emph {et~al.}(2020)\citenamefont {Li},
  \citenamefont {Goryachev}, \citenamefont {adn Michael E~Tobar}, \citenamefont
  {Zhao}, \citenamefont {Adhikari},\ and\ \citenamefont {Chen}}]{pt705}%
  \BibitemOpen
  \bibfield  {author} {\bibinfo {author} {\bibnamefont {Li}, \bibfnamefont
  {Xiang}}, \bibinfo {author} {\bibfnamefont {Maxim}\ \bibnamefont
  {Goryachev}}, \bibinfo {author} {\bibfnamefont {Yiqiu~Ma}\ \bibnamefont {adn
  Michael E~Tobar}}, \bibinfo {author} {\bibfnamefont {Chunnong}\ \bibnamefont
  {Zhao}}, \bibinfo {author} {\bibfnamefont {Rana~X}\ \bibnamefont {Adhikari}},
  and\ \bibinfo {author} {\bibfnamefont {Yanbei}\ \bibnamefont {Chen}}}
  (\bibinfo {year} {2020}),\ \bibfield  {title} {\enquote {\bibinfo {title}
  {Broadband sensitivity improvement via coherent quantum feedback with pt
  symmetry},}\ }\href {https://doi.org/10.48550/arXiv.2012.00836} {\bibfield
  {journal} {\bibinfo  {journal} {arXiv}\ }\textbf {\bibinfo {volume}
  {quant-ph}},\ 10.48550/arXiv.2012.00836}\BibitemShut {NoStop}%
\bibitem [{\citenamefont {Li}\ \emph {et~al.}(2023{\natexlab{b}})\citenamefont
  {Li}, \citenamefont {Chen}, \citenamefont {Abbasi}, \citenamefont {Murch},\
  and\ \citenamefont {Whaley}}]{pt633}%
  \BibitemOpen
  \bibfield  {author} {\bibinfo {author} {\bibnamefont {Li}, \bibfnamefont
  {Zeng-Zhao}}, \bibinfo {author} {\bibfnamefont {Weijian}\ \bibnamefont
  {Chen}}, \bibinfo {author} {\bibfnamefont {Maryam}\ \bibnamefont {Abbasi}},
  \bibinfo {author} {\bibfnamefont {Kater~W}\ \bibnamefont {Murch}}, and\
  \bibinfo {author} {\bibfnamefont {K~Birgitta}\ \bibnamefont {Whaley}}}
  (\bibinfo {year} {2023}{\natexlab{b}}),\ \bibfield  {title} {\enquote
  {\bibinfo {title} {Speeding up entanglement generation by proximity to
  higher-order exceptional points},}\ }\href
  {https://doi.org/10.1103/PhysRevLett.131.100202} {\bibfield  {journal}
  {\bibinfo  {journal} {Physical Review Letters}\ }\textbf {\bibinfo {volume}
  {131}},\ 10.1103/PhysRevLett.131.100202}\BibitemShut {NoStop}%
\bibitem [{\citenamefont {Limonov}\ \emph {et~al.}(2017)\citenamefont
  {Limonov}, \citenamefont {Rybin}, \citenamefont {Poddubny},\ and\
  \citenamefont {Kivshar}}]{pt537}%
  \BibitemOpen
  \bibfield  {author} {\bibinfo {author} {\bibnamefont {Limonov}, \bibfnamefont
  {Mikhail~F}}, \bibinfo {author} {\bibfnamefont {Mikhail~V}\ \bibnamefont
  {Rybin}}, \bibinfo {author} {\bibfnamefont {Alexander~N}\ \bibnamefont
  {Poddubny}}, and\ \bibinfo {author} {\bibfnamefont {Yuri~S}\ \bibnamefont
  {Kivshar}}} (\bibinfo {year} {2017}),\ \bibfield  {title} {\enquote {\bibinfo
  {title} {Fano resonances in photonics},}\ }\href
  {https://doi.org/10.1038/nphoton.2017.142} {\bibfield  {journal} {\bibinfo
  {journal} {Nature Photonics}\ }\textbf {\bibinfo {volume} {11}},\ \bibinfo
  {pages} {543--554}}\BibitemShut {NoStop}%
\bibitem [{\citenamefont {Lin}\ \emph {et~al.}(2011)\citenamefont {Lin},
  \citenamefont {Ramezani}, \citenamefont {Eichelkraut}, \citenamefont
  {Kottos}, \citenamefont {Cao},\ and\ \citenamefont
  {Christodoulides}}]{pt193}%
  \BibitemOpen
  \bibfield  {author} {\bibinfo {author} {\bibnamefont {Lin}, \bibfnamefont
  {Zin}}, \bibinfo {author} {\bibfnamefont {Hamidreza}\ \bibnamefont
  {Ramezani}}, \bibinfo {author} {\bibfnamefont {Toni}\ \bibnamefont
  {Eichelkraut}}, \bibinfo {author} {\bibfnamefont {Tsampikos}\ \bibnamefont
  {Kottos}}, \bibinfo {author} {\bibfnamefont {Hui}\ \bibnamefont {Cao}}, and\
  \bibinfo {author} {\bibfnamefont {Demetrios~N}\ \bibnamefont
  {Christodoulides}}} (\bibinfo {year} {2011}),\ \bibfield  {title} {\enquote
  {\bibinfo {title} {Unidirectional invisibility induced by
  $\cal{PT}$-symmetric periodic structures},}\ }\href
  {https://doi.org/10.1103/PhysRevLett.106.213901} {\bibfield  {journal}
  {\bibinfo  {journal} {Physical Review Letters}\ }\textbf {\bibinfo {volume}
  {106}},\ \bibinfo {pages} {213901}}\BibitemShut {NoStop}%
\bibitem [{\citenamefont {Lipatov}\ \emph {et~al.}(1979)\citenamefont
  {Lipatov}, \citenamefont {Bukhvostov},\ and\ \citenamefont {Malkov}}]{pt604}%
  \BibitemOpen
  \bibfield  {author} {\bibinfo {author} {\bibnamefont {Lipatov}, \bibfnamefont
  {L~N}}, \bibinfo {author} {\bibfnamefont {A~P}\ \bibnamefont {Bukhvostov}},
  and\ \bibinfo {author} {\bibfnamefont {E~I}\ \bibnamefont {Malkov}}}
  (\bibinfo {year} {1979}),\ \bibfield  {title} {\enquote {\bibinfo {title}
  {Large-order estimates for perturbation theory of a yang-mills field coupled
  to a scalar field},}\ }\href {https://doi.org/10.1103/PhysRevD.19.2974}
  {\bibfield  {journal} {\bibinfo  {journal} {Physics Review D}\ }\textbf
  {\bibinfo {volume} {19}},\ 10.1103/PhysRevD.19.2974}\BibitemShut {NoStop}%
\bibitem [{\citenamefont {Liu}\ \emph {et~al.}(2016)\citenamefont {Liu},
  \citenamefont {Zhang}, \citenamefont {\c{S}ahin Kaya~\"{O}zdemir},
  \citenamefont {Peng}, \citenamefont {Jing}, \citenamefont {L\"{u}},
  \citenamefont {Chun-Wen~Li}, \citenamefont {Nori},\ and\ \citenamefont
  {xi~Liu}}]{pt509}%
  \BibitemOpen
  \bibfield  {author} {\bibinfo {author} {\bibnamefont {Liu}, \bibfnamefont
  {Zhong-Peng}}, \bibinfo {author} {\bibfnamefont {Jing}\ \bibnamefont
  {Zhang}}, \bibinfo {author} {\bibnamefont {\c{S}ahin Kaya~\"{O}zdemir}},
  \bibinfo {author} {\bibfnamefont {Bo}~\bibnamefont {Peng}}, \bibinfo {author}
  {\bibfnamefont {Hui}\ \bibnamefont {Jing}}, \bibinfo {author} {\bibfnamefont
  {Xin-You}\ \bibnamefont {L\"{u}}}, \bibinfo {author} {\bibfnamefont
  {Lan~Yang}\ \bibnamefont {Chun-Wen~Li}}, \bibinfo {author} {\bibfnamefont
  {Franco}\ \bibnamefont {Nori}}, and\ \bibinfo {author} {\bibfnamefont
  {Yu}~\bibnamefont {xi~Liu}}} (\bibinfo {year} {2016}),\ \bibfield  {title}
  {\enquote {\bibinfo {title} {Metrology with $\mathcal{PT}$-symmetric
  cavities: Enhanced sensitivity near the $\mathcal{PT}$-phase transition},}\
  }\href {https://doi.org/PhysRevLett.117.110802} {\bibfield  {journal}
  {\bibinfo  {journal} {Physical Review Letters}\ }\textbf {\bibinfo {volume}
  {117}},\ \bibinfo {pages} {110802}}\BibitemShut {NoStop}%
\bibitem [{\citenamefont {Loeffel}\ and\ \citenamefont {Martin}(1970)}]{pt616}%
  \BibitemOpen
  \bibfield  {author} {\bibinfo {author} {\bibnamefont {Loeffel}, \bibfnamefont
  {J~J}}, and\ \bibinfo {author} {\bibfnamefont {A}~\bibnamefont {Martin}}}
  (\bibinfo {year} {1970}),\ \bibfield  {title} {\enquote {\bibinfo {title}
  {Propri\'et\'es analytiques des niveaux de l'oscillateur anharmonique et
  convergence des approximants de {P}ad\'e},}\ }in\ \href
  {http://cds.cern.ch/record/944354} {\emph {\bibinfo {booktitle} {Ecole
  d'Et\'e de Physique Th\'eorique : Carg\`ese Lectures in Physics}}},\
  Vol.~\bibinfo {volume} {5},\ \bibinfo {editor} {edited by\ \bibinfo {editor}
  {\bibfnamefont {D}~\bibnamefont {Bessis}}},\ \bibinfo {note}
  {{CERN-TH-1167}}\BibitemShut {NoStop}%
\bibitem [{\citenamefont {Loeffel}\ \emph {et~al.}(1969)\citenamefont
  {Loeffel}, \citenamefont {Martin}, \citenamefont {Simon},\ and\ \citenamefont
  {Wightman}}]{pt597}%
  \BibitemOpen
  \bibfield  {author} {\bibinfo {author} {\bibnamefont {Loeffel}, \bibfnamefont
  {J~J}}, \bibinfo {author} {\bibfnamefont {A}~\bibnamefont {Martin}}, \bibinfo
  {author} {\bibfnamefont {B}~\bibnamefont {Simon}}, and\ \bibinfo {author}
  {\bibfnamefont {A~S}\ \bibnamefont {Wightman}}} (\bibinfo {year} {1969}),\
  \bibfield  {title} {\enquote {\bibinfo {title} {Pade approximants and the
  anharmonic oscillator},}\ }\href
  {https://doi.org/10.1016/0370-2693(69)90087-2} {\bibfield  {journal}
  {\bibinfo  {journal} {Physics Letters B}\ }\textbf {\bibinfo {volume} {30}},\
  10.1016/0370-2693(69)90087-2}\BibitemShut {NoStop}%
\bibitem [{\citenamefont {Longhi}(2010)}]{pt202}%
  \BibitemOpen
  \bibfield  {author} {\bibinfo {author} {\bibnamefont {Longhi}, \bibfnamefont
  {S}}} (\bibinfo {year} {2010}),\ \bibfield  {title} {\enquote {\bibinfo
  {title} {$\cal{PT}$-symmetric laser absorber},}\ }\href
  {https://doi.org/10.1103/PhysRevA.82.031801} {\bibfield  {journal} {\bibinfo
  {journal} {Physical Review A}\ }\textbf {\bibinfo {volume} {82}},\ \bibinfo
  {pages} {031801(R)}}\BibitemShut {NoStop}%
\bibitem [{\citenamefont {Longhi}(2015)}]{pt217}%
  \BibitemOpen
  \bibfield  {author} {\bibinfo {author} {\bibnamefont {Longhi}, \bibfnamefont
  {S}}} (\bibinfo {year} {2015}),\ \bibfield  {title} {\enquote {\bibinfo
  {title} {Bound states in the continuum in $\cal{PT}$-symmetric optical
  lattices},}\ }\href {https://doi.org/10.1364/OL.39.001697} {\bibfield
  {journal} {\bibinfo  {journal} {Optics Letters}\ }\textbf {\bibinfo {volume}
  {39}},\ \bibinfo {pages} {1697--1700}}\BibitemShut {NoStop}%
\bibitem [{\citenamefont {Longhi}\ and\ \citenamefont
  {Della~{V}alle}(2014)}]{pt218}%
  \BibitemOpen
  \bibfield  {author} {\bibinfo {author} {\bibnamefont {Longhi}, \bibfnamefont
  {Stefano}}, and\ \bibinfo {author} {\bibfnamefont {Giuseppe}\ \bibnamefont
  {Della~{V}alle}}} (\bibinfo {year} {2014}),\ \bibfield  {title} {\enquote
  {\bibinfo {title} {Optical lattices with exceptional points in the
  continuum},}\ }\href {https://doi.org/10.1103/PhysRevA.89.052132} {\bibfield
  {journal} {\bibinfo  {journal} {Physical Review A}\ }\textbf {\bibinfo
  {volume} {89}},\ \bibinfo {pages} {052132}}\BibitemShut {NoStop}%
\bibitem [{\citenamefont {Lv}\ \emph {et~al.}(2022)\citenamefont {Lv},
  \citenamefont {Zhang}, \citenamefont {Zhai},\ and\ \citenamefont
  {Zhou}}]{pt666}%
  \BibitemOpen
  \bibfield  {author} {\bibinfo {author} {\bibnamefont {Lv}, \bibfnamefont
  {Chenwei}}, \bibinfo {author} {\bibfnamefont {Ren}\ \bibnamefont {Zhang}},
  \bibinfo {author} {\bibfnamefont {Zhengzheng}\ \bibnamefont {Zhai}}, and\
  \bibinfo {author} {\bibfnamefont {Qi}~\bibnamefont {Zhou}}} (\bibinfo {year}
  {2022}),\ \bibfield  {title} {\enquote {\bibinfo {title} {Curving the space
  by non-hermiticity},}\ }\href {https://doi.org/10.1038/s41467-022-29774-8}
  {\bibfield  {journal} {\bibinfo  {journal} {Nature Communications}\ }\textbf
  {\bibinfo {volume} {13}},\ 10.1038/s41467-022-29774-8}\BibitemShut {NoStop}%
\bibitem [{\citenamefont {Makris}\ \emph {et~al.}(2008)\citenamefont {Makris},
  \citenamefont {El-Ganainy}, \citenamefont {Christodoulides},\ and\
  \citenamefont {Musslimani}}]{pt184}%
  \BibitemOpen
  \bibfield  {author} {\bibinfo {author} {\bibnamefont {Makris}, \bibfnamefont
  {K~G}}, \bibinfo {author} {\bibfnamefont {R}~\bibnamefont {El-Ganainy}},
  \bibinfo {author} {\bibfnamefont {D~N}\ \bibnamefont {Christodoulides}}, and\
  \bibinfo {author} {\bibfnamefont {Z~H}\ \bibnamefont {Musslimani}}} (\bibinfo
  {year} {2008}),\ \bibfield  {title} {\enquote {\bibinfo {title} {Beam
  dynamics in $\cal{PT}$ symmetric optical lattices},}\ }\href
  {https://doi.org/10.1103/PhysRevLett.100.103904} {\bibfield  {journal}
  {\bibinfo  {journal} {Physical Review Letters}\ }\textbf {\bibinfo {volume}
  {100}},\ \bibinfo {pages} {103904}}\BibitemShut {NoStop}%
\bibitem [{\citenamefont {Makris}\ \emph {et~al.}(2010)\citenamefont {Makris},
  \citenamefont {El-Ganainy}, \citenamefont {Christodoulides},\ and\
  \citenamefont {Musslimani}}]{pt187}%
  \BibitemOpen
  \bibfield  {author} {\bibinfo {author} {\bibnamefont {Makris}, \bibfnamefont
  {Konstantinos~G}}, \bibinfo {author} {\bibfnamefont {Ramy}\ \bibnamefont
  {El-Ganainy}}, \bibinfo {author} {\bibfnamefont {Demetrios~N}\ \bibnamefont
  {Christodoulides}}, and\ \bibinfo {author} {\bibfnamefont {Z~H}\ \bibnamefont
  {Musslimani}}} (\bibinfo {year} {2010}),\ \bibfield  {title} {\enquote
  {\bibinfo {title} {$\cal{PT}$-symmetric optical lattices},}\ }\href
  {https://doi.org/10.1103/PhysRevA.81.063807} {\bibfield  {journal} {\bibinfo
  {journal} {Physical Review A}\ }\textbf {\bibinfo {volume} {81}},\ \bibinfo
  {pages} {063807}}\BibitemShut {NoStop}%
\bibitem [{\citenamefont {Mannheim}(2011)}]{pt642}%
  \BibitemOpen
  \bibfield  {author} {\bibinfo {author} {\bibnamefont {Mannheim},
  \bibfnamefont {Philip~D}}} (\bibinfo {year} {2011}),\ \bibfield  {title}
  {\enquote {\bibinfo {title} {Making the case for conformal gravity},}\ }\href
  {https://doi.org/10.1007/s10701-011-9608-6} {\bibfield  {journal} {\bibinfo
  {journal} {Foundations of Physics}\ }\textbf {\bibinfo {volume} {42}},\
  10.1007/s10701-011-9608-6}\BibitemShut {NoStop}%
\bibitem [{\citenamefont {Mannheim}(2017)}]{pt643}%
  \BibitemOpen
  \bibfield  {author} {\bibinfo {author} {\bibnamefont {Mannheim},
  \bibfnamefont {Philip~D}}} (\bibinfo {year} {2017}),\ \bibfield  {title}
  {\enquote {\bibinfo {title} {Mass generation, the cosmological constant
  problem, conformal symmetry, and the higgs boson},}\ }\href
  {https://doi.org/10.1016/j.ppnp.2017.02.001} {\bibfield  {journal} {\bibinfo
  {journal} {Progress in Particle and Nuclear Physics}\ }\textbf {\bibinfo
  {volume} {94}},\ 10.1016/j.ppnp.2017.02.001}\BibitemShut {NoStop}%
\bibitem [{\citenamefont {Mannheim}(2018)}]{pt639}%
  \BibitemOpen
  \bibfield  {author} {\bibinfo {author} {\bibnamefont {Mannheim},
  \bibfnamefont {Philip~D}}} (\bibinfo {year} {2018}),\ \bibfield  {title}
  {\enquote {\bibinfo {title} {Unitarity of loop diagrams for the ghostlike
  $1/(k^2-m_1^2)-1/(k^2-m^2_2)$ propagator},}\ }\href
  {https://doi.org/10.1103/physrevd.98.045014} {\bibfield  {journal} {\bibinfo
  {journal} {Physical Review D}\ }\textbf {\bibinfo {volume} {98}},\
  10.1103/physrevd.98.045014}\BibitemShut {NoStop}%
\bibitem [{\citenamefont {Mavromatos}\ \emph {et~al.}(2023)\citenamefont
  {Mavromatos}, \citenamefont {Sarkar},\ and\ \citenamefont {Soto}}]{pt622}%
  \BibitemOpen
  \bibfield  {author} {\bibinfo {author} {\bibnamefont {Mavromatos},
  \bibfnamefont {N~E}}, \bibinfo {author} {\bibfnamefont {Sarben}\ \bibnamefont
  {Sarkar}}, and\ \bibinfo {author} {\bibfnamefont {A}~\bibnamefont {Soto}}}
  (\bibinfo {year} {2023}),\ \bibfield  {title} {\enquote {\bibinfo {title}
  {Schwinger-dyson equations and mass generation for an axion theory with a
  pt-symmetric yukawa fermion interaction},}\ }\href
  {https://doi.org/10.1016/j.nuclphysb.2022.116048} {\bibfield  {journal}
  {\bibinfo  {journal} {Nuclear Physics B}\ }\textbf {\bibinfo {volume}
  {986}},\ 10.1016/j.nuclphysb.2022.116048}\BibitemShut {NoStop}%
\bibitem [{\citenamefont {Mavromatos}\ \emph {et~al.}(2022)\citenamefont
  {Mavromatos}, \citenamefont {Sarkar},\ and\ \citenamefont {Soto}}]{pt660}%
  \BibitemOpen
  \bibfield  {author} {\bibinfo {author} {\bibnamefont {Mavromatos},
  \bibfnamefont {N~E}}, \bibinfo {author} {\bibfnamefont {Sarben}\ \bibnamefont
  {Sarkar}}, and\ \bibinfo {author} {\bibfnamefont {Alex}\ \bibnamefont
  {Soto}}} (\bibinfo {year} {2022}),\ \bibfield  {title} {\enquote {\bibinfo
  {title} {Pt-symmetric fermionic field theories with axions: Renormalization
  and dynamical mass generation},}\ }\href
  {https://doi.org/10.1103/physrevd.106.015009} {\bibfield  {journal} {\bibinfo
   {journal} {Physical Review D}\ }\textbf {\bibinfo {volume} {106}},\
  10.1103/physrevd.106.015009}\BibitemShut {NoStop}%
\bibitem [{\citenamefont {Mavromatos}\ and\ \citenamefont
  {Soto}(2021)}]{pt659}%
  \BibitemOpen
  \bibfield  {author} {\bibinfo {author} {\bibnamefont {Mavromatos},
  \bibfnamefont {Nick~E}}, and\ \bibinfo {author} {\bibfnamefont {Alex}\
  \bibnamefont {Soto}}} (\bibinfo {year} {2021}),\ \bibfield  {title} {\enquote
  {\bibinfo {title} {Dynamical majorana neutrino masses and axions ii:
  Inclusion of anomaly terms and axial background},}\ }\href
  {https://doi.org/10.1016/j.nuclphysb.2020.115275} {\bibfield  {journal}
  {\bibinfo  {journal} {Nuclear Physics B}\ }\textbf {\bibinfo {volume}
  {962}},\ 10.1016/j.nuclphysb.2020.115275}\BibitemShut {NoStop}%
\bibitem [{\citenamefont {Mezincescu}(2000)}]{pt377}%
  \BibitemOpen
  \bibfield  {author} {\bibinfo {author} {\bibnamefont {Mezincescu},
  \bibfnamefont {G~Andrei}}} (\bibinfo {year} {2000}),\ \bibfield  {title}
  {\enquote {\bibinfo {title} {Some properties of eigenvalues and
  eigenfunctions of the cubic oscillator with imaginary coupling constant},}\
  }\href {https://doi.org/10.1088/0305-4470/33/27/308} {\bibfield  {journal}
  {\bibinfo  {journal} {Journal of Physics A: Mathematical and General}\
  }\textbf {\bibinfo {volume} {33}},\ 10.1088/0305-4470/33/27/308}\BibitemShut
  {NoStop}%
\bibitem [{\citenamefont {Monticone}\ \emph {et~al.}(2016)\citenamefont
  {Monticone}, \citenamefont {Valagiannopoulos},\ and\ \citenamefont
  {Al\'{u}}}]{pt596}%
  \BibitemOpen
  \bibfield  {author} {\bibinfo {author} {\bibnamefont {Monticone},
  \bibfnamefont {Francesco}}, \bibinfo {author} {\bibfnamefont
  {Constantinos~A}\ \bibnamefont {Valagiannopoulos}}, and\ \bibinfo {author}
  {\bibfnamefont {Andrea}\ \bibnamefont {Al\'{u}}}} (\bibinfo {year} {2016}),\
  \bibfield  {title} {\enquote {\bibinfo {title} {Parity-time symmetric
  nonlocal metasurfaces: All-angle negative refraction and volumetric
  imaging},}\ }\href {https://doi.org/10.1103/PhysRevX.6.041018} {\bibfield
  {journal} {\bibinfo  {journal} {Physical Review X}\ }\textbf {\bibinfo
  {volume} {6}},\ 10.1103/PhysRevX.6.041018}\BibitemShut {NoStop}%
\bibitem [{\citenamefont {Mostafazadeh}(2002{\natexlab{a}})}]{pt140}%
  \BibitemOpen
  \bibfield  {author} {\bibinfo {author} {\bibnamefont {Mostafazadeh},
  \bibfnamefont {Ali}}} (\bibinfo {year} {2002}{\natexlab{a}}),\ \bibfield
  {title} {\enquote {\bibinfo {title} {Pseudo-{H}ermiticity versus $\cal{PT}$
  symmetry: The necessary condition for the reality of the spectrum of a
  non-{H}ermitian {H}amiltonian},}\ }\href {https://doi.org/10.1063/1.1418246}
  {\bibfield  {journal} {\bibinfo  {journal} {Journal of Mathematical Physics}\
  }\textbf {\bibinfo {volume} {43}},\ \bibinfo {pages} {205--214}}\BibitemShut
  {NoStop}%
\bibitem [{\citenamefont {Mostafazadeh}(2002{\natexlab{b}})}]{pt385}%
  \BibitemOpen
  \bibfield  {author} {\bibinfo {author} {\bibnamefont {Mostafazadeh},
  \bibfnamefont {Ali}}} (\bibinfo {year} {2002}{\natexlab{b}}),\ \bibfield
  {title} {\enquote {\bibinfo {title} {Pseudo-{H}ermiticity versus
  $\mathcal{PT}$-symmetry. {II}. {A} complete characterization of
  non-{H}ermitian {H}amiltonians with a real spectrum},}\ }\href
  {https://doi.org/10.1063/1.1461427} {\bibfield  {journal} {\bibinfo
  {journal} {Journal of Mathematical Physics}\ }\textbf {\bibinfo {volume}
  {43}},\ 10.1063/1.1461427}\BibitemShut {NoStop}%
\bibitem [{\citenamefont {Mostafazadeh}(2002{\natexlab{c}})}]{pt386}%
  \BibitemOpen
  \bibfield  {author} {\bibinfo {author} {\bibnamefont {Mostafazadeh},
  \bibfnamefont {Ali}}} (\bibinfo {year} {2002}{\natexlab{c}}),\ \bibfield
  {title} {\enquote {\bibinfo {title} {Pseudo-{H}ermiticity versus
  $\mathcal{PT}$-symmetry {III}: {E}quivalence of pseudo-{H}ermiticity and the
  presence of antilinear symmetries},}\ }\href
  {https://doi.org/10.1063/1.1489072} {\bibfield  {journal} {\bibinfo
  {journal} {Journal of Mathematical Physics}\ }\textbf {\bibinfo {volume}
  {43}},\ 10.1063/1.1489072}\BibitemShut {NoStop}%
\bibitem [{\citenamefont {Mostafazadeh}(2003{\natexlab{a}})}]{pt181}%
  \BibitemOpen
  \bibfield  {author} {\bibinfo {author} {\bibnamefont {Mostafazadeh},
  \bibfnamefont {Ali}}} (\bibinfo {year} {2003}{\natexlab{a}}),\ \bibfield
  {title} {\enquote {\bibinfo {title} {Exact $\cal{PT}$-symmetry is equivalent
  to {H}ermiticity},}\ }\href {https://doi.org/10.1088/0305-4470/36/25/312}
  {\bibfield  {journal} {\bibinfo  {journal} {Journal of Physics A:
  Mathematical and General}\ }\textbf {\bibinfo {volume} {36}},\ \bibinfo
  {pages} {7081--7092}}\BibitemShut {NoStop}%
\bibitem [{\citenamefont {Mostafazadeh}(2003{\natexlab{b}})}]{pt389}%
  \BibitemOpen
  \bibfield  {author} {\bibinfo {author} {\bibnamefont {Mostafazadeh},
  \bibfnamefont {Ali}}} (\bibinfo {year} {2003}{\natexlab{b}}),\ \bibfield
  {title} {\enquote {\bibinfo {title} {Pseudo-{H}ermiticity and generalized
  $\mathcal{PT}$- and $\mathcal{CPT}$-symmetries},}\ }\href
  {https://doi.org/10.1063/1.1539304} {\bibfield  {journal} {\bibinfo
  {journal} {Journal of Mathematical Physics}\ }\textbf {\bibinfo {volume}
  {44}},\ 10.1063/1.1539304}\BibitemShut {NoStop}%
\bibitem [{\citenamefont {Mostafazadeh}(2005)}]{pt391}%
  \BibitemOpen
  \bibfield  {author} {\bibinfo {author} {\bibnamefont {Mostafazadeh},
  \bibfnamefont {Ali}}} (\bibinfo {year} {2005}),\ \bibfield  {title} {\enquote
  {\bibinfo {title} {Pseudo-{H}ermitian description of $\mathcal{PT}$-symmetric
  systems defined on a complex contour},}\ }\href
  {https://doi.org/10.1088/0305-4470/38/14/011} {\bibfield  {journal} {\bibinfo
   {journal} {Journal of Physics A: Mathematical and General}\ }\textbf
  {\bibinfo {volume} {38}},\ 10.1088/0305-4470/38/14/011}\BibitemShut {NoStop}%
\bibitem [{\citenamefont {Mostafazadeh}(2009)}]{pt141}%
  \BibitemOpen
  \bibfield  {author} {\bibinfo {author} {\bibnamefont {Mostafazadeh},
  \bibfnamefont {Ali}}} (\bibinfo {year} {2009}),\ \bibfield  {title} {\enquote
  {\bibinfo {title} {Spectral singularities of complex scattering potentials
  and infinite reflection and transmission coefficients at real energies},}\
  }\href {https://doi.org/10.1103/PhysRevLett.102.220402} {\bibfield  {journal}
  {\bibinfo  {journal} {Physical Review Letters}\ }\textbf {\bibinfo {volume}
  {102}},\ \bibinfo {pages} {220402}}\BibitemShut {NoStop}%
\bibitem [{\citenamefont {Mostafazadeh}(2014)}]{pt142}%
  \BibitemOpen
  \bibfield  {author} {\bibinfo {author} {\bibnamefont {Mostafazadeh},
  \bibfnamefont {Ali}}} (\bibinfo {year} {2014}),\ \bibfield  {title} {\enquote
  {\bibinfo {title} {Unidirectionally invisible potentials as local building
  blocks of all scattering potentials},}\ }\href
  {https://doi.org/10.1103/PhysRevA.90.023833} {\bibfield  {journal} {\bibinfo
  {journal} {Physical Review A}\ }\textbf {\bibinfo {volume} {90}},\ \bibinfo
  {pages} {023833}}\BibitemShut {NoStop}%
\bibitem [{\citenamefont {Mostafazadeh}\ and\ \citenamefont
  {Batal}(2004)}]{pt380}%
  \BibitemOpen
  \bibfield  {author} {\bibinfo {author} {\bibnamefont {Mostafazadeh},
  \bibfnamefont {Ali}}, and\ \bibinfo {author} {\bibfnamefont {Ahmet}\
  \bibnamefont {Batal}}} (\bibinfo {year} {2004}),\ \bibfield  {title}
  {\enquote {\bibinfo {title} {Physical aspects of pseudo-{H}ermitian and
  $\mathcal{PT}$-symmetric quantum mechanics},}\ }\href
  {https://doi.org/10.1088/0305-4470/37/48/009} {\bibfield  {journal} {\bibinfo
   {journal} {Journal of Physics A: Mathematical and General}\ }\textbf
  {\bibinfo {volume} {37}},\ 10.1088/0305-4470/37/48/009}\BibitemShut {NoStop}%
\bibitem [{\citenamefont {Mudute-Ndumbe}\ and\ \citenamefont
  {Graefe}(2020)}]{pt679}%
  \BibitemOpen
  \bibfield  {author} {\bibinfo {author} {\bibnamefont {Mudute-Ndumbe},
  \bibfnamefont {Steve}}, and\ \bibinfo {author} {\bibfnamefont {Eva-Maria}\
  \bibnamefont {Graefe}}} (\bibinfo {year} {2020}),\ \bibfield  {title}
  {\enquote {\bibinfo {title} {A non-hermitian pt-symmetric kicked top},}\
  }\href {https://doi.org/10.1088/1367-2630/abb27a} {\bibfield  {journal}
  {\bibinfo  {journal} {New Journal of Physics}\ }\textbf {\bibinfo {volume}
  {22}},\ 10.1088/1367-2630/abb27a}\BibitemShut {NoStop}%
\bibitem [{\citenamefont {Musslimani}\ \emph {et~al.}(2008)\citenamefont
  {Musslimani}, \citenamefont {Makris}, \citenamefont {El-Ganainy},\ and\
  \citenamefont {Christodoulides}}]{pt183}%
  \BibitemOpen
  \bibfield  {author} {\bibinfo {author} {\bibnamefont {Musslimani},
  \bibfnamefont {Z~H}}, \bibinfo {author} {\bibfnamefont {K~G}\ \bibnamefont
  {Makris}}, \bibinfo {author} {\bibfnamefont {R}~\bibnamefont {El-Ganainy}},
  and\ \bibinfo {author} {\bibfnamefont {D~N}\ \bibnamefont {Christodoulides}}}
  (\bibinfo {year} {2008}),\ \bibfield  {title} {\enquote {\bibinfo {title}
  {Optical solitons in $\cal{PT}$ periodic potentials},}\ }\href
  {https://doi.org/10.1103/PhysRevLett.100.030402} {\bibfield  {journal}
  {\bibinfo  {journal} {Physical Review Letters}\ }\textbf {\bibinfo {volume}
  {100}},\ \bibinfo {pages} {030402}}\BibitemShut {NoStop}%
\bibitem [{\citenamefont {Naghiloo}\ \emph {et~al.}(2019)\citenamefont
  {Naghiloo}, \citenamefont {Abbasi}, \citenamefont {Joglekar},\ and\
  \citenamefont {Murch}}]{pt629}%
  \BibitemOpen
  \bibfield  {author} {\bibinfo {author} {\bibnamefont {Naghiloo},
  \bibfnamefont {M}}, \bibinfo {author} {\bibfnamefont {M}~\bibnamefont
  {Abbasi}}, \bibinfo {author} {\bibfnamefont {Yogesh~N}\ \bibnamefont
  {Joglekar}}, and\ \bibinfo {author} {\bibfnamefont {K~W}\ \bibnamefont
  {Murch}}} (\bibinfo {year} {2019}),\ \bibfield  {title} {\enquote {\bibinfo
  {title} {Quantum state tomography across the exceptional point in a single
  dissipative qubit},}\ }\href {https://doi.org/10.1038/s41567-019-0652-z}
  {\bibfield  {journal} {\bibinfo  {journal} {Nature Physics}\ }\textbf
  {\bibinfo {volume} {15}},\ 10.1038/s41567-019-0652-z}\BibitemShut {NoStop}%
\bibitem [{\citenamefont {Nanayakkara}(2004{\natexlab{a}})}]{pt429}%
  \BibitemOpen
  \bibfield  {author} {\bibinfo {author} {\bibnamefont {Nanayakkara},
  \bibfnamefont {A}}} (\bibinfo {year} {2004}{\natexlab{a}}),\ \bibfield
  {title} {\enquote {\bibinfo {title} {Classical motion of complex 2-{D}
  non-{H}ermitian {H}amiltonian systems},}\ }\href
  {https://doi.org/10.1023/B:CJOP.0000014374.61647.55} {\bibfield  {journal}
  {\bibinfo  {journal} {Czechoslovak Journal of Physics}\ }\textbf {\bibinfo
  {volume} {54}},\ 10.1023/B:CJOP.0000014374.61647.55}\BibitemShut {NoStop}%
\bibitem [{\citenamefont {Nanayakkara}(2004{\natexlab{b}})}]{pt430}%
  \BibitemOpen
  \bibfield  {author} {\bibinfo {author} {\bibnamefont {Nanayakkara},
  \bibfnamefont {A}}} (\bibinfo {year} {2004}{\natexlab{b}}),\ \bibfield
  {title} {\enquote {\bibinfo {title} {Classical trajectories of 1{D} complex
  non-{H}ermitian {H}amiltonian systems},}\ }\href
  {https://doi.org/10.1088/0305-4470/37/15/002} {\bibfield  {journal} {\bibinfo
   {journal} {Journal of Physics A: Mathematical and General}\ }\textbf
  {\bibinfo {volume} {37}},\ 10.1088/0305-4470/37/15/002}\BibitemShut {NoStop}%
\bibitem [{\citenamefont {Nasari}\ \emph {et~al.}(2022)\citenamefont {Nasari},
  \citenamefont {Lopez-Galmiche}, \citenamefont {Lopez-Aviles}, \citenamefont
  {Schumer}, \citenamefont {Hassan}, \citenamefont {Zhong}, \citenamefont
  {Rotter}, \citenamefont {LiKamWa}, \citenamefont {Christodoulides},\ and\
  \citenamefont {Khajavikhan}}]{pt663}%
  \BibitemOpen
  \bibfield  {author} {\bibinfo {author} {\bibnamefont {Nasari}, \bibfnamefont
  {Hadiseh}}, \bibinfo {author} {\bibfnamefont {Gisela}\ \bibnamefont
  {Lopez-Galmiche}}, \bibinfo {author} {\bibfnamefont {Helena~E}\ \bibnamefont
  {Lopez-Aviles}}, \bibinfo {author} {\bibfnamefont {Alexander}\ \bibnamefont
  {Schumer}}, \bibinfo {author} {\bibfnamefont {Absar~U}\ \bibnamefont
  {Hassan}}, \bibinfo {author} {\bibfnamefont {Qi}~\bibnamefont {Zhong}},
  \bibinfo {author} {\bibfnamefont {Stefan}\ \bibnamefont {Rotter}}, \bibinfo
  {author} {\bibfnamefont {Patrick}\ \bibnamefont {LiKamWa}}, \bibinfo {author}
  {\bibfnamefont {Demetrios~N}\ \bibnamefont {Christodoulides}}, and\ \bibinfo
  {author} {\bibfnamefont {Mercedeh}\ \bibnamefont {Khajavikhan}}} (\bibinfo
  {year} {2022}),\ \bibfield  {title} {\enquote {\bibinfo {title} {Observation
  of chiral state transfer without encircling an exceptional point},}\ }\href
  {https://doi.org/10.1038/s41586-022-04542-2} {\bibfield  {journal} {\bibinfo
  {journal} {Nature}\ }\textbf {\bibinfo {volume} {605}},\
  10.1038/s41586-022-04542-2}\BibitemShut {NoStop}%
\bibitem [{\citenamefont {Ohlsson}(2016)}]{pt572}%
  \BibitemOpen
  \bibfield  {author} {\bibinfo {author} {\bibnamefont {Ohlsson}, \bibfnamefont
  {Tommy}}} (\bibinfo {year} {2016}),\ \bibfield  {title} {\enquote {\bibinfo
  {title} {Non-{H}ermitian neutrino oscillations in matter with $\mathcal{PT}$
  symmetric {H}amiltonians},}\ }\href
  {https://doi.org/10.1209/0295-5075/113/61001} {\bibfield  {journal} {\bibinfo
   {journal} {Europhysics Letters}\ }\textbf {\bibinfo {volume} {113}},\
  \bibinfo {pages} {61001}}\BibitemShut {NoStop}%
\bibitem [{\citenamefont {\"{O}zdemir}\ \emph {et~al.}(2019)\citenamefont
  {\"{O}zdemir}, \citenamefont {Rotter}, \citenamefont {Nori},\ and\
  \citenamefont {Yang}}]{pt699}%
  \BibitemOpen
  \bibfield  {author} {\bibinfo {author} {\bibnamefont {\"{O}zdemir},
  \bibfnamefont {\c{S}~K}}, \bibinfo {author} {\bibfnamefont {S}~\bibnamefont
  {Rotter}}, \bibinfo {author} {\bibfnamefont {F}~\bibnamefont {Nori}}, and\
  \bibinfo {author} {\bibfnamefont {L}~\bibnamefont {Yang}}} (\bibinfo {year}
  {2019}),\ \bibfield  {title} {\enquote {\bibinfo {title} {Parity-time
  symmetry and exceptional points in photonics},}\ }\href
  {https://doi.org/10.1038/s41563-019-0304-9} {\bibfield  {journal} {\bibinfo
  {journal} {Nature Materials}\ }\textbf {\bibinfo {volume} {18}},\
  10.1038/s41563-019-0304-9}\BibitemShut {NoStop}%
\bibitem [{\citenamefont {Parisi}(1977)}]{pt603}%
  \BibitemOpen
  \bibfield  {author} {\bibinfo {author} {\bibnamefont {Parisi}, \bibfnamefont
  {G}}} (\bibinfo {year} {1977}),\ \bibfield  {title} {\enquote {\bibinfo
  {title} {Asymptotic estimates in perturbation theory},}\ }\href
  {https://doi.org/10.1016/0370-2693(77)90168-X} {\bibfield  {journal}
  {\bibinfo  {journal} {Physics Letters B}\ }\textbf {\bibinfo {volume} {66}},\
  10.1016/0370-2693(77)90168-X}\BibitemShut {NoStop}%
\bibitem [{\citenamefont {Park}\ \emph {et~al.}(2021)\citenamefont {Park},
  \citenamefont {Hwang}, \citenamefont {Choi},\ and\ \citenamefont
  {Yang}}]{pt686}%
  \BibitemOpen
  \bibfield  {author} {\bibinfo {author} {\bibnamefont {Park}, \bibfnamefont
  {Sungjoon}}, \bibinfo {author} {\bibfnamefont {Yoonseok}\ \bibnamefont
  {Hwang}}, \bibinfo {author} {\bibfnamefont {Hong~Chul}\ \bibnamefont {Choi}},
  and\ \bibinfo {author} {\bibfnamefont {Bohm-Jung}\ \bibnamefont {Yang}}}
  (\bibinfo {year} {2021}),\ \bibfield  {title} {\enquote {\bibinfo {title}
  {Topological acoustic triple point},}\ }\href
  {https://doi.org/10.1038/s41467-021-27158-y} {\bibfield  {journal} {\bibinfo
  {journal} {Nature Communications}\ }\textbf {\bibinfo {volume} {12}},\
  10.1038/s41467-021-27158-y}\BibitemShut {NoStop}%
\bibitem [{\citenamefont {Peng}\ \emph {et~al.}(2014)\citenamefont {Peng},
  \citenamefont {\c{S}ahin Kaya~\"{O}zdemir}, \citenamefont {Lei},
  \citenamefont {Monifi}, \citenamefont {Gianfreda}, \citenamefont {Long},
  \citenamefont {Fan}, \citenamefont {Nori}, \citenamefont {Bender},\ and\
  \citenamefont {Yang}}]{pt499}%
  \BibitemOpen
  \bibfield  {author} {\bibinfo {author} {\bibnamefont {Peng}, \bibfnamefont
  {Bo}}, \bibinfo {author} {\bibnamefont {\c{S}ahin Kaya~\"{O}zdemir}},
  \bibinfo {author} {\bibfnamefont {Fuchuan}\ \bibnamefont {Lei}}, \bibinfo
  {author} {\bibfnamefont {Faraz}\ \bibnamefont {Monifi}}, \bibinfo {author}
  {\bibfnamefont {Mariagiovanna}\ \bibnamefont {Gianfreda}}, \bibinfo {author}
  {\bibfnamefont {Gui~Lu}\ \bibnamefont {Long}}, \bibinfo {author}
  {\bibfnamefont {Shanhui}\ \bibnamefont {Fan}}, \bibinfo {author}
  {\bibfnamefont {Franco}\ \bibnamefont {Nori}}, \bibinfo {author}
  {\bibfnamefont {Carl~M}\ \bibnamefont {Bender}}, and\ \bibinfo {author}
  {\bibfnamefont {Lan}\ \bibnamefont {Yang}}} (\bibinfo {year} {2014}),\
  \bibfield  {title} {\enquote {\bibinfo {title} {Parity-time-symmetric
  whispering-gallery microcavities},}\ }\href
  {https://doi.org/10.1038/nphys2927} {\bibfield  {journal} {\bibinfo
  {journal} {Nature Physics}\ }\textbf {\bibinfo {volume} {10}},\
  10.1038/nphys2927}\BibitemShut {NoStop}%
\bibitem [{\citenamefont {Petermann}(1953)}]{pt590}%
  \BibitemOpen
  \bibfield  {author} {\bibinfo {author} {\bibnamefont {Petermann},
  \bibfnamefont {A}}} (\bibinfo {year} {1953}),\ \bibfield  {title} {\enquote
  {\bibinfo {title} {Divergence of perturbation expansion},}\ }\href
  {https://doi.org/10.1103/PhysRev.89.1160} {\bibfield  {journal} {\bibinfo
  {journal} {Physical Review}\ }\textbf {\bibinfo {volume} {89}},\
  10.1103/PhysRev.89.1160}\BibitemShut {NoStop}%
\bibitem [{\citenamefont {del Pino}\ \emph {et~al.}(2022)\citenamefont {del
  Pino}, \citenamefont {Slim},\ and\ \citenamefont {Verhagen}}]{pt664}%
  \BibitemOpen
  \bibfield  {author} {\bibinfo {author} {\bibnamefont {del Pino},
  \bibfnamefont {Javier}}, \bibinfo {author} {\bibfnamefont {Jesse~J}\
  \bibnamefont {Slim}}, and\ \bibinfo {author} {\bibfnamefont {Ewold}\
  \bibnamefont {Verhagen}}} (\bibinfo {year} {2022}),\ \bibfield  {title}
  {\enquote {\bibinfo {title} {Non-hermitian chiral phononics through
  optomechanically induced squeezing},}\ }\href
  {https://doi.org/10.1038/s41586-022-04609-0} {\bibfield  {journal} {\bibinfo
  {journal} {Nature}\ }\textbf {\bibinfo {volume} {606}},\
  10.1038/s41586-022-04609-0}\BibitemShut {NoStop}%
\bibitem [{\citenamefont {Qin}\ \emph {et~al.}(2021)\citenamefont {Qin},
  \citenamefont {Fu}, \citenamefont {Glasser},\ and\ \citenamefont
  {Yahalom}}]{pt619}%
  \BibitemOpen
  \bibfield  {author} {\bibinfo {author} {\bibnamefont {Qin}, \bibfnamefont
  {Hong}}, \bibinfo {author} {\bibfnamefont {Yichen}\ \bibnamefont {Fu}},
  \bibinfo {author} {\bibfnamefont {Alexander~S}\ \bibnamefont {Glasser}}, and\
  \bibinfo {author} {\bibfnamefont {Asher}\ \bibnamefont {Yahalom}}} (\bibinfo
  {year} {2021}),\ \bibfield  {title} {\enquote {\bibinfo {title} {Spontaneous
  and explicit parity-time-symmetry breaking in drift-wave instabilities},}\
  }\href {https://doi.org/10.1103/physreve.104.015215} {\bibfield  {journal}
  {\bibinfo  {journal} {Physical Review E}\ }\textbf {\bibinfo {volume}
  {104}},\ 10.1103/physreve.104.015215}\BibitemShut {NoStop}%
\bibitem [{\citenamefont {Qin}\ \emph {et~al.}(2019)\citenamefont {Qin},
  \citenamefont {Zhang}, \citenamefont {Glasser},\ and\ \citenamefont
  {Xiao}}]{pt617}%
  \BibitemOpen
  \bibfield  {author} {\bibinfo {author} {\bibnamefont {Qin}, \bibfnamefont
  {Hong}}, \bibinfo {author} {\bibfnamefont {Ruili}\ \bibnamefont {Zhang}},
  \bibinfo {author} {\bibfnamefont {Alexander~S}\ \bibnamefont {Glasser}}, and\
  \bibinfo {author} {\bibfnamefont {Jianyuan}\ \bibnamefont {Xiao}}} (\bibinfo
  {year} {2019}),\ \bibfield  {title} {\enquote {\bibinfo {title}
  {Kelvin-helmholtz instability is the result of parity-time symmetry
  breaking},}\ }\href {https://doi.org/10.1063/1.5088498} {\bibfield  {journal}
  {\bibinfo  {journal} {Physics of Plasmas}\ }\textbf {\bibinfo {volume}
  {26}},\ 10.1063/1.5088498}\BibitemShut {NoStop}%
\bibitem [{\citenamefont {Rechtsman}\ \emph {et~al.}(2012)\citenamefont
  {Rechtsman}, \citenamefont {Zeuner}, \citenamefont {T\"{u}nnermann},
  \citenamefont {Nolte}, \citenamefont {Segev},\ and\ \citenamefont
  {Szameit}}]{pt548}%
  \BibitemOpen
  \bibfield  {author} {\bibinfo {author} {\bibnamefont {Rechtsman},
  \bibfnamefont {Mikael~C}}, \bibinfo {author} {\bibfnamefont {Julia~M}\
  \bibnamefont {Zeuner}}, \bibinfo {author} {\bibfnamefont {Andreas}\
  \bibnamefont {T\"{u}nnermann}}, \bibinfo {author} {\bibfnamefont {Stefan}\
  \bibnamefont {Nolte}}, \bibinfo {author} {\bibfnamefont {Mordechai}\
  \bibnamefont {Segev}}, and\ \bibinfo {author} {\bibfnamefont {Alexander}\
  \bibnamefont {Szameit}}} (\bibinfo {year} {2012}),\ \bibfield  {title}
  {\enquote {\bibinfo {title} {Strain-induced pseudomagnetic field and photonic
  {L}andau levels in dielectric structures},}\ }\href
  {https://doi.org/10.1038/nphoton.2012.302} {\bibfield  {journal} {\bibinfo
  {journal} {Nature Photonics}\ }\textbf {\bibinfo {volume} {84}},\ \bibinfo
  {pages} {153--158}}\BibitemShut {NoStop}%
\bibitem [{\citenamefont {Regensburger}\ \emph {et~al.}(2012)\citenamefont
  {Regensburger}, \citenamefont {Bersch}, \citenamefont {Miri}, \citenamefont
  {Onishchukov}, \citenamefont {Christodoulides},\ and\ \citenamefont
  {Peschel}}]{pt527}%
  \BibitemOpen
  \bibfield  {author} {\bibinfo {author} {\bibnamefont {Regensburger},
  \bibfnamefont {Alois}}, \bibinfo {author} {\bibfnamefont {Christoph}\
  \bibnamefont {Bersch}}, \bibinfo {author} {\bibfnamefont {Mohammad-Ali}\
  \bibnamefont {Miri}}, \bibinfo {author} {\bibfnamefont {Georgy}\ \bibnamefont
  {Onishchukov}}, \bibinfo {author} {\bibfnamefont {Demetrios~N}\ \bibnamefont
  {Christodoulides}}, and\ \bibinfo {author} {\bibfnamefont {Ulf}\ \bibnamefont
  {Peschel}}} (\bibinfo {year} {2012}),\ \bibfield  {title} {\enquote {\bibinfo
  {title} {Parity-time synthetic photonic lattices},}\ }\href
  {https://doi.org/10.1038/nature11298} {\bibfield  {journal} {\bibinfo
  {journal} {Nature}\ }\textbf {\bibinfo {volume} {488}},\ \bibinfo {pages}
  {167--171}}\BibitemShut {NoStop}%
\bibitem [{\citenamefont {Regge}(1960)}]{pt613}%
  \BibitemOpen
  \bibfield  {author} {\bibinfo {author} {\bibnamefont {Regge}, \bibfnamefont
  {T}}} (\bibinfo {year} {1960}),\ \bibfield  {title} {\enquote {\bibinfo
  {title} {Bound states, shadow states and mandelstam representation},}\ }\href
  {https://doi.org/10.1007/BF02733035} {\bibfield  {journal} {\bibinfo
  {journal} {Il Nuovo Cimento}\ }\textbf {\bibinfo {volume} {18}},\
  10.1007/BF02733035}\BibitemShut {NoStop}%
\bibitem [{\citenamefont {Ren}\ \emph {et~al.}(2017{\natexlab{a}})\citenamefont
  {Ren}, \citenamefont {Hodaei}, \citenamefont {Harari}, \citenamefont
  {Hassan}, \citenamefont {Chow}, \citenamefont {Soltani}, \citenamefont
  {Christodoulides},\ and\ \citenamefont {Khajavikhan}}]{pt493}%
  \BibitemOpen
  \bibfield  {author} {\bibinfo {author} {\bibnamefont {Ren}, \bibfnamefont
  {J}}, \bibinfo {author} {\bibfnamefont {H}~\bibnamefont {Hodaei}}, \bibinfo
  {author} {\bibfnamefont {G}~\bibnamefont {Harari}}, \bibinfo {author}
  {\bibfnamefont {A~U}\ \bibnamefont {Hassan}}, \bibinfo {author}
  {\bibfnamefont {W}~\bibnamefont {Chow}}, \bibinfo {author} {\bibfnamefont
  {M}~\bibnamefont {Soltani}}, \bibinfo {author} {\bibfnamefont
  {D}~\bibnamefont {Christodoulides}}, and\ \bibinfo {author} {\bibfnamefont
  {M}~\bibnamefont {Khajavikhan}}} (\bibinfo {year} {2017}{\natexlab{a}}),\
  \bibfield  {title} {\enquote {\bibinfo {title} {Ultrasensitive micro-scale
  parity-time-symmetric ring laser gyroscope},}\ }\href
  {https://doi.org/10.1364/OL.42.001556} {\bibfield  {journal} {\bibinfo
  {journal} {Optics Letters}\ }\textbf {\bibinfo {volume} {42}},\
  10.1364/OL.42.001556}\BibitemShut {NoStop}%
\bibitem [{\citenamefont {Ren}\ \emph {et~al.}(2017{\natexlab{b}})\citenamefont
  {Ren}, \citenamefont {Hodaei}, \citenamefont {Harari}, \citenamefont
  {Hassan}, \citenamefont {Chow}, \citenamefont {Soltani}, \citenamefont
  {Christodoulides},\ and\ \citenamefont {Khajavikhan}}]{pt510}%
  \BibitemOpen
  \bibfield  {author} {\bibinfo {author} {\bibnamefont {Ren}, \bibfnamefont
  {J}}, \bibinfo {author} {\bibfnamefont {H}~\bibnamefont {Hodaei}}, \bibinfo
  {author} {\bibfnamefont {G}~\bibnamefont {Harari}}, \bibinfo {author}
  {\bibfnamefont {A~U}\ \bibnamefont {Hassan}}, \bibinfo {author}
  {\bibfnamefont {W}~\bibnamefont {Chow}}, \bibinfo {author} {\bibfnamefont
  {M}~\bibnamefont {Soltani}}, \bibinfo {author} {\bibfnamefont
  {D}~\bibnamefont {Christodoulides}}, and\ \bibinfo {author} {\bibfnamefont
  {M}~\bibnamefont {Khajavikhan}}} (\bibinfo {year} {2017}{\natexlab{b}}),\
  \bibfield  {title} {\enquote {\bibinfo {title} {Ultrasensitive micro-scale
  parity-time-symmetric ring laser gyroscope},}\ }\href
  {https://doi.org/10.1364/OL.42.001556} {\bibfield  {journal} {\bibinfo
  {journal} {Optics Letters}\ }\textbf {\bibinfo {volume} {42}},\ \bibinfo
  {pages} {1556--1559}}\BibitemShut {NoStop}%
\bibitem [{\citenamefont {Rotter}(2009)}]{pt672}%
  \BibitemOpen
  \bibfield  {author} {\bibinfo {author} {\bibnamefont {Rotter}, \bibfnamefont
  {Ingrid}}} (\bibinfo {year} {2009}),\ \bibfield  {title} {\enquote {\bibinfo
  {title} {A non-hermitian hamilton operator and the physics of open quantum
  systems},}\ }\href {https://doi.org/10.1088/1751-8113/42/15/153001}
  {\bibfield  {journal} {\bibinfo  {journal} {Journal of Physics A:
  Mathematical and Theoretical}\ }\textbf {\bibinfo {volume} {42}},\
  10.1088/1751-8113/42/15/153001}\BibitemShut {NoStop}%
\bibitem [{\citenamefont {Rubinstein}\ \emph {et~al.}(2007)\citenamefont
  {Rubinstein}, \citenamefont {Sternberg},\ and\ \citenamefont {Ma}}]{pt693}%
  \BibitemOpen
  \bibfield  {author} {\bibinfo {author} {\bibnamefont {Rubinstein},
  \bibfnamefont {J}}, \bibinfo {author} {\bibfnamefont {P}~\bibnamefont
  {Sternberg}}, and\ \bibinfo {author} {\bibfnamefont {Q}~\bibnamefont {Ma}}}
  (\bibinfo {year} {2007}),\ \bibfield  {title} {\enquote {\bibinfo {title}
  {Bifurcation diagram and pattern formation of phase slip centers in
  superconducting wires driven with electric currents},}\ }\href
  {https://doi.org/10.1103/physrevlett.99.167003} {\bibfield  {journal}
  {\bibinfo  {journal} {Physical Review Letters}\ }\textbf {\bibinfo {volume}
  {99}},\ 10.1103/physrevlett.99.167003}\BibitemShut {NoStop}%
\bibitem [{\citenamefont {R\"{u}ter}\ \emph {et~al.}(2010)\citenamefont
  {R\"{u}ter}, \citenamefont {Makris}, \citenamefont {El-Ganainy},
  \citenamefont {Christodoulides}, \citenamefont {Segev},\ and\ \citenamefont
  {Kip}}]{pt188}%
  \BibitemOpen
  \bibfield  {author} {\bibinfo {author} {\bibnamefont {R\"{u}ter},
  \bibfnamefont {Christian~E}}, \bibinfo {author} {\bibfnamefont
  {Konstantinos~G}\ \bibnamefont {Makris}}, \bibinfo {author} {\bibfnamefont
  {Ramy}\ \bibnamefont {El-Ganainy}}, \bibinfo {author} {\bibfnamefont
  {Demetrios~N}\ \bibnamefont {Christodoulides}}, \bibinfo {author}
  {\bibfnamefont {Mordechai}\ \bibnamefont {Segev}}, and\ \bibinfo {author}
  {\bibfnamefont {Detlef}\ \bibnamefont {Kip}}} (\bibinfo {year} {2010}),\
  \bibfield  {title} {\enquote {\bibinfo {title} {Observation of parity-time
  symmetry in optics},}\ }\href {https://doi.org/10.1038/nphys1515} {\bibfield
  {journal} {\bibinfo  {journal} {Nature Physics}\ }\textbf {\bibinfo {volume}
  {6}},\ \bibinfo {pages} {192--195}}\BibitemShut {NoStop}%
\bibitem [{\citenamefont {Sakhdari}\ \emph {et~al.}(2017)\citenamefont
  {Sakhdari}, \citenamefont {Farhat},\ and\ \citenamefont {Chen}}]{pt494}%
  \BibitemOpen
  \bibfield  {author} {\bibinfo {author} {\bibnamefont {Sakhdari},
  \bibfnamefont {Maryam}}, \bibinfo {author} {\bibfnamefont {Mohamed}\
  \bibnamefont {Farhat}}, and\ \bibinfo {author} {\bibfnamefont {Pai-Yen}\
  \bibnamefont {Chen}}} (\bibinfo {year} {2017}),\ \bibfield  {title} {\enquote
  {\bibinfo {title} {$\mathcal{PT}$-symmetric metasurfaces: wave manipulation
  and sensing using singular points},}\ }\href
  {https://doi.org/10.1088/1367-2630/aa6bb9} {\bibfield  {journal} {\bibinfo
  {journal} {New Journal of Physics}\ }\textbf {\bibinfo {volume} {19}},\
  10.1088/1367-2630/aa6bb9}\BibitemShut {NoStop}%
\bibitem [{\citenamefont {Schindler}\ \emph {et~al.}(2011)\citenamefont
  {Schindler}, \citenamefont {Li}, \citenamefont {Zheng}, \citenamefont
  {Ellis},\ and\ \citenamefont {Kottos}}]{pt488}%
  \BibitemOpen
  \bibfield  {author} {\bibinfo {author} {\bibnamefont {Schindler},
  \bibfnamefont {Joseph}}, \bibinfo {author} {\bibfnamefont {Ang}\ \bibnamefont
  {Li}}, \bibinfo {author} {\bibfnamefont {Mei~C}\ \bibnamefont {Zheng}},
  \bibinfo {author} {\bibfnamefont {F~M}\ \bibnamefont {Ellis}}, and\ \bibinfo
  {author} {\bibfnamefont {Tsampikos}\ \bibnamefont {Kottos}}} (\bibinfo {year}
  {2011}),\ \bibfield  {title} {\enquote {\bibinfo {title} {Experimental study
  of active lrc circuits with $\mathcal{PT}$ symmetries},}\ }\href
  {https://doi.org/10.1103/PhysRevA.84.040101} {\bibfield  {journal} {\bibinfo
  {journal} {Physical Review A}\ }\textbf {\bibinfo {volume} {84}},\
  10.1103/PhysRevA.84.040101}\BibitemShut {NoStop}%
\bibitem [{\citenamefont {Schindler}\ \emph
  {et~al.}(2022{\natexlab{a}})\citenamefont {Schindler}, \citenamefont
  {Schindler},\ and\ \citenamefont {Ogilvie}}]{pt698}%
  \BibitemOpen
  \bibfield  {author} {\bibinfo {author} {\bibnamefont {Schindler},
  \bibfnamefont {M}}, \bibinfo {author} {\bibfnamefont {S}~\bibnamefont
  {Schindler}}, and\ \bibinfo {author} {\bibfnamefont {M}~\bibnamefont
  {Ogilvie}}} (\bibinfo {year} {2022}{\natexlab{a}}),\ \bibfield  {title}
  {\enquote {\bibinfo {title} {Finite-density qcd, $\mathcal{PT}$ symmetry, and
  dual algorithms},}\ }in\ \href {https://doi.org/10.22323/1.396.0417} {\emph
  {\bibinfo {booktitle} {Proceedings of Science: The 38th International
  Symposium on Lattice Field Theory (LATTICE2021)}}},\ Vol.\ \bibinfo {volume}
  {396},\ p.\ \bibinfo {pages} {9 pages}\BibitemShut {NoStop}%
\bibitem [{\citenamefont {Schindler}\ \emph
  {et~al.}(2022{\natexlab{b}})\citenamefont {Schindler}, \citenamefont
  {Schindler},\ and\ \citenamefont {Ogilvie}}]{pt697}%
  \BibitemOpen
  \bibfield  {author} {\bibinfo {author} {\bibnamefont {Schindler},
  \bibfnamefont {M}}, \bibinfo {author} {\bibfnamefont {S}~\bibnamefont
  {Schindler}}, and\ \bibinfo {author} {\bibfnamefont {M}~\bibnamefont
  {Ogilvie}}} (\bibinfo {year} {2022}{\natexlab{b}}),\ \bibfield  {title}
  {\enquote {\bibinfo {title} {Finite-density qcd, $\mathcal{PT}$ symmetry, and
  exotic phases},}\ }in\ \href {https://doi.org/10.22323/1.396.0555} {\emph
  {\bibinfo {booktitle} {Proceedings of Science: The 38th International
  Symposium on Lattice Field Theory (LATTICE2021)}}},\ Vol.\ \bibinfo {volume}
  {396},\ p.\ \bibinfo {pages} {9 pages}\BibitemShut {NoStop}%
\bibitem [{\citenamefont {Schindler}\ \emph {et~al.}(2020)\citenamefont
  {Schindler}, \citenamefont {Schindler}, \citenamefont {Medina},\ and\
  \citenamefont {Ogilvie}}]{pt695}%
  \BibitemOpen
  \bibfield  {author} {\bibinfo {author} {\bibnamefont {Schindler},
  \bibfnamefont {Moses~A}}, \bibinfo {author} {\bibfnamefont {Stella~T}\
  \bibnamefont {Schindler}}, \bibinfo {author} {\bibfnamefont {Leandro}\
  \bibnamefont {Medina}}, and\ \bibinfo {author} {\bibfnamefont {Michael~C}\
  \bibnamefont {Ogilvie}}} (\bibinfo {year} {2020}),\ \bibfield  {title}
  {\enquote {\bibinfo {title} {Universality of pattern formation},}\ }\href
  {https://doi.org/10.1103/physrevd.102.114510} {\bibfield  {journal} {\bibinfo
   {journal} {Physical Review D}\ }\textbf {\bibinfo {volume} {102}},\
  10.1103/physrevd.102.114510}\BibitemShut {NoStop}%
\bibitem [{\citenamefont {Schindler}\ \emph {et~al.}(2021)\citenamefont
  {Schindler}, \citenamefont {Schindler},\ and\ \citenamefont
  {Ogilvie}}]{pt696}%
  \BibitemOpen
  \bibfield  {author} {\bibinfo {author} {\bibnamefont {Schindler},
  \bibfnamefont {Moses~A}}, \bibinfo {author} {\bibfnamefont {Stella~T}\
  \bibnamefont {Schindler}}, and\ \bibinfo {author} {\bibfnamefont {Michael~C}\
  \bibnamefont {Ogilvie}}} (\bibinfo {year} {2021}),\ \bibfield  {title}
  {\enquote {\bibinfo {title} {Pt symmetry, pattern formation, and
  finite-density qcd},}\ }\href
  {https://doi.org/10.1088/1742-6596/2038/1/012022} {\bibfield  {journal}
  {\bibinfo  {journal} {Journal of Physics Conference Series}\ }\textbf
  {\bibinfo {volume} {2038}},\ 10.1088/1742-6596/2038/1/012022}\BibitemShut
  {NoStop}%
\bibitem [{\citenamefont {Schindler}\ and\ \citenamefont
  {Bender}(2018)}]{pt376}%
  \BibitemOpen
  \bibfield  {author} {\bibinfo {author} {\bibnamefont {Schindler},
  \bibfnamefont {Stella~T}}, and\ \bibinfo {author} {\bibfnamefont {Carl~M}\
  \bibnamefont {Bender}}} (\bibinfo {year} {2018}),\ \bibfield  {title}
  {\enquote {\bibinfo {title} {Winding in non-{H}ermitian systems},}\ }\href
  {https://doi.org/10.1088/1751-8121/aa9faf} {\bibfield  {journal} {\bibinfo
  {journal} {Journal of Physics A: Mathematical and Theoretical}\ }\textbf
  {\bibinfo {volume} {51}},\ 10.1088/1751-8121/aa9faf}\BibitemShut {NoStop}%
\bibitem [{\citenamefont {Schnabel}\ \emph {et~al.}(2017)\citenamefont
  {Schnabel}, \citenamefont {Cartarius}, \citenamefont {Main}, \citenamefont
  {Wunner},\ and\ \citenamefont {Heiss}}]{pt669}%
  \BibitemOpen
  \bibfield  {author} {\bibinfo {author} {\bibnamefont {Schnabel},
  \bibfnamefont {Jan}}, \bibinfo {author} {\bibfnamefont {Holger}\ \bibnamefont
  {Cartarius}}, \bibinfo {author} {\bibfnamefont {J{\"o}rg}\ \bibnamefont
  {Main}}, \bibinfo {author} {\bibfnamefont {G{\"u}nter}\ \bibnamefont
  {Wunner}}, and\ \bibinfo {author} {\bibfnamefont {Walter~Dieter}\
  \bibnamefont {Heiss}}} (\bibinfo {year} {2017}),\ \bibfield  {title}
  {\enquote {\bibinfo {title} {Pt-symmetric waveguide system with evidence of a
  third-order exceptional point},}\ }\href
  {https://doi.org/10.1103/physreva.95.053868} {\bibfield  {journal} {\bibinfo
  {journal} {Physical Review A}\ }\textbf {\bibinfo {volume} {95}},\
  10.1103/physreva.95.053868}\BibitemShut {NoStop}%
\bibitem [{\citenamefont {Schurig}\ \emph {et~al.}(2006)\citenamefont
  {Schurig}, \citenamefont {Mock}, \citenamefont {Justice}, \citenamefont
  {Cummer}, \citenamefont {Pendry}, \citenamefont {Starr},\ and\ \citenamefont
  {Smith}}]{pt233}%
  \BibitemOpen
  \bibfield  {author} {\bibinfo {author} {\bibnamefont {Schurig}, \bibfnamefont
  {D}}, \bibinfo {author} {\bibfnamefont {J~J}\ \bibnamefont {Mock}}, \bibinfo
  {author} {\bibfnamefont {B~J}\ \bibnamefont {Justice}}, \bibinfo {author}
  {\bibfnamefont {S.~A}\ \bibnamefont {Cummer}}, \bibinfo {author}
  {\bibfnamefont {J~B}\ \bibnamefont {Pendry}}, \bibinfo {author}
  {\bibfnamefont {A~F}\ \bibnamefont {Starr}}, and\ \bibinfo {author}
  {\bibfnamefont {D~R}\ \bibnamefont {Smith}}} (\bibinfo {year} {2006}),\
  \bibfield  {title} {\enquote {\bibinfo {title} {Metamaterial electromagnetic
  cloak at microwave frequencies},}\ }\href
  {https://doi.org/10.1126/science.1133628} {\bibfield  {journal} {\bibinfo
  {journal} {Science}\ }\textbf {\bibinfo {volume} {314}},\ \bibinfo {pages}
  {977--980}}\BibitemShut {NoStop}%
\bibitem [{\citenamefont {Simon}\ and\ \citenamefont {Dicke}(1970)}]{pt17}%
  \BibitemOpen
  \bibfield  {author} {\bibinfo {author} {\bibnamefont {Simon}, \bibfnamefont
  {Barry}}, and\ \bibinfo {author} {\bibfnamefont {A}~\bibnamefont {Dicke}}}
  (\bibinfo {year} {1970}),\ \bibfield  {title} {\enquote {\bibinfo {title}
  {Coupling constant analyticity for the anharmonic oscillator},}\ }\href
  {https://doi.org/10.1016/0003-4916(70)90240-X} {\bibfield  {journal}
  {\bibinfo  {journal} {Annals of Physics}\ }\textbf {\bibinfo {volume}
  {58}}~(\bibinfo {number} {1}),\ \bibinfo {pages} {76--136}}\BibitemShut
  {NoStop}%
\bibitem [{\citenamefont {Smilga}(2009)}]{pt676}%
  \BibitemOpen
  \bibfield  {author} {\bibinfo {author} {\bibnamefont {Smilga}, \bibfnamefont
  {A~V}}} (\bibinfo {year} {2009}),\ \bibfield  {title} {\enquote {\bibinfo
  {title} {Exceptional points in quantum and classical dynamics},}\ }\href
  {https://doi.org/10.1088/1751-8113/42/9/095301} {\bibfield  {journal}
  {\bibinfo  {journal} {Journal of Physics A: Mathematical and Theoretical}\
  }\textbf {\bibinfo {volume} {42}},\
  10.1088/1751-8113/42/9/095301}\BibitemShut {NoStop}%
\bibitem [{\citenamefont {Soley}\ \emph {et~al.}(2023)\citenamefont {Soley},
  \citenamefont {Bender},\ and\ \citenamefont {Stone}}]{pt595}%
  \BibitemOpen
  \bibfield  {author} {\bibinfo {author} {\bibnamefont {Soley}, \bibfnamefont
  {Micheline~B}}, \bibinfo {author} {\bibfnamefont {Carl~M}\ \bibnamefont
  {Bender}}, and\ \bibinfo {author} {\bibfnamefont {A~Douglas}\ \bibnamefont
  {Stone}}} (\bibinfo {year} {2023}),\ \bibfield  {title} {\enquote {\bibinfo
  {title} {Experimentally realizable $\mathcal{PT}$ phase transitions in
  reflectionless quantum scattering},}\ }\href
  {https://doi.org/10.1103/PhysRevLett.130.250404} {\bibfield  {journal}
  {\bibinfo  {journal} {Physical Review Letters}\ }\textbf {\bibinfo {volume}
  {130}},\ 10.1103/PhysRevLett.130.250404}\BibitemShut {NoStop}%
\bibitem [{\citenamefont {Sounas}\ \emph {et~al.}(2015)\citenamefont {Sounas},
  \citenamefont {Fleury},\ and\ \citenamefont {Al\'{u}}}]{pt237}%
  \BibitemOpen
  \bibfield  {author} {\bibinfo {author} {\bibnamefont {Sounas}, \bibfnamefont
  {Dimitrios~L}}, \bibinfo {author} {\bibfnamefont {Romain}\ \bibnamefont
  {Fleury}}, and\ \bibinfo {author} {\bibfnamefont {Andrea}\ \bibnamefont
  {Al\'{u}}}} (\bibinfo {year} {2015}),\ \bibfield  {title} {\enquote {\bibinfo
  {title} {Unidirectional cloaking based on metasurfaces with balanced loss and
  gain},}\ }\href {https://doi.org/10.1103/PhysRevApplied.4.014005} {\bibfield
  {journal} {\bibinfo  {journal} {Physical Review Applied}\ }\textbf {\bibinfo
  {volume} {4}},\ \bibinfo {pages} {014005}}\BibitemShut {NoStop}%
\bibitem [{\citenamefont {Streater}\ and\ \citenamefont
  {Wightman}(2000)}]{pt566}%
  \BibitemOpen
  \bibfield  {author} {\bibinfo {author} {\bibnamefont {Streater},
  \bibfnamefont {Raymond~F}}, and\ \bibinfo {author} {\bibfnamefont {Arthur~S}\
  \bibnamefont {Wightman}}} (\bibinfo {year} {2000}),\ \href@noop {} {\emph
  {\bibinfo {title} {{PCT}, Spin and Statistics, and All That}}}\ (\bibinfo
  {publisher} {Princeton},\ \bibinfo {address} {New Jersey})\BibitemShut
  {NoStop}%
\bibitem [{\citenamefont {Suchkov}\ \emph {et~al.}(2016)\citenamefont
  {Suchkov}, \citenamefont {Sukhorukov}, \citenamefont {Huang}, \citenamefont
  {Dmitriev}, \citenamefont {Lee},\ and\ \citenamefont {Kivshar}}]{pt227}%
  \BibitemOpen
  \bibfield  {author} {\bibinfo {author} {\bibnamefont {Suchkov}, \bibfnamefont
  {Sergey~V}}, \bibinfo {author} {\bibfnamefont {Andrey~A}\ \bibnamefont
  {Sukhorukov}}, \bibinfo {author} {\bibfnamefont {Jiahao}\ \bibnamefont
  {Huang}}, \bibinfo {author} {\bibfnamefont {Sergey~V}\ \bibnamefont
  {Dmitriev}}, \bibinfo {author} {\bibfnamefont {Chaohong}\ \bibnamefont
  {Lee}}, and\ \bibinfo {author} {\bibfnamefont {Yuri~S}\ \bibnamefont
  {Kivshar}}} (\bibinfo {year} {2016}),\ \bibfield  {title} {\enquote {\bibinfo
  {title} {Nonlinear switching and solitons in $\cal{PT}$-symmetric photonic
  systems},}\ }\href {https://doi.org/10.1002/lpor.201500227} {\bibfield
  {journal} {\bibinfo  {journal} {Laser and Photonics Reviews}\ }\textbf
  {\bibinfo {volume} {10}},\ \bibinfo {pages} {177--213}}\BibitemShut {NoStop}%
\bibitem [{\citenamefont {Symanzik}(1975)}]{pt586}%
  \BibitemOpen
  \bibfield  {author} {\bibinfo {author} {\bibnamefont {Symanzik},
  \bibfnamefont {K}}} (\bibinfo {year} {1975}),\ \bibfield  {title} {\enquote
  {\bibinfo {title} {Renormalization problem in nonrenormalizable massless
  $\phi^4$ theory},}\ }\href {https://doi.org/10.1007/BF01609868} {\bibfield
  {journal} {\bibinfo  {journal} {Communications in Mathematical Physics}\
  }\textbf {\bibinfo {volume} {45}},\ 10.1007/BF01609868}\BibitemShut {NoStop}%
\bibitem [{\citenamefont {Szameit}\ \emph {et~al.}(2011)\citenamefont
  {Szameit}, \citenamefont {Rechtsman}, \citenamefont {Bahat-Treidel},\ and\
  \citenamefont {Segev}}]{pt547}%
  \BibitemOpen
  \bibfield  {author} {\bibinfo {author} {\bibnamefont {Szameit}, \bibfnamefont
  {Alexander}}, \bibinfo {author} {\bibfnamefont {Mikael~C}\ \bibnamefont
  {Rechtsman}}, \bibinfo {author} {\bibfnamefont {Omri}\ \bibnamefont
  {Bahat-Treidel}}, and\ \bibinfo {author} {\bibfnamefont {Mordechai}\
  \bibnamefont {Segev}}} (\bibinfo {year} {2011}),\ \bibfield  {title}
  {\enquote {\bibinfo {title} {$\mathcal{PT}$-symmetry in honeycomb photonic
  lattices},}\ }\href {https://doi.org/10.1103/physreva.84.021806} {\bibfield
  {journal} {\bibinfo  {journal} {Physical Review A}\ }\textbf {\bibinfo
  {volume} {84}},\ \bibinfo {pages} {021806}}\BibitemShut {NoStop}%
\bibitem [{\citenamefont {Thirring}(1953)}]{pt589}%
  \BibitemOpen
  \bibfield  {author} {\bibinfo {author} {\bibnamefont {Thirring},
  \bibfnamefont {Walter}}} (\bibinfo {year} {1953}),\ \bibfield  {title}
  {\enquote {\bibinfo {title} {On the divergence of perturbation theory for
  quantized fields},}\ }\href {https://doi.org/10.5169/seals-112398} {\bibfield
   {journal} {\bibinfo  {journal} {Helvetica Physica Acta}\ }\textbf {\bibinfo
  {volume} {26}},\ 10.5169/seals-112398}\BibitemShut {NoStop}%
\bibitem [{\citenamefont {Turok}(2014)}]{pt585}%
  \BibitemOpen
  \bibfield  {author} {\bibinfo {author} {\bibnamefont {Turok}, \bibfnamefont
  {Neil}}} (\bibinfo {year} {2014}),\ \bibfield  {title} {\enquote {\bibinfo
  {title} {On quantum tunneling in real time},}\ }\href
  {https://doi.org/10.1088/1367-2630/16/6/063006} {\bibfield  {journal}
  {\bibinfo  {journal} {New Journal of Physics}\ }\textbf {\bibinfo {volume}
  {16}},\ 10.1088/1367-2630/16/6/063006}\BibitemShut {NoStop}%
\bibitem [{\citenamefont {Ushveridze}(1994)}]{pt157}%
  \BibitemOpen
  \bibfield  {author} {\bibinfo {author} {\bibnamefont {Ushveridze},
  \bibfnamefont {A~G}}} (\bibinfo {year} {1994}),\ \href@noop {} {\emph
  {\bibinfo {title} {Quasi-exactly solvable models in quantum mechanics}}}\
  (\bibinfo  {publisher} {IOP},\ \bibinfo {address} {Bristol})\BibitemShut
  {NoStop}%
\bibitem [{\citenamefont {V\'azquez-{C}andanedo}\ \emph
  {et~al.}(2014)\citenamefont {V\'azquez-{C}andanedo}, \citenamefont
  {Hern\'andez-{H}errej\'{o}n}, \citenamefont {Izrailev},\ and\ \citenamefont
  {Christodoulides}}]{pt220}%
  \BibitemOpen
  \bibfield  {author} {\bibinfo {author} {\bibnamefont {V\'azquez-{C}andanedo},
  \bibfnamefont {O}}, \bibinfo {author} {\bibfnamefont {J~C}\ \bibnamefont
  {Hern\'andez-{H}errej\'{o}n}}, \bibinfo {author} {\bibfnamefont {F~M}\
  \bibnamefont {Izrailev}}, and\ \bibinfo {author} {\bibfnamefont {D~N}\
  \bibnamefont {Christodoulides}}} (\bibinfo {year} {2014}),\ \bibfield
  {title} {\enquote {\bibinfo {title} {{Gain- or loss-induced localization in
  one-dimensional $\cal{PT}$-symmetric tight-binding models}},}\ }\href
  {https://doi.org/10.1103/PhysRevA.89.013832} {\bibfield  {journal} {\bibinfo
  {journal} {Physical Review A}\ }\textbf {\bibinfo {volume} {89}},\ \bibinfo
  {pages} {013832}}\BibitemShut {NoStop}%
\bibitem [{\citenamefont {Veneziano}(1968)}]{pt626}%
  \BibitemOpen
  \bibfield  {author} {\bibinfo {author} {\bibnamefont {Veneziano},
  \bibfnamefont {G}}} (\bibinfo {year} {1968}),\ \bibfield  {title} {\enquote
  {\bibinfo {title} {Construction of a crossing-simmetric, {R}egge-behaved
  amplitude for linearly rising trajectories},}\ }\href
  {https://doi.org/10.1007/BF02824451} {\bibfield  {journal} {\bibinfo
  {journal} {Il Nuovo Cimento A}\ }\textbf {\bibinfo {volume} {57}},\
  10.1007/BF02824451}\BibitemShut {NoStop}%
\bibitem [{\citenamefont {Weidemann}\ \emph {et~al.}(2022)\citenamefont
  {Weidemann}, \citenamefont {Kremer}, \citenamefont {Longhi},\ and\
  \citenamefont {Szameit}}]{pt665}%
  \BibitemOpen
  \bibfield  {author} {\bibinfo {author} {\bibnamefont {Weidemann},
  \bibfnamefont {Sebastian}}, \bibinfo {author} {\bibfnamefont {Mark}\
  \bibnamefont {Kremer}}, \bibinfo {author} {\bibfnamefont {Stefano}\
  \bibnamefont {Longhi}}, and\ \bibinfo {author} {\bibfnamefont {Alexander}\
  \bibnamefont {Szameit}}} (\bibinfo {year} {2022}),\ \bibfield  {title}
  {\enquote {\bibinfo {title} {Topological triple phase transition in
  non-hermitian floquet quasicrystals},}\ }\href
  {https://doi.org/10.1038/s41586-021-04253-0} {\bibfield  {journal} {\bibinfo
  {journal} {Nature}\ }\textbf {\bibinfo {volume} {601}},\
  10.1038/s41586-021-04253-0}\BibitemShut {NoStop}%
\bibitem [{\citenamefont {Weigert}(2003)}]{pt370}%
  \BibitemOpen
  \bibfield  {author} {\bibinfo {author} {\bibnamefont {Weigert}, \bibfnamefont
  {S}}} (\bibinfo {year} {2003}),\ \bibfield  {title} {\enquote {\bibinfo
  {title} {Completeness and orthonormality in $\mathcal{PT}$-symmetric quantum
  systems},}\ }\href {https://doi.org/10.1103/PhysRevA.68.062111} {\bibfield
  {journal} {\bibinfo  {journal} {Physical Review A}\ }\textbf {\bibinfo
  {volume} {68}},\ 10.1103/PhysRevA.68.062111}\BibitemShut {NoStop}%
\bibitem [{\citenamefont {Weimann}\ \emph {et~al.}(2016)\citenamefont
  {Weimann}, \citenamefont {Kremer}, \citenamefont {Plotnik}, \citenamefont
  {Lumer}, \citenamefont {Nolte}, \citenamefont {Makris}, \citenamefont
  {Segev}, \citenamefont {Rechtsman},\ and\ \citenamefont {Szameit}}]{pt543}%
  \BibitemOpen
  \bibfield  {author} {\bibinfo {author} {\bibnamefont {Weimann}, \bibfnamefont
  {S}}, \bibinfo {author} {\bibfnamefont {M}~\bibnamefont {Kremer}}, \bibinfo
  {author} {\bibfnamefont {Y}~\bibnamefont {Plotnik}}, \bibinfo {author}
  {\bibfnamefont {Y}~\bibnamefont {Lumer}}, \bibinfo {author} {\bibfnamefont
  {S}~\bibnamefont {Nolte}}, \bibinfo {author} {\bibfnamefont {K~G}\
  \bibnamefont {Makris}}, \bibinfo {author} {\bibfnamefont {M}~\bibnamefont
  {Segev}}, \bibinfo {author} {\bibfnamefont {M~C}\ \bibnamefont {Rechtsman}},
  and\ \bibinfo {author} {\bibfnamefont {A}~\bibnamefont {Szameit}}} (\bibinfo
  {year} {2016}),\ \bibfield  {title} {\enquote {\bibinfo {title}
  {Topologically protected bound states in photonic parity-time-symmetric
  crystals},}\ }\href {https://doi.org/10.1038/nmat4811} {\bibfield  {journal}
  {\bibinfo  {journal} {Nature Materials}\ }\textbf {\bibinfo {volume} {16}},\
  \bibinfo {pages} {433--438}}\BibitemShut {NoStop}%
\bibitem [{\citenamefont {West}\ \emph {et~al.}(2010)\citenamefont {West},
  \citenamefont {Kottos},\ and\ \citenamefont {Prosen}}]{pt680}%
  \BibitemOpen
  \bibfield  {author} {\bibinfo {author} {\bibnamefont {West}, \bibfnamefont
  {Carl~T}}, \bibinfo {author} {\bibfnamefont {Tsampikos}\ \bibnamefont
  {Kottos}}, and\ \bibinfo {author} {\bibfnamefont {Toma{\v{z}}}\ \bibnamefont
  {Prosen}}} (\bibinfo {year} {2010}),\ \bibfield  {title} {\enquote {\bibinfo
  {title} {Pt-symmetric wave chaos},}\ }\href
  {https://doi.org/10.1103/physrevlett.104.054102} {\bibfield  {journal}
  {\bibinfo  {journal} {Physical Review Letters}\ }\textbf {\bibinfo {volume}
  {104}},\ 10.1103/physrevlett.104.054102}\BibitemShut {NoStop}%
\bibitem [{\citenamefont {Wong}\ \emph {et~al.}(2016)\citenamefont {Wong},
  \citenamefont {Xu}, \citenamefont {Kim}, \citenamefont {O'Brien},
  \citenamefont {Wang}, \citenamefont {Feng},\ and\ \citenamefont
  {Zhang}}]{pt550}%
  \BibitemOpen
  \bibfield  {author} {\bibinfo {author} {\bibnamefont {Wong}, \bibfnamefont
  {Zi~Jing}}, \bibinfo {author} {\bibfnamefont {Ye-Long}\ \bibnamefont {Xu}},
  \bibinfo {author} {\bibfnamefont {Jeongmin}\ \bibnamefont {Kim}}, \bibinfo
  {author} {\bibfnamefont {Kevin}\ \bibnamefont {O'Brien}}, \bibinfo {author}
  {\bibfnamefont {Yuan}\ \bibnamefont {Wang}}, \bibinfo {author} {\bibfnamefont
  {Liang}\ \bibnamefont {Feng}}, and\ \bibinfo {author} {\bibfnamefont {Xiang}\
  \bibnamefont {Zhang}}} (\bibinfo {year} {2016}),\ \bibfield  {title}
  {\enquote {\bibinfo {title} {Lasing and anti-lasing in a single cavity},}\
  }\href {https://doi.org/10.1038/nphoton.2016.216} {\bibfield  {journal}
  {\bibinfo  {journal} {Nature Photonics}\ }\textbf {\bibinfo {volume} {10}},\
  \bibinfo {pages} {796--801}}\BibitemShut {NoStop}%
\bibitem [{\citenamefont {Wu}\ \emph {et~al.}(1957)\citenamefont {Wu},
  \citenamefont {Ambler}, \citenamefont {Hayward}, \citenamefont {Hoppes},\
  and\ \citenamefont {Hudson}}]{pt581}%
  \BibitemOpen
  \bibfield  {author} {\bibinfo {author} {\bibnamefont {Wu}, \bibfnamefont
  {C~S}}, \bibinfo {author} {\bibfnamefont {E}~\bibnamefont {Ambler}}, \bibinfo
  {author} {\bibfnamefont {R~W}\ \bibnamefont {Hayward}}, \bibinfo {author}
  {\bibfnamefont {D~D}\ \bibnamefont {Hoppes}}, and\ \bibinfo {author}
  {\bibfnamefont {R~P}\ \bibnamefont {Hudson}}} (\bibinfo {year} {1957}),\
  \bibfield  {title} {\enquote {\bibinfo {title} {Experimental test of parity
  conservation in beta decay},}\ }\href
  {https://doi.org/10.1103/PhysRev.105.1413} {\bibfield  {journal} {\bibinfo
  {journal} {Physical Review}\ }\textbf {\bibinfo {volume} {105}},\
  10.1103/PhysRev.105.1413}\BibitemShut {NoStop}%
\bibitem [{\citenamefont {Yang}\ \emph {et~al.}(2022)\citenamefont {Yang},
  \citenamefont {Xie}, \citenamefont {Li}, \citenamefont {Zhang}, \citenamefont
  {Peng}, \citenamefont {Wang}, \citenamefont {Li}, \citenamefont {Li},
  \citenamefont {Chen},\ and\ \citenamefont {Gao}}]{pt667}%
  \BibitemOpen
  \bibfield  {author} {\bibinfo {author} {\bibnamefont {Yang}, \bibfnamefont
  {Yumeng}}, \bibinfo {author} {\bibfnamefont {Xinrong}\ \bibnamefont {Xie}},
  \bibinfo {author} {\bibfnamefont {Yuanzhen}\ \bibnamefont {Li}}, \bibinfo
  {author} {\bibfnamefont {Zijian}\ \bibnamefont {Zhang}}, \bibinfo {author}
  {\bibfnamefont {Yiwei}\ \bibnamefont {Peng}}, \bibinfo {author}
  {\bibfnamefont {Chi}\ \bibnamefont {Wang}}, \bibinfo {author} {\bibfnamefont
  {Erping}\ \bibnamefont {Li}}, \bibinfo {author} {\bibfnamefont {Ying}\
  \bibnamefont {Li}}, \bibinfo {author} {\bibfnamefont {Hongsheng}\
  \bibnamefont {Chen}}, and\ \bibinfo {author} {\bibfnamefont {Fei}\
  \bibnamefont {Gao}}} (\bibinfo {year} {2022}),\ \bibfield  {title} {\enquote
  {\bibinfo {title} {Radiative anti-parity-time plasmonics},}\ }\href
  {https://doi.org/10.1038/s41467-022-35447-3} {\bibfield  {journal} {\bibinfo
  {journal} {Nature Communications}\ }\textbf {\bibinfo {volume} {13}},\
  10.1038/s41467-022-35447-3}\BibitemShut {NoStop}%
\bibitem [{\citenamefont {Zezyulin}\ and\ \citenamefont
  {Konotop}(2012)}]{pt226}%
  \BibitemOpen
  \bibfield  {author} {\bibinfo {author} {\bibnamefont {Zezyulin},
  \bibfnamefont {D~A}}, and\ \bibinfo {author} {\bibfnamefont {V~V}\
  \bibnamefont {Konotop}}} (\bibinfo {year} {2012}),\ \bibfield  {title}
  {\enquote {\bibinfo {title} {Nonlinear modes in finite-dimensional
  $\cal{PT}$-symmetric systems},}\ }\href
  {https://doi.org/10.1103/PhysRevLett.108.213906} {\bibfield  {journal}
  {\bibinfo  {journal} {Physical Review Letters}\ }\textbf {\bibinfo {volume}
  {108}},\ \bibinfo {pages} {213906}}\BibitemShut {NoStop}%
\bibitem [{\citenamefont {Zhang}\ \emph {et~al.}(2022)\citenamefont {Zhang},
  \citenamefont {Yang}, \citenamefont {Sheng},\ and\ \citenamefont
  {Wu}}]{pt685}%
  \BibitemOpen
  \bibfield  {author} {\bibinfo {author} {\bibnamefont {Zhang}, \bibfnamefont
  {Qiankun}}, \bibinfo {author} {\bibfnamefont {Cheng}\ \bibnamefont {Yang}},
  \bibinfo {author} {\bibfnamefont {Jiteng}\ \bibnamefont {Sheng}}, and\
  \bibinfo {author} {\bibfnamefont {Haibin}\ \bibnamefont {Wu}}} (\bibinfo
  {year} {2022}),\ \bibfield  {title} {\enquote {\bibinfo {title} {Dissipative
  coupling-induced phonon lasing},}\ }\href
  {https://doi.org/10.1073/pnas.2207543119} {\bibfield  {journal} {\bibinfo
  {journal} {Proceedings of the National Academy of Sciences of the United
  States of America}\ }\textbf {\bibinfo {volume} {119}},\
  10.1073/pnas.2207543119}\BibitemShut {NoStop}%
\bibitem [{\citenamefont {Zhang}\ \emph {et~al.}(2017)\citenamefont {Zhang},
  \citenamefont {Wu},\ and\ \citenamefont {Zhang}}]{pt558}%
  \BibitemOpen
  \bibfield  {author} {\bibinfo {author} {\bibnamefont {Zhang}, \bibfnamefont
  {Weixuan}}, \bibinfo {author} {\bibfnamefont {Tong}\ \bibnamefont {Wu}}, and\
  \bibinfo {author} {\bibfnamefont {Xiangdong}\ \bibnamefont {Zhang}}}
  (\bibinfo {year} {2017}),\ \bibfield  {title} {\enquote {\bibinfo {title}
  {Tailoring eigenmodes at spectral singularities in graphene-based {PT}
  systems},}\ }\href {https://doi.org/10.1038/s41598-017-11231-y} {\bibfield
  {journal} {\bibinfo  {journal} {Scientific Reports}\ }\textbf {\bibinfo
  {volume} {7}},\ \bibinfo {pages} {11407}}\BibitemShut {NoStop}%
\bibitem [{\citenamefont {Zhu}\ \emph {et~al.}(2014)\citenamefont {Zhu},
  \citenamefont {Ramezani}, \citenamefont {Shi}, \citenamefont {Zhu},\ and\
  \citenamefont {Zhang}}]{pt242}%
  \BibitemOpen
  \bibfield  {author} {\bibinfo {author} {\bibnamefont {Zhu}, \bibfnamefont
  {Xuefeng}}, \bibinfo {author} {\bibfnamefont {Hamidreza}\ \bibnamefont
  {Ramezani}}, \bibinfo {author} {\bibfnamefont {Chengzhi}\ \bibnamefont
  {Shi}}, \bibinfo {author} {\bibfnamefont {Jie}\ \bibnamefont {Zhu}}, and\
  \bibinfo {author} {\bibfnamefont {Xiang}\ \bibnamefont {Zhang}}} (\bibinfo
  {year} {2014}),\ \bibfield  {title} {\enquote {\bibinfo {title}
  {$\cal{PT}$-symmetric acoustics},}\ }\href
  {https://doi.org/10.1103/PhysRevX.4.031042} {\bibfield  {journal} {\bibinfo
  {journal} {Physical Review X}\ }\textbf {\bibinfo {volume} {4}},\ \bibinfo
  {pages} {031042}}\BibitemShut {NoStop}%
\bibitem [{\citenamefont {Znojil}(2001)}]{pt383}%
  \BibitemOpen
  \bibfield  {author} {\bibinfo {author} {\bibnamefont {Znojil}, \bibfnamefont
  {M}}} (\bibinfo {year} {2001}),\ \bibfield  {title} {\enquote {\bibinfo
  {title} {$\mathcal{PT}$-symmetric square well},}\ }\href
  {https://doi.org/10.1016/S0375-9601(01)00301-2} {\bibfield  {journal}
  {\bibinfo  {journal} {Physics Letters A}\ }\textbf {\bibinfo {volume}
  {285}},\ 10.1016/S0375-9601(01)00301-2}\BibitemShut {NoStop}%
\bibitem [{\citenamefont {Znojil}\ and\ \citenamefont {Tater}(2001)}]{pt61}%
  \BibitemOpen
  \bibfield  {author} {\bibinfo {author} {\bibnamefont {Znojil}, \bibfnamefont
  {M}}, and\ \bibinfo {author} {\bibfnamefont {M}~\bibnamefont {Tater}}}
  (\bibinfo {year} {2001}),\ \bibfield  {title} {\enquote {\bibinfo {title}
  {Complex {C}alogero model with real energies},}\ }\href
  {https://doi.org/10.1088/0305-4470/34/8/321} {\bibfield  {journal} {\bibinfo
  {journal} {Journal of Physics A: Mathematical and General}\ }\textbf
  {\bibinfo {volume} {34}},\ \bibinfo {pages} {1793--1803}}\BibitemShut
  {NoStop}%
\bibitem [{\citenamefont {Znojil}(2006)}]{pt160}%
  \BibitemOpen
  \bibfield  {author} {\bibinfo {author} {\bibnamefont {Znojil}, \bibfnamefont
  {Miloslav}}} (\bibinfo {year} {2006}),\ \bibfield  {title} {\enquote
  {\bibinfo {title} {Coupled-channel version of the $\cal{PT}$-symmetric square
  well},}\ }\href {https://doi.org/10.1088/0305-4470/39/2/014} {\bibfield
  {journal} {\bibinfo  {journal} {Journal of Physics A: Mathematical and
  General}\ }\textbf {\bibinfo {volume} {39}},\ \bibinfo {pages}
  {441--455}}\BibitemShut {NoStop}%
\end{thebibliography}%
\end{document}